\documentclass[11pt]{article}

\usepackage{graphicx}
\usepackage{amssymb}

\def\baselinestretch{1.2}

\begin{document}

\begin{center}

{\Large \bf Oscillations of neutrinos and mesons in quantum field
theory} \vskip 0.8cm

{\large Mikael Beuthe\footnote{Electronic address:mbeuthe@oma.be}}\\

{\it \small Institut de Physique Th\'eorique, Universit\'e
Catholique de Louvain, B-1348 Louvain-la-Neuve,
Belgium\footnote{Present address: Observatoire
Royal de Belgique, Avenue Circulaire 3, B-1180 Bruxelles, Belgium}}\\

\end{center}

\vskip 5mm \small

\begin{abstract}
This report deals with the quantum field theory of particle
oscillations in vacuum. We first review the various controversies
regarding quantum-mechanical derivations of the oscillation
formula, as well as the different field-theoretical approaches
proposed to settle them. We then clear up the contradictions
between the existing field-theoretical treatments by a thorough
study of the external wave packet model. In particular, we show
that the latter includes stationary models as a subcase. In
addition, we compute explicitly decoherence terms, which destroy
interferences, in order to prove that the coherence length can be
increased without bound by more accurate energy measurements. We
show that decoherence originates not only in the width and in the
separation of wave packets, but also in their spreading through
space-time. In this review, we neither assume the relativistic
limit nor the stability of oscillating particles, so that the
oscillation formula derived with field-theoretical methods can be
applied not only to neutrinos but also to neutral $K$ and $B$
mesons. Finally, we discuss oscillations of correlated particles
in the same framework.
\end{abstract}



\tableofcontents \listoffigures

\normalsize \setlength{\topmargin}{-1cm}
\setlength{\textheight}{23cm}
\newpage

\section{Introduction}

Nearly fifty years ago, the $\theta$ puzzle led Gell-Mann and Pais
\cite{Gellmann55} to discover that interaction (or flavor)
eigenstates need not coincide with particles having well-defined
masses and lifetimes. The concept of flavor-mixing was born. Pais
and Piccioni \cite{Pais55} went one step further and suggested
that the propagation of a flavor eigenstate, as a coherent
superposition of mass eigenstates, leads to the partial conversion
of the initial flavor into other flavors. The flavor-mixing
transition probability oscillates in space with a wavelength
depending on the mass differences.  The theory of flavor-mixing
and particle oscillations was soon vindicated in a series of
experiments on the kaon system
\cite{Lande56,Fitch61,Good61,Kabir}. Similar oscillations were
observed in 1987 with $B$ mesons \cite{Albrecht87}.

By analogy, Pontecorvo proposed in 1957 the hypothesis of
neutrino-antineutrino oscillations
\cite{Pontecorvo57,Pontecorvo58}. Neutrino flavor-mixing, strictly
speaking, was suggested somewhat later \cite{Maki62}, and the
two-flavor oscillation case, as well as its application to solar
neutrinos, was examined at the end of the sixties by Pontecorvo
and Gribov \cite{Pontecorvo68,Gribov69}. Neutrino oscillations
are, however, much more difficult to observe than meson
oscillations, because of the small masses and the weak
interactions of the neutrinos. They remained for thirty years a
theoretical subject (see for example the reviews
\cite{Bilenky78,Bilenky87}), with the noteworthy exception of the
solar neutrino experiment in the Homestake Mine
\cite{Davis68,Davis94,Cleveland98,Bahcall}. In the 90's, the
persistent deficit in solar neutrinos (in comparison with the
predictions of solar models \cite{Bahcall,Bahcall01,Couvidat02})
observed at Homestake was confirmed by the experiments
Kamiokande-II \cite{Hirata89,Fukuda96}, SAGE
\cite{Abazov91,Abdurashitov99}, GALLEX
\cite{Anselmann92,Hampel99}, GNO \cite{Altmann00} and
Super-Kamiokande \cite{Fukuda98a,Fukuda01}. Moreover the Sudbury
Neutrino Observatory recently announced the detection of a
nonelectronic component in the solar neutrino flux
\cite{Ahmad01,Ahmad02}, the total flux being compatible with the
predictions of solar models. All these observations can be
explained by neutrino mixing and oscillations
\cite{Barger02,Bandyopadhyay02b,Bahcall02,Fukuda02,Nakamura02} On
another length scale, the observation of an angular dependence of
the atmospheric muon-neutrino flux by the Super-Kamiokande
Collaboration \cite{Fukuda98b,Fukuda00,Kajita01,Jung01} gave a
huge impetus to the neutrino oscillation hypothesis (note that a
global muon-neutrino depletion had already been observed in other
atmospheric neutrino experiments
\cite{Hirata88,Fukuda94,Becker-Szendy92,Ambrosio98,Allison99}).
Preliminary results from the long-baseline experiment K2K
\cite{Ahn01,Ishii02} confirm these results by showing a depletion
in an artificially produced muon-neutrino beam. Finally,
controversial evidence for neutrino oscillations has appeared in
the laboratory experiment LSND
\cite{Athanassopoulos96,Athanassopoulos98} but has not been
confirmed (nor totally excluded) by the KARMEN experiment
\cite{Armbruster95,Armbruster02,Church02}.

Particle oscillations are an interference phenomenon which allows
the measurement of extremely small quantities, such as masses too
small to be measurable by other means (mass differences between
conjugate mesons \cite{Fitch61,Good61,Albrecht87} or neutrinos
\cite{Fisher99,Kayser02}), or CP violation (measured nearly forty
years ago in neutral meson systems \cite{Christenson64,Branco} and
yet to be measured in the leptonic sector \cite{Dick99}). Thus, it
is a privileged tool in the discovery of new physics. Nowadays,
$B$ meson and neutrino oscillations have a huge importance in the
determination of the unknown parameters of the Standard Model and
its minimal extension, in particular the mixing matrix elements
(including the CP violation phase) and the neutrino masses.
Besides serving as a delicate probe in particle physics, particle
oscillations are a good testing ground for quantum mechanics.
Einstein-Podolsky-Rosen correlations
\cite{Inglis61,Day61,Lipkin68} have been studied theoretically
\cite{Bertlmann98,Bertlmann99,Foadi99,Pompili00} as well as
experimentally \cite{Apostolakis98} in the neutral $K$ and $B$
systems. Bell inequalities for neutral kaons are the subject of
active research \cite{Ancochea99,Bertlmann01,Genovese01,Bramon02}.
Furthermore, fundamental issues of quantum mechanics are raised by
oscillation experiments. The determination of the conditions under
which the coherence between the different mass eigenstates is
maintained is crucial for the phenomenology of neutrino
oscillations \cite{Kiers96,Malyshkin00}. The question of the
correlation between neutral mesons, in processes where $B^0\bar
B^0$ is included in the final state, is important for an estimate
of the oscillation frequency \cite{Gronau01}.

Considering the practical importance of the oscillations, the
theoretical framework describing them should be rock-solid.
Surprisingly, the validity of the standard treatment of
oscillations, resorting to plane waves in a simple
quantum-mechanical model, was not seriously questioned until
neutrino oscillations finally gained credibility in the 90's. Two
pioneering articles should be noted. First, a quantum field model
of kaon oscillations was proposed by Sachs in 1963 \cite{Sachs63},
but was soon forgotten, until the use of Sachs' reciprocal basis
was recently revived for a rigorous quantum-mechanical treatment
of meson oscillations \cite{Branco,Alvarez-Gaume99}. Second,
Kayser showed in 1981 that neutrino oscillations are in
contradiction with the plane wave assumption of the standard
treatment of oscillations, and he proposed to modelize the
oscillating particles with wave packets \cite{Kayser81}. His
model, however, was not widely accepted and numerous articles have
continued to appear to this day, discussing the ambiguities of the
plane wave standard treatment, such as the proper choice of the
frame \cite{Lowe96} or the equality of the energies or momenta of
the different mass eigenstates \cite{Giunti01b}.

Although Kayser's wave packet model is a serious improvement on
the standard plane wave treatment, it also suffers from serious
defects: flavor states are ill-defined \cite{Giunti92a}; the
nonrelativistic limit is problematic if the mass eigenstates are
not nearly degenerate \cite{Giunti92a}; the question about the
equality (or not) of the energy of the different mass eigenstates
is not answered \cite{Grossman97,Stodolsky98}; the possibility
remains of obtaining nonstandard oscillation formulas
\cite{Srivastava95b}; the relation between the coherence length,
beyond which oscillations vanish, and the conditions of production
and detection is vague (problem of the determination of the wave
packet size \cite{Kiers96}); finally, it is not adequate for the
oscillations of unstable particles.

In order to solve these problems, a quantum field treatment of
neutrino oscillations, similar to Sachs's model, has been proposed
by Giunti, Kim, Lee and Lee \cite{Giunti93}. The model consists in
treating the oscillating particle as an internal line of a Feynman
diagram, propagating between a source and a detector, which are
represented by ingoing and outgoing external wave packets. This
approach has been followed by other authors, but no agreement has
yet been reached on the correct assumptions and calculation
methods:
\begin{enumerate}

\item There is disagreement on whether the interfering mass
eigenstates have the same energy \cite{Grimus96} or whether their
energy is spread; the existence of propagating wave packets,
associated to the mass eigenstates, is also at stake
\cite{Grimus99}.

\item Whereas oscillations vanish if the 3-momentum is measured
too accurately, it is not clear whether the same thing happens
when the energy is perfectly measured
\cite{Kiers98,Giunti98a,Giunti98b}.

\item Nonstandard formulas for neutrino oscillations have been
derived in specific cases, either showing a strong dependence of
the flavor-mixing transition probability on the neutrino mass
ratios \cite{Shtanov98}, or having a `plane wave' behavior
\cite{Ioannisian99}, or leading to oscillations of charged leptons
in pion decay \cite{Srivastava98}.
\end{enumerate}
Moreover, this formalism cannot be applied, as it is, to meson
oscillations or to unstable neutrinos. The existence of neutrino
mass makes indeed not only neutrino mixing possible but also
neutrino decay \cite{Mohapatra,Boehm,Kim,Lindner01}. Although it
cannot alone explain the data, neutrino decay has a long history
as a possible explanation for the solar neutrino anomaly
\cite{Bahcall72,Pakvasa72,Bahcall86,Berezhiani92a,Berezhiani92b,Acker92,Acker94,Ma00,Choubey00b,
Bandyopadhyay01,Joshipura02,Bandyopadhyay02a,Beacom02}. Moreover
this hypothesis was compatible until recently with atmospheric
neutrino data
\cite{Barger99a,Lipari99,Fogli99,Barger99b,Choubey00a,Ma00},
although it now seems ruled out \cite{Kayser02}. Neutrino decay
has also been discussed with respect to supernovae
\cite{Frieman88,Lindner02}.

In this report, we intend to reconcile the different
field-theoretical treatments in order to solve the above problems.
Our first aim is to argue in favor of a field theory approach of
oscillations. We want to show that it provides, on the one hand,
the best available derivation of the oscillation formula and
gives, on the other hand, a good explanation of the
characteristics of the oscillation process (origin of the
oscillations, observability conditions, boundary conditions). Our
second aim is to strengthen the phenomenological basis of the
neutrino oscillation formula through its unification with the
mesonic oscillation formula. We thus extend the formalism to the
unstable case, so that the same formula can be applied to K and B
mesons, to neutrinos and to unstable neutrinos.

In section~\ref{section1}, we discuss the quantum-mechanical
approaches, so as to understand the problems addressed in most of
the articles and the different solutions adopted in the
literature. The existing field-theoretical treatments are reviewed
in section~\ref{section2}, and their contradictions are summarized
in five questions, which will serve as a guideline in the
calculations of sections \ref{section3} to \ref{section5}; these
questions will be answered in section \ref{answers} and
\ref{section9}. In sections \ref{section3} and \ref{section4}, we
explain all the assumptions behind the external wave packet model.
The relativistic case and the nonrelativistic (but nearly mass
degenerate) cases are treated in a unified way, and the
oscillating particle is allowed to be unstable.

Oscillations of stable particles are analyzed in
section~\ref{section5}. With the aim of reconciling the different
points of view in the literature, we discuss different methods of
computation, which allow to piece together a complete physical
interpretation in section~\ref{section6}. We shall see that this
formalism lends itself to a wave packet interpretation of the
oscillating particle and can thus take up the physical insights of
Kayser's intermediate wave packet model. The dispersion (or
spreading in space-time of the amplitude) is taken into account,
as it is important for mesons and might be relevant for solar and
supernova neutrinos. We can then show that the existence of the
coherence length is due not only to the separation of wave
packets, but also to their dispersion. We compute the explicit
dependence of the coherence length on the conditions of production
and detection, to check that it increases without bound with the
accuracy of energy measurements. We also show in which sense the
oscillation formula can be seen as a superposition of mass
eigenstates with the same energy, in order to clear up the
misunderstanding of stationary boundary conditions. In addition,
we analyze in section~\ref{section6} the nonstandard oscillation
formulas derived within field-theoretical models and we disprove
them.

Oscillations of unstable particles are examined in
section~\ref{section7}. Nonexponential corrections to mixing are
computed for the first time in quantum field theory, and they are
shown to be negligible. We clearly state the correspondence
between the field-theoretical treatment and the Wigner-Weisskopf
and reciprocal basis formalisms. Some more topics are discussed in
section~\ref{section8}: the normalization of the transition
probability, the size of the external wave packets and the
instability of the source. Finally, we examine correlated
oscillations (DA$\Phi$NE, KEKB and BABAR experiments) in
section~\ref{section9}, including an analysis of the possible
energy-momentum correlation at the source. In the same framework,
oscillations of recoil particles are shown to be nonexistent.

Let us mention four restrictions. First, we shall only consider
free propagation in vacuum. Interactions with background matter
can have a dramatic effect on mixing, as demonstrated by the
Mikheyev-Smirnov-Wolfenstein (or MSW) effect
\cite{Wolfenstein78,Mikheyev85,Mikheyev86,Kuo89}. In practice, the
MSW oscillations are washed out by averaging over the energy
spectrum \cite{Petcov89,Malyshkin00}, so that the question of
intrinsic decoherence discussed in the present report is not very
important in that case. It is true that a field-theoretical
formalism for oscillations with matter interactions should be
developed to deal with dense astrophysical environments
\cite{Raffelt,Prakash01,Cardall01,Dolgov02}, but this goes far
beyond the limits of this review. Second, we shall not explain the
nonperturbative quantum field theory of particle oscillations, as
proposed by Blasone, Vitiello and co-workers
\cite{Blasone95,Blasone99a}. This approach is indeed of a
completely different nature than the perturbative
field-theoretical treatments discussed in this report. We shall
however comment on their results in section \ref{review}. Third,
we shall not examine exotic mechanisms, such as oscillations of
massless neutrinos due to a violation of the Equivalence Principle
\cite{Raychaudhuri02}, to resonant spin-flavor precession
\cite{Akhmedov00,Barranco02}, to flavor-changing (or
non-universal) interactions \cite{Fornengo02} or to dissipation
\cite{Lisi00,Benatti01} (comparative reviews can be found in
\cite{Lipari99,Pakvasa00,Gago02}). Finally, we consider neither
the gravitational effects on oscillations \cite{Wudka01,Linet02}
nor the possibility of extra dimensions
\cite{Dvali99,Davoudiasl02}.

\section{Oscillations in quantum mechanics}
\label{section1}

We first discuss the concept of particle mixing as it is a
requirement for oscillations. We then present the standard
derivation of the oscillation formula, stressing the problematic
choice of basis for unstable particles, as well as the ambiguities
arising in the conversion of an oscillation in time into an
oscillation in space. This section ends with a review of the
attempts to solve these problems with more sophisticated
quantum-mechanical models.

\subsection{Mixing in quantum mechanics}
\label{mixing}

The possibility of particle mixing arises from the fundamental
principles of quantum theory. The measurement of an observable
causes the initial state to jump into an eigenstate of the
operator associated to this observable. Thus the act of
measurement determines a basis of physical states, made up of all
the eigenstates of the corresponding operator. The different bases
associated to the different operators are linked by linear
transformations; a state expanded in a given basis can be seen as
a {\it mixing} of the eigenstates of that basis. While one gets
used to the quantum superposition principle as long as wave
functions correspond to rather abstract properties of a system, it
is always a bit shocking to apply it to particles, which are
associated in our mind with classical concepts. The two-slit
experiment with particles is the best known example, but particle
mixing, appearing when an interacting eigenstate is a
superposition of different mass eigenstates, is hardly less
unsettling.

The textbook case is given by the neutral kaon system: the decay
pattern of a $K^0$ into pions is well explained by the
superposition of two mass eigenstates, a CP-even state decaying
quickly into two pions, and a CP-odd state decaying more slowly
into three pions. Thus a $K^0$ can be seen as a mixing of a $K_S$
({\it K short}) and a $K_L$ ({\it K long}), having well-defined
masses and widths (the so-called mass eigenstate basis). Whereas
pionic decays do not allow to identify the final strangeness
($K^0$ or $\bar K^0$), it is possible to do so in the decays of
kaons into $\pi e\nu$, with the result that one observes a
nonnegligible probability to detect a transition from the initial
$K^0$ into a $\bar K^0$. This is the phenomenon referred to as
{\it particle oscillation}. The flavor-mixing transition
probability oscillates indeed with the propagation distance,
unless the oscillation is averaged by some mechanism, in which
case the transition probability simply follows the exponential
decay law for each mass eigenstate.

Particle oscillations can be compared to a 2-slit interference
experiment, which requires firstly that several paths must be
available to a particle, and secondly that these paths should not
be distinguishable by the experimental conditions. In particle
oscillations, the different mass eigenstates are the analogue of
the different paths, since they are eigenstates of the space-time
evolution operators ${\bf \hat P}$ and $\hat H$. Of course, their
other quantum numbers should be equal. The equivalent of the
second requirement is that the masses should not be
distinguishable by the experimental apparatus. Under these
conditions, the probability to observe the propagating particle in
a given state, other than a mass eigenstate, oscillates in space
(and in time) with a wavelength inversely proportional to the mass
difference, because of the interference between the mass
eigenstate amplitudes. Thus particle oscillations occur {\it if}
the initial and final states are not mass eigenstates {\it and if}
it is impossible to ascertain which mass eigenstate has
contributed as an intermediate state. This last assumption implies
that mass differences should be much smaller than the typical
energies involved in the processes, with the result that the
oscillation wavelength is macroscopically large and the particle
oscillates over macroscopic distances.

Of course, oscillations are only observable in processes where the
propagation distance is of the order of, or much larger than, the
oscillation wavelength. Otherwise it is perfectly adequate to work
with interaction (or flavor) eigenstates, as it has been done
until now with neutrinos in accelerator experiments. Note that one
often talks about particle oscillations even if the oscillation
term is averaged by some mechanism, such as an integration over
the energy spectrum, though this situation could be more
appropriately called {\it incoherent mixing}.

Let us be more precise about the basis pertaining to the
oscillation phenomenon. The Hamiltonian should be such that it can
be split into a propagation Hamiltonian $H_{propag}$, describing
the free propagation of the particle, and an interaction
Hamiltonian $H_{int}$, describing the interactions producing the
propagating particle. These two subsets of the Hamiltonian can be
distinguished if there exists a quantum number, called {\it
flavor}, conserved by $H_{int}$, but violated by $H_{propag}$. The
appropriate basis for the production of the particle is made up of
eigenstates of both $H_{int}$ and the flavor operator. It is
called the {\it flavor basis}. The basis relevant to the
propagation of the particle is the one in which $H_{propag}$ is
diagonal and is called the {\it propagation basis}, or {\it mass
basis}. This basis does not coincide with the flavor basis, since
the flavor operator does not commute with $H_{propag}$.

The simplest example is given by stable neutrinos, in which case
the Hamiltonian $H_{propag}$ includes the mass matrix, whereas the
Hamiltonian $H_{int}$ includes the weak interactions of the
neutrinos with the bosons $Z$ and $W$. The flavor is here the
leptonic number, which can be electronic, muonic or tauic and
maybe sterile. This number is conserved by the weak interactions,
but is violated by a non-degenerate mass matrix which mixes
neutrinos of different flavors.

As regards unstable particles, the Wigner-Weisskopf method
\cite{Lee57,Kabirww,Nachtmann} is used to build a non-Hermitian
effective Hamiltonian $H_{propag}$, with the aim of describing
both propagation and decay. $H_{int}$ includes the interactions
involved in the production of the oscillating particles. In the
case of neutral $K$ (or $B$) mesons, $H_{propag}$ includes, on the
one hand, the degenerate effective mass matrix (due to Yukawa,
strong and electromagnetic interactions) and, on the other hand,
the weak interactions causing the decay and the flavor-mixing
transitions $K^0\leftrightarrow\bar K^0$ (or
$B^0\leftrightarrow\bar B^0$). $H_{int}$ is an effective
Hamiltonian describing the strong interactions of the mesons; it
can be built with chiral perturbation theory. The flavor is here
the strangeness (or bottomness) with values $\pm1$, which is
conserved by strong interactions but violated by weak
interactions. The non-Hermiticity of $H_{propag}$ means that the
neutral meson system does not evolve in isolation, because of its
possible decay.

The relationship between the propagation and flavor basis can be
written as
\begin{equation}
   |\nu_\alpha (0) \!> = \sum_{j} \, U_{\alpha j} \, |\nu_j (0) \!> \, ,
   \label{chgtbase}
\end{equation}
where $U$ is the matrix diagonalizing $H_{propag}$, if this
Hamiltonian has been expressed from the start in the flavor basis.
The states \mbox{$|\nu_j (0) \!>$} are the propagation (or mass)
eigenstates, at time $t=0$, with well-defined masses and widths,
belonging to the propagation basis, whereas the states
\mbox{$|\nu_\alpha (0) \!>$} are the flavor eigenstates.

The properties of the matrix $U$ depend on the Hamiltonian
$H_{propag}$. In the case of a stable oscillating particle, the
matrix $U$ arises from the diagonalization of the mass matrix and
is unitary. In the case of an unstable oscillating particle,
$H_{propag}$ is necessarily non-Hermitian, as explained above, so
that the time evolution is nonunitary, i.e.\ the oscillating
particle is allowed to decay. The matrix $U$ diagonalizing that
Hamiltonian is generally not unitary, although it can be unitary
if some symmetry makes the Hamiltonian normal\footnote{$H$ is
normal if $[H,H^\dagger$]=0.} \cite{Byron}.

The computation of an amplitude requires the knowledge of the
scalar product values of the eigenstates. In the flavor basis, the
following orthogonality properties hold:
\begin{equation}
   <\! \nu_\beta(0) \, | \, \nu_\alpha(0) \!> = \delta_{\alpha\beta} \, ,
   \label{orthogsaveur}
\end{equation}
since $H_{int}$ is Hermitian and commutes with the flavor
operator. In the propagation basis, $H_{propag}$ is Hermitian only
if the oscillating particle is stable. Thus the scalar product of
the flavor eigenstates must be defined by transforming the flavor
scalar products (\ref{orthogsaveur}):
\begin{equation}
   <\! \nu_i(0) \, | \, \nu_j(0) \!>
   = \sum_\gamma \,  U^{-1}_{j\gamma} \, U^{-1 \, \dagger}_{\gamma i} \, .
    \label{orthogmasse}
\end{equation}
These scalar products become orthogonal if the matrix $U$ is
unitary, which is true if the particle is stable or, more
generally, if $H_{propag}$ is normal. In the example of neutral
kaons, the approximation of CP symmetry makes $H_{propag}$ normal
and $U$ unitary. Since the violation of this symmetry is of the
order of $10^{-3}$, the right-hand side of Eq.~(\ref{orthogmasse})
is of the same order in the neutral kaon system. This lack of
orthogonality implies that the states \mbox{$| \, \nu_i(0) \!>$}
can be normalized in different ways. The normalization factors do
not matter as long as they do not appear in the final result. A
normalization problem arises however in the case of unstable
particles if we try to compute an amplitude involving a mass
eigenstate in the initial or final state. For example, the
amplitude \mbox{$<\!\pi\pi\, |H_{total}| \, K_L\!>$} depends on
the normalization chosen for the state \mbox{$| \, K_L\!>$}. In
that case, the normalization ambiguity is of the order of the
direct CP violation parameter $\epsilon'$, so that it cannot be
neglected \cite{Beuthe98}. This problem does not arise in the
field-theoretical approach, where the kaons are considered as
intermediate states, which are not directly observed.

It must be stressed that the orthogonality properties in the
flavor basis are valid only if all mass eigenstates are
kinematically allowed \cite{Bilenky93,Pakvasa00,Czakon01}. For
example, let us suppose that there are four stable neutrinos such
that $m_i=0$ for $i=1,2,3$ and $m_4\!\gg\!1\,$GeV. Under 1~GeV,
the flavor scalar products become
\begin{equation}
   <\! \nu_\beta(0) \, | \, \nu_\alpha(0) \!> =
   \delta_{\alpha\beta} - U_{\alpha4} \, U^*_{\beta4} \, .
   \label{nonunitary}
\end{equation}
This difficulty is a first indication of the problems arising in
the definition of a flavor basis. Is it really possible to
interpret a flavor eigenstate as an observable particle? In
quantum mechanics, each stable particle can be associated with an
irreducible representation of the Poincar\'e group, with a given
mass and a given spin. In that framework, the flavor eigenstates
do not correspond to particles, unless they are degenerate in
mass. The problem is not easier to solve in quantum field theory.
Whereas creation and annihilation operators of mass eigenstates
are well-defined, the transformation (\ref{chgtbase}) does not
yield creation and annihilation operators of flavor states that
satisfy canonical commutation relations \cite{Giunti92a}. One
could thus wonder whether flavor eigenstates really exist or not.
A phenomenological argument in favor of their existence is that,
for obvious practical reasons, no one has yet devised an
experiment involving only one mass eigenstate of a system of
oscillating particles\footnote{This might have been possible for
neutral kaons, had the CP symmetry been exact in that system,
making the propagation basis coincide with the CP basis. Decay
channels respecting CP ($K\to2\pi,3\pi$) would have then allowed a
nonambiguous determination of the propagating eigenstate.}. On the
contrary, oscillating particles are produced and detected in a
superposition of mass eigenstates formally equivalent to a flavor
eigenstate. In that sense, flavor eigenstates are observed,
whereas mass eigenstates are not.

The case of an unstable particle differs from the stable case in
two respects. First, the decay widths of the propagating
eigenstates can vary greatly, allowing the isolation of a nearly
pure beam of the longest-lived eigenstate (so the phenomenologist
is tempted to declare this state `observable'). Second, there is
no state corresponding to an unstable particle in the Hilbert
space of physical states. These particles are best described by
S-matrix poles in quantum field theory (so the theoretician is
tempted to declare these unstable states `not observable').

\subsection{Oscillations with plane waves}
\label{oscilwithplane}

Many articles and books give derivations of the oscillation
formula with plane wave states (see for example
Refs.~\cite{Bilenky78,Mohapatra}). In this section and in the
next, we examine this method in its fine details, emphasizing
subtle points and doubtful assumptions.

Following the rules of quantum mechanics, the transition
probability between an initial state of flavor $\alpha$ at time
$t=0$ and position $x=0$ and a final state of flavor $\beta$ at
time $t$ and position ${\bf x}$ is given by
$$
   {\cal A}_{\alpha \to \beta}(t,{\bf x})
   = <\! \nu_\beta(0) \, |
     \exp(-i \, \hat H_{propag} t + i \, {\bf \hat P \cdot x})
     | \, \nu_\alpha(0) \!> \, ,
$$
where $\hat H_{propag}$ is the Hamiltonian operator and ${\bf \hat
P}$ is the generator of translations. The states
\mbox{$|\,\nu_{\alpha,\beta}(0)\!>$} are taken to be
energy-momentum eigenstates, so that a plane wave treatment will
follow. This amplitude can be computed with the help of the change
of basis (\ref{chgtbase}):
\begin{eqnarray}
   {\cal A}_{\alpha \to \beta}(t,{\bf x})
   &=& <\! \nu_\beta(0) \, |
       \sum_{j} \, U_{\alpha j} \,
       e^{ -i \phi_j -\gamma_j }
       | \, \nu_j(0) \!>
   \nonumber \\
   &=& <\! \nu_\beta(0) \, |
       \sum_{j,\rho} \, U_{\alpha j} \,
       e^{ -i \phi_j - \gamma_j }
       U_{j \rho}^{-1}
       | \, \nu_\rho(0) \!> \, ,
   \label{ampliQM}
\end{eqnarray}
where the phase is given by
\begin{equation}
   \phi_j = E_j t - {\bf p}_j \cdot {\bf x} \, ,
   \label{phasephij}
\end{equation}
and the decay term  is defined by
\begin{equation}
   \gamma_j = \frac{m_j\Gamma_j t}{2E_j} \, .
   \label{decayterm}
\end{equation}
The symbols $E_j$, ${\bf p}_j$, $m_j$ and $\Gamma_j$ stand for the
energy, momentum, mass and width of the mass eigenstate
\mbox{$|\,\nu_j(0)\!>$}. The factor ${\bf p}_j \cdot{\bf x}$ is
often dropped, either because the mass eigenstate is assumed to be
in its rest frame (the time $t$ is then the proper time), or
because of an assumption of equal momenta for all mass eigenstates
(the factor ${\bf p}_j \cdot{\bf x}$ then does not appear in the
probability). These assumptions cannot be justified and neither
will be assumed in the following derivation.

If the particle is unstable, $H_{propag}$ is an effective
Hamiltonian, computed with the Wigner-Weisskopf method in the rest
frame of the decaying particle. Hence, the phase depends on the
proper time, $\phi_j=m_j\tau$, so that a boost is necessary to
obtain the expressions (\ref{phasephij}) and (\ref{decayterm})
valid in the laboratory frame. However the concept of a rest frame
has no meaning in the case of a superposition of different mass
eigenstates, which may have different velocities. Thus the choice
of the boost is ambiguous because it is not clear whether a common
boost should be used for the different mass eigenstates, or
whether each mass eigenstate is boosted by a different velocity.
This difficulty is another indication of the problems arising in
the correspondence between a flavor state and a particle. It will
be seen in section \ref{timespaceconversion} that the choice of
the boost has a big impact on the value of the oscillation length.

With the help of the orthogonality property (\ref{orthogsaveur}),
the amplitude (\ref{ampliQM}) can be rewritten as
\begin{equation}
   {\cal A}_{\alpha \to \beta}(t,{\bf x}) =
   \sum_j \, U_{\alpha j} \,
   e^{ -i \phi_j - \gamma_j } \,
   U_{j \beta}^{-1} \, .
   \label{oscMQ}
\end{equation}

In the case of $K$ and $B$ mesons, the oscillation formula
(\ref{oscMQ}) is sometimes written in the following way
\cite{NachtmannCP}:
\begin{equation}
   {\cal A}_{\alpha \to \beta}(t,{\bf x}) =
   \sum_{i,j} \, U_{\alpha j} \,
   e^{ -i \phi_j - \gamma_j }
   U_{i \beta}^\dagger \,
   <\! \nu_i(0) \, | \, \nu_j(0) \!> \, .
   \label{amplitrad}
\end{equation}
The equivalence of this expression with Eq.~(\ref{oscMQ}) can be
checked with the scalar products (\ref{orthogmasse}).

Another way to obtain the transition amplitude (\ref{oscMQ}) for
neutral mesons is to define a {\it reciprocal basis}
\cite{Sachs63,Enz65,Alvarez-Gaume99,Branco,Silva00}. This method
consists in defining two mass bases: the {\it kets} belong to the
{\it in} basis and are the right-eigenvectors of the Hamiltonian,
whereas the {\it bras} belong to the {\it out} basis (or
reciprocal basis) and are the left-eigenvectors of the
Hamiltonian. Their Hermitian conjugate vectors are not used to
write amplitudes. The Hamiltonian can then be expressed in a
diagonal form:
\begin{equation}
   \hat H_{propag} = \sum_j
   | \, \nu_j(0) \!>_{in} \, \lambda_j \; {}_{out}\!\!<\! \nu_j(0) \, | \, ,
   \label{Hreciprocal}
\end{equation}
where the $\lambda_j=m_j-i\Gamma_j/2$ are the complex eigenvalues
of the Hamiltonian in the rest frame of the oscillating particle
(in the mass-degenerate limit). The new bases are related to the
flavor basis by
\begin{eqnarray}
   | \, \nu_\alpha(0) \!>
   &=& \sum_j \, U_{\alpha j} \, | \, \nu_j(0) \!>_{in} \; ,
   \nonumber \\
   <\! \nu_\alpha(0) \, |
   &=& \sum_j \, {}_{out}\!<\! \nu_j(0) \, | \, U_{j\alpha}^{-1} \; .
   \label{doublebase}
\end{eqnarray}
With these notations, a kind of orthogonality property is
restored:
$$
   {}_{out}\!<\! \nu_i(0) \, | \, \nu_j(0) \!>_{in} = \delta_{ij} \, .
$$
The oscillation formula obtained with these new bases is identical
to Eq.~(\ref{oscMQ}), since it is simply another way to decompose
$\hat H_{propag}$.

The physical interpretation of the reciprocal basis is not clear
and its use is not really necessary, since the same result can be
obtained without it. Actually, the reciprocal basis method is a
technical trick which is best understood from a field-theoretical
viewpoint. After all, the new bases were first introduced
\cite{Sachs63} in a quantum field theory approach as left- and
right-eigenvectors of the propagator representing the oscillating
particle. As emphasized by Enz and Lewis \cite{Enz65}, they should
always appear through Eq.~(\ref{Hreciprocal}), underlining their
unphysical and intermediate character. For example, the quantity
\mbox{$|<\!\pi\pi\, |H_{total}| \, K_L\!>|^2$} is not an
observable, as mentioned in section \ref{mixing}.

In the special case of a Hermitian $H_{propag}$, $U$ is unitary
and $\gamma_j=0$. The amplitude then reads
$$
   {\cal A}_{\alpha \to \beta}(t,{\bf x}) = \sum_{j}
    U_{\alpha \, j} \;
    e^{ -i \, E_j t + i \, {\bf p}_j \cdot {\bf x} } \;
    U_{\beta \, j}^{*} \, .
$$
This is the expression commonly used for neutrinos, except that
the term ${\bf p}_j \cdot {\bf x}$ is often dropped for the
reasons mentioned after Eq.~(\ref{decayterm}).

In the general case of a non-Hermitian Hamiltonian $H_{propag}$,
the transition probability is given by the squared modulus of the
oscillation amplitude (\ref{oscMQ}):
\begin{equation}
   {\cal P}_{\alpha \to \beta}(t,{\bf x})
    = \sum_{i,j} \, U_{\alpha \, i}   \,  U_{i \, \beta}^{-1}    \,
                       U_{\alpha \, j}^* \, (U^{-1})_{j \, \beta}^* \,
         e^{ -i (\phi_i-\phi_j) - \gamma_i - \gamma_j } \, .
   \label{standardproba}
\end{equation}
The particle oscillation phenomenon is manifest in the oscillatory
behavior of the interference terms in the transition probability
(\ref{standardproba}), with a phase \mbox{$\phi_i-\phi_j$}
depending on the point $(t,{\bf x})$ of detection.

Regarding antiparticle oscillations, the CPT theorem applied to
the amplitude (\ref{oscMQ}) gives the relationship
$$
   {\cal A}_{\bar\alpha \to \bar\beta}(t,{\bf x})
   = {\cal A}_{\beta \to \alpha}(t,{\bf x})  \, ,
$$
so that the oscillation formula for antiparticles is obtained from
Eq.~(\ref{standardproba}) by the exchange of $\alpha$ and $\beta$.
Note that CP violation arises from terms in
Eq.~(\ref{standardproba}) breaking the
$\alpha\!\leftrightarrow\!\beta$ symmetry, since it appears
through a difference between $|{\cal A}_{\bar\alpha \to
\bar\beta}(t,{\bf x})|^2$ and $|{\cal A}_{\alpha \to \beta}(t,{\bf
x})|^2$.

\subsection{Time to space conversion of the oscillating phase}
\label{timespaceconversion}

\subsubsection{Standard oscillation phase}

There has been some controversy about the conversion of the
oscillation phase $\phi_i-\phi_j$, appearing in
Eq.~(\ref{standardproba}), into a measurable quantity. Since the
propagation time is not measured in oscillation
experiments\footnote{However time measurements are important
according to Okun \cite{Okun00a} in an experiment performed at
IHEP (Serpukhov) \cite{Anikeev98}.}, a prescription is needed to
get rid of the time dependence in the phase difference.

The numerous prescriptions proposed in the literature are somewhat
confusing. They can be classified by expanding the phase around an
average energy or momentum. Since the oscillating particle is
on-shell, the energy $E_j$ can be expressed in function of the
momentum $p_j$. The phase $\phi_j$ can then be expanded around an
average momentum $p$, not very different from $p_j$ or $p_k$, and
an average mass $m$. Although the mass $m$ can be very different
from $m_j$ or $m_k$ in the ultra-relativistic case, the expansion
is correct as long as the mass difference $\delta m_j^2=m_j^2-m^2$
is small with respect to the energy. The momentum difference
$\delta p_j=p_j-p$ is expected to be of the same order than
$\delta m_j^2$. In one spatial dimension, the expansion of the
phase reads, to first order in $\delta m_j^2$ and $\delta p_j$,
\begin{eqnarray}
   \phi_j &=& \sqrt{p_j^2+m_j^2} \, t_j - p_j \, x \nonumber \\
   &=& Et-px \;+\; \frac{\delta m_j^2}{2E} t \;+\; (vt-x) \delta p_j
             \;+\; E \delta t_j \, ,
   \label{expansionplanephase}
\end{eqnarray}
with the average energy and velocity defined by $E=\sqrt{p^2+m^2}$
and $v=p/E$, respectively. A different time $t_j$ has been allowed
for each mass eigenstate, and expanded around an average time $t$,
with \mbox{$\delta t_j=t_j-t$}. Of course, the following arguments
will only be correct to first order in \mbox{$\delta m_j^2/2E$},
but it is useless to argue about further orders in a flawed
approach such as the plane wave treatment.

All prescriptions leading to the standard oscillation formula set
$\delta t_j=0$, i.e.\ they impose that interference only takes
place for equal propagation times (and lengths) for the different
mass eigenstates. This {\it equal time prescription} has been
explicitly stated \cite{Kayser95,Kayser96,Kayser97a,Lowe96}, in
reaction against articles proposing different detection times
\cite{Srivastava95a,Srivastava95b}. It has also been legitimated
by an {\it equal velocity prescription}
\cite{Srivastava95b,DeLeo00}, which is seen to be equivalent to
the previous prescription with the help of the classical relation
$t_j=x/v_j$. However the equal velocity condition leads to
$\frac{E_j}{E_i}=\frac{m_j}{m_i}$, which is very unlikely for
neutrinos \cite{Okun00b}.

Note that imposing equal times $t_i=t_j$, in the laboratory frame,
also means imposing equal proper times $\tau_i=\tau_j$ if the
classical relation $\tau_j=\sqrt{t_j^2-x^2}$ is used. The last
relation implies that a boost of velocity $v_j=x/t_j$ is used to
go from the rest frame of the mass eigenstate $m_j$ to the
laboratory frame. Thus the question of the choice of the correct
boost boils down to the question of the equality of propagation
times in the laboratory frame.

A second prescription is needed in order to obtain the standard
oscillation formula. It could be called the {\it classical
propagation condition}: $vt-x=0$. It imposes that the term $(vt-x)
\delta p_j$, appearing in Eq.~(\ref{expansionplanephase}), is
negligible in comparison with $\delta m_j^2t/2E$. This condition
can be weakened to $|vt-x|\ll t$. Since plane waves are
delocalized in space-time, this condition cannot be justified
without a more sophisticated treatment, for example with wave
packets.

These two prescriptions are sufficient to derive the following
formula:
\begin{equation}
   \phi_i-\phi_j
   \cong \frac{\delta m_{ij}^2 \, |{\bf x}|}{2|{\bf p}|}
   = 2\pi \frac{|{\bf x}|}{L^{osc}_{ij}} \, ,
   \label{standardphase}
\end{equation}
obtained from a three-dimensional generalization of
Eq.~(\ref{expansionplanephase}). The {\it oscillation length}
$L^{osc}_{ij}$ is defined by
\begin{equation}
   L_{ij}^{osc} = \frac{4\pi |{\bf p}|}{\delta m_{ij}^2} \, ,
   \label{longoscMQ}
\end{equation}
where $\delta m_{ij}^2=m_i^2-m_j^2$ is assumed to be positive. The
classical propagation condition also justifies the substitution
$t\to E|{\bf x}|/|{\bf p}|$ in $\gamma_{i,j}$ (see
Eq.~(\ref{decayterm})).

The transition probability (\ref{standardproba}) together with the
phase (\ref{standardphase}) form the standard oscillation formula
used to fit the experimental data (see \cite{Groom02} for an
application to two-flavor neutrino oscillations). Of course, the
probability (\ref{standardproba}) should first be averaged over
the energy spectrum and over the region of production and
detection (see section \ref{incoherenteffects}). If the
oscillation is completely washed out by these averaging
procedures, the transition probability (\ref{standardproba}) can
be simplified by the substitution
$e^{-i(\phi_i-\phi_j)}\to\delta_{ij}$.

Although the equal time prescription and the classical propagation
condition are sufficient to obtain the standard oscillation phase
(\ref{standardphase}), additional prescriptions leading to the
same result are commonly found in the literature:
\begin{enumerate}

\item The {\it equal momentum prescription} is the most common:
$\delta p_i=\delta p_j=0$. As seen above, this assumption is not
necessary. Moreover, it is impossible to impose experimental
conditions such that the momentum uncertainty is zero, since the
oscillations are destroyed by a momentum measurement more accurate
than the mass difference. Thus, this prescription is groundless.

\item The {\it equal energy prescription} has been recently
advocated by Lipkin: $\delta E_i=\delta E_j=0$ \cite{Lipkin95}. It
has the advantage of avoiding the classical propagation condition
since it leads to $\delta p_j=-\delta m_j^2/2p$, so that the time
dependence completely drops from the phase difference
$\phi_j-\phi_k$ (at least if the equal time prescription is
assumed), and the standard oscillation phase (\ref{standardphase})
is directly obtained.

In principle, oscillation experiments are feasible with a zero
energy uncertainty, since spatial oscillations are not expected to
vanish in that case (note that a quantum field treatment is
necessary to prove it). In practice, the energy uncertainty is far
from being negligible and is often of the same order of magnitude
as the momentum uncertainty. Thus the equal energy prescription is
only justified if an extremely small uncertainty on the energy is
imposed by the physical properties of the process itself. Do we
have any theoretical reason to expect that this uncertainty is
smaller than the mass difference? No convincing arguments
supporting that assumption have been given until now. For
example\footnote{Lipkin gives at least three reasons for the
equality of energies: the strict energy conservation discussed
here \cite{Lipkin95,Lipkin99}, the flavor-energy factorization
\cite{Grossman97,Lipkin99} and the stationarity resulting from a
time average \cite{Lipkin95,Lipkin99}. The last two arguments are
discussed in section \ref{intermwavepackets}.}, Lipkin computes
the energy-momentum uncertainties $\delta p_K$, $\delta E_K$ of
the kaon in the process $\pi p^-\!\to\! \Lambda K^0$
\cite{Lipkin99}. If the proton is at rest in a lattice, its
momentum uncertainty, due to the Debye temperature of the crystal,
can be estimated at $\delta p_p\!\sim\!10^3\,$eV. Lipkin estimates
$\delta p_K$ to be of the same order of magnitude, whereas he
neglects $\delta E_K\!\sim\!(\delta p_p)^2/m_p$. However $\delta
E_K$ is still much larger than the mass difference $\delta
m_K\!\sim\!10^{-6}\,$eV, which is the most important mass scale in
the experiment. Moreover the pion momentum uncertainty gives a
first order contribution to $\delta E_K$ and should not be
neglected. Even if one has shown that $\delta E\ll\delta p$, there
is a long way to go to show that $\delta E\ll\delta m$.

Another argument against the equality of energies is that it holds
only in one particular frame. For example, if the energies of the
different mass eigenstates are equal in the decay of a pion at
rest (\mbox{$\pi\!\to\!\mu\nu$}), the energy difference becomes
approximately equal to the momentum difference if the pion is
relativistic \cite{Giunti01b}. Thus the equal energy prescription
should be shown to be true in the laboratory frame for any
experimental conditions.

However there is a much more reasonable way of looking at the
equal energy prescription. It consists in seeing it as the result
of a time average washing out the interference between wave packet
components having different energies. There is then no need to
prove that the energy uncertainty is zero. This argument will be
considered in section \ref{intermwavepackets}.

\item {\it Energy-momentum conservation at the production}: first
proposed by Winter \cite{Winter81}, this recurring prescription
\cite{Boehm,Goldman96,Lowe96,Srivastava95a,Srivastava95b,Giunti01a,Giunti01b,Tsukerman01}
allows to compute explicitly $\delta p_j$. It has often been used
to show that neither the momenta nor the energies of the different
mass eigenstates are equal. In the example of the pion decay at
rest (\mbox{$\pi\!\to\!\mu\nu$}), the energies and momenta of the
muon and neutrino can be computed exactly if the energy-momentum
of the pion is perfectly known. However this knowledge is usually
not available: when the energy-momentum spread of the source is
much larger than the mass difference $\delta m_{ij}^2$, it is
meaningless to compute the exact values of the energies and
momenta to order $\delta m_{ij}^2$. A more detailed examination of
this question requires wave packets instead of plane waves, or
even better, quantum field theory.
\end{enumerate}

\subsubsection{Non-standard oscillation phase}
\label{nonstandard}

Controversial prescriptions leading to nonstandard oscillations
formulas involve {\it different propagation times} $\delta t_i\neq
\delta t_j\neq0$ or, equivalently, different proper times
$\tau_i\neq\tau_j$. Let us parametrize $p_j$ by
\begin{equation}
   p_j = p + (\rho-1) \frac{\delta m_j^2}{2 p} \, ,
\label{expansionpjQM}
\end{equation}
where $\rho$ is a dimensionless number of order unity. The
corresponding energy and velocity can be written as
\begin{eqnarray}
  E_j &=& E + \rho \frac{\delta m_j^2}{2E} \, ,
  \nonumber
  \\
  v_j &=& v
  + \left(\rho(1-v^2)-1 \right)\frac{\delta m_j^2}{2pE} \, .
  \label{expansionvjQM}
\end{eqnarray}
The momenta are equal if $\rho=1$ whereas the energies are equal
if $\rho=0$. As explained above, it has also been proposed to
determine $\rho$ through energy-momentum conservation at
production, leading to a value of $\rho$ of order unity. In any
case, the time difference is computed with the help of the
classical relation $t_j=x/v_j$ and reads
$$
   \delta t_j = (E^2 - \rho \, m^2) \,\frac{\delta m_j^2x}{2p^3E} \, .
$$
Inserting this value in Eq.~(\ref{expansionplanephase}) and using
$t=x/v$, we obtain
\begin{equation}
   \phi_j = Et-px \;+\;
   \left( 1+\rho + \frac{1-\rho}{v^2} \right) \frac{\delta m_j^2x}{2p} \, .
   \label{phaseEPconserv}
\end{equation}
With the equal momentum prescription ($\rho=1$), the corresponding
oscillation length will be smaller by a factor $2$ than the
standard value given in Eq.~(\ref{longoscMQ}) \cite{Lipkin95}.
With the equal energy prescription ($\rho=0$), the oscillation
length will be smaller by a factor $1+v^{-2}$ than in
Eq.~(\ref{longoscMQ}). Thus the equal energy prescription may also
lead to a nonstandard oscillation length, contrary to what was
claimed in \cite{Lipkin95}. With the energy-momentum conservation
prescription, the oscillation length will be smaller than in
Eq.~(\ref{longoscMQ}) by a factor depending on the value of
$\rho$, which depends on the energy \cite{Srivastava95a}. These
formulas can also be applied to the case of correlated
oscillations, such as $\phi(1020)\!\to\!K^0\bar K^0$ or
$\Upsilon(4s)\!\to\!B^0\bar B^0$. In the center-of-mass frame of
the resonance, the different time prescription leads to an
oscillation length smaller by a factor 2 than the standard value,
since the equality $|p_i|=|p_j|$ valid in that frame leads to
$\rho=1$ \cite{Srivastava95b}.

Another disturbing consequence of the different time prescription
is the oscillation of recoil particles, for example $\Lambda$ in
$\pi^- p\!\to\!\Lambda K^0$ \cite{Srivastava95a}, or the muon in
$\pi\!\to\!\mu\nu$ \cite{Srivastava95c,Srivastava98}. This is
easily seen by applying Eq.~(\ref{expansionplanephase}) to the
recoil particle. Although $\delta m_j^2=0$ (the recoil particle
has only one mass eigenstate), $\phi_i\neq\phi_j$ because $\delta
t_i\neq\delta t_j$. The oscillation of particles having only one
mass eigenstate is unacceptable since it leads to non-conservation
of the detection probability of this particle. This is not the
case when there are several mass eigenstates, as the sum of the
detection probabilities of the different mass eigenstates is
always equal to 1 for a given propagation distance.

The treatment of neutrino oscillations in $\pi\to\mu\nu$ proposed
by Field \cite{Field01} also resorts to the different time
prescription. This author claims that the different neutrino mass
eigenstates are detected at the same space-time point but are
produced at different space-time points. Following Field, the
oscillation phase would not only receive a contribution from the
neutrino path difference but also from the path difference of the
source which decays at different times. Field computes the first
contribution with the energy-momentum conservation prescription,
which gives for a source at rest an oscillation phase larger than
the standard result by a factor 2 (see Eq.~(\ref{phaseEPconserv})
with $v=1$). The second contribution, which only appears in
Field's article, is computed in the same way. The oscillation
phase obtained by Field therefore differs from the standard result
(\ref{standardphase}). Field's method also leads to the prediction
of muon oscillations because of the use of the different time
prescription\footnote{Field's criticism of wave packets models
will be examined in section \ref{intermwavepackets}.}.

Another example of the different time prescription can be found in
Malyshkin and Kulsrud's analysis \cite{Malyshkin00} of the time
variations of the solar neutrino flux, which leads to a result
different from Parke's formula \cite{Parke86}. As we do not
consider here oscillations in matter, this new oscillation formula
will not be discussed further, although our comments on the
different time prescription also apply to that case.

The different time prescription has been strongly criticized by
several authors within the plane wave framework
\cite{Lowe96,Kayser95,Kayser96,Kayser97a,Lipkin95,Lipkin99}. Their
argument\footnote{A more convincing (though indirect) argument has
been given by Giunti \cite{Giunti02a}: the use of the different
time prescription in the double-slit experiment leads to a wrong
interference pattern.}, which can be stated as `interference only
occurs between states taken at the same space-time point', does
not hold when examined in a wave packet or field-theoretical
model. Kiers, Nussinov and Weiss \cite{Kiers96,Kiers98} have
indeed shown that the coherent character of the detection process
allows wave functions at different space-time points to interfere.
In particular, a long coherent measurement in time may be used to
revive oscillations, even after the mass eigenstate wave packets
have completely separated spatially (see section
\ref{intermwavepackets}). The question of the correct time
prescription is thus subtler than it seems at first sight. The
crux of the matter is to take into account the production and
detection processes. In this way, each increment in the phase
associated to the propagating particle, due to a slightly
different production (respectively detection) point, is cancelled
by a decrement in the phase of the wave packet of the source
(respectively detector) \cite{Giunti01a}. For example, this
cancellation can be implemented in the intermediate wave packet
model (see section \ref{intermwavepackets}) by computing the total
amplitude as an overlap of the propagating wave packet with the
source and detector wave packets
\cite{Giunti98b,Giunti02a,Giunti02b}. Note however that this
mechanism is much more natural in a quantum field model such as
the external wave packet model (see section~\ref{section3}), where
the amplitude is integrated over all possible microscopic
production and detection points, with the result that the phase
depends only on average (i.e.\ macroscopic) production and
detection points.

In conclusion, neither the equal time prescription nor the
different time prescription can be justified in the plane wave
approach, although the choice of the prescription has an important
effect on value of the oscillation length. Moreover other
prescriptions, such as the classical propagation condition and the
equal energy prescription, cannot be understood within the plane
waves formalism. A wave packet or quantum field treatment is thus
inescapable. Let us also insist on the dubiousness of the
arguments using energy-momentum conservation. In most cases, they
are invalidated by the energy-momentum spectrum of the source:
different energy-momentum components of the source can contribute
to different mass eigenstates.

\subsection{Problems with the plane wave treatment}
\label{planewaveproblems}

In section \ref{timespaceconversion}, it has been shown that the
plane wave treatment of particle oscillations cannot deal in a
satisfactory way with the time dependence of the oscillating
phase. Besides, this approach implies a perfectly well-known
energy-momentum and an infinite uncertainty on the space-time
localization of the oscillating particle. Oscillations are
destroyed under these assumptions \cite{Kayser81}. On the one
hand, the perfect knowledge of the energy-momentum allows to
determine which mass eigenstate propagates. On the other hand, the
spatial delocalization makes impossible the measurement of the
oscillation length. A correct oscillation formula should include
observability conditions in such a way that the oscillation term
vanishes if either
\begin{enumerate}
   \item
   the energy-momentum uncertainty is smaller than the mass
   difference between the interfering mass eigenstates, or
   \item
   the oscillation length is of the same order, or smaller, than
   the uncertainty on the position of the source or of the
   detection point of the oscillating particle.
\end{enumerate}

Another kind of problem is not specific to the plane wave
treatment, but affects all approaches where the oscillating
particle is considered to be directly observable. On the one hand,
flavor eigenstates are ill-defined for stable particles, because
we do not know how to define creator and annihilation operators of
flavor states satisfying canonical (anti)commutation relations
\cite{Giunti92a}. On the other hand, mass eigenstates are
ill-defined for unstable particles, since they are in general not
orthogonal (except if the Hamiltonian is normal, see section
\ref{mixing}). The solution to these problems simply consists in
considering the oscillating particle as an intermediate state.
Actually this stand reflects well the experimental situation,
where the oscillating particle is not directly observed. One
rather detects the particles in interaction with the oscillating
state, both at the source and at the detector. The flavor
transition probability should thus be computed with observable
particles as initial and final states
\cite{Kobzarev82,Giunti92a,Giunti93,Rich93}.

Finally, we should mention two other problems regarding unstable
particles. These difficulties arise because of the nonrelativistic
Wigner-Weisskopf method used to compute the effective Hamiltonian.
First, interference between different mass eigenstates is
forbidden in nonrelativistic quantum mechanics (Bargmann
superselection rule \cite{Bargmann54,Kaempffer,Galindo}). The
argument is the following. The invariance of the Schr\"odinger
equation under Galilean transformations determines the
transformation law of a quantum state: it is multiplied by a phase
factor depending on the mass and space-time position of the state.
Thus different mass eigenstates transform differently, so that the
relative phase in a superposition of such eigenstates is not
conserved under Galilean transformations. Therefore a coherent
superposition of different mass eigenstates is forbidden. Second,
unstable particles cannot be consistently described in
nonrelativistic quantum mechanics for the same reason at the
origin of the Bargmann superselection rule: transitions between
different mass eigenstates are forbidden \cite{Kaempffer}.
Unstable states cannot be considered as asymptotic states. Thus
they do not appear in the Hilbert space of physical states and
must be treated in quantum field theory where they appear as
complex poles of the full propagator \cite{Veltman63}.

\subsection{Intermediate wave packets and other improvements}
\label{intermwavepackets}

Some of the problems of the plane wave treatment are solved by the
{\it intermediate wave packet model}, in which a wave packet is
associated with each propagating mass eigenstate. Note that this
model is usually discussed with respect to neutrinos, i.e.\ in the
relativistic limit. Nussinov was the first to put forward the
existence of wave packets as the cause of a coherence length
beyond which oscillations vanish \cite{Nussinov76}. Oscillations
with wave packets were then studied in detail by Kayser
\cite{Kayser81}. The oscillation formula was later explicitly
computed with Gaussian wave packets by Giunti, Kim and Lee
\cite{Giunti91,Giunti98b,Giunti02a}.

The intermediate wave packet model shows that oscillations vanish
if $\sigma_x\gtrsim L^{osc}_{ij}$, i.e.\ if the uncertainty over
the position is larger than the oscillation length (Fig.~1). For
minimal uncertainty wave packets, this condition can be rewritten
as $\sigma_p\lesssim\delta m^2_{ij}/2p$, i.e.\ oscillations are
forbidden if the momentum spread of the wave packets is smaller
than the mass difference between the interfering eigenstates. Thus
oscillations are destroyed by energy-momentum measurements aiming
to determine which mass eigenstate propagates.
\begin{figure}
\begin{center}
\includegraphics[width=5cm]{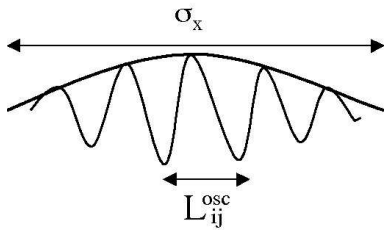}
\caption{Decoherence due to a wave packet width $\sigma_x$ larger
than the oscillation length.}
\end{center}
\end{figure}

The other main result of the model is the existence of a coherence
length beyond which oscillations vanish
\cite{Nussinov76,Kayser81,Giunti91}. Its usual explanation is that
wave packets associated to different mass eigenstates have
different group velocities. Hence, wave packets progressively
separate, and interference disappears when they do not overlap
anymore (Fig.~2). We shall see in section \ref{subdecoh} that
dispersion is also at the origin of the coherence length.
\begin{figure}
\begin{center}
\includegraphics[width=10cm]{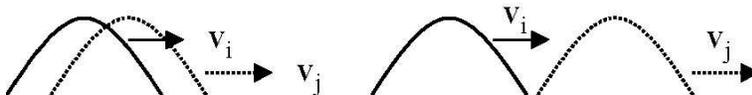}
\caption{Decoherence due to the non-overlapping of the different
mass eigenstates wave packets.}
\end{center}
\end{figure}

In the intermediate wave packet model, the classical propagation
condition, $|vt-x|\ll t$, is automatically implemented by the
space-time localization of the wave packet. However the question
of the equal or different time prescription is not solved, since
the times of production and detection are not specified by the
formalism. As regards the question of the equality of energies or
momenta, it is in principle no longer relevant, since the
energy-momentum is spread out. In fact, the oscillation length
depends only on the zeroth order values (in a $\delta
m_{ij}^2$-expansion) of the average energies and momenta, because
of the cancellation between the time and space parts of the phase
(see Eqs.~(\ref{expansionplanephase}) and (\ref{expansionpjQM})
with $p_j$ equal to the average momentum). The time dependence of
the probability is dealt with by averaging over time
\cite{Giunti91}. It is thus not necessary to worry about the
equality of energies or momenta.

Nevertheless there is still much argument about the equality of
the energy spectra, or the equality of the average energies,
momenta or velocities. For example, Grossman and Lipkin
\cite{Grossman97} imposed a same energy spectrum for all mass
eigenstates, so that the probability of finding a particle with
the wrong flavor vanishes for all times at the position of the
source. It can be objected that this boundary condition is
automatically satisfied without imposing an identical energy
spectrum, since the probability to detect the oscillating particle
at the source becomes negligible once the wave packet has left it.
Equality of average velocities has been proposed by Takeuchi,
Tazaki, Tsai and Yamazaki \cite{Takeuchi99,Takeuchi01}, but is
very unlikely for neutrinos, as noted by Okun and Tsukerman
\cite{Okun00b}, because the ratio of the average energies
$E_i/E_j$ should then be equal to the mass ratio $m_i/m_j$.

Actually there is another way of looking at this question.
Starting with arbitrary wave packets, it is simple to check that
interference occurs only between wave packet components with the
same energy, at least if the oscillation process is strictly
stationary. This line of thought has been advocated by Sudarsky
{\it et al.} \cite{Sudarsky91} (within a field-theoretical model),
by Lipkin \cite{Lipkin95,Lipkin99} (with wave packets) and by
Stodolsky \cite{Stodolsky98} (within a density matrix formalism).
Let us explain it with the intermediate wave packet model. In one
dimension, the wave function corresponding to the mass eigenstate
$m_i$ can be written as
$$
   \psi_i(x,t) = \int dE \, A_i(E) \, e^{-i E(t-t_0) + i k_i (x-x_0)} \, ,
$$
where $k_i=\sqrt{E^2-m_i^2}$ and $(t_0,x_0)$ are the coordinates
of the source. The flavor-mixing transition probability is a
linear superposition of terms $\psi_i(x,t)\psi_j^*(x,t)$. If the
time of emission is unknown, the probability is averaged over
$t_0$, so that it becomes a superposition of terms like
\begin{equation}
   \int dt_0 \, \psi_i(x,t) \, \psi_j^*(x,t)
   = 2\pi \int dE \, A_i(E) A_j^*(E) \, e^{i(k_i-k_j)(x-x_0)} \, .
   \label{incohsum}
\end{equation}
This equation makes clear that interference occurs only between
wave packet components with the same energy. The time-averaged
transition probability in the wave packet model is equivalent to
an incoherent sum over energy eigenstates (`incoherent' means a
sum in the probability, not in the amplitude).
Eq.~(\ref{incohsum}) can thus be seen as a sum over plane waves of
energy $E$ and momentum $k_i$, weighed by the factor $A_i(E)
A_j^*(E)$. The equivalence between a wave packet and a plane wave
decomposition of a stationary beam was already observed by Dicke
and Wittke \cite{Dicke} in connection with electron-interference
experiments, and by Comsa \cite{Comsa83} in connection with
neutron-interferometry. The same issue was recently raised by
Kiers, Nussinov and Weiss \cite{Kiers96} and by Stodolsky
\cite{Stodolsky98}, this time in connection with neutrinos.

Of course the time integral in Eq.~(\ref{incohsum}) only yields a
delta function if the time interval is infinite, i.e.\ if the
process is strictly stationary. In practice, there is always some
available information on the time of emission. If the time
integral is limited to the interval $\Delta T$, the delta function
becomes a narrow peak of width $1/\Delta T$, which can be
neglected as long as it is small in comparison with the mass
difference $\delta m^2_{ij}/2E$. This condition can be written as
$\Delta T\gg T^{osc}_{ij}$, where $T^{osc}_{ij}$ is the
oscillation time.

A few more results have been obtained with wave packets. Kiers,
Nussinov and Weiss have used the equivalence between wave packets
and plane waves, explained above, to show that it is always
possible to increase the coherence length by more accurate energy
measurements at the detector \cite{Kiers96}. After Giunti and Kim
\cite{Giunti98b} showed how to take into account the detector's
momentum uncertainty in the intermediate wave packet model, Giunti
\cite{Giunti02a} did the same with the detector's energy
uncertainty, so as to reproduce Kiers, Nussinov and Weiss'
prediction. Wave packets have also been applied by Nauenberg to
study correlated oscillations of neutrinos or mesons
\cite{Nauenberg99}, and to the propagation of neutrinos in matter
\cite{Giunti92b,Peltoniemi00}. Finally, it is worth mentioning
that the wave packet approach is explained in great detail in Kim
and Pevsner's book \cite{Kimbis}, is discussed by Fukugita and
Yanagida \cite{Fukugita94} and has been reviewed by
Zra\makebox[0pt][l]{/}lek \cite{Zralek98}.

The intermediate wave packet model has been recently discussed by
De Leo, Nishi and Rotelli \cite{DeLeo02}, who recover the standard
oscillation formula under two conditions. The first condition,
which they call `minimal slippage', is equivalent to the existence
of the coherence length discussed above, i.e.\ mass eigenstate
wave packets should overlap at detection otherwise there is
decoherence. The second condition, which they call `non
instantaneous creation' or `pure flavor creation event-wise',
means that interference occurs between wave packet components
corresponding to different initial wave packet points, so that the
standard oscillation phase (\ref{standardphase}) is obtained. This
mechanism, which was already explained by Giunti and Kim
\cite{Giunti01a} (see end of section \ref{nonstandard}), is
automatically included in quantum field treatments where the
oscillating particle is coupled to a source and detector (see
section~\ref{section3}). In the same article, De Leo, Nishi and
Rotelli argue that the oscillation formula might depend on the
wave packet shape, for example if the wave packet is a
superposition of two well-separated Gaussian peaks, each having
its own phase factor. In that case, there would be a succession of
coherence/decoherence/coherence/decoherence regions as the mass
eigenstates wave packets shift one with respect to the other.
Moreover, the constant phase difference between the two Gaussian
peaks crops up in the oscillation phase in the second coherence
regime. While surprising phenomena indeed arise for such special
wave packet shapes, it seems difficult to set up an experiment in
which a specific wave packet shape is maintained for all the
particles within the produced flux. In practice we should average
over all wave packet shapes, with the result that only the typical
width of the wave packet will have an effect on the final
oscillation formula. All other effects due to the different shapes
of the wave packets cancel each other out. Besides, the analysis
of oscillations in a faraway region (such as the second coherence
regime mentioned above) should take into account the fact that
dispersion wipes out oscillations at large distances (see section
\ref{subdecoh}).

The representation of the oscillating particle by a minimal
uncertainty wave packet has been criticized by Field
\cite{Field01}, who claims that that oscillating particles follow
classical space-time trajectories and that the oscillation phase
is due to the propagation time difference between different mass
eigenstates (see section \ref{timespaceconversion}). One of his
main arguments is that the neutrino momentum spread in
$\pi\to\mu\nu$ is of the order of decay width $\Gamma_\mu$ of the
muon, so that neutrino oscillations should be completed suppressed
by decoherence in wave packet models. Moreover, the Heisenberg
relation is violated if one admits that the neutrino spatial
uncertainty is of the order of the cm. These arguments do not hold
since the neutrino momentum spread reflects the momentum
uncertainties of the pion source and of the emitted muon, which
are much larger than $\Gamma_\mu$. Finally, Field criticizes
models where the neutrino wave packet is coupled to a wave packet
source, by claiming that the initial quantum state is not the same
for all amplitudes, so that interference is forbidden. This
criticism is not justified, since it makes no sense to cut the
source wave packet into disjointed parts which are then
interpreted as different initial states.

Although they solve some of the problems of the plane wave
approach, intermediate wave packets are not satisfying for many
reasons:
\begin{enumerate}

\item The question of the existence of a Fock space for the weak
eigenstates remains.

\item Oscillating particles are not, and cannot be, directly
observed. It would be more convincing to write a transition
probability between observable particles, involved in the
production and detection of the oscillating particle.

\item The wave packet shapes, sizes and normalizations are
arbitrary. In particular, the amplitudes of production and
detection are not taken into account. This is not a problem in the
relativistic limit or for nearly degenerate masses, in which cases
these amplitudes can be factorized from the sum over the mass
eigenstates. However this factorization is not possible for
nonrelativistic particles (unless they are nearly degenerate in
mass).

\item The coherence length depends on the difference between the
group velocities of the wave packets, the exact values of which
are unknown in the nonrelativistic
case\footnote{Eq.~(\ref{expansionvjQM}) can be interpreted as
giving the group velocity of a mass eigenstate wave packet. In the
relativistic limit ($v\to1$), the $\rho$-dependence drops from
this equation, so that the velocity can be determined up to order
$\delta m_{ij}^2/2p$ without needing to know $\rho$.}. There is no
reason to believe that the group velocities could be determined,
to order $\delta m_{ij}^2/2p$, by energy-momentum conservation at
the source, contrary to what is claimed in \cite{Giunti91}.

\item The coherence length increases with the precision of the
energy-momentum measurements at the source and at the detector.
Whereas the momentum precision is limited by the condition
$\sigma_p\gtrsim\delta m_{ij}^2/2p$, there is, a priori, no such
limitation on the energy accuracy and, therefore, no bound on the
coherence length. Although it has been claimed \cite{Giunti98b}
that such a bound exists, the intermediate wave packet model
cannot settle that question, because the energy uncertainty has to
be inserted by hand into the model.

\item Experiments measure a particle flux, not a particle density.
The transition probability must thus be converted to a flux
density, involving inverse velocity factors. These enhance
nonrelativistic mass eigenstates and strongly skew the probability
\cite{Giunti91}. This effect is due to the neglect of the
amplitudes of production and detection and to an analysis
restricted to one dimension.

\item Although it has not been controversial in the wave packet
model, why should the times appearing in the different wave
packets be identical? A justification of this equal time
prescription is necessary to rule out the nonstandard oscillation
formulas discussed in section \ref{timespaceconversion}.

\item It is not possible to analyze in the wave packet model the
corrections to the oscillation probability due to the possible
instability of the propagating particle.

\item The influence of the source instability on the observability
of the oscillations cannot be studied in that framework.
\end{enumerate}

A quantum-mechanical model of neutrino oscillations solving the
first three problems has been put forward by Rich \cite{Rich93}.
This author considers the oscillating particle as an intermediate
state and computes the probability transition between initial and
final observable states with second-order time-dependent
perturbation theory of standard quantum mechanics. Since this
model takes into account the production and detection processes,
it has the other advantage of not requiring the equal time
prescription. The spirit of Rich's model is similar to the
stationary boundary condition models resorting to
field-theoretical methods (see section \ref{review}), but it has
the disadvantage of being nonrelativistic.

Ancochea, Bramon, Mu\~noz-Tapia and Nowakowski have tried to solve
the difficulty of converting the probability density for neutral
kaons into a flux \cite{Ancochea96}. In their nonrelativistic wave
packet model, they construct flavor probability currents
associated to a Schr\"odinger equation. A problem arising in this
approach is that the flavor currents are not conserved
\cite{Zralek98}. Moreover, no one knows how to extend this method
to the relativistic case.

Sassaroli has proposed an hybrid model for neutrino oscillations,
going half-way to quantum field theory, in which a coupled system
of two Dirac equations is quantized \cite{Sassaroli99}. However
boundary conditions cannot be applied consistently, unless lepton
flavor wave functions are considered as observable and the
relativistic limit is taken. This difficulty was already noted by
Giunti, Kim and Lee \cite{Giunti92a}.

The review of intermediate wave packet models would not be
complete without mentioning its latest and most sophisticated
version, as proposed by Giunti \cite{Giunti02b}. Instead of
representing the oscillating particle by a superposition of
arbitrary wave packets, Giunti computes the form of the
intermediate wave packet from basic principles. More precisely, it
consists in creating the intermediate wave packet from quantum
field interactions between the external wave packets involved in
the production process of the oscillating particle. The wave
packet then evolves in space-time, before interacting with quantum
field interactions at the detector. This {\it interacting wave
packet model} is very close in method, spirit and results to the
external wave packet model at the core of the present report. It
will be further discussed in section \ref{poleintegrations}.

\section{Oscillations in quantum field theory}
\label{section2}

\subsection{Review of the literature}
\label{review}

Few authors deny that the most rigorous treatment of oscillations
is done in the quantum field theory framework. However, although
the quantum field computations in the literature all reproduce in
some limit the naive quantum-mechanical formula given by
Eqs.~(\ref{standardproba}) and (\ref{standardphase}), there is not
yet an agreement in which respect they differ from the naive
formula. There are two reasons for the lack of agreement between
the existing quantum field derivations of the oscillation formula:
first, the different authors use different physical assumptions,
and second, they use different approximation schemes to compute
the transition probability.

The field-theoretical approach to particle oscillations is quite
old. Already in 1963, Sachs \cite{Sachs63} applied S-matrix
methods to neutral kaon interferences\footnote{Quoting Sachs, from
\cite{Sachs63}: `The question of whether it is the momenta or the
energies of these particles that are to be taken equal in the
treatment of interference phenomena has often been raised. [Our]
method provides a clear answer to this question since it is based
on an analysis of the phenomena in terms of wave packets'. Nearly
40 years after, new articles continue to appear, wondering about
the equality of energies or momenta.}, using a model developed
earlier with Jacob \cite{Jacob61} for unstable particles. Let us
call this model, as applied to non-oscillating particles, the {\it
Jacob-Sachs model}. Its application to systems of mixed particles
will be called the {\it external wave packet model}. In this
model, the particle to be studied is represented by its
propagator; it propagates between a source and a detector, where
wave packets representing the external particles are in
interaction. Much later, Sudarsky, Fischbach, Talmadge, Aronson
and Cheng \cite{Sudarsky91} studied the influence of a spatially
varying potential on the neutral kaon system. They resort to a
one-dimensional model similar to Sachs' but do not specify the
contour of integration, so that the finiteness of their final
expressions is not guaranteed. Neither Sachs nor Sudarsky {\it et
al.} studied the observability conditions of oscillations. Another
simplified model was proposed by Beuthe, L\'{o}pez Castro and
Pestieau \cite{Beuthe98}, with the aim of modelizing experiments
at CPLEAR and DA$\Phi$NE. Their model is not satisfying since they
use external wave packets localized in time but not in space,
which does not correspond to actual experiments.

As regards neutrinos, Kobzarev, Martemyanov, Okun and Shchepkin
analyzed neutrino oscillations with a bare-bones quantum field
model \cite{Kobzarev82}: the source and detector are infinitely
heavy nuclei, so that the propagation distance is perfectly known,
whereas the propagation time is left undetermined. No constraints
on oscillations are discussed in that article. Next came an
important article by Giunti, Kim and Lee \cite{Giunti92a}, showing
that it is impossible to build a Fock space for flavor states,
because the mixing of the ladder operators for mass eigenstates
does not yield flavor ladder operators satisfying canonical
(anti)commutation relations. This observation strikes a blow to
the quantum-mechanical wave packet approach, which should not be
mistaken for the quantum field model with external wave packets,
as it is sometimes the case in the literature. Nonrelativistic
corrections to the neutrino propagation are explicitly computed in
Ref.~\cite{Giunti92a} in a few examples.

In a pioneering article, Giunti, Kim, Lee and Lee \cite{Giunti93}
studied neutrino oscillations within a Gaussian external wave
packet model. They derive a localization condition (no
oscillations if $L^{osc} \lesssim \sigma_x$) and a coherence
length beyond which oscillations vanish. These conditions agree
with those obtained in the quantum-mechanical picture
\cite{Nussinov76,Kayser81,Giunti91}. A later paper by Giunti, Kim
and Lee \cite{Giunti98a} contains essentially identical results as
in Ref.~\cite{Giunti93}, but with more generality. Cardall used
the same model \cite{Cardall00}, paying greater attention to the
normalization of the event rate and to the spin structure. A model
close in spirit to the external wave packet model was proposed by
Kiers and Weiss \cite{Kiers98}. These authors couple the
oscillating neutrino with localized source and detector, which are
idealized by oscillators. They show how the coherence length
increases with the energy precision at the detector. Note that
they had already predicted this phenomenon using elementary
quantum mechanics \cite{Kiers96}.

Another kind of model arises from the use of {\it stationary
boundary conditions}, leading to a unique value for the energy of
the oscillating particle. The simplest example is the Kobzarev
{\it et al.} model mentioned above \cite{Kobzarev82}. Going a bit
further, Grimus and Stockinger proposed a model with external
particles represented either by bound states or by plane waves.
They obtain a localization condition but no coherence length. In
two other papers, Grimus, Mohanty and Stockinger
\cite{Grimus99,Grimus00} studied the influence of an unstable
source on the oscillations, with the model of the previous paper
modified by a Wigner-Weisskopf approximation for the unstable
source. Ioannisian and Pilaftsis analyzed neutrino oscillations
\cite{Ioannisian99} within a scalar version of the
Grimus-Stockinger model. They claim to have found a novel form of
neutrino oscillations at short distance, which they call `plane
wave oscillations'. Stationary boundary conditions were also used
by Cardall and Chung \cite{Cardall99} to study the MSW effect in
quantum field theory. Note that a quantum field derivation of the
MSW effect has been proposed by Mannheim \cite{Mannheim88}, who
represents the interaction with matter with an effective
potential, whereas this potential has been derived with finite
temperature field theory by N\"otzold and Raffelt \cite{Notzold88}
and by Pal and Pham \cite{Pal89}.

Campagne \cite{Campagne97} studied a neutrino source decaying in
flight, such as a relativistic pion, with field-theoretical
methods. He sidesteps external wave packets by limiting
arbitrarily the interactions regions with the help of step
functions, but it is only a trick to replace complicated
interactions at the source and detector by simple stationary
boundary conditions. Decay in flight has also been studied by
Dolgov \cite{Dolgov00}, within a simplified external wave packet
model modified by a Wigner-Weisskopf approximation.

Another approach ({\it source-propagator models}) consists in
using the propagator in configuration space coupled to a source
but not to a detector. Srivastava, Widom and Sassaroli chose this
method to modelize correlated oscillations of two kaons
\cite{Srivastava95b} and neutrino oscillations
\cite{Srivastava98}. Since external wave packets are absent,
time-space conversion problems cannot be avoided and lead to
nonstandard oscillation lengths or recoil oscillations (see
section \ref{timespaceconversion}). Shtanov also used a
source-propagator model, and claims to have found a strong
dependence of the oscillation formula on the neutrino masses if
the source and detector are very well localized in space-time
\cite{Shtanov98}.

In a completely different line of thought, Blasone, Vitiello and
other researchers have attempted to define a Fock space of weak
eigenstates and to derive a nonperturbative oscillation formula.
The main results of these studies are summarized in
\cite{Blasone01b}. Note that a previous formalism developed in
\cite{Blasone95,Alfinito95,Binger99,Hannabuss00} lead to an
oscillation formula not invariant under reparametrization
\cite{Fujii99,Blasone99b,Fujii01,Blasone01c} and was replaced by a
more satisfying theory in \cite{Blasone99a,Blasone01a}, which was
further developed in \cite{Blasone01b,Ji01,Ji02,Blasone02}. This
new theory aims at defining flavor creation and annihilation
operators, satisfying canonical (anti)commutation relations, by
means of Bogoliubov transformations. As a result, new oscillation
formulas are obtained for fermions and bosons, with the
oscillation frequency depending surprisingly not only on the
difference but also on the sum of the energies of the different
mass eigenstates. Apart from the speculative nature of the
enterprise, the drawbacks of the approach are the dependence on
time, not on space, of the oscillation formula (Lorentz covariance
is broken), as well as the lack of observability conditions.
Although these studies are very interesting from a fundamental
point of view, it is not obvious whether the new features of the
Blasone-Vitiello oscillation formulas are observable in practice.
Since these new oscillation formulas tends to the standard
oscillation formula (\ref{standardproba}) in the relativistic
limit or if the mass eigenstates are nearly degenerate, we can
focus on the case of a nonrelativistic oscillating particle having
very distinct mass eigenstates. In that case, $p\sim\delta
m^2/2E$, so that either $\sigma_p\lesssim\delta m^2/2E$ or
$p\lesssim\sigma_p$. Under these conditions, the quantum theory of
measurement says that interference between the different mass
eigenstates vanishes. Once the oscillation terms have been
averaged to zero, the Blasone-Vitiello formulas do not differ
anymore from the standard oscillation formula
(\ref{standardproba}). Therefore, the Blasone-Vitiello formalism
does not seem to be relevant to the phenomenology of oscillations
on macroscopic distances. This observation does not detract from
the theoretical worth of that approach.

All the above models, whether for mesons or for neutrinos, can be
grouped in four categories: external wave packet models,
stationary boundary conditions models, source-propagator models
and Blasone-Vitiello models. In the following sections, the
connection between external wave packet models and stationary
boundary condition models will be studied and the
Ioannisian-Pilaftsis and Shtanov models will be analyzed in
detail\footnote{Some of our results have been published in
Ref.~\cite{Beuthe02}.}.

Finally let us mention again the interacting wave packet model
recently proposed by Giunti \cite{Giunti02b} which was described
at the end of section \ref{intermwavepackets}, since it bridges
the gap between the intermediate wave packet model and quantum
field treatments. Moreover its results are equivalent to those
obtained in the external wave packet model.

\subsection{Five questions}
\label{fivequestions}

The contradictions between the existing quantum field derivations
of the oscillation formula can be summarized into five questions:
\begin{enumerate}

\item Whereas the external wave packet model allows to associate
intermediate `wave packets' to the oscillating particle, it is not
possible to do so in models using stationary boundary conditions.
Moreover, the coherence length is finite in the former case, but
infinite in the latter. Is it possible to see the oscillation
formula derived in the models using stationary boundary conditions
\cite{Kobzarev82,Grimus96,Grimus99,Grimus00,Ioannisian99,Cardall99},
as a particular case of the oscillation formula derived in the
models using external wave packets
\cite{Giunti93,Giunti98a,Cardall00}?

\item Kiers, Nussinov and Weiss \cite{Kiers96} have shown in a
quantum-mechanical model that the value of the coherence length
depends on the accuracy of the energy-momentum measurements at the
detector. This effect has been confirmed by quantum field theory
calculations \cite{Giunti98a,Kiers98}, but it is not clear whether
oscillations survive a perfect measurement of the energy
\cite{Giunti98a,Giunti98b}. Does a perfect knowledge of the energy
lead to an infinite coherence length or is there decoherence
anyway?

\item For a source strongly localized in space and time, Shtanov
has derived an oscillation formula where each oscillating
exponential $exp(-i\phi_j)$ is multiplied by a prefactor depending
on the mass $m_j$. Unless the masses $m_j$ are nearly degenerate,
these prefactors strongly modify the standard oscillation formula
given by Eqs.~(\ref{standardproba})-(\ref{standardphase}). Is this
result correct?

\item Ioannisian and Pilaftsis \cite{Ioannisian99} claim to have
found a novel form of neutrino oscillations (`plane wave
oscillations'), if the spatial spread of the source and detector
is of macroscopic size. Does such behavior exist?

\item As in the quantum-mechanical treatment of section
\ref{timespaceconversion}, the correlation between an oscillating
particle and other particles (or {\it recoil particles}) at the
source has been said to modify the oscillation length with respect
to Eq.~(\ref{standardphase}) and to bring about oscillations of
the recoil particles \cite{Srivastava95b,Srivastava98}. Is this
assertion true?

\end{enumerate}

In order to answer these questions, we shall use a model such that
all specific models used in the articles cited above can be
recovered in some limit. Moreover, to extend the usual treatment
of a stable relativistic oscillating particle, our model will also
allow the oscillating particle to be nonrelativistic and/or
unstable and will take into account the dispersion. Answers to
questions~1 to 4 can be found in section \ref{answers} while
question~5 is treated in section~\ref{section9}.

\section{The external wave packet model}
\label{section3}

The numerous problems arising in the plane wave approach of
oscillations (see section \ref{planewaveproblems}) and in the
intermediate wave packet method (see section
\ref{intermwavepackets}) show that the oscillating particle cannot
be treated in isolation. The oscillation process must be
considered globally: the oscillating states become intermediate
states, not directly observed, which propagate between a source
and a detector. This idea is easily implemented in quantum field
theory, where intermediate states are represented by internal
lines of Feynman diagrams. Quantum field theory has the advantage
of providing a relativistic treatment from the start, which is
required to study the mixings of relativistic and nonrelativistic
particles. It also allows to describe unstable particles in a
consistent way. The oscillating particle is represented by a
relativistic propagator, which determines the space-time evolution
and the possible decay of the particle. Boosts and the
consideration of specific rest frames become pointless. Since
interactions are included in the amplitude, equal or different
time prescriptions are not needed anymore. Particles interacting
with the oscillating particles at the source and at the detector
are described by wave packets. This {\it external wave packet
model} was first proposed by Sachs \cite{Sachs63} for kaons and by
Giunti, Kim, Lee and Lee \cite{Giunti93} for neutrinos.

As appears clearly from the review of section \ref{review}, there
is an ongoing controversy in the literature about whether the
boundary conditions (i.e.\ the source and the detector) should be
taken as stationary or not. In other words, can the time
independence of most oscillation experiments be translated into
the assumption that the energy of the process is perfectly
defined? Although it is true that most sources are stationary from
a macroscopic point of view (for example, the flux of solar
neutrinos is steady), there is no reason to think that it should
be the case from a microscopic point of view. Whereas it can be
reasonable to make a stationary approximation in a
quantum-mechanical model where we do not have any information on
the microscopic processes (for example a density matrix model as
in \cite{Stodolsky98}), the same approximation is very dubious in
quantum field theory, which describes the interactions of
individual particles. After all, perturbative quantum field theory
applies to one-particle propagation processes, in which a
stationary source and detector are the exception, rather than the
rule. As emphasized by Cardall \cite{Cardall00}, the Sun is
certainly not stationary at the atomic scale, and neither is a
detector composed of bound state particles. For example, a water
Cerenkov detector sees charged lepton wave packets with finite
energy and time spread. The finite character of the spread is
partly due to the limited coherence time of the bound state
particle that has interacted with the incoming neutrino. Boundary
conditions can be considered as stationary at the microscopic
level when the energy uncertainty at the source or detector is
smaller than the inverse propagation distance, i.e.\ $\sigma_{e\scriptscriptstyle
P,D}\lesssim1/L$. This constraint is extremely stringent. In the
example of atmospheric neutrinos, the process can be considered as
stationary if $\sigma_{e\scriptscriptstyle P,D}\lesssim10^{-19}\,$MeV, which is
not satisfied in current oscillation experiments.

Although it seems difficult to argue that the energy uncertainty
is smaller than the mass difference, it is possible to take up
another stand regarding stationary boundary conditions. In the
same way as in section \ref{intermwavepackets}, it consists in
arguing that interference occurs between wave packet components
with the same energy, because of the time average on the
transition probability. As it is unrealistic to consider in
isolation an interference for a given energy, we should take care
to integrate the probability (computed with stationary boundary
conditions) over the wave packet energy width. At that point a
question arises: is this energy width determined by the source or
by the detector? There is thus information to be gained by working
with non-stationary boundary conditions. Moreover, we shall show
that stationary boundary conditions can be imposed in the external
wave packet model by assigning zero velocities to some states,
whereas other states are represented by plane waves.

In section \ref{jacobsachsmodel}, we describe the Jacob-sachs
model which is the prototype for the external wave packet model.
We then examine the diagonalization of mixed propagators in
section \ref{mixedpropagators}. Next, we compute the transition
amplitude of the process associated to particle oscillations in
section \ref{flavormixing}. In section \ref{simplest}, we discuss
the simplest model in which oscillations can be consistently
described. Finally, we present in section \ref{overlapgaussian}
the external wave packet model in its Gaussian version.

\subsection{The Jacob-Sachs model}
\label{jacobsachsmodel}

The first version of the external wave packet model was developed
by Jacob and Sachs \cite{Jacob61} to describe the propagation of a
non-oscillating unstable particle.

\subsubsection{The process}
\label{theprocess}

The propagating process of a particle between a source and a
detector (extended in space-time and indicated by dotted circles)
is symbolized by Fig.~3.
\begin{figure}
\begin{center}
\includegraphics[width=10cm]{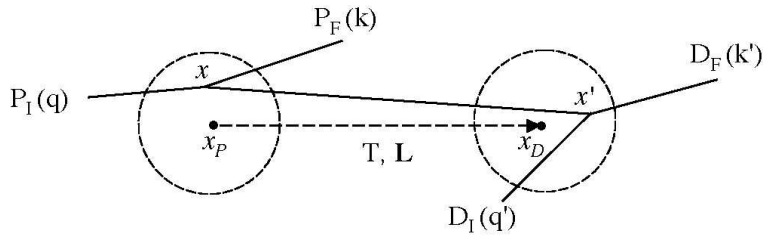}
\caption{Propagation of a particle between a source and a
detector, centered in $x_{\scriptscriptstyle P}$ and $x_{\scriptscriptstyle D}$.}
\end{center}
\end{figure}

The arrows indicate the momentum flow. $P_I$ represents the set of
incoming particles, of total momentum $q$, arriving in the
production region (or source), which is centered around the point
$( t_{\scriptscriptstyle P},{\bf x}_{\scriptscriptstyle P} )$. $P_F$ represents the set of
outgoing particles, of total momentum $k$, coming from the
production region, with the exception of the intermediate particle
whose propagation is studied. $D_I$, $D_F$ and $( t_{\scriptscriptstyle D},{\bf
x}_{\scriptscriptstyle D} )$ are defined similarly, but apply to the detection
process. If the intermediate particle decays, $D_I(q')$ can be
seen as an outgoing state. The interaction points at production
and detection are noted $x$ and $x'$, respectively.

The external particles are assumed to be stable, whereas the
intermediate particle is allowed to be unstable. Although it is
possible to describe external unstable particles by considering a
more global process where all unstable particles are considered as
intermediate states, the technical difficulties involved in
computing the corresponding amplitudes are enormous. All
computations will be carried out for a scalar propagating particle
since it can be argued (see section \ref{flavormixing}) that the
spin has no impact on the characteristics of oscillations.

With the aim of localizing the production region around $(t_{\scriptscriptstyle
P},{\bf x}_{\scriptscriptstyle P} )$, the in- and outgoing particles at point $x$
are represented by wave packets which overlap (in configuration
space) around $(t_{\scriptscriptstyle P},{\bf x}_{\scriptscriptstyle P} )$. As the
energy-momenta are often measured, the wave packets are supposed
to be well-localized in momentum space, around their average
momentum. The detection region is localized in the same way. Note
that there is no difficulty in replacing certain wave packets by
bound states or by plane waves, if needed. For example, it is
practical to use plane waves if an emitted particle is not
observed \cite{Grimus96}.

If the dotted circles are removed, the above picture can be
interpreted as a Feynman diagram. The internal line represents
either a particle or an antiparticle. The experimental conditions
are chosen such that a quasi-real particle propagates on a
macroscopic distance, transferring positive energy from $x$ to
$x'$. The energy-momenta of the initial and final states are such
that the S-matrix element, corresponding to the above process, is
evaluated at the singularity of the propagator of the intermediate
particle. If $x'^0-x^0$ is a macroscopic time, it will be seen
that the intermediate particle (positive energy component)
contributes to the propagation, but not the intermediate
antiparticle (negative energy component).

\subsubsection{Wave packets}
\label{sectionwavepackets}

A wave packet representing a given state $|\,\psi \!\!>$ of mass
$m$ can be expressed in momentum space \cite{Peskin} as
$$
   |\,\psi \!> = \int [d{\bf k}] \, \psi({\bf k}) \, |\,{\bf k} \!> \, ,
$$
where $\psi({\bf k})$ is the wave function in momentum space taken
at time $t=0$, and $|\,{\bf k}\!>$ belongs to the basis of
one-particle states of momentum ${\bf k}$, also taken at time
$t=0$. In the interaction picture, the time-dependence of the
basis and of the wave function cancel each other so that the state
$|\,\psi \!>$ is independent of time. That is why we may take
$\psi({\bf k})$ and $|\,{\bf k}\!>$ at time $t=0$. The following
notation is used
$$
   [d{\bf k}] = \frac{d {\bf k}}{(2\pi)^3 } \,
   \frac{1}{ \sqrt{2 \, E({\bf k}) } } \, ,
$$
where $E({\bf k}) = \sqrt{ {\bf k}^2 + m^2}$. The normalization of
free states is chosen as follows \cite{Peskin}:
$$
   <\! {\bf k} \, | \, {\bf p} \!>
   = 2 E({\bf k}) \, (2\pi)^3 \, \delta^{(3)}( {\bf p} \! - \! {\bf k} ) \, ,
$$
so that
\begin{equation}
   <\! \psi \,|\,\psi \!> = 1
   \quad \mbox {if} \quad
   \int \frac{d {\bf k}}{(2\pi)^3 } \,
   |\psi({\bf k})|^2 =1 \, .
   \label{norm}
\end{equation}
In configuration space, the wave function reads
\begin{equation}
   \widetilde{\psi}({\bf x},t)
   = \int \frac{d {\bf k}}{(2\pi)^3 } \, \psi({\bf k}) \,
   e^{ - i E({\bf k}) t  + i \, {\bf k} \cdot {\bf x} } \, .
   \label{wpconfig}
\end{equation}
It is here necessary to use the time-dependent wave function,
i.e.\ $\psi({\bf k})\,exp(-iE({\bf k})t)$, since the wave function
is considered alone.

If the wave packet represents a particle with an approximately
known momentum ${\bf K}$, the wave function $\psi({\bf k})$ is
sharply peaked at \mbox{${\bf k}={\bf K}$}. The configuration
space wave function \mbox{$\widetilde{\psi}( {\bf x},t \!=\! 0 )$}
has a maximum at the point \mbox{${\bf x} \!=\!{\bf 0}$} if
\mbox{$\psi({\bf K} \!+\! {\bf k'}) = \psi({\bf K} \!-\! {\bf
k'})$}. On this assumption, the wave function will be noted
\mbox{$\psi({\bf k},{\bf K})$}.

Wave packets centered in ${\bf x}_0$ at time $t_0$ are built with
the help of the space-time translation operator $\exp (i \hat{P}
\!\cdot\! x_0)$, where $x_0 \!=\! (t_0,{\bf x}_0)$. If a wave
packet in momentum space is given by
\begin{equation}
   \Psi({\bf k},{\bf K},{\bf x}_0,t_0)
   = \psi({\bf k},{\bf K}) \,
   e^{  i E({\bf k}) t_0 - i \, {\bf k} \cdot {\bf x}_0 } \, ,
   \label{paquetonde}
\end{equation}
the corresponding wave packet in configuration space will be
peaked at the point ${\bf x}_0$ at time $t_0$:
$$
   \widetilde{\Psi}({\bf x},t,{\bf K},{\bf x}_0,t_0)
   = \int \frac{d {\bf k}}{(2\pi)^3 } \; \psi({\bf k},{\bf K}) \,
      e^{ - i E({\bf k})(t-t_0)  +  i\, {\bf k} \cdot ( {\bf x} - {\bf
x}_0 ) } \, .
$$
Without loss of generality, let us choose to work with only one
particle in $P_I(q)$, in $P_F(k)$, in $D_I(q')$ and in $D_F(k')$.
The extension to a larger number of initial and final particles is
straightforward and would only complicate the notation. The wave
packets are built such that those involved in the production are
centered at ${\bf x}_{\scriptscriptstyle P}$ at time $t_{\scriptscriptstyle P}$, whereas those
involved in the detection are centered at ${\bf x}_{\scriptscriptstyle D}$ at
time $t_{\scriptscriptstyle D}$. They are noted \cite{Giunti93,Giunti98a}
\begin{eqnarray*}
   |\, P_I \!> &=& \int [d{\bf q}] \,
   \Psi_{ P_{\scriptscriptstyle I} } \left( {\bf q},{\bf Q},{\bf x}_{\scriptscriptstyle P},t_{\scriptscriptstyle P} \right)
   |\, P_I ({\bf q}) \!>
   \\
   |\, P_F \!> &=& \int [d{\bf k}] \,
   \Psi_{ P_{\scriptscriptstyle F} } \left( {\bf k},{\bf K},{\bf x}_{\scriptscriptstyle P},t_{\scriptscriptstyle P} \right)
   |\, P_F ({\bf k}) \!>
   \\
   |\, D_I \!> &=& \int [d{\bf q'}] \,
   \Psi_{ D_{\scriptscriptstyle I} } \left( {\bf q'},{\bf Q'},{\bf x}_{\scriptscriptstyle D},t_{\scriptscriptstyle D} \right)
   |\, D_I ({\bf q'}) \!>
   \\
   |\, D_F \!> &=& \int [d{\bf k'}] \,
   \Psi_{ D_{\scriptscriptstyle F} } \left( {\bf k'},{\bf K'},{\bf x}_{\scriptscriptstyle D},t_{\scriptscriptstyle D} \right)
   |\, D_F ({\bf k'}) \!> \, .
\end{eqnarray*}
Gaussian wave packets are discussed in section
\ref{sectiongaussianwp}.

\subsubsection{The propagation amplitude}
\label{propagationampli}

The general formula of the connected amplitude corresponding to
Fig.~3 is given by
$$
   {\cal A} =
   <\! P_F,D_F \,| \,
   \hat T \left( \exp \left( - i \int d^4x \, {\cal H}_I \right)
\right) - {\bf 1}
   |\, P_I,D_I \!> \, ,
$$
where ${\cal H}_I$ is the interaction Lagrangian for the
intermediate particle and $\hat T$ is the time ordering operator.
Let $g$ be the coupling constant in ${\cal H}_I$. Expanding the
amplitude to order $g^2$, and inserting the wave packets
expressions, we obtain
\begin{equation}
   {\cal A} =
   \int [d{\bf q}] \, \Psi_{ P_{\scriptscriptstyle I} }
   \int [d{\bf k}] \, \Psi_{ P_{\scriptscriptstyle F} }^*
   \int [d{\bf q'}] \, \Psi_{ D_{\scriptscriptstyle I} }
   \int [d{\bf k'}] \, \Psi_{ D_{\scriptscriptstyle F} }^* \;
   {\cal A}_{planewave}(q,k,q',k')
   \label{generalampli}
\end{equation}
with
\begin{equation}
   {\cal A}_{planewave}(q,k,q',k') =
   \int d^4x \, M_P(q,k) \, e^{ -i (q-k) \cdot x }
   \int d^4x' \, M_D(q',k') \, e^{ -i (q'-k') \cdot x' }
   \; G(x'-x)
\end{equation}
where $M_P(q,k)$ and $M_D(q',k')$ are the interaction amplitudes
at production and detection. The propagator $G(x'-x)$ has been
obtained by field contraction and reads
\begin{equation}
   G(x'-x) =
   \int \frac{d^4p}{(2\pi)^4} \, e^{ -ip \cdot (x'-x) } \, G(p^2) \, .
   \label{proppartic}
\end{equation}
where $G(p^2)= i (p^2 - M_0^2 +i\epsilon)^{-1}$ is the free
propagator in momentum space and $M_0$ is the bare mass of the
propagating particle. It is assumed that renormalization has been
carried out so that $M_0$ can be replaced by the physical mass
$m$. The external particles are on shell:
$$
   q^0 = E_{ P_{\scriptscriptstyle I}}({\bf q}) = \sqrt{ {\bf q}^2 + m^2_{P_{\scriptscriptstyle I} } } \, ,
$$
and so on.

We have supposed that the particle ($p^0 \!>\! 0$) propagates from
$x$ to $x'$ and the antiparticle ($p^0 \!<\! 0$) from $x'$ to $x$.
If interactions at the source and at the detector are such that
the particle propagates from $x'$ to $x$ and the antiparticle from
$x$ to $x'$, the contraction of the fields would have yielded the
propagator
$$
   \overline{G}(x'-x) =
   \int \frac{d^4p}{(2\pi)^4} \, e^{ ip \cdot (x'-x) } \, G(p^2) \, .
   \label{propantipartic}
$$
In that case, the different phase sign would select the
antiparticle pole in the subsequent contour integral.

If the propagating particle is unstable, the {\it complete}
propagator should replace the free propagator in the amplitude.
For this purpose, the amplitude is expanded to all orders and one
sums over all diagrams corresponding to one-particle-irreducible
insertions (1PI self-energy) in the propagator
\cite{Brown,Peskin}. This sum is a geometrical series in the 1PI
self-energy $-i\Pi(p^2)$ and yields the same formula as above,
except that $G(p^2)$ represents now the complete propagator:
\begin{equation}
   G(p^2) = \frac{i}{p^2 - M^2_0 - \Pi(p^2) + i\epsilon} \, .
   \label{fullperturbpropag}
\end{equation}
It is convenient to proceed to a change of variables in
Eq.~(\ref{generalampli}):
$$
   x \to x + x_{\scriptscriptstyle P} \qquad \mbox{and} \qquad x' \to x' + x_{\scriptscriptstyle D} \, ,
$$
where $x_{\scriptscriptstyle P} = (t_{\scriptscriptstyle P},{\bf x}_{\scriptscriptstyle P})$ and $x_D = (t_{\scriptscriptstyle
D},{\bf x}_{\scriptscriptstyle D})$. The amplitude (\ref{generalampli}) becomes
\begin{equation}
   {\cal A} = \int \frac{d^4p}{(2\pi)^4} \, \psi(p^0,{\bf p}) \, G(p^2) \,
    e^{ -ip \cdot (x_D-x_P) } \, ,
  \label{ampli}
\end{equation}
where the {\it overlap function} $\psi(p^0,{\bf p})$ represents
the overlap of the incoming and outgoing wave packets, both at the
source and at the detector. It is defined by
\begin{eqnarray}
   \psi(p^0,{\bf p}) =
   \int d^4x \, e^{ipx} \int d^4x' \,e^{-ipx'}
   \int [d{\bf q}]  \, \psi_{\scriptscriptstyle P_{in}}   ({\bf q},{\bf Q}) \, e^{-iqx}
   \int [d{\bf k}]  \, \psi_{\scriptscriptstyle P_{out}}^* ({\bf k},{\bf K}) \, e^{ikx}
   \nonumber \\  \times \
   \int [d{\bf q'}] \, \psi_{\scriptscriptstyle D_{in}}   ({\bf q'},{\bf Q'}) \, e^{-iq'x'}
   \int [d{\bf k'}] \, \psi_{\scriptscriptstyle D_{out}}^* ({\bf k'},{\bf K'}) \, e^{ik'x'} \,
   M_P(q,k) \, M_D(q',k') \, .
   \label{recouvrement}
\end{eqnarray}
All external particles are on-shell. Note that the overlap
function is independent of $x_{\scriptscriptstyle P}$ and $x_{\scriptscriptstyle D}$. The
integrals over $x$ and $x'$ in Eq.~(\ref{recouvrement}) yield
delta functions, which impose energy-momentum conservation at the
source and the detector. Remark also that the overlap function
depends on the directions of incoming and outgoing momenta.

Most computations in field theory apply to microscopic processes,
where in- and outgoing states can be approximated by plane waves.
In that case, the wave functions  $\psi_{ P_{\scriptscriptstyle I} }$, $\psi_{
P_{\scriptscriptstyle F} }$, $\psi_{ D_{\scriptscriptstyle I} }$ and $\psi_{ D_{\scriptscriptstyle F} }$ become
delta functions, so that the overlap function $\psi(p^0,{\bf p})$
is proportional to
$$
   \psi(p^0,{\bf p}) \sim
   \delta^{(4)}(K+K'-Q-Q') \, \delta^{(4)}(p-Q+K) \,
   M_P(Q,K) \, M_D(Q',K') \, ,
$$
where $Q =\left( \sqrt{ {\bf Q}^2 + m_{P_{\scriptscriptstyle I}}^2 },{\bf Q}
\right)$ and so on. The momentum integral in Eq.~(\ref{ampli}) is
then easy to perform and yields
$$
   {\cal A} \sim \delta^{(4)}(K+K'-Q-Q') \, G\left( (Q-K)^2 \right) \,
   e^{-i (Q-K) \cdot ( x_D - x_P ) } \, .
$$
This expression cannot lead to oscillations in a superposition of
several amplitudes, since the phase of the exponential does not
depend on the mass of the intermediate state. Thus, oscillations
vanish in the plane wave limit.

In the case of an amplitude corresponding to a macroscopic
process, such as the propagation on a macroscopic distance, the
overlap function is not a delta function so that the momentum
integral in Eq.~(\ref{ampli}) is not trivial (see section
\ref{simplest} for a simple exact model). The main contribution
comes from the pole associated to the propagating particle, so
that the phase of the exponential will depend on the mass.

\subsection{Mixed propagators}
\label{mixedpropagators}

In section \ref{mixing}, mixing was defined in quantum mechanics
as the result of the noncoincidence of the flavor basis and the
mass basis. The definition of mixing is similar in field theory,
except that it applies to fields, not to physical states. This
difference allows to bypass the problems arising in the definition
of flavor and mass bases.

The complete Lagrangian is split in a propagation Lagrangian
$L_{propag}$ and an interaction Lagrangian $L_{int}$. These two
subsets of the Lagrangian can be distinguished if there is a {\it
flavor} transformation which is a symmetry of $L_{int}$ but not of
$L_{propag}$. {\it Particle mixing} occurs if the propagator built
from $L_{propag}$, and representing the creation of a particle of
flavor $\alpha$ at point $x$ and the annihilation of a particle of
flavor $\beta$ at point $x'$, is not diagonal, i.e.\ not zero for
$\alpha\neq\beta$. The Lagrangian $L_{propag}$ always includes the
kinetic and the mass terms. If the particle is unstable,
$L_{propag}$ also contains the interaction causing the decay.

In the simplest case, illustrated by the light neutrinos $\nu_e$,
$\nu_\mu$ and $\nu_\tau$, the propagating particle is stable and
the form of its fundamental Lagrangian is known (although the mass
matrix might have its origin in an effective Lagrangian).
$L_{propag}$ contains the mass matrix, generated by Yukawa
interactions, whereas $L_{int}$ includes the weak interactions. In
the flavor basis (called the gauge eigenstates basis for
neutrinos), the mixed propagator is nondiagonal. It is well-known
\cite{Mohapatra,Bilenky99} that $L_{propag}$ can be diagonalized
by a unitary transformation on the fields\footnote{The matrix $V$
corresponds to the matrix $U^t$ of section \ref{mixing}.}:
\begin{equation}
   \nu_\alpha = \sum_j \, V^\dagger_{\alpha j} \, \nu_j \, ,
   \label{relchamp}
\end{equation}
where Greek and Latin indices refer to the flavor and to the mass
basis, respectively. It could however turn out that only a subset
of $V$ is used to describe the mixing in an oscillation
experiment. One reason could be that some of the mixed states are
too heavy to be produced (see Eq.~(\ref{nonunitary})). In this
case, the mixing matrix for the light states can be nonunitary
\cite{Bilenky93,Pakvasa00,Czakon01}. Another reason could be that
some of the mixed states are sterile, i.e.\ they only interact
through mass mixing. In that case, the mixing matrix for the
non-sterile states is rectangular and does not satisfy unitarity
\cite{Schechter80}.

The propagator is defined by the time-ordered two-point function:
$$
   G_{\beta\alpha}(x'-x) =
   <\! 0 \,|T \left( \nu_\beta(x') \: \nu^*_\alpha(x) \right) |\, 0 \!> \; .
$$
Since Wick's theorem applies to fields in the mass basis, we must
substitute Eq.~(\ref{relchamp}) in this equation. In the scalar
case, field contraction yields
\begin{equation}
   G_{\beta\alpha}(x'-x) =
   \sum_j \, V^\dagger_{\beta j} \,
   G_{D,jj}(x'-x)
    \, V_{j\alpha} \, ,
   \label{stablenondiag}
\end{equation}
where $G_{D,jj}(x'-x)$ is the free propagator (with $D$ for
`diagonal') of a scalar particle of mass $m_j$:
$$
   G_{D,jj}(x'-x) =
   \int \, \frac{d^4p}{(2\pi)^4} \, e^{-i p (x'-x)} \,
   \frac{i}{p^2 -m_j^2 +i \epsilon} \, .
$$
The derivation of the mixed propagator for a spin one-half
particle is similar.

The treatment of unstable particles is more involved. The complete
Lagrangian is again split in $L_{int}$ and $L_{propag}$, but the
interactions causing the decay should be included in $L_{propag}$.
The free propagator is then replaced by the complete propagator,
obtained by an infinite sum on the self-energy $-i\Pi(p^2)$
\cite{Baulieu82}:
\begin{equation}
   i \, G^{-1}(p^2) = p^2 \, {\bf 1} - M^2_0 - \Pi(p^2) \, ,
   \label{fullpropag}
\end{equation}
in an obvious matrix notation. The propagation eigenstates are
characterized by the complex poles of the complete propagator
$z_j=m_j^2-im_j\Gamma_j$ or, equivalently, by the zeros of the
inverse propagator. It is always possible to diagonalize the
propagator if its eigenvalues are distinct:
\begin{equation}
   G(p^2) = V^{-1}(p^2) \, G_D(p^2) \, V(p^2) \, ,
   \label{propdiag}
\end{equation}
where $G_D(p^2)$ is the diagonal matrix the elements of which are
given by
\begin{equation}
   G_{D,jj}(p^2) = \frac{i}{p^2-m_j^2+im_j\Gamma_j} \, .
   \label{unstablediag}
\end{equation}
The matrix $V$ is not necessarily unitary, since the self-energy
matrix for unstable particles is usually not normal (see section
\ref{mixing}). Moreover, it depends on the energy. If the mass
eigenstates are nearly degenerate, the self-energy can be
approximated by its value at $p^2=m^2$ and the matrix $V$ becomes
constant. In section \ref{nonexponential}, we show that the energy
dependence of the self-energy generates corrections to the
amplitude in inverse powers of the propagation distance $L$. These
corrections are due to production thresholds of multi-particles
states \cite{Schwinger60,Jacob61,Brown}.

\subsection{The flavor-mixing amplitude}
\label{flavormixing}

In order to derive the flavor-mixing amplitude, the Jacob-Sachs
model of section \ref{jacobsachsmodel} has to be modified to take
into account the different flavors. The process corresponding to
the production at the source of a particle of flavor $\alpha$, and
the detection of a particle of flavor $\beta$ at a detector placed
at a macroscopic distance from the source, can still be symbolized
by the Fig.~3 of section \ref{theprocess}, except that the
intermediate particle should be considered as a superposition of
different mass eigenstates. The initial flavor $\alpha$ is tagged,
for example, by the outgoing state $P_F(k)$, whereas the final
flavor $\beta$ can be tagged by the outgoing state $D_F(k')$. If
it is impossible to identify the flavor at the detector (ex:
$K^0,\bar K^0 \to \pi^+\pi^-$), one should sum over the different
flavors.

If the energy dependence of the matrix $V$ diagonalizing the
propagator can be neglected (see Eq.~(\ref{propdiag})), the
amplitude corresponding to the global process  can be expressed as
a linear combination of amplitudes ${\cal A}_j$ corresponding to
the propagation of different mass eigenstates:
\begin{equation}
   {\cal A}(\alpha \!\to\! \beta,T,{\bf L})
   = \sum_j V_{\beta j}^{-1} \, {\cal A}_j \, V_{j\alpha} \, ,
   \label{amplitot}
\end{equation}
where the average propagation time $T$ is defined by
\mbox{$T=x_{\scriptscriptstyle D}^0-x_{\scriptscriptstyle P}^0$} and the average propagation
distance by \mbox{${\bf L}={\bf x}_{\scriptscriptstyle D}-{\bf x}_{\scriptscriptstyle P}$}. The
partial amplitude ${\cal A}_j$ has the same form as the
propagation amplitude of an isolated particle, given in
Eq.~(\ref{ampli}):
\begin{equation}
   {\cal A}_j = \int \frac{d^4p}{(2\pi)^4} \;
   \psi(p^0,{\bf p}) \; G_{D,jj}(p^2) \;
   e^{-i p^0 T + i \, {\bf p} \cdot {\bf L} } \, .
   \label{defAj}
\end{equation}
The overlap function $\psi(p^0,{\bf p})$ is defined by
Eq.~(\ref{recouvrement}) and the propagator for the $j$th mass
eigenstate by Eq.~(\ref{unstablediag}).

As noted in section \ref{theprocess}, it will be enough for our
purpose to work with a scalar oscillating particle. When the spin
is taken into account, three cases must be distinguished:
\begin{enumerate}

\item If the particle is relativistic and the interactions at the
source and detector are chiral, only one helicity eigenstate
contributes to the propagation. The spin structure can then be
factorized and the computation can proceed with a scalar
propagator. The two mentioned requirements are satisfied for light
neutrino oscillations, since the energy threshold of the detectors
is much higher than the light neutrino mass scale, and because of
the chirality of the Standard Model flavor-changing interactions.

\item If the particle is nonrelativistic, the spin structure is
approximately equal for the different helicities provided that the
mass eigenstates are nearly degenerate: $m_i\cong m_j\equiv m$.
The helicity factors can then be factorized from the sum over the
mass eigenstates:
$$
  {\cal P}^{spin}_{\alpha \to \beta}(T,{\bf L})
  \sim \left( \sum_s |H(s,m)|^2 \right) \,
  |{\cal A}^{scalar}_{\alpha \to \beta}(T,{\bf L})|^2
$$
where $H(s,m)$ includes all helicity dependent factors (with the
index $s$ referring to the spins of the external particles).
However the source and detector contributions to
$\sum_s|H(s,m)|^2$ cannot be disentangled if the particle is not
relativistic, so that the whole process cannot be factorized into
a product of source/propagation/detection probabilities (this
point was emphasized Refs.~\cite{Giunti93,Cardall99}). As the mass
eigenstates are nearly degenerate, it seems natural to define the
oscillation probability as the modification introduced by the mass
difference:
$$
  {\cal P}_{\alpha \to \beta}(T,{\bf L}) =
  \frac{{\cal P}^{spin}_{\alpha \to \beta}(T,{\bf L})}
  {{\cal P}^{spin}_{\alpha \to \beta}(T,{\bf L})|_{m_i=m_j}}
  \, ,
$$
so that the calculation of the oscillation can be done as if the
oscillating particle were scalar. The discussion is very similar
for a relativistic particle with nonchiral interactions.

\item If the particle is nonrelativistic and the mass eigenstates
have very different masses, each specific process has to be
computed separately, with the influences of the amplitudes of
production and detection carefully taken into account for each
mass eigenstate. This situation is not really relevant to
oscillations, but rather to incoherent mixing, since oscillations
are averaged to zero in such experimental conditions. In other
words, the contributions of the Feynman diagrams corresponding to
the different mass eigenstates are summed in the probability, not
in the amplitude.
\end{enumerate}
In this report, we always assume the condition $\delta m_{ij}^2\ll
E^2$, under which oscillations are observable, so that the third
case is not considered. Some explicit examples of the third case
have been given by Giunti, Kim and Lee \cite{Giunti92a}, and by
Kiers and Weiss \cite{Kiers98}. The conditions of factorization of
the transition probability have been discussed by Cardall and
Chung \cite{Cardall99}.

\subsection{The simplest consistent model}
\label{simplest}

A simple model, in which the flavor-changing amplitude
(Eqs.~(\ref{amplitot})-(\ref{defAj})) can be computed exactly, has
been proposed by Kobzarev, Martemyanov, Okun and Shchepkin
\cite{Kobzarev82}. In this model, a charged lepton, represented by
a plane wave, collides with an infinitely heavy nucleus, situated
in ${\bf x}_{\scriptscriptstyle P}$. The neutrino produced at that point has a
definite energy equal to the energy of the incident lepton. At
point ${\bf x}_{\scriptscriptstyle D}$, the neutrino collides with a second
infinitely heavy nucleus and, as a result of this collision, a
charged lepton is emitted  with an energy equal to the neutrino
energy.

Let us see what these assumptions mean in the wave packet notation
of section \ref{sectionwavepackets}. There are two ingoing states
at the source, the lepton and the nucleus. The lepton is
represented by a plane wave, so that
$$
   \psi_{ P_{\scriptscriptstyle I,lept} }({\bf q}_l,{\bf Q}_l) \sim
   \delta^{(3)}({\bf q}_l - {\bf Q}_l) \, .
$$
The uncertainty on the momentum of the nucleus is infinite, so
that the ingoing and outgoing momentum wave functions are
constant:
$$
   \psi_{ P_{\scriptscriptstyle I,nucl} }({\bf q}_n,{\bf Q}_n) \sim const
   \hspace{1cm} \mbox{and} \hspace{1cm}
   \psi_{ P_{\scriptscriptstyle F,nucl} }({\bf k}_n,{\bf K}_n) \sim const \, .
$$
The situation is similar at the detector, except that there are
one ingoing and two outgoing states.

The overlap function (\ref{recouvrement}) is easily computed:
\begin{equation}
   \psi(p^0,{\bf p}) \sim \delta (p^0-E_{in}) \, \delta (p^0-E_{out}) \, ,
   \label{overlapsimplest}
\end{equation}
where $E_{in}$ is the energy of the incoming lepton at the source
and $E_{out}$ is the energy of the outgoing lepton at the
detector. The partial amplitude (\ref{defAj}) becomes
$$
   {\cal A}_j \sim \delta (E_{in}-E_{out})
   \int d^3p \; G_{D,jj}(E_{in},{\bf p}) \; e^{i {\bf p} \cdot {\bf L} } \, ,
$$
where ${\bf L}={\bf x}_{\scriptscriptstyle D} - {\bf x}_{\scriptscriptstyle P}$. The Fourier
transform of the propagator with respect to its momentum can be
computed by contour integration and yields
\begin{equation}
   {\cal A}_j \sim \frac{1}{L} \, \delta (E_{in}-E_{out}) \, e^{i p_j L} \, ,
   \label{amplisimplest}
\end{equation}
where $p_j = \sqrt{ E_{in}^2 - m_j^2 }$ and $L=|{\bf L}|$.

The transition probability between an initial state of flavor
$\alpha$ and a final state of flavor $\beta$ is given by the
squared modulus of the amplitude (\ref{amplitot}) with ${\cal
A}_j$ given by Eq.~(\ref{amplisimplest}). Oscillations between
mass eigenstates $m_i$ and $m_j$ arise from interference terms
${\cal A}_i{\cal A}_j^*\sim e^{ i (p_i -p_j) \, L }$. Thus, the
$ij$-interference term oscillates with a frequency equal to
\begin{equation}
   (p_i -p_j) \, L \cong \frac{\delta m_{ij}^2}{2 p_m} \, L \, ,
   \label{phasesimplest}
\end{equation}
where $\delta m_{ij}^2 \equiv m_i^2-m_j^2$ and $p_m =
\sqrt{E_{in}^2-m^2}$, with $m$ the mass in the degenerate limit.
The phase (\ref{phasesimplest}) is equal to the standard
oscillation phase (\ref{standardphase}) calculated in the
quantum-mechanical plane wave treatment.

This very simple model has the advantage of being consistent,
unlike the plane wave approach in quantum mechanics. It could be
recommended as a pedagogical tool. It is however not sufficient
for a thorough study of the oscillation phenomenon. First, it
cannot describe unstable particles, since they cannot decay into
an infinitely heavy state. Moreover, the approximation of plane
waves and stationary states is too strong to allow the study of
the observability conditions of oscillations. Another drawback of
this model is that the amplitude is independent of the direction
of ${\bf L}$. Last, a spatial localization of the source or the
detector more precise than the Compton wave length of the lightest
external particle is unphysical \cite{Itzykson}.

\subsection{The overlap function for Gaussian wave packets}
\label{overlapgaussian}

In most cases, it is not possible to compute exactly the
flavor-changing amplitude (\ref{amplitot}). The technical
difficulty lies in the energy-momentum integration in
Eq.~(\ref{defAj}). The two exceptions are, on the one hand, the
Kobzarev {\it et al.} model of section \ref{simplest} and, and the
other hand, the limit in which all the external wave packets are
plane waves (in which case oscillations vanish as shown in section
\ref{propagationampli}). To proceed further, we have to be more
specific about the shape of the overlap function
(\ref{recouvrement}). In order to answer the questions on the
coherence length, it is useful to work with an overlap function
which depends explicitly on the energy and 3-momentum
uncertainties at the source and at the detector. This can be done
by approximating the in- and outgoing particles with Gaussian wave
packets, as first proposed by Giunti, Kim, Lee and Lee
\cite{Giunti93,Giunti98a}. The general case of arbitrary wave
packets can then be analyzed as a superposition of Gaussian wave
packets. Note that the Gaussian approximation is used by most
authors since it allows analytical integrations.

\subsubsection{Gaussian wave packets}
\label{sectiongaussianwp}

General wave packets were defined in section
\ref{sectionwavepackets}. The wave function of Gaussian wave
packet can be written at time $t=0$ in momentum space as
\begin{equation}
   \psi_\chi({\bf p},{\bf p}_\chi) =
   \left( \frac{2\pi}{ \sigma_{p\chi}^2 } \right)^{3/4}
   \exp \left( - \frac{({\bf p}-{\bf p}_\chi)^2}{ 4\sigma_{p\chi}^2 }
        \right) \, .
   \label{gaussianwp}
\end{equation}
We chose to set the initial phase to zero. This wave packet is
thus centered in ${\bf x}={\bf 0}$ at time $t=0$ (see
Eq.~(\ref{paquetonde})). Recall that the normalization is given by
Eq.~(\ref{norm}). The width $\sigma_{p\chi}$ is the momentum
uncertainty of the wave packet, as can be checked by computing the
variance of the operator $\hat p^a$:
$$
  <\! (\delta p^a)^2 \!>_\chi
  = <\! \chi \,|(\hat p^a- \bar p^a_\chi)^2 |\,\chi \!>
  = \sigma_{p\chi}^2 \, ,
$$
with the average momentum given by
$$
  \bar p^a_\chi = <\! \chi \,|\hat p^a|\,\chi \!> = p^a_\chi \, .
$$
It is also useful to define a width $\sigma_{x\chi}$ in
configuration space by $\sigma_{p\chi}\sigma_{x\chi} = 1/2$.

The wave function is given in configuration space by
Eq.~(\ref{wpconfig}). This integral is Gaussian if the energy is
expanded to second order around the average momentum ${\bf
p}_\chi$:
\begin{equation}
  \sqrt{{\bf p}^2 + m_\chi^2}
  \cong E_\chi + {\bf v}_\chi \cdot ({\bf p}-{\bf p}_\chi)
    + \frac{1}{2E_\chi} \, \left( ({\bf p}-{\bf p}_\chi)^2
       - \left( {\bf v}_\chi \cdot ({\bf p}-{\bf p}_\chi) \right)^2 \right) \, ,
\end{equation}
where $E_\chi=\sqrt{{\bf p}_\chi^2 + m_\chi^2}$ and ${\bf
v}_\chi={\bf p}_\chi/E_\chi$. The wave packet is thus given in
configuration space by
\begin{equation}
  \tilde \psi_\chi({\bf x},t)
  = \frac{(2\pi\sigma_{x\chi}^2)^{-3/4}}{\sqrt{\det \Sigma}}
   \exp \left(
        - E_\chi t + i {\bf p}_\chi \cdot {\bf x}
        - \frac{({\bf x}-{\bf v}_\chi t) \Sigma^{-1} ({\bf x}-{\bf v}_\chi t)}
               { 4\sigma_{x\chi}^2 }
        \right) \, ,
   \label{wpgaussianconfig}
\end{equation}
where the matrix notation is implicit and with the matrix $\Sigma$
defined by
\begin{equation}
  \Sigma^{ij}=
  \delta^{ij}
  +(\delta^{ij}-v_\chi^i v_\chi^j)\frac{2it\sigma_p^2}{E_\chi} \, .
  \label{defSigma}
\end{equation}
The spatial uncertainty of the wave packet can be computed in
configuration space with Eq.~(\ref{wpgaussianconfig}):
\begin{equation}
  <\! (\delta x^a)^2 \!>_\chi
  = <\! \chi \,|(\hat x^a- \bar x^a_\chi)^2 |\,\chi \!>
  = \sigma_{x\chi}^2 \, \left[ (Re \Sigma^{-1})^{-1} \right]^{aa} \, ,
  \label{gaussianwpdispersion}
\end{equation}
where the average position is given by
$$
  \bar x^a_\chi = <\! \chi \,|\hat x^a|\,\chi \!> = v^a_\chi t \, .
$$
If the $z$ axis is chosen along ${\bf v}_\chi$, we obtain
\begin{eqnarray}
  \left[ (Re \Sigma^{-1})^{-1} \right]^{xx}
  &=& \left[ (Re \Sigma^{-1})^{-1} \right]^{yy}
  = 1 + \frac{4\sigma_{p\chi}^4 \, t^2}{E_\chi^2} \, ,
  \label{thresholdtrans}
  \\
  \left[ (Re \Sigma^{-1})^{-1} \right]^{zz}
  &=& 1 + \frac{4m_\chi^4\sigma_{p\chi}^4 \, t^2}{E_\chi^6} \, .
  \label{thresholdlong}
\end{eqnarray}
These equations show that the wave packet begins to spread in
directions transverse to ${\bf v}_\chi$ at $t\cong
E_\chi/2\sigma_{p\chi}^2$, whereas the spreading along ${\bf
v}_\chi$ begins at the later time $t\cong
E_\chi^3/2m_\chi^2\sigma_{p\chi}^2$. In this report, we refer to
the time-dependence of Eq.~(\ref{gaussianwpdispersion}) as the
{\it dispersion} of the wave packet. Although dispersion will be
neglected in the external wave packets, we shall see that the
propagation amplitude associated to the oscillating particle
spreads in the same way.

In order to study the stationary limit ${\bf v}_\chi\to0$, it is
of great interest to compute the energy uncertainty of the wave
packet (as proposed in Ref.~\cite{Giunti02b}). The average energy
$\bar E_\chi$ and the average squared energy should both be
computed to order $\sigma_p^2$. In momentum space, the average
energy is given by
\begin{equation}
  \bar E_\chi = <\chi|\hat E|\chi> = E_\chi
  + (3- {\bf v}_\chi^2)\frac{\sigma_p^2}{2E_\chi}\, .
  \label{averageEmom}
\end{equation}
The second term comes from the second order term in the expansion
of the energy operator around the average momentum. Remarkably the
average energy $\bar E_\chi$ is different from $E_\chi=\sqrt{{\bf
p}_\chi^2 + m_\chi^2}$. The average squared energy can be computed
exactly:
\begin{equation}
  <\chi|\hat E^2|\chi>
  = E_\chi^2 + 3\sigma_p^2 \, ,
  \label{averageE2mom}
\end{equation}
so that the squared energy uncertainty is given to order $\sigma_p^2$ by
\begin{equation}
  <(\delta E)^2>_\chi
  = <\! \chi \,|(\hat E- \bar E_\chi)^2 |\,\chi \!>
  = {\bf v}_\chi^2 \, \sigma_p^2 \, .
  \label{energyuncertainty}
\end{equation}

In configuration space, the same results are obtained only if the
dispersion (i.e.\ time spread) of the wave packet is taken into
account. We compute to order $\sigma_p^2$ (additional terms appear
at higher order) the average energy and squared energy:
\begin{eqnarray}
  <\chi|\hat E|\chi>   &=& E_\chi -\frac{i}{4}
  Tr\left(\frac{\partial\Sigma}{\partial t}\right) \, ,
  \label{averageEconf}\\
  <\chi|\hat E^2|\chi> &=& E_\chi^2 + {\bf v}_\chi^2 \sigma_p^2 -
  \frac{iE_\chi}{2} \,
  Tr\left(\frac{\partial\Sigma}{\partial t}\right) \, .
  \label{averageE2conf}
\end{eqnarray}
With the help of Eq.~(\ref{defSigma}), we can check that
Eqs.~(\ref{averageEconf}) and (\ref{averageE2conf}) are equal to
Eqs.~(\ref{averageEmom}) and (\ref{averageE2mom}), respectively
(but only if dispersion is not neglected). Thus the squared energy
uncertainty is the same in both representations as expected
(similar computations for the momentum uncertainty show that its
value is the same in momentum and configuration space even if
dispersion is neglected).

Note that the energy uncertainty coincides at first order with the
naive expectation:
\begin{equation}
  \delta E
  \cong \sqrt{({\bf p}_\chi+\delta {\bf p})^2+m^2} -\sqrt{{\bf p}_\chi^2+m^2}
  \cong {\bf v}_\chi \cdot \delta {\bf p} \, .
\end{equation}
It could be objected that the squared energy uncertainty does not
vanish to order $\sigma_p^4/E_\chi^2$ in the limit ${\bf
v}_\chi\to0$. However external wave packet models are always built
with non-spreading wave packets, i.e. one sets
$\Sigma^{ij}=\delta^{ij}$. If this approximation is taken
seriously, we can forget about the original expression in momentum
space. In that case, the average energy and squared energy
computed in configuration space are shifted to $E_\chi$ and
$E_\chi^2+ {\bf v}_\chi^2\sigma_p^2$ respectively with no
correction of higher order in $\sigma_p^2$ (see
Eqs.(\ref{averageEconf})-(\ref{averageE2conf})), so that the
energy uncertainty is still given by
Eq.~(\ref{energyuncertainty}). Thus the energy uncertainty of
non-spreading wave packets vanishes in the stationary limit ${\bf
v}_\chi\to0$ to all orders in $\sigma_p^2$.

This result is of great interest to study the stationary limit of
the external wave packet model. For example, the bound state wave
function used by Grimus and Stockinger \cite{Grimus96} coincides
with the wave function of a non-spreading wave packet at rest,
i.e.\ Eq.~(\ref{wpgaussianconfig}) with ${\bf v}_\chi=0$ and
$\Sigma={\bf 1}$. Besides, plane waves can be obtained by taking
$\sigma_{x\chi}\to\infty$ in Eq.~(\ref{wpgaussianconfig}).

\subsubsection{Gaussian overlap function}
\label{gaussianoverlap}

We now compute the overlap function (\ref{recouvrement}) with
non-spreading external wave packets as done in
Ref.~\cite{Giunti98a}. The factors $M_P(q,k)$ and $M_D(q',k')$
multiplying the exponential vary slowly over the width of the wave
packet and can be approximated by their value at the average
momentum. They can be factorized outside the sum over the mass
eigenstates since it is assumed that the neutrinos are either
relativistic or nearly degenerate in mass. The wave packets
$\psi_{\scriptscriptstyle P_{in,out}}$ and $\psi_{\scriptscriptstyle D_{in,out}}$ have the form
given in Eq.~(\ref{gaussianwp}). Expanding the energy to first
order around the average momenta, the momentum integrations in
Eq.~(\ref{recouvrement}) give results similar to
Eq.~(\ref{wpgaussianconfig}) but with $\Sigma={\bf 1}$ and
additional factors $\sqrt{2E}$ coming from the integration
measure. For example we obtain
\begin{equation}
   \int [d{\bf q}]  \, \psi_{\scriptscriptstyle P_{in}}   ({\bf q},{\bf Q}) \, e^{-iqx} =
   N_{\scriptscriptstyle P_{in}} \, \exp \left(
       -i E_{\scriptscriptstyle P_{in}}({\bf Q}) t + i{\bf Q} \cdot {\bf x}
       - \frac{ ({\bf x} - {\bf v}_{\scriptscriptstyle P_{in}} t)^2}
              {4\sigma_{x\scriptscriptstyle P_{in}}^2}
                           \right) \, ,
   \label{packetin}
\end{equation}
where $N_{\scriptscriptstyle P_{in}} =(2\pi\sigma_{x\scriptscriptstyle P_{in}}^2)^{-3/4}(2E_{\scriptscriptstyle
P_{in}}({\bf Q}))^{-1/2}$ is a normalization constant.

Doing the same for the other wave packets, we can write the
overlap function as
\begin{equation}
   \psi(p^0,{\bf p}) = N \, \psi_{\scriptscriptstyle P}(p^0,{\bf p}) \,
   \psi_{\scriptscriptstyle D}^*(p^0,{\bf p}) \, ,
   \label{overlapPD}
\end{equation}
with
$$
\psi_{\scriptscriptstyle P}(p^0,{\bf p}) = \int d^4x \exp \left( i \left( p^0 -
E_{\scriptscriptstyle P} \right) t - i \left( {\bf p} - {\bf p}_{\scriptscriptstyle P} \right)
\cdot {\bf x} - \frac{ {\bf x}^2 - 2{\bf v}_{\scriptscriptstyle P} \cdot {\bf x}
\, t + \Sigma_{\scriptscriptstyle P} t^2 }
        {4\sigma_{x\scriptscriptstyle P}^2} \right) \, ,
$$
where $E_{\scriptscriptstyle P}=E_{\scriptscriptstyle P_{in}} - E_{\scriptscriptstyle P_{out}}$, ${\bf p}_{\scriptscriptstyle
P}={\bf Q} - {\bf K}$. The function $\psi_{\scriptscriptstyle D}(p^0,{\bf p})$ is
defined in the same way, with the index $P$ replaced by $D$,
except for the energy-momentum which is defined so as to be
positive: $E_{\scriptscriptstyle D}=E_{\scriptscriptstyle D_{out}} - E_{\scriptscriptstyle D_{in}}$ and ${\bf
p}_{\scriptscriptstyle D}={\bf K}' - {\bf Q}'$. The constant $N$ includes the
normalization constants as well as the factors $M_{P,D}$ evaluated
at the maxima of the wave packets.

A new width $\sigma_{p\scriptscriptstyle P}$ has been defined by $\sigma_{p\scriptscriptstyle
P}\sigma_{x\scriptscriptstyle P} = 1/2$, with
$$
   \frac{1}{\sigma_{x\scriptscriptstyle P}^2} = \frac{1}{\sigma_{x\scriptscriptstyle P_{in}}^2}
   + \frac{1}{\sigma_{x\scriptscriptstyle P_{out}}^2} \, .
$$
$\sigma_{p\scriptscriptstyle P}$ can be interpreted as the {\it momentum
uncertainty at the source}. The width $\sigma_{x\scriptscriptstyle P}$ is mainly
determined by the external particle with the smallest space width.
This is expected since the production region depends on the
overlap in space-time of the external wave packets.

The symbol ${\bf v}_{\scriptscriptstyle P}$ is defined by
$$
   {\bf v}_{\scriptscriptstyle P} = \sigma_{x\scriptscriptstyle P}^2 \left(
   \frac{ {\bf v}_{\scriptscriptstyle P_{in}} }{\sigma_{x\scriptscriptstyle P_{in}}^2} +
   \frac{ {\bf v}_{\scriptscriptstyle P_{out}} }{\sigma_{x\scriptscriptstyle P_{out}}^2}
   \right) \, .
$$
It can be interpreted as the velocity of the production region,
approximately equal to the velocity of the particle with the
smallest spatial spread (unless the velocities of the different
in- and outgoing particles are very different).

The symbol $\Sigma_{\scriptscriptstyle P}$, satisfying $0\leq\Sigma_{\scriptscriptstyle P}\leq
1$, is defined by
$$
   \Sigma_{\scriptscriptstyle P} = \sigma_{x\scriptscriptstyle P}^2 \left(
   \frac{ {\bf v}_{\scriptscriptstyle P_{in}}^2 }{\sigma_{x\scriptscriptstyle P_{in}}^2} +
   \frac{ {\bf v}_{\scriptscriptstyle P_{out}}^2 }{\sigma_{x\scriptscriptstyle P_{out}}^2}
   \right) \, .
$$
Recall in all above definitions that there might be more than one
in- and outgoing state in the production and detection process.

Integrating over ${\bf x}$ and $t$, we obtain
\begin{equation}
   \psi_{\scriptscriptstyle P}(p^0,{\bf p}) =
   \pi^2 \sigma_{p \scriptscriptstyle P}^{-3} \, \sigma_{e\scriptscriptstyle P}^{-1}\,
   \exp \left(
              - f_{\scriptscriptstyle P}(p^0,{\bf p})
        \right) \, ,
   \label{overlapP}
\end{equation}
with
\begin{equation}
   f_{\scriptscriptstyle P}(p^0,{\bf p}) =
     \frac{ \left( {\bf p}-{\bf p}_{\scriptscriptstyle P} \right)^2}{4\sigma_{p \scriptscriptstyle P}^2}
   + \frac{ \left(
                 p^0 - E_{\scriptscriptstyle P} -({\bf p} - {\bf p}_{\scriptscriptstyle P}) \cdot {\bf v}_{\scriptscriptstyle P}
           \right)^2}
         {4\sigma_{e\scriptscriptstyle P}^2} \, ,
   \label{overlapfP}
\end{equation}
where
\begin{equation}
  \sigma_{e\scriptscriptstyle P}^2 =
  \sigma_{p \scriptscriptstyle P}^2 \,
  \left( \Sigma_{\scriptscriptstyle P} - {\bf v}_{\scriptscriptstyle P}^2 \right)
  \leq \sigma_{p \scriptscriptstyle P}^2 \, .
\end{equation}
The quantity $\sigma_{e\scriptscriptstyle P}$ can be interpreted as the {\it
energy uncertainty at the source}, or also as the inverse of the
time of overlap of wave packets during the production process.
Indeed, we can show that
$$
   \sigma_{e\scriptscriptstyle P}^2 = \sum_{\alpha<\beta}
   \frac{ \sigma_{x\scriptscriptstyle P}^2 }{ 4\sigma_{x\alpha}^2 \sigma_{x\beta}^2 } \,
   \left( {\bf v}_\alpha - {\bf v}_\beta \right)^2 \, ,
$$
where the sum is over all wave packets involved in the production
process. This sum is dominated by the term including the two
smallest wave packets in configuration space (unless their
velocities are nearly equal). If $\sigma_{x1}$ is the smallest
width and $\sigma_{x2}$ the second smallest, we obtain
\begin{equation}
   \sigma_{e\scriptscriptstyle P} \sim \frac{ |{\bf v}_1 - {\bf v}_2| }{ \sigma_{x2} }
   \sim \frac{1}{T^{overlap}_{\scriptscriptstyle P}} \, .
   \label{timeoverlap}
\end{equation}
where $T^{overlap}_{\scriptscriptstyle P}$ is defined as the duration of the
production process. Thus, $\sigma_{e\scriptscriptstyle P}$ can be interpreted as
the energy uncertainty at the source, since it is proportional to
the inverse of the time of overlap of the external wave packets at
the source. The quantities $\sigma_{x\scriptscriptstyle D}$, $\sigma_{p\scriptscriptstyle D}$,
${\bf v}_{\scriptscriptstyle D}$, $\Sigma_{\scriptscriptstyle D}$, $\sigma_{e\scriptscriptstyle D}$,
$T^{overlap}_{\scriptscriptstyle D}$ have similar definitions and properties.

Note that stationary boundary conditions are recovered by setting
${\bf v}_{\scriptscriptstyle P,D}=0$ and $\sigma_{e\scriptscriptstyle P,D}=0$, with
$\sigma_{p\scriptscriptstyle P,D}$ different from zero. Besides, we shall need to
know the stationary limit of $|{\bf v}_{\scriptscriptstyle P,D}|/\sigma_{e\scriptscriptstyle
P,D}$. Since ${\bf v}_{\scriptscriptstyle P}$ is the velocity of the production
region and $\sigma_{e\scriptscriptstyle P}^{-1}$ is the duration of the
production process, the ratio $|{\bf v}_{\scriptscriptstyle P}|/\sigma_{e\scriptscriptstyle P}$
is bounded by the size $S_{\scriptscriptstyle P}$ of the macroscopic region of
production. The argument is similar for the ratio $|{\bf v}_{\scriptscriptstyle
D}|/\sigma_{e\scriptscriptstyle D}$. Thus we assume the following constraints:
\begin{equation}
  \frac{|{\bf v}_{\scriptscriptstyle P,D}|}{\sigma_{e\scriptscriptstyle P,D}}
  \lesssim S_{\scriptscriptstyle P,D} \, .
  \label{statlim}
\end{equation}
These bounds are very conservative, since we shall see that
stationary models such as those found in
Refs.~\cite{Grimus96,Ioannisian99} are recovered by setting $|{\bf
v}_{\scriptscriptstyle P,D}|/\sigma_{e\scriptscriptstyle P,D}=0$. In the example of the
Grimus-Stockinger model \cite{Grimus96}, an initial stationary
neutron (${\bf v}_{\scriptscriptstyle P_{in},n}=0$) decays into a stationary
proton (${\bf v}_{\scriptscriptstyle P_{out},pr}=0$), a `plane-wave' electron
($\sigma_{x\scriptscriptstyle P_{out},el}=\infty$) and the intermediate
antineutrino. At detection, the antineutrino collides with a
stationary electron (${\bf v}_{\scriptscriptstyle D_{in}}=0$) and the outgoing
antineutrino and electron are represented as plane waves
($\sigma_{x\scriptscriptstyle D_{out},\nu}=\sigma_{x\scriptscriptstyle D_{out},el}=\infty$). As
argued at the end of section \ref{sectiongaussianwp}, all these
stationary states are limiting cases of wave packets.

As the propagation distance is macroscopic, only processes
satisfying global conservation of energy-momentum have a
nonnegligible probability of occurring. Since our aim is not to
prove this well-known fact, we impose that
\begin{equation}
   {\bf p}_{\scriptscriptstyle P} = {\bf p}_{\scriptscriptstyle D} \equiv {\bf p}_0
   \hspace{1cm} \mbox{and} \hspace{1cm}
   E_{\scriptscriptstyle P} = E_{\scriptscriptstyle D} \equiv E_0 \, ,
   \label{conserv}
\end{equation}
but we still have ${\bf v}_{\scriptscriptstyle P}\neq{\bf v}_{\scriptscriptstyle D}$. These
approximations allow to do expansions around ${\bf p}_0$ and
$E_0$. An associated velocity can be defined by ${\bf v}_0={\bf
p}_0/E_0$.

\section{Propagation amplitude and dispersion}
\label{section4}

\subsection{Introduction}

In section~\ref{section3}, we described the external wave packet
model and we applied to flavor-mixing transitions. All the
information relevant to the source and detector was included in
what we called the overlap function (Eq.~(\ref{recouvrement})). We
computed this overlap function, on the one hand, in the
pedagogical case of infinitely heavy source and detector
(Eq.~(\ref{overlapsimplest})) and, on the other hand, in the more
general case of Gaussian external wave packets
(Eqs.~(\ref{overlapPD})-(\ref{overlapfP})). It is now possible to
evaluate the transition amplitude.

Although the flavor-mixing amplitude, as defined in
Eqs.~(\ref{amplitot}) -(\ref{defAj}), is the common starting point
of most field-theoretical models found in the literature, the
obtained oscillation formulas do not always agree. In
section~\ref{section2}, several conflicting results were
mentioned, namely stationary states versus non-stationary states,
the possible existence of plane wave oscillations and of mass
prefactors. Actually, it will be seen in sections \ref{section5}
and \ref{section6} that these apparent contradictions vanish if we
proceed to a careful evaluation of the integrals present in
Eq.~(\ref{defAj}). In the present section, the possible methods of
evaluating the amplitude (\ref{defAj}) are compared. We shall see
that the choice of the integration method depends on the
propagation distance, so that the distance range can be divided in
three regimes. We then evaluate the amplitude in the three
scenarios.

\subsection{Pole integration and wave packet correspondence}
\label{poleintegrations}

Since the experimental conditions are such that the propagating
particle is on-shell, the main contribution to the transition
amplitude (\ref{defAj}) comes from the pole of the propagator.
Although integrals on $p^0$ and ${\bf p}$ can, in principle, be
done in any order, it is practical to integrate first on the
energy, i.e. to do the pole integration on the energy, in order to
make possible an interpretation of the amplitude in terms of
propagating wave packets. In section \ref{shortcut}, we shall
discuss another method of integration (for stable oscillating
particles) that has the advantages of being shorter and of
clearing up the meaning of the stationary limit. However, this
last method neither lends itself to a wave packet picture nor can
be easily extended to unstable oscillating particles.

The crucial step in the pole integration is the choice of an
appropriate contour in the complex plane. The contour should be
carefully chosen as the analytic continuation of most overlap
functions diverges at infinity in the complex plane. The
integration on the energy can be performed with the help of the
Jacob-Sachs theorem \cite{Jacob61}. This theorem is based on the
assumption that the energy spectrum of all incident particles is
limited to a finite range. Thus the overlap function $\psi(E,{\bf
p})$ is distinct from zero only for $p^2=E^2-{\bf p}^2$ within
certain bounds (with $E>0$). On this interval, $\psi(E,{\bf p})$
is taken to be infinitely differentiable. In that case, the
Jacob-Sachs theorem says that the asymptotic value ($T\to\infty$)
of the energy integral in Eq.~(\ref{defAj}) is given by its
residue at the pole below the real axis. Thus, the evaluation of
the partial amplitude ${\cal A}_j$ with the Jacob-Sachs theorem
yields
\begin{equation}
   {\cal A}_j \cong \frac{\pi}{(2\pi)^4} \,  \int
   \frac{d^3p}{E_j({\bf p})} \;
   \psi(E_j({\bf p}),{\bf p}) \;\,
   e^{ - i \phi_j({\bf p}) - \gamma_j({\bf p}) } \, ,
   \label{jacobsachsinteg}
\end{equation}
where
\begin{eqnarray}
  E_j({\bf p}) &=& \sqrt{ m^2_j + {\bf p}^2 }
  \label{defEj}
  \\
  \phi_j({\bf p}) &=& E_j({\bf p}) \, T - {\bf p} \cdot {\bf L}
  \\
  \gamma_j({\bf p}) &=& \frac{m_j \Gamma_j}{2E_j({\bf p})} \, T \, .
  \label{defgam}
\end{eqnarray}

For Gaussian external wave packets, the overlap function is given
by Eqs.~(\ref{overlapPD})-(\ref{overlapfP}). In principle, this
function should be cut off outside the energy range determined by
experimental conditions so as to satisfy the conditions of the
Jacob-Sachs theorem. However these corrections are very small and
will be neglected in the computations. The reader is referred to
the Appendix for more details on the validity conditions of this
theorem.

It is interesting to note that the amplitude
(\ref{jacobsachsinteg}) is mathematically equivalent to the
amplitude obtained in the intermediate wave packet model
\cite{Kayser81}, in which the mass eigenstates are directly
represented by wave packets. The overlap function $\psi(E_j({\bf
p}),{\bf p})$ corresponds to the wave function of the jth mass
eigenstate. Thus it makes sense, in an external wave packet model,
to talk about mass eigenstate `wave packets' associated with the
propagating particle.

This wave packet picture brings up a problem, as the overlap
function takes into account not only the properties of the source,
but also of the detector. This is unusual for a wave packet
interpretation and not satisfying for causality. This point was
recently clarified in an interesting article by Giunti
\cite{Giunti02b} which was already mentioned in section
\ref{intermwavepackets}. In this paper, Giunti proposes a
sophisticated version of the intermediate wave packet model, in
which the wave packet of the oscillating particle is explicitly
computed with field-theoretical methods in terms of external wave
packets. The intermediate wave packet depends only on the function
$\psi_{\scriptscriptstyle P}(p^0,{\bf p})$ (see Eq.~(\ref{overlapP})). The
translation of this wave packet to the detection point yields a
factor $exp[-i\phi_j({\bf p})]$. At the detector, its interaction
with external wave packets gives rise to the function $\psi_{\scriptscriptstyle
D}(p^0,{\bf p})$. Remarkably, the total amplitude is exactly the
same as Eq.~(\ref{jacobsachsinteg}). However the propagating wave
packet now depends only on the properties of the source through
the function $\psi_{\scriptscriptstyle P}(p^0,{\bf p})$. The only drawbacks of
this model are that it needs the equal time prescription and that
it is not applicable to unstable oscillating particles.

\subsection{Three propagation regimes}
\label{threepropagation}

\subsubsection{Laplace's and stationary phase methods}
\label{alternative}

Unfortunately, the integration over the 3-momentum in
Eq.~(\ref{jacobsachsinteg}) cannot be done analytically. Resorting
to the explicit form (\ref{overlapPD})-(\ref{overlapfP}) of the
overlap function valid for Gaussian external wave packets, we see
that the integral (\ref{jacobsachsinteg}) can be approximated by
means of an asymptotic expansion for which two kinds of large
parameters can be used. On the one hand, $\sigma_{p\scriptscriptstyle P,D}^{-2}$
and $\sigma_{e\scriptscriptstyle P,D}^{-2}$ are large parameters appearing in the
overlap function (\ref{overlapPD})-(\ref{overlapfP}). They suggest
a second order expansion of the integrand around the maximum ${\bf
p}_j$ of the overlap function, followed by a Gaussian integration:
this is called {\it Laplace's method} \cite{Erdelyi,Bender}. On
the other hand, $T$ and ${\bf L}$ are large parameters appearing
in the phase. They suggest a second order expansion of the
integrand around the stationary point ${\bf p}_{cl,j}$ of the
phase, followed by a Gaussian integration: this is called the {\it
method of stationary phase} \cite{Erdelyi,Bender}. The competition
between these two asymptotic behaviors implies a detailed study of
the oscillation of the phase around the average momentum ${\bf
p}_j$ (Fig.~4). The expansion of the phase in
Eq.~(\ref{jacobsachsinteg}) should be compared with the expansion
of the overlap function. Although both methods are expected to
lead roughly to the same answer in the case of the propagation of
a single particle, it should be checked whether the delicate
compensation mechanism resulting in the oscillation phase is
independent of the method chosen.
\begin{figure}
\begin{center}
\includegraphics[width=9cm]{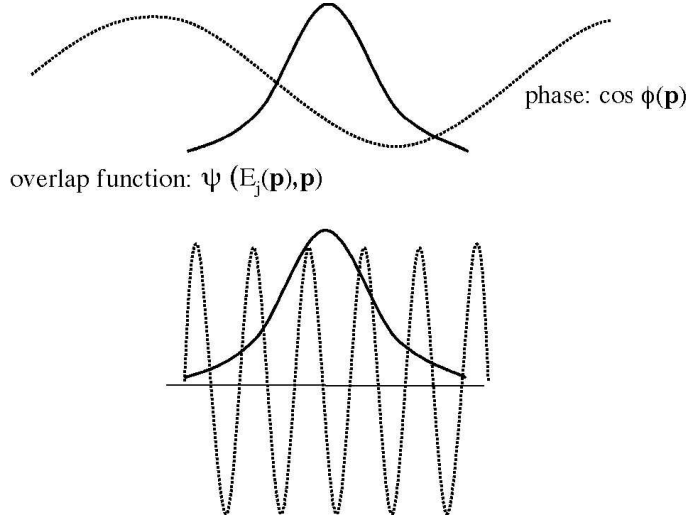}
\caption{Laplace's method is preferable if the phase varies slowly
(upper drawing), whereas the method of stationary phase is
preferable if the phase varies rapidly (lower drawing). The
overlap function (resp. the cosine of the phase) is indicated on
both drawings by a continuous (resp. dotted) line.}
\end{center}
\end{figure}

In section \ref{nodispersionregime}, it will be shown that
Laplace's method is preferable if the dispersion (i.e. spreading
in time) of the amplitude is negligible. This is certainly the
case sufficiently near the source but is not true anymore beyond
some threshold. From the discussion of section
\ref{sectiongaussianwp}, it is expected that the spreading of the
amplitude first begins in directions transversal to the
propagation axis, before becoming significant along this axis. It
is thus appropriate to divide the propagation range into three
regimes: the {\it no-dispersion regime}, the {\it
transversal-dispersion regime} and the {\it
longitudinal-dispersion regime}. Laplace's method will be used in
the first regime in all directions and in the second regime along
the propagation axis, whereas the method of stationary phase will
be used in the second regime along transversal directions and in
the third regime in all directions.

As the phase of an amplitude corresponding to a propagating
particle oscillates in $pL$, the interference between two
amplitudes corresponding to two mass eigenstates provides a phase
of the form $\epsilon pL$, where $\epsilon=\delta
m_{ij}^2/2p_0^2$, with $\delta m_{ij}^2=m_i^2-m_j^2$. When the
conditions of production and detection of the oscillating particle
are studied with the help of Gaussian wave packets, a new
parameter is introduced, namely $\sigma_p^2$. Therefore, the
parameter $\epsilon$ can also appear with other dimensionless
coefficients like $(p/\sigma_p)^2$ and $(\sigma_pL)^2$. These new
coefficients are smaller than the original coefficient $pL$ if the
distance $L$ is macroscopic. The new contributions will be
neglected in the phase, since they are much smaller than the
oscillating phase $\epsilon pL$. On the other hand, they should
not be neglected in the modulus of the amplitude, where they are
the leading terms in the $\epsilon$-expansion and yield
observability conditions of oscillations.

We shall thus calculate the transition probability to ${\cal
O}(\epsilon^2)$ in the real part of the argument of the
exponential and to ${\cal O}(\epsilon)$ in the phase. Gaussian
integrations will be consistent with the $\epsilon$-expansion if
the extremum is computed to ${\cal O}(\epsilon)$, the real part of
the argument of the exponential to ${\cal O}(\epsilon^2)$, the
phase to ${\cal O}(\epsilon)$, the first derivatives to ${\cal
O}(\epsilon)$ and the second derivatives to ${\cal
O}(\epsilon^0)$.

\subsubsection{No-dispersion regime}
\label{nodispersionregime}

As discussed above, the integration over the 3-momentum in
Eq.~(\ref{jacobsachsinteg}) is done by Laplace's method if the
phase varies slowly over the width of the overlap function, i.e.
if the particle is sufficiently near the source. The aim of this
section is to determine the extension of this no-dispersion regime
and to compute the amplitude ${\cal A}_j$ in this range.

The first step consists in analyzing the form of the overlap
function (\ref{overlapPD})-(\ref{overlapfP}). The influence of the
decay term $\gamma_j({\bf p})$ (Eq.~(\ref{defgam})) will be
neglected in this section since we are near the source. The study
of the overlap function amounts to study the argument of the
exponential, i.e. of the function $f_j({\bf p})$ defined by
\begin{equation}
  f_j({\bf p}) = f_{\scriptscriptstyle P}(E_j({\bf p}),{\bf p}) +
  f_{\scriptscriptstyle D}(E_j({\bf p}),{\bf p}) \, ,
  \label{definfj}
\end{equation}
where $f_{\scriptscriptstyle P,D}$ are defined by Eq.~(\ref{overlapfP}) (with the
constraint (\ref{conserv})) and $E_j({\bf p})$ by
Eq.~(\ref{defEj}). It is convenient to expand all results in small
mass differences around ($E_0$, ${\bf p}_0$, $m_0$, ${\bf v}_0$),
where $m_0^2=E_0^2-{\bf p}_0^2$ and ${\bf v}_0={\bf p}_0/E_0$. The
expansion parameter is noted $\epsilon$ and refers collectively to
all $\delta m_j^2/2E_0^2$, where $\delta m_j^2=m_j^2-m_0^2$.  The
value ${\bf p}_j$ minimizing $f_j({\bf p})$ is given to ${\cal
O}(\epsilon)$ by
\begin{equation}
  {\bf p}_j = {\bf p}_0 +
  \left(  \alpha {\bf u}_{\scriptscriptstyle P} + \beta  {\bf u}_{\scriptscriptstyle D} \right) \,
  \frac{\delta m_j^2}{2E_0} \, ,
  \label{momentumtilde}
\end{equation}
where
$$
  {\bf u}_{\scriptscriptstyle P,D} =
  \frac{{\bf v}_0-{\bf v}_{\scriptscriptstyle P,D}}{2\sigma_{e\scriptscriptstyle P,D}}\, .
$$
The associated energy $E_j=\sqrt{{\bf p}_j^2+m_j^2}$ can be
expanded to ${\cal O}(\epsilon)$ as
\begin{equation}
  E_j = E_0 + \tilde\rho \, \frac{\delta m_j^2}{2E_0} \, ,
  \label{energytilde}
\end{equation}
where $\tilde\rho=1+\alpha{\bf v}_0\cdot{\bf u}_{\scriptscriptstyle P}+\beta{\bf
v}_0\cdot{\bf u}_{\scriptscriptstyle D}$. The values of the dimensionless
coefficients $\alpha$ and $\beta$ can be computed but their
explicit expressions will not be needed. It is sufficient to know
that $\tilde\rho\to0$ in the stationary limit. A velocity ${\bf
v}_j={\bf p}_j/E_j$ is also defined for future use.

We are now going to approximate the overlap function as a Gaussian
and compute its three characteristic widths. At the extremum ${\bf
p}_j$, the Hessian matrix of $f_j({\bf p})$ reads to ${\cal
O}(\epsilon^0)$
\begin{eqnarray*}
  \Sigma^{ab} &\equiv&
  \frac{1}{2} \,
  \frac{\partial^2 f_j}{\partial p^a\partial p^b}({\bf p}_j)
  \nonumber \\
  &=& \frac{\delta^{ab}}{4\sigma_p^2}
    + u_{\scriptscriptstyle P}^a u_{\scriptscriptstyle P}^b + u_{\scriptscriptstyle D}^a u_{\scriptscriptstyle D}^b \, ,
\end{eqnarray*}
where the momentum width $\sigma_p$ is defined by
\begin{equation}
   \frac{1}{\sigma_p^2}=
   \frac{1}{\sigma_{p\scriptscriptstyle P}^2} + \frac{1}{\sigma_{p\scriptscriptstyle D}^2} \, ,
   \label{definitionsigmap}
\end{equation}
and is approximately equal to the smallest width among the
production and detection momentum widths. The associated width
$\sigma_x$ in configuration space is defined by
$\sigma_p\sigma_x=1/2$. The matrix $\Sigma^{ab}$ determines the
range of ${\bf p}$ values for which the overlap function
$\psi(E_j({\bf p}),{\bf p})$ is not negligible. As $\Sigma^{ab}$
is symmetric, it can be diagonalized by an orthogonal coordinate
transformation. The eigenvalues of $\Sigma^{ab}$ are
\begin{eqnarray}
  \sigma_x^2 &=& \frac{1}{4\sigma_p^2} \, ,
  \nonumber
  \\
  \sigma_{x\pm}^2 &=& \frac{1}{4\sigma_p^2}
  + \frac{1}{2} \left( {\bf u}_{\scriptscriptstyle P}^2 + {\bf u}_{\scriptscriptstyle D}^2 \right)
  \pm \frac{1}{2}
    \sqrt{\left( {\bf u}_{\scriptscriptstyle P}^2 + {\bf u}_{\scriptscriptstyle D}^2 \right)^2
          - 4 \left( {\bf u}_{\scriptscriptstyle P} \times {\bf u}_{\scriptscriptstyle D} \right)^2}
  \; .
  \label{defsigmapm}
\end{eqnarray}
The eigenvector associated with $\sigma_x^2$ is in the direction
of ${\bf u}_{\scriptscriptstyle P} \times {\bf u}_{\scriptscriptstyle D}$, whereas the
eigenvectors associated with $\sigma_{x\pm}^2$ belong to the plane
defined by ${\bf u}_{\scriptscriptstyle P}$ and ${\bf u}_{\scriptscriptstyle D}$. In the limit
$|{\bf u}_{\scriptscriptstyle P}|\gg|{\bf u}_{\scriptscriptstyle D}|$ (resp. $|{\bf u}_{\scriptscriptstyle
P}|\ll|{\bf u}_{\scriptscriptstyle D}|$), the eigenvalues $\sigma_x^2$ and
$\sigma_{x-}^2$ become degenerate and the eigenvector associated
with $\sigma_{x+}^2$ becomes aligned with ${\bf u}_{\scriptscriptstyle P}$ (resp.
${\bf u}_{\scriptscriptstyle D}$). This is also the case in the limit of parallel
${\bf u}_{\scriptscriptstyle P}$ and ${\bf u}_{\scriptscriptstyle D}$. These limits are relevant
to the case of stationary boundary conditions which are examined
below.

Let us choose coordinate axes $({\bf e}_x,{\bf e}_y,{\bf e}_z)$
coinciding with the normalized eigenvectors associated with
$(\sigma_x^2,\sigma_{x-}^2,\sigma_{x+}^2)$ respectively. The
quantities $(\sigma_p^2,\sigma_{p-}^2,\sigma_{p+}^2)$  (with
$\sigma_{p\pm}\sigma_{x\pm}=1/2$) can be interpreted as the
momentum widths of the overlap function, since they give
constraints on the range of ${\bf p}$ values for which the overlap
function is non-negligible:
\begin{eqnarray}
  |p^x-p_j^x| &\lesssim& \sigma_p \;\; ,
  \nonumber \\
  |p^y-p_j^y| &\lesssim& \sigma_{p-} \; ,
  \nonumber \\
  |p^z-p_j^z| &\lesssim& \sigma_{p+} \; .
  \label{widthconstraint}
\end{eqnarray}

The case of the stationary limit is of special interest since this
assumption is used by several authors
\cite{Kobzarev82,Grimus96,Grimus99,Ioannisian99,Cardall99}. Recall
that stationary boundary conditions are obtained in the external
wave packet model by taking ${\bf v}_{\scriptscriptstyle P,D}\to0$ and
$\sigma_{e\scriptscriptstyle P,D}\to0$ with $|{\bf v}_{\scriptscriptstyle
P,D}|\lesssim\sigma_{e\scriptscriptstyle P,D}S_{\scriptscriptstyle P,D}$ (see
Eq.~(\ref{statlim})). In this limit, the velocity ${\bf v}_0$
becomes aligned with the axis ${\bf e}_z$,
\begin{equation}
  v_0^{x,y} \sim v_{\scriptscriptstyle P,D}^{x,y} \to 0  \, ,
  \label{statvo}
\end{equation}
and two eigenvalues become degenerate whereas the third diverges:
\begin{eqnarray}
  \sigma_{x-}^2 &\to& \sigma_x^2 \, ,
  \nonumber \\
  \sigma_{x+}^2 &\to&
  \frac{1}{4\sigma_p^2}
   + {\bf u}_{\scriptscriptstyle P}^2 + {\bf u}_{\scriptscriptstyle D}^2
   \to \infty  \, ,
  \label{statwidth}
\end{eqnarray}
In other words, the transversal widths (i.e. in directions
orthogonal to ${\bf p}_0$) are given by $\sigma_p$ in the
stationary limit, whereas the longitudinal width (i.e. in the
direction of ${\bf p}_0$) is given by $\sigma_{p+}\to0$.

The second step consists in comparing the expansion of the overlap
function with the expansion of the phase $\phi_j({\bf p})$ around
${\bf p}_j$, which reads
\begin{equation}
  \phi_j({\bf p}) \cong \phi_j({\bf p}_j)
  + ({\bf v}_j T - {\bf L}) ({\bf p} - {\bf p}_j)
  + \frac{T}{2E_0} (p^a - p_j^a)R^{ab}(p^b - p_j^b)
  \, ,
  \label{expansionphase}
\end{equation}
where $R^{ab}=\delta^{ab}-v_0^av_0^b$. The second derivatives have
been evaluated to ${\cal O}(\epsilon^0)$.

Laplace's method will be preferred to the method of stationary
phase if the phase $\phi_j({\bf p})$ varies slowly over the `bump'
of the overlap function. As
$\sigma_p\geq\sigma_{p-}\geq\sigma_{p+}$, the variation of the
phase will become important in the first place along the direction
$p^x$, then in the direction $p^y$ and finally in the direction
$p^z$. The criterion for the use of Laplace's method in all three
directions $p^{x,y,z}$ will thus be determined by considering the
largest momentum width $\sigma_p$. The insertion of the
constraints (\ref{widthconstraint}) into the phase
(\ref{expansionphase}) yields first order conditions for a slowly
varying phase,
\begin{eqnarray}
  |v_j^x T - L^x| \sigma_p \; &\lesssim& 1 \, , \nonumber \\
  |v_j^y T - L^y| \sigma_{p-} &\lesssim& 1 \, , \nonumber \\
  |v_j^z T - L^z| \sigma_{p+} &\lesssim& 1 \, ,
  \label{cond1}
\end{eqnarray}
as well as a second order condition,
\begin{equation}
  \frac{T}{E_0} \, \sigma_p^2 \lesssim 1 \, ,
  \label{cond2}
\end{equation}
where we have used the property $q^aR^{ab}q^b\leq{\bf q}^2$ and
the fact that $\sigma_p$ is the largest width. For a given $T$
satisfying Eq.~(\ref{cond2}), it is always possible to find a
range of ${\bf L}$ values so that conditions (\ref{cond1}) are
satisfied. For other ${\bf L}$ values, the amplitude is negligible
as will be checked on the result (see Eq.~(\ref{Fdiag})). Thus the
criterion allowing to choose the integration method is determined
by Eq.~(\ref{cond2}): the integration on ${\bf p}$ will be done by
Laplace's method if $T\lesssim E_0/\sigma_p^2$ or, equivalently
(with $L=|{\bf L}|$ and $p_0=|{\bf p}_0|$), if
\begin{equation}
   L\lesssim \frac{p_0}{\sigma_p^2} \, ,
  \label{ipcondition}
\end{equation}
since conditions (\ref{cond1}) impose the relation ${\bf L}\cong
{\bf v}_0T$ as long as $\sigma_{x+}\ll L$.

For $\sigma_{x+}\gtrsim L$ (stationary limit), we now show that
condition (\ref{ipcondition}) is directly obtained without going
through condition (\ref{cond2}). The overlap function imposes in
that limit that $|{\bf p}|=\sqrt{E_0^2-m_j^2}$, so that we are
left with an angular integration with the angular part of the
integrand given by
\begin{equation}
  \exp\left( \frac{{\bf p}\cdot{\bf p}_0}{2\sigma_p^2}
            + i {\bf p}\cdot{\bf L}
      \right) \, .
  \label{angular}
\end{equation}
Condition (\ref{ipcondition}) shows that the angular variation of
the phase in (\ref{angular}) is slow with respect to the angular
variation of the overlap function, in which case Laplace's method
will give good results. Therefore condition (\ref{ipcondition})
constitutes a good criterion for the use of Laplace's method
whether the stationary limit is taken or not.

Condition (\ref{ipcondition}) is usually not verified in
oscillation experiments, because $L/\sigma_x\gg p_0/\sigma_p$ in
most cases. This condition is the same than the one under which
Ioannisian and Pilaftsis \cite{Ioannisian99} obtain `plane wave'
oscillations.

If condition (\ref{ipcondition}) is satisfied, the evaluation of
the integral (\ref{jacobsachsinteg}) can be done by Laplace's
method and yields
\begin{equation}
  {\cal A}_j = N \sigma_p\sigma_{p-}\sigma_{p+} \,
  \exp \Big(
      - i E_j T + i {\bf p}_j \cdot {\bf L}
      - f_j({\bf p}_j)
      - F_j(T) - \gamma_j({\bf p}_j) \Big) \, ,
  \label{nodispersion}
\end{equation}
where $N$ absorbs numerical constants. The function $F_j(T)$ is
defined by
\begin{equation}
  F_j(T)= \frac{1}{4}
          ({\bf v}_j T - {\bf L})^t
          \left( \Sigma + i\frac{T}{2E_0} R \right)^{-1}
          ({\bf v}_j T - {\bf L}) \, ,
  \label{definF}
\end{equation}
where $\Sigma^{ab}$ and $R^{ab}$ are considered as matrices. In
the framework of the wave packet interpretation developed after
Eq.~(\ref{jacobsachsinteg}), the function $\exp(- F_j(T))$ plays
the part of the space-time envelope of the wave packet associated
with the jth mass eigenstate. The elements of the matrix
$Re(\Sigma+i\frac{T}{2E_0}R)^{-1}$ constrain the extent of the
wave packet envelope in space-time. As $T$ increases, the wave
packet spreads because of the $i\frac{T}{E_0}R$ term. Thus the
dispersion of the wave packet is due to the second order term in
the expansion of the phase $\phi_j({\bf p}_j)$. Therefore,
condition (\ref{cond2}) or, equivalently, condition
(\ref{ipcondition}) means that dispersion has not yet begun in any
direction, transversal or longitudinal. For that reason, the range
of ${\bf L}$ values defined by $L\lesssim p_0/\sigma_p^2$ is
called the {\it no-dispersion regime}. Note that this threshold
was already met in the analysis of Gaussian wave packets in
section \ref{sectiongaussianwp} (see Eq.~(\ref{thresholdtrans})).
Of course, the above interpretation is not valid for
$\sigma_{x+}\gtrsim L$, in which case the propagation time $T$
becomes indeterminate and dispersion loses its meaning.

Now that the origin of dispersion has been clarified, the term in
Eq.~(\ref{definF}) including $R$ can be neglected with respect to
$\Sigma$. Moreover, we choose to approximate ${\bf v}_j$ by ${\bf
v}_0$ in $F_j(T)$. In comparison with the oscillation formulas
that will be derived in the transversal- and
longitudinal-dispersion regimes, this approximation will lead to
the absence of the coherence-length term, since this term
exclusively arises, when the dispersion is neglected, from the
velocity difference ${\bf v}_i-{\bf v}_j$. Dropping the index $j$,
the wave packet envelope in Eq.~(\ref{nodispersion}) can then be
written in the coordinate system diagonalizing $\Sigma$ as
\begin{equation}
  F(T)=    \frac{(v_0^x T - L^x)^2}{4\sigma_x^2}
         + \frac{(v_0^y T - L^y)^2}{4\sigma_{x-}^2}
         + \frac{(v_0^z T - L^z)^2}{4\sigma_{x+}^2} \, ,
  \label{Fdiag}
\end{equation}
which shows that the conditions (\ref{cond1}) assumed for
Laplace's method are required to obtain a non-negligible amplitude
${\cal A}_j$. In other words, ${\bf p}_0$ and ${\bf L}$ should be
nearly parallel in order to have an amplitude significantly
different from zero. We shall see that this constraint explicitly
appears in the final oscillation formula
(Eq.~(\ref{probanodisp})).

\subsubsection{Transversal-dispersion regime}
\label{transverse}

Let us now assume that condition (\ref{ipcondition}) is not
satisfied, i.e.\ $L\gtrsim p_0/\sigma_p^2$. In that case,
Laplace's method cannot be used to integrate on all three
components $p^{x,y,z}$ in the amplitude (\ref{jacobsachsinteg})
since dispersion becomes significant. However, the spreading of
the amplitude is not identical in all directions. More
specifically, the onset of dispersion in the direction ${\bf p}_0$
can be delayed by two factors. First, the matrix element $R^{ab}$
present in Eq.~(\ref{expansionphase}) leads to a relativistic
contraction (of $1-{\bf v}_0^2$) in the direction ${\bf p}_0$ of
the dispersion of the amplitude (see Eq.~(\ref{definF})). Second,
the momentum width along ${\bf p}_0$ is given for $\sigma_{e\scriptscriptstyle
P,D}\ll\sigma_x$ (i.e. in the stationary limit (\ref{statwidth}))
by a vanishing $\sigma_{p+}$. Thus Laplace's method is valid for a
longer time $T$ in the direction ${\bf p}_0$ than in directions
transverse to this vector. For this reason, the choice of the
integration method in the direction ${\bf p}_0$ will be postponed
for a short while, whereas the method of stationary phase will be
preferred  for momentum integrations in directions transverse to
${\bf p}_0$.

Let the $z$ axis be along ${\bf L}$, i.e.\ ${\bf L}=L\,{\bf e}_z$.
As in section \ref{nodispersionregime}, the examination of the
amplitude (\ref{jacobsachsinteg}) shows that the quick variation
of the phase averages the amplitude to zero unless ${\bf p}_0$ and
${\bf L}$ are nearly parallel. The method of stationary phase can
thus be applied in directions $p^{x,y}$, the stationary points of
which are given by $p^x=p^y=0$.  The result of the method of
stationary phase for the transverse momenta in
(\ref{jacobsachsinteg}) can be written as follows:
\begin{equation}
   {\cal A}_j = \frac{Ng({\bf l})}{T-i\mu} \,
   \int dp \;
   \exp \left(
              - i \phi_j(p) - f_j(p) - \gamma_j(p)
        \right) \, ,
   \label{longitudinalamplitude}
\end{equation}
where $p\equiv p^z$ and
\begin{eqnarray}
   \phi_j(p)&=& \sqrt{p^2+m_j^2} \, T - pL \, ,
   \label{phasej}
   \\
   f_j(p) &=& f_{j\scriptscriptstyle P}(p) + f_{j\scriptscriptstyle D}(p) \, ,
   \label{definitionfj}
\end{eqnarray}
with
\begin{equation}
   f_{j\scriptscriptstyle P}(p) =
   \frac{ (p-p_0)^2}{4\sigma_{p \scriptscriptstyle P}^2}
   +\frac{ \left( \sqrt{p^2+m_j^2} - E_0 -( p-p_0) v_{\scriptscriptstyle P} \right)^2}
          { 4\sigma_{e\scriptscriptstyle P}^2 } \, ,
   \label{definitionfjP}
\end{equation}
where $v_{\scriptscriptstyle P}= v_{\scriptscriptstyle P}^z$ and $p_0=p_0^z$. The average energy
$E_0$ has been redefined so as to absorb a factor $p_0^x v_{\scriptscriptstyle
P}^x + p_0^y v_{\scriptscriptstyle P}^y$. The definition of $f_{j\scriptscriptstyle D}(p)$ is
similar. Finally, $\gamma_j(p)$ expresses the possible decay of
the oscillating particle:
\begin{equation}
   \gamma_j(p) = \frac{m_j \Gamma_j T}{2 \sqrt{p^2+m_j^2}} \, .
   \label{defgammaj}
\end{equation}

The function $g({\bf l})$ comes from the transversal part of the
overlap function. It expresses the geometrical constraint between
the direction of observation ${\bf l}={\bf L}/L$ and the momentum
${\bf p}_0$:
\begin{equation}
   g({\bf l})
   = \exp \left(
                 - \frac{({\bf p}_0\times{\bf l})^2}{4\sigma_p^2}
          \right) \, ,
   \label{geom}
\end{equation}
where $\sigma_p$ is defined by Eq.~(\ref{definitionsigmap}). The
function $g({\bf l})$ restricts the propagation to a cone of axis
${\bf p}_0$ and angle $arcsin(\sigma_p/p_0)$ (see Fig.~6 in
section \ref{osclargedist}).

Numerical constants have been included in $N$, as well as 4-volume
factors $\sigma_{p\scriptscriptstyle P}^{-3}\sigma_{e\scriptscriptstyle P}^{-1}$ and
$\sigma_{p\scriptscriptstyle D}^{-3}\sigma_{e\scriptscriptstyle D}^{-1}$, and an energy factor
$E_0$. The constant $\mu=E_0/2\sigma_p^2$ comes from the overlap
function and acts as a cut-off for small $T$. As $T$ is
macroscopic, the prefactor $1/(T-i\mu)$ can be approximated by
$1/T$ so as to give a prefactor $1/L^2$ in the transition
probability. This expected geometrical decrease is seen to
originate in the transverse dispersion of the wave packet
corresponding to the oscillating particle. As in section
\ref{nodispersionregime}, it is convenient to define a reference
mass $m_0$ and a velocity $v_0$ by
\begin{eqnarray}
   m_0^2 &=& E_0^2-p_0^2 \, , \label{defm0} \\
   v_0 &=& p_0/E_0 \label{defv0} \, .
\end{eqnarray}

As before, the choice of the method to perform the longitudinal
momentum integration (\ref{longitudinalamplitude}) is done by
comparing the expansions of the phase and of the overlap function
around the value $p_j$ for which $f_j(p)+\gamma_j(p)$ is extremal
(the decay term is not neglected in this section). The explicit
value of $p_j$ will be computed in section \ref{osclongdisp} for
the stable case and in section \ref{sectionunstable} for the
unstable case.

The expansions of $f_j(p)$ and $\phi_j(p)$ are given by
\begin{eqnarray}
  f_j(p) &\cong& f_j(p_j) + \frac{(p-p_j)^2}{4\sigma_{peff}^2} \, ,
  \nonumber
  \\
  \phi_j(p)&\cong& \phi_j(p_j) + (v_jT-L)(p-p_j)
  + \frac{m_j^2T}{2E_0^3} (p-p_j)^2 \, ,
  \label{expansionphij}
\end{eqnarray}
where $E_j=\sqrt{p_j^2+m_j^2}$ and $v_j=p_j/E_j$.

The {\it effective width} $\sigma_{peff}$ and $\sigma_{xeff}$ are
defined by
\begin{eqnarray}
   && \frac{1}{\sigma_{peff}^2} =
   \frac{1}{\sigma_{p\scriptscriptstyle P}^2} + \frac{1}{\sigma_{p\scriptscriptstyle D}^2}
   + \frac{(v_0-v_{\scriptscriptstyle P})^2}{\sigma_{e\scriptscriptstyle P}^2}
   + \frac{(v_0-v_{\scriptscriptstyle D})^2}{\sigma_{e\scriptscriptstyle D}^2} \, ,
   \nonumber \\
   && \sigma_{peff}\sigma_{xeff} = \frac{1}{2} \, .
   \label{sigmaeff}
\end{eqnarray}
Note that $\sigma_{peff}$ has the same form as the asymptotic
value of $\sigma_{p+}$ (see Eq.~(\ref{statwidth})).

With the help of the wave packet correspondence discussed at the
end of section \ref{poleintegrations}, the effective width can be
interpreted as the energy-momentum width of the oscillation
process, since it is the width of the overlap function. It is
dominated by the smallest among the energy uncertainties (recall
that $\sigma_{e\scriptscriptstyle P,D}\leq \sigma_{p\scriptscriptstyle P,D}$). The effective
width $\sigma_{xeff}$ is then approximately equal either to the
production or to the detection time uncertainty, depending on
which one is the largest.

Laplace's method is preferable if the phase (\ref{expansionphij})
varies slowly over the width $\sigma_{peff}$, i.e. if the two
following conditions are satisfied
\begin{eqnarray}
  |v_jT-L|2\sigma_{peff} &\lesssim& 1 \, ,
  \label{constraint1}
  \\
  \frac{m_j^2T}{2E_0^3} \, 4\sigma_{peff}^2 &\lesssim& 1 \, .
  \label{constraint2}
\end{eqnarray}
As in section \ref{nodispersionregime}, the first order constraint
(\ref{constraint1}) will be included in the result of Laplace's
method. Thus the criterion allowing to choose between Laplace and
stationary phase methods is given by Eq.~(\ref{constraint2}). In
other words, it is better to use Laplace's method if $T$ is
smaller than a {\it dispersion time} $T^{disp}_j$ defined by
\begin{equation}
  T^{disp}_j = \frac{E_0^3}{2m_j^2\sigma_{peff}^2} \, .
  \label{dispersiontime}
\end{equation}
The term `dispersion time' is justified by the fact that it is the
time at which the longitudinal dispersion of the amplitude becomes
important, more precisely twice the initial size. A {\it
dispersion length} $L^{disp}_j$ can be defined by $L^{disp}_j=v_0
\, T^{disp}_j$. The distance range $p_0/\sigma_p^2\lesssim
L\lesssim L^{disp}_j$ will be called the {\it
tranversal-dispersion regime}. For $L\gtrsim L^{disp}_j$, the
stationary phase method is more accurate: this distance range will
be called the {\it longitudinal-dispersion regime}. Note that this
threshold was already met in the analysis of Gaussian wave packets
in section \ref{sectiongaussianwp} (see
Eq.~(\ref{thresholdlong})). Various estimates of the dispersion
length are discussed in section \ref{displengthsection}, showing
that the concept of dispersion length is relevant to
nonrelativistic particles such as $K$ and $B$ mesons, as well as
to supernova neutrinos, and possibly to solar neutrinos.

In the transversal-dispersion regime, the evaluation of the
amplitude (\ref{longitudinalamplitude}) as a Gaussian integral
around $p_j$ gives
\begin{equation}
  {\cal A}_j =
  \frac{Ng({\bf l}) \sigma_{peff}}{ T\sqrt{1+i T/T^{disp}_j} } \,
  \exp \left(
  -i E_j T + i p_j L
  - f_j(p_j) - \gamma_j(p_j)
  - \frac{1}{1+iT/T^{disp}_j} \frac{(v_jT-L)^2}{4\sigma_{xeff}^2}
  \right) \, ,
  \label{transdisp}
\end{equation}
where $N$ absorbs numerical constants and $p_j$ is the value for
which $f_j(p)+\gamma_j(p)$ is extremal.

The amplitude (\ref{transdisp}) behaves as a wave packet of group
velocity $v_j$ and space-time extent
$(1+(T/T_j^{disp})^2)^\frac{1}{2}\,\sigma_{xeff}$. If the
longitudinal dispersion is neglected ($T_j^{disp}=\infty$), the
amplitude (\ref{transdisp}) is similar to Eq.~(18) of
Ref.~\cite{Giunti98a}.

\subsubsection{Longitudinal-dispersion regime}

At sufficiently large distance, dispersion becomes significant and
all neutrinos propagating freely enter into the
longitudinal-dispersion regime. In this regime ($L\gtrsim
L_j^{disp}$), it has been argued that the integral
(\ref{longitudinalamplitude}) should be evaluated with the method
of stationary phase. The stationary point of the phase $\phi_j(p)$
is given by
\begin{equation}
  p_{cl,j} = m_j \frac{v_{cl}}{\sqrt{1-v_{cl}^2}} \, ,
  \label{statpoint}
\end{equation}
where $v_{cl}=L/T$. It can be interpreted as the classical
momentum of a particle of mass $m_j$, travelling at the classical
velocity $v_{cl}$. Of course, a stationary point exists only for
$T \geq L $. Otherwise Laplace's method must be used, but the
amplitude is nearly zero in that case anyway. The evaluation of
the amplitude (\ref{longitudinalamplitude}) as a Gaussian integral
around $p_{cl,j}$ gives
\begin{equation}
  {\cal A}_j =
  \frac{Ng({\bf l})\sigma_{peff}}{ T\sqrt{1+i T/T^{disp}_j} } \,
  \exp \left( - i m_j \, \sqrt{T^2-L^2} - f_j(p_{cl,j}) - \gamma_j(p_{cl,j})
              + \sigma_{peff}^2 \,
                \frac{ \left( f_j'(p_{cl,j}) \right)^2 }{1+i T/T^{disp}_j}
       \right)  \, ,
  \label{longdisp}
\end{equation}
where $f_j(p)$ is defined by Eq.~(\ref{definitionfj}), $f_j'(p)$
refers to its derivative and $N$ absorbs numerical constants. The
wave packet interpretation of the amplitude (\ref{longdisp}) is
not obvious but the shape of the associated wave packet can be
studied by an expansion around the maximum of the amplitude.

\subsection{Estimates of the dispersion thresholds}

In section \ref{threepropagation}, it was shown that the
propagation range could be divided into three regimes separated by
two thresholds (Fig.~5). The first threshold is determined by the
no-dispersion condition (\ref{ipcondition}) and separates the
no-dispersion regime from the transversal-dispersion regime at
$L^{nodisp}=p_0/\sigma_p^2$. The second threshold (see
Eq.~(\ref{dispersiontime})) is given by the dispersion length
$L^{disp}_j = p_0E_0^2/2m_j^2\sigma_{peff}^2$ and separates the
transversal-dispersion regime from the longitudinal-dispersion
regime. It is not straightforward to estimate the value of these
thresholds, mainly because of our ignorance of the external wave
packet sizes.
\begin{figure}
\begin{center}
\includegraphics[width=10cm]{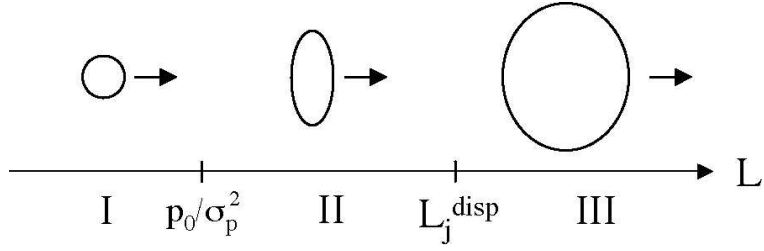}
\caption{The propagation range can be divided into three regimes
according to the dispersion.}
\end{center}
\end{figure}

\subsubsection{No-dispersion condition}

The first threshold is rarely of the order of a macroscopic
distance. A far-fetched example where the propagation distance
could be smaller than the no-dispersion threshold
$L^{nodisp}=p_0/\sigma_p^2$ is given by Ioannisian and Pilaftsis
\cite{Ioannisian99}: atmospheric neutrinos are detected in the
no-dispersion regime if $\sigma_x\gtrsim 10^{-2}\,$cm (with
$L\!\sim\!1000\,$km and $p\!\sim\!1\,$GeV). Although such a
macroscopic size of the neutrino `wave packet' is not totally
excluded (see section \ref{displengthsection}), it is unlikely to
be verified. However this case  will be examined in detail in
section \ref{oscwithoutdisp} both for reason of completeness and
because it was argued in Ref.~\cite{Ioannisian99} that the
corresponding oscillation formula would differ significantly from
the standard result.

\subsubsection{Dispersion length}
\label{displengthsection}

This threshold is a priori more interesting than the first, as it
is of the order of a macroscopic distance. As said above, the main
difficulty lies in the lack of knowledge of the external wave
packet sizes. In the case of neutrinos, another uncertainty arises
because of the dependence of the dispersion length on the absolute
mass scale, instead of a dependence on the mass differences. Some
examples are given below, partly based on the wave packet size
discussion of section \ref{coherenteffects}. We should always
remember that $\sigma_{peff}$ is determined by the smallest width
among the production and detection energy or momentum widths.

First, the neutrinos in the experiment LSND
\cite{Athanassopoulos96,Athanassopoulos98} have a momentum around
$30\,$MeV. The neutrino mass can be taken to be $1\,$eV, and the
production and detection widths should be around $0.01\,$MeV
\cite{Grimus00}. The corresponding dispersion length, around
$10^7\,$m, is much longer than the $30\,$m length scale of the
experiment. Second, atmospheric neutrinos have a momentum around
$1\,$GeV and let us take a mass of $1\,$eV as a bound. An estimate
of the source width can be obtained from the inverse pion and muon
lifetimes and is given by $\sigma_{p\scriptscriptstyle P}\sim10^{-14}\,$MeV in
the case of a pion source. The detector width can be estimated at
$\sigma_{x\scriptscriptstyle D}\sim10^{-10}\,$m \cite{Rich93}, i.e.\
$\sigma_{p\scriptscriptstyle D}\sim10^{-3}\,$MeV, and is dominated by the source
width. The corresponding dispersion length, larger than
$10^{33}\,$km, is completely irrelevant to the experiment. It is
however doubtful that the source width is truly estimated by the
inverse source lifetime (see the discussion of section
\ref{coherenteffects}). Third, solar neutrinos have a momentum
around $1\,$MeV and let us take a mass of $1\,$eV as a bound.
Various estimates of the source width have been given, between
$10^{-7}\,$MeV and $10^{-3}\,$MeV
\cite{Nussinov76,Krauss85,Loeb89,Kimbis,Kiers96,Malyshkin00}. The
detector width can be estimated as above at $\sigma_{p\scriptscriptstyle
D}\sim10^{-3}\,$MeV. The dispersion length, varying between
$10^{2}\,$km and $10^{10}\,$km for a $1\,$eV neutrino mass
(depending on the chosen source width), could be relevant to the
experiment, but is much larger for smaller neutrino masses.

As regards mesons, kaons in the CPLEAR experiment have a momentum
around $550\,$MeV \cite{Fry00}. If the width is guessed to be
$0.01\,$MeV, the dispersion length, around $10^{-6}\,$m, is much
smaller than the length scale of the experiment. More generally,
the ratio between the dispersion length and the decay length
(defined by $L^{decay}=p/m\Gamma$) is given by
$$
   \frac{L^{disp}}{L^{decay}} = \frac{E^2}{\sigma_{peff}^2} \,
   \frac{\Gamma}{2m} \, ,
$$
so that dispersion occurs for all quasi-stable particles
($\Gamma\ll m$), and in particular for $B$ mesons.

In conclusion, if the estimates above are truthful, dispersion
cannot be ignored for $K$ and $B$ mesons, and might be important
for solar neutrinos (and certainly for supernova neutrinos). It is
then necessary to develop a formalism taking into account the
dispersion, in which case the use of the stationary phase method
cannot be avoided.

\section{Oscillations of stable particles}
\label{section5}

In section~\ref{section4}, the choice of the method of evaluation
of the amplitude (\ref{defAj}) was seen to depend on the distance
range within which the oscillating particle detected. Three
distance ranges were distinguished, depending on whether the
amplitude had begun to spread in the transverse and longitudinal
directions (with respect to the direction of propagation). The
amplitude was then evaluated in the three possible cases, yielding
formulas (\ref{nodispersion}), (\ref{transdisp}) and
(\ref{longdisp}) corresponding to the no-dispersion, transversal-
and longitudinal-dispersion regimes respectively.

In this section\footnote{The reader not interested in all
computational details can skip this section. Its results are
summarized in section~\ref{section6}.}, we compute the transition
probability for stable oscillating particles. The transition
probability is then obtained by squaring the amplitude and
averaging over the macroscopic propagation time $T$, which is not
observed in experiments:
\begin{equation}
   {\cal P}_{\alpha \to \beta}({\bf L}) \sim
   \sum_{i,j} V_{i\alpha} \, V_{\beta i}^{-1} \,
          V_{j\alpha}^* \, V_{\beta j}^{-1 \, *} \,
   \int dT \, {\cal A}_i \, {\cal A}_j^* \, .
   \label{transitionproba}
\end{equation}
We shall come back to the proportionality factor in sections
\ref{section6} and \ref{normalization}.

The oscillation formula is derived in the three different
propagation regimes in sections \ref{oscwithoutdisp},
\ref{osctransdisp} and \ref{osclongdisp}. A shorter method of
integration, valid for stable oscillating particles in the second
and third regimes, is then discussed in section \ref{shortcut}.

\subsection{Oscillations without dispersion}
\label{oscwithoutdisp}

\subsubsection{Amplitude}

For $L\lesssim p_0/\sigma_p^2$, the momentum integration in
Eq.~(\ref{jacobsachsinteg}) is done by Laplace's method, yielding
for stable oscillating particles (see Eq.~(\ref{nodispersion})
with $\gamma_j({\bf p}_j)=0$) the following amplitude:
\begin{equation}
  {\cal A}_j = N \sigma_p\sigma_{p-}\sigma_{p+} \,
  \exp \left(
      - i E_j T + i {\bf p}_j \cdot {\bf L}
      - \left( \frac{\delta m_j^2}{4\tilde\sigma_mE_0} \right)^2
      - F(T) \right) \, ,
  \label{nodispersionbis}
\end{equation}
where ${\bf p}_j$ is given by Eq.~(\ref{momentumtilde}), and
\begin{equation}
  F(T)=    \frac{(v_0^x T - L^x)^2}{4\sigma_x^2}
         + \frac{(v_0^y T - L^y)^2}{4\sigma_{x-}^2}
         + \frac{(v_0^z T - L^z)^2}{4\sigma_{x+}^2} \, ,
  \label{Fdiagbis}
\end{equation}
with $\sigma_x$, $\sigma_{x\pm}$ defined by
Eqs.~(\ref{defsigmapm}). The parameter $\tilde\sigma_m$, which has
the dimension of a width, is related to $f_j({\bf p}_j)$ (see
Eq.~(\ref{definfj})) by
$$
  f_j({\bf p}_j) =
  \left( \frac{\delta m_j^2}{4\tilde\sigma_mE_0} \right)^2 \, .
  \label{sigmatilde}
$$

\subsubsection{Probability}
\label{sectionprobanodisp}

The last step towards the oscillation formula consists in
computing the time average of the transition probability, which is
a Gaussian integral on $T$:
\begin{equation}
  \int dT \; {\cal A}_i{\cal A}_j^* = N_{\tilde g} \,
  \exp \left(
      - i\phi_{ij}(T_0)
      - \frac{(\delta m_i^2)^2+(\delta m_j^2)^2}{16\tilde\sigma_m^2E_0^2}
      - \frac{(E_i-E_j)^2}{4F''(T_0)} - 2F(T_0) \right) \, ,
  \label{averagenodisp}
\end{equation}
where $T_0$ is the solution of $F'(T_0)=0$, while the phase
$\phi_{ij}(T_0)$ is given by
\begin{equation}
  \phi_{ij}(T_0) = (E_i-E_j) T_0 - ({\bf p}_i-{\bf p}_j) \cdot {\bf L}
  \, .
  \label{phaseprobanodisp}
\end{equation}
Non-exponential factors are included in $N_{\tilde g}$. The
function $F(T)$ can be rewritten as
\begin{equation}
  F(T)= \frac{(\tilde{\bf v}_0 T - \tilde{\bf L})^2}{4\sigma_x^2}
  \, ,
\end{equation}
where
\begin{eqnarray*}
  \tilde{\bf v}_0 &=& \sigma_x\sqrt{\Sigma^{-1}} \, {\bf v}_0 \, ,
  \\
  \tilde{\bf L} &=& \sigma_x\sqrt{\Sigma^{-1}} \, {\bf L} \, ,
\end{eqnarray*}
with
$\Sigma^{-1}=diag(\sigma_x^{-2},\sigma_{x-}^{-2},\sigma_{x+}^{-2})$.
It is then easy to compute
\begin{eqnarray}
  T_0 &=& \frac{\tilde{\bf v}_0\cdot\tilde{\bf L}}{\tilde{\bf v}_0^2}
  \, , \label{valueTO} \\
  F(T_0) &=&
  \frac{(\tilde{\bf v}_0\times\tilde{\bf L})^2}{4\sigma_x^2\,\tilde{\bf v}_0^2}  \, ,
  \label{directioncond} \\
  F''(T_0) &=& \frac{\tilde{\bf v}_0^2}{2\sigma_x^2} \, .
  \label{secderiv}
\end{eqnarray}
Actually, the time interval of integration $\Delta T$ is finite.
The approximation $\Delta T\to\infty$ is good if $\Delta T$ is
larger than the width of the overlap function given by
Eq.~(\ref{secderiv}):
\begin{equation}
  |{\bf v}_0| \Delta T \gtrsim \tilde\sigma_{xeff} \, ,
  \label{deltaTnodisp}
\end{equation}
where
\begin{equation}
  \tilde\sigma_{xeff}= \frac{|{\bf v}_0|}{|\tilde{\bf v}_0|} \, \sigma_x \, .
  \label{defsigmatilde}
\end{equation}

The insertion of Eq.~(\ref{directioncond}) in
Eq.~(\ref{averagenodisp}) shows that $F(T_0)$ plays the role of a
directional constraint since $\int dT{\cal A}_i{\cal A}_j^*$ is
non-negligible if $\tilde{\bf L}$ is nearly parallel to
$\tilde{\bf v}_0$. In order to express this constraint in terms of
non-tilded quantities, two cases should be distinguished:
$\sigma_{x+}\sim\sigma_x$ and $\sigma_{x+}\gg\sigma_x$.

For $\sigma_{x+}\sim\sigma_x$, $\int dT{\cal A}_i{\cal A}_j^*$ is
non-negligible if ${\bf L}$ is nearly parallel to ${\bf v}_0$:
\begin{equation}
  {\bf L} = \frac{{\bf v}_0}{|{\bf v}_0|} \, L
  + {\cal O}(\sigma_{x+})  \, .
  \label{substi}
\end{equation}
With the substitution (\ref{substi}), the value of $T_0$ given by
Eq.~(\ref{valueTO}) becomes $T_0=L/|{\bf v}_0|+{\cal
O}(\sigma_{x+})$, so that the phase (\ref{phaseprobanodisp})
becomes
\begin{equation}
  \phi_{ij}(T_0)
  = \frac{\delta m_{ij}^2}{2p_0} (L + {\cal O}(\sigma_{x+})) \, ,
  \label{phaseT0}
\end{equation}
which is equal to the standard oscillation phase
(\ref{standardphase}) if $\sigma_{x+}\ll L$.

If $\sigma_{x+}\gg\sigma_x$, $\tilde{\bf v}_0$ and $\tilde L^z$
tend to zero so that $F(T_0)$ should directly be studied as a
function of ${\bf v}_0$ and ${\bf L}$. General conclusions for
arbitrary $\sigma_{x\pm}$ can be drawn from the study of the
quadratic form in $(L^x,L^y,L^z)$ associated with $F(T_0)=1$. This
analysis shows that there is an eigenvalue $s_3=0$ corresponding
to an eigenvector along ${\bf v}_0$. The two other eigenvalues
$s_{1,2}$ are positive (with $s_2\leq s_1$) , so that the surface
$F(T_0)=1$ in $(L^x,L^y,L^z)$-space is a cylinder of elliptical
section with an axis along ${\bf v}_0$. This geometrical picture
can be interpreted as imposing that the components of ${\bf L}$
orthogonal to ${\bf v}_0$ should be smaller than $\sqrt{1/s_2}$,
whereas there is no constraint at all on the component of ${\bf
L}$ along ${\bf v}_0$. In other words, the probability is maximal
within a cylinder of axis ${\bf v}_0$.

In the stationary limit ($\sigma_{x+}\to\infty$), the lengthy
expressions of the non-zero eigenvalues become
\begin{eqnarray}
  s_1 &\to& \frac{1}{4\sigma_x^2} \, ,
  \nonumber \\
  s_2 &\to& \frac{1}{4\sigma_x^2} \,
  \frac{\sigma_x^2\,(v_0^z)^2}
       {\sigma_{x+}^2(v_0^x)^2+\sigma_{x+}^2(v_0^y)^2+\sigma_{x}^2(v_0^z)^2} \, .
  \label{s12stat}
\end{eqnarray}
The properties of the stationary limit, namely
Eqs.~(\ref{statlim}), (\ref{statvo}) and (\ref{statwidth}), lead
to the bound $\sqrt{1/s_2}\lesssim S_{\scriptscriptstyle P,D}$. In the stationary
limit, the components $L^{x,y}$ (which are then orthogonal to
${\bf v}_0$) should thus be smaller than $S_{\scriptscriptstyle P,D}$. The
time-averaged probability is therefore always negligible in
directions other than along the average momentum ${\bf p}_0$. Thus
the evaluation of the phase $\phi_{ij}(T_0)$ yields
\begin{equation}
  \phi_{ij}(T_0)
  = \frac{\delta m_{ij}^2}{2p_0} (L + {\cal O}(S_{\scriptscriptstyle P,D})) \, ,
  \label{phaseT0stat}
\end{equation}
which is equal to the standard oscillation phase
(\ref{standardphase}) if $S_{\scriptscriptstyle P,D}\ll L$.

With the results (\ref{averagenodisp}), (\ref{directioncond}),
(\ref{secderiv}) and (\ref{phaseT0}) (or (\ref{phaseT0stat})), the
flavor-mixing transition probability (\ref{transitionproba}) can
be written for a propagation distance satisfying $S_{\scriptscriptstyle P,D}\ll L
\lesssim p_0/\sigma_p^2$ as
\begin{eqnarray}
  && {\cal P}_{\alpha \to \beta}({\bf L}) \sim N_{\tilde g} \,
  \exp \left( - \frac{(\tilde{\bf v}_0\times\tilde{\bf L})^2}
                     {2\sigma_x^2\,\tilde{\bf v}_0^2}
       \right)
  \sum_{i,j} V_{i\alpha} \, V_{i\beta}^* \, V_{j\alpha}^* \, V_{j\beta}
  \nonumber \\
  && \hspace{2cm} \times
  \exp \left(
      - 2\pi i\frac{L}{L_{ij}^{osc}}
      - \frac{(\delta m_i^2)^2+(\delta m_j^2)^2}{16\tilde\sigma_m^2E_0^2}
      - 2\pi^2\left( \frac{\tilde\rho\tilde\sigma_{xeff}}{L_{ij}^{osc}} \right)^2
      \right) \, ,
  \label{probanodisp}
\end{eqnarray}
where $\tilde \rho$ and $\tilde\sigma_{xeff}$ are defined by
Eqs.~(\ref{energytilde}) and (\ref{defsigmatilde}), respectively.
$N_{\tilde g}$ includes all prefactors independent of $L$ as well
as normalization factors. The oscillation length $L_{ij}^{osc}$ is
defined by
\begin{equation}
   L^{osc}_{ij} = \frac{4\pi |{\bf p}_0|}{\delta m_{ij}^2} \, .
   \label{osclengthnodisp}
\end{equation}
As discussed above, the probability is maximal within a cylinder
of axis ${\bf v}_0$ and radius smaller than $S_{\scriptscriptstyle P,D}$. There
is no decrease in $1/L^2$ since the spatial spreading of the `wave
packet' has not yet begun.

\subsection{Oscillations with transversal dispersion}
\label{osctransdisp}

\subsubsection{Amplitude}
\label{laplaceregimelongitud}

For $p_0/\sigma_p^2\lesssim L\lesssim L^{disp}_j$, the momentum
integration in Eq.~(\ref{jacobsachsinteg}) is done by the method
of stationary phase in the transverse directions and by Laplace's
method in the longitudinal direction, yielding in the stable case
(see Eq.~(\ref{transdisp}) with $\gamma_j(p_j)=0$)
\begin{equation}
  {\cal A}_j =
  \frac{Ng({\bf l}) \sigma_{peff}}{ T\sqrt{1+i T/T^{disp}_j} } \,
  \exp \left(
  -i E_j T + i p_j L
  - f_j(p_j)
  - \frac{1}{1+iT/T^{disp}_j} \frac{(v_jT-L)^2}{4\sigma_{xeff}^2}
  \right) \, ,
  \label{amplitudelaplace}
\end{equation}
with $f_j(p)$ and $\sigma_{peff}$ defined by
Eqs.~(\ref{definitionfj}) and (\ref{sigmaeff}), respectively. We
had not yet determined the value $p_j$ for which $f_j(p)$ is
extremal. The computation of $p_j$ to ${\cal O}(\epsilon)$ yields
\begin{equation}
  p_j=p_0 + (\rho-1)\, \frac{\delta m_j^2}{2p_0} \, ,
  \label{expansionpj}
\end{equation}
where the dimensionless number $\rho$ is defined by
\begin{equation}
   \rho = \sigma_{peff}^2
   \left(
   \frac{1}{\sigma_p^2}
   - \frac{v_{\scriptscriptstyle P}(v_0-v_{\scriptscriptstyle P})}{\sigma_{e\scriptscriptstyle P}^2}
   - \frac{v_{\scriptscriptstyle D}(v_0-v_{\scriptscriptstyle D})}{\sigma_{e\scriptscriptstyle D}^2}
   \right) \, ,
   \label{definitionrho}
\end{equation}
with $\sigma_p$ defined by Eq.~(\ref{definitionsigmap}). The
associated energy $E_j=\sqrt{p_j^2+m_j^2}$ and velocity
$v_j=p_j/E_j$ are given to ${\cal O}(\epsilon)$ by
\begin{eqnarray}
  E_j &=& E_0 + \rho \frac{\delta m_j^2}{2E_0} \, ,
  \label{expansionEj}
  \\
  v_j &=& v_0
  + \left(\rho(1-v_0^2)-1 \right)\frac{\delta m_j^2}{2p_0E_0} \, .
  \label{expansionvj}
\end{eqnarray}
The parameter $\rho$ has been defined so as to be in
correspondence with the notation of Ref.~\cite{Giunti98a}. The
symbol $\omega$ appearing in that article is related to our
notation by $\omega=\sigma_p^2/\sigma_{peff}^2$. Note that the
authors of Ref.~\cite{Giunti98a} do not compute $\rho$ explicitly
and also take the relativistic limit $v_0=1$. The explicit value
of $\rho$ is very interesting to know, since $\rho=0$ in the case
of stationary boundary conditions, in which case all mass
eigenstates have the same energy $E_0$.

The value of $f_j(p)$ at order $\epsilon^2$ is given by
\begin{equation}
   f_j(p_j) = \frac{1}{4\sigma_m^2} \left(
   \frac{\delta m_j^2}{2E_0}
   \right)^2 \;+\; {\cal O}(\epsilon^3) \, ,
   \label{valuefjpj}
\end{equation}
where $\sigma_m$ is defined by
\begin{equation}
   \frac{1}{\sigma_m^2} = \sigma_{peff}^2
   \left(
   \frac{1}{\sigma_p^2} \,
   \left( \frac{1}{\sigma_{e\scriptscriptstyle P}^2} + \frac{1}{\sigma_{e\scriptscriptstyle D}^2} \right)
   + \frac{(v_{\scriptscriptstyle P}-v_{\scriptscriptstyle D})^2}{\sigma_{e\scriptscriptstyle P}^2\sigma_{e\scriptscriptstyle D}^2}
   \right) \, .
   \label{omegam}
\end{equation}
$\sigma_m$ will be called the {\it mass width}, as it imposes a
constraint between the masses $m_{i,j}$ and the ingoing
energy-momentum (see section \ref{massconstraint}).

\subsubsection{Probability}
\label{sectionprobatrans}

As in section \ref{sectionprobanodisp}, the last step towards the
oscillation formula consists in computing the time average of the
transition probability. Actually, the computation can be
simplified by considering separately two cases. The examination of
Eq.~(\ref{valuefjpj}) shows indeed that the amplitude ${\cal A}_j$
(Eq.~(\ref{amplitudelaplace})) is negligible unless $|\delta
m_j^2/2E_0| \lesssim \sigma_m$. Together with the same condition
on ${\cal A}_i$, it puts a constraint on the mass difference:
$$
    \left( \frac{\delta m_{ij}^2}{2E_0} \right)^2 \lesssim 2\sigma_m^2
   \hspace{1cm} \mbox{where} \hspace{1cm} \delta m_{ij}^2 =m_i^2-m_j^2 \, .
$$
There are two possible cases:
\begin{enumerate}
   \item the masses are nearly degenerate;
   \item if not, the masses must be very small in comparison with the energy
   $E_0$, i.e.\ the oscillating particles are relativistic.
\end{enumerate}

We first consider the case of {\it nearly degenerate masses}. If
$|m_i-m_j| \ll m_i,m_j$, it is possible to make the approximation
$T^{disp}_i \cong T^{disp}_j$ and to work with only one dispersion
time defined by  $T^{disp}=E_0^3/2\tilde m_0^2\sigma_{peff}^2$,
with $\tilde m_0$ the mass in the degenerate limit. The integrand
in (\ref{transitionproba}) can be written as
$$
   {\cal A}_i {\cal A}_j^* =
   \frac{N^2g^2({\bf l})\sigma_{peff}^2}{T^2\sqrt{1+(T/T^{disp})^2}} \,
   \exp \left( - \frac{(\delta m_i^2)^2+(\delta m_j^2)^2}{16\sigma_m^2E_0^2}
              -i \phi_{ij}(T,L) - f_{ij}(T,L)
        \right) \, ,
$$
with the phase given by
\begin{equation}
   \phi_{ij}(T,L) = (E_i-E_j) T - (p_i-p_j) L
   - \frac{T}{T^{disp}} \, \frac{1}{1+(T/T^{disp})^2} \,
   \frac{ (v_iT-L)^2 - (v_jT-L)^2 }{4\sigma_{xeff}^2} \, ,
   \label{defphaseij}
\end{equation}
and the function $f_{ij}(T,L)$ given by
\begin{equation}
   f_{ij}(T,L) = \frac{1}{1+(T/T^{disp})^2} \,
   \frac{(v_iT-L)^2+(v_jT-L)^2}{4\sigma_{xeff}^2} \, .
   \label{definitionfij}
\end{equation}
The prefactor $N^2$ absorbs numerical constants.

The time integral in Eq.~(\ref{transitionproba}) can be evaluated
with Laplace's method. We want to compute the transition
probability to order $\epsilon^2$ in the real part of argument of
the exponential, and to order $\epsilon$ in the phase. This again
implies computing the minimum $T_{ij}$ of $f_{ij}(T,L)$ to order
$\epsilon$, $f_{ij}(T_{ij},L)$ to order $\epsilon^2$, its second
derivative to order $\epsilon^0$, and the first derivative of the
phase to order $\epsilon$. The minimum of $f_{ij}(T,L)$ is reached
to order $\epsilon^0$ for $T=L/v_0$, and to order $\epsilon$ for
\begin{equation}
   T_{ij} = \frac{L}{v_0}
   \left( 1 - \frac{v_i +  v_j - 2v_0}{2v_0} \right)
   \;+\; {\cal O}(\epsilon^2) \, ,
   \label{defTij}
\end{equation}
If the expansion (\ref{expansionvj}) is used, the minimum of
$f_{ij}(T,L)$ reads
\begin{eqnarray}
   f_{ij}(T_{ij},L) &=&
   \frac{L^2}{1+\ell^2} \, \frac{(v_i-v_j)^2}{8v_0^2\sigma_{xeff}^2}
   \nonumber \\
   &=& \frac{E_0^4}{8\sigma_{peff}^2 \tilde m_0^4} \, \frac{\ell^2}{1+\ell^2} \,
   (\rho(1-\tilde v_0^2) - 1)^2 \, \left( \frac{\delta m_{ij}^2}{2p_0} \right)^2
   \;+\; {\cal O}(\epsilon^3) \, ,
   \label{minfijlaplace}
\end{eqnarray}
where $\ell= L/(v_0T^{disp})$ and $1-\tilde v_0^2=\tilde
m_0^2/E_0^2$. The velocity $v_0$ has been replaced by $\tilde v_0$
in $v_{i,j}$ because $|\rho(v_0^2-\tilde v_0^2)|\ll 1$. The value
of the second derivative of $f_{ij}(T,L)$ reads
\begin{equation}
   \frac{1}{2} \frac{d^2f_{ij}}{dT^2}(T_{ij},L) =
   \frac{2 v_0^2 \sigma_{peff}^2}{1+\ell^2} \;+\; {\cal O}(\epsilon)\, .
   \label{fijlaplaceprime}
\end{equation}
The value of the phase at $T=T_{ij}$ is
\begin{equation}
   \phi_{ij}(T_{ij},L) = \frac{\delta m_{ij}^2}{2 p_0} \, L
   \;+\; {\cal O}(\epsilon^2) \, .
   \label{laplacephase}
\end{equation}
The derivative of the phase is given to order $\epsilon$ by
\begin{eqnarray}
   \frac{d\phi_{ij}}{dT}(T_{ij},L) &=&
   E_i - E_j \,-\, \frac{2 \ell }{1+\ell^2} \, \sigma_{peff}^2 \, (v_i-v_j) \, L
   \nonumber \\
   &=& \frac{\rho(1-\tilde v_0^2) + \ell^2}{1+\ell^2} \,
     \frac{E_0 \delta m_{ij}^2}{2 \tilde m_0^2} \;+\; {\cal O}(\epsilon^2) \, ,
   \label{laplacephaseprime}
\end{eqnarray}
where the expansions (\ref{expansionEj})-(\ref{expansionvj}) have
been used.  As above, the velocity $v_0$ has been replaced by
$\tilde v_0$ in $v_{i,j}$ because $|\rho(v_0^2-\tilde v_0^2)|\ll
1$. The second derivative of the phase is of order $\epsilon$ and
thus does not contribute to the transition probability to order
$\epsilon^2$, at least in the argument of the exponential.

The approximation of the time average integration by Laplace's
method can now be done and yields
\begin{eqnarray}
   && \int dT \, {\cal A}_i \; {\cal A}_j^* =
   v_0N^2\sigma_{peff} \,
   \frac{g^2({\bf l})}{L^2}
   \nonumber \\ && \hspace{1.5cm} \times
   \exp \left(
   - 2 \pi i \frac{L}{L^{osc}_{ij}}
   - \frac{(\delta m_i^2)^2+(\delta m_j^2)^2}{16\sigma_m^2E_0^2}
   - 2 \pi^2 \left( \frac{\rho \,  \sigma_{xeff}}{L^{osc}_{ij}} \right)^2
   - \left( \frac{L}{L^{coh}_{ij}} \right)^2
        \right) \, ,
   \label{average}
\end{eqnarray}
where $N^2$ absorbs numerical constants. The oscillation length
$L^{osc}_{ij}$ for the masses $m_i$ and $m_j$ is given by
\begin{equation}
   L^{osc}_{ij} = \frac{4\pi p_0}{\delta m_{ij}^2} \, .
   \label{osclength}
\end{equation}
Without loss of generality, $\delta m_{ij}^2$ is taken to be
positive. The coherence length $L^{coh}_{ij}$ is defined by
\begin{equation}
   L^{coh}_{ij} = \frac{1}{\sqrt{2}\pi} \, \frac{p_0}{\sigma_{peff}}
                  L^{osc}_{ij} \, .
   \label{cohlength}
\end{equation}
It is assumed that the time interval $\Delta T$ used to average is
larger than the width of the overlap function:
\begin{equation}
  v_0\Delta T \gtrsim \sigma_{xeff} \hspace{1cm} (\ell\ll1) \, .
  \label{deltaTtransverse}
\end{equation}
Moreover, the time separation $T^{sep}$ between the wave packet
peaks is supposed to be smaller than the time interval $\Delta T$.
This assumption is true as long as the distance is not hugely
larger than the oscillation length, since Eq.~(\ref{defTij}) shows
that $T^{sep}$ is of the order of
\begin{equation}
   T^{sep} = |T^{ii}-T^{jj}| \sim \frac{L}{L^{osc}_{ij}} \,
\frac{1}{v_0p_0} \, ,
   \label{defTsep}
\end{equation}
where $1/v_0p_0$ is of the order of a microscopic time. The
violation of the condition $\Delta T \gtrsim T^{sep}$ gives rise
to the interesting possibility of detecting separate pulses,
corresponding to the different mass eigenstates
\cite{Kayser81,Reinartz85}. For example, a time-dependence of the
neutrino burst from the supernova SN1987A has been searched for
(see \cite{Bahcall,Raffelt} for reviews), but only upper mass
limits have been derived.

Let us now consider relativistic particles with {\it very
different masses}. Without loss of generality, we suppose that
$m_i\gg m_j$.  As $T^{disp}_i \ll T^{disp}_j$, the approximation
of taking only one dispersion time for the two mass eigenstates is
not valid anymore. Therefore, we shall suppose that $L \lesssim
L^{disp}_i$ and show that the decoherence sets in before the
dispersion length $L^{disp}_i$ is reached. The calculation of the
time average proceeds as in the nearly mass degenerate case, with
the approximations $\ell_i^2 \ll 1$ (with $\ell_i=L/L_i^{disp}$)
and $\rho \ll \gamma_0^2$ (the particle is relativistic), so that
the same result is obtained (see Eq.~(\ref{average})). Now, the
coherence length $L^{coh}_{ij}$ is shorter than the dispersion
length $L^{disp}_i$ if $\delta m_{ij}^2/m_i^2 \gtrsim
\sigma_{peff}/E_0$. This condition is always true for very
different masses, since in that case $\delta m_{ij}^2/m_i^2 \cong
1$. Therefore, the interference becomes negligible before the
dispersion of the heaviest mass eigenstate begins, so that the
result (\ref{average}) is also valid for very different masses.

In conclusion, the flavor-mixing transition probability
(\ref{transitionproba}) for a propagation distance ${\bf
L}=L\,{\bf l}$ satisfying $p_0/\sigma_p^2\lesssim L\lesssim
min(L^{disp}_i,L^{disp}_j)$ has the same form whatever the mass
values and reads
\begin{eqnarray}
   && {\cal P}_{\alpha \to \beta}({\bf L}) \sim
   v_0N^2\sigma_{peff} \,
   \frac{g^2({\bf l})}{L^2} \,
   \sum_{i,j} V_{i\alpha} \, V_{\beta i}^{-1} \,
          V_{j\alpha}^* \, V_{\beta j}^{-1 \, *} \,
   \nonumber \\
   && \hspace{1.5cm} \times \exp \left( -2 \pi i \frac{L}{L^{osc}_{ij}}
   - \frac{(\delta m_i^2)^2+(\delta m_j^2)^2}{16\sigma_m^2E_0^2}
   - 2 \pi^2 \left( \frac{\rho \, \sigma_{xeff}}{L^{osc}_{ij}} \right)^2
   - \left( \frac{L}{L^{coh}_{ij}} \right)^2
               \right) \, .
   \label{averagelaplace}
\end{eqnarray}

\subsection{Oscillations with longitudinal dispersion}
\label{osclongdisp}

\subsubsection{Amplitude}

For $L\gtrsim L^{disp}_j$, the momentum integration in
Eq.~(\ref{jacobsachsinteg}) is done by the method of stationary
phase in all directions, yielding in the stable case (see
Eq.~(\ref{longdisp}) with $\gamma_j(p_{cl,j})=0$):
\begin{equation}
  {\cal A}_j =
  \frac{Ng({\bf l})\sigma_{peff}}{ T\sqrt{1+i T/T^{disp}_j} } \,
  \exp \left( - i m_j \, \sqrt{T^2-L^2} - f_j(p_{cl,j})
              + \sigma_{peff}^2 \,
                \frac{ \left( f_j'(p_{cl,j}) \right)^2 }{1+i T/T^{disp}_j}
       \right)  \, ,
  \label{amplitudephasestat}
\end{equation}
where $f_j(p)$ and $\sigma_{peff}$ are defined by
Eqs.~(\ref{definitionfj}) and (\ref{sigmaeff}), respectively. The
stationary point is given by Eq.~(\ref{statpoint}):
\begin{equation}
   p_{cl,j} = \frac{m_j v_{cl}}{\sqrt{1-v_{cl}^2}}
   = \frac{m_j L}{\sqrt{T^2-L^2}}\, ,
   \label{pclass}
\end{equation}
The corresponding energy is given by
\begin{equation}
   E_{cl,j} = \sqrt{p_{cl,j}^2+m_j^2} = \frac{m_j T}{\sqrt{T^2-L^2}} \, .
   \label{Eclass}
\end{equation}
The value of $f_j(p_{cl,j})$ is given by $f_j(p_{cl,j})=f_{j\scriptscriptstyle
P}(p_{cl,j})+f_{j\scriptscriptstyle D}(p_{cl,j})$ with
\begin{equation}
   f_{j\scriptscriptstyle P,D}(p_{cl,j}) =
   \frac{\left( p_{cl,j} - p_0 \right)^2}{4\sigma_{p\scriptscriptstyle P,D}^2} \,+\,
   \frac{\left( E_{cl,j} -E_0 -( p_{cl,j} -p_0 ) v_{\scriptscriptstyle P,D} \right)^2}
        {4\sigma_{e\scriptscriptstyle P,D}^2} \, ,
   \label{definitionfjPD}
\end{equation}
while the value of $f_j'(p_{cl,j})$ is given by
$f_j'(p_{cl,j})=f_{j\scriptscriptstyle P}'(p_{cl,j})+f_{j\scriptscriptstyle D}'(p_{cl,j})$, with
$$
   f_{j\scriptscriptstyle P,D}'(p_{cl,j}) =
   \frac{ p_{cl,j} - p_0 }{2\sigma_{p\scriptscriptstyle P,D}^2} \,+\,
   (v_{cl} - v_{\scriptscriptstyle P,D}) \,
   \frac{E_{cl,j} - E_0 - (p_{cl,j}- p_0) v_{\scriptscriptstyle P,D}}
        {2\sigma_{e\scriptscriptstyle P,D}^2} \, .
$$

\subsubsection{Probability}

As in sections \ref{sectionprobanodisp} and
\ref{sectionprobatrans}, the last step towards the oscillation
formula consists in computing the time average of the transition
probability. Examination of the term $f_j(p_{cl,j})$ in
Eq.~(\ref{amplitudephasestat}) shows that the amplitude ${\cal
A}_j$ is nearly zero unless $|p_{cl,j} - p_0| \lesssim
\sigma_{p\scriptscriptstyle P,D}$ (see Eq.~(\ref{definitionfjPD})). This
condition means that the interference term ${\cal A}_i{\cal
A}_j^*$ will be negligible unless $|p_{cl,i} - p_{cl,j}| \lesssim
\sigma_{p\scriptscriptstyle P,D}$, that is $|\delta m_{ij}|/m \lesssim
\sigma_{p\scriptscriptstyle P,D}/p$ (with $m$ referring to $m_i$ or $m_j$). In
other words, the interference term is negligible if the masses are
not nearly degenerate. This result is in agreement with the
conclusion of section \ref{sectionprobatrans}: if the masses $m_i$
and $m_j$ are very different, decoherence sets in before the
dispersion length is reached.

It is thus possible to work with only one dispersion time
$T^{disp} = T^{disp}_i \cong T^{disp}_j$. It is understood that
for noninterference terms, i.e.\ $\int dT {\cal A}_j{\cal A}_j^*$,
the dispersion time $T^{disp}$ will be taken to be $T^{disp}_j$.
The integral to be computed is $\int dT {\cal A}_i{\cal A}_j^*$,
with
\begin{equation}
   {\cal A}_i{\cal A}_j^* =
   \frac{N^2g^2({\bf l})\sigma_{peff}^2}{T^2\sqrt{1+(T/T^{disp})^2}} \,
   \exp \left( - i\tilde \phi_{ij}(T,L) -\tilde f_{ij}(T,L) \right) \, ,
   \label{interferencephasestat}
\end{equation}
with the phase given by
\begin{equation}
   \tilde \phi_{ij}(T,L) = \delta m_{ij} \, \sqrt{T^2-L^2}
   + \sigma_{peff}^2 \, \frac{T}{T^{disp}} \,
     \frac{ \Big(f_i'(p_{cl,i}) \Big)^2 - \left(f_j'(p_{cl,j}) \right)^2 }
          {1+(T/T^{disp})^2} \, ,
   \label{defphaseijtilde}
\end{equation}
and the function $\tilde f_{ij}(T,L)$ defined by
$$
   \tilde f_{ij}(T,L) = f_i(p_{cl,i}) + f_j(p_{cl,j})
   - \sigma_{peff}^2 \,
     \frac{ \Big(f_i'(p_{cl,i}) \Big)^2 + \left(f_j'(p_{cl,j}) \right)^2 }
          {1+(T/T^{disp})^2} \, .
$$

We would like to evaluate $\int dT {\cal A}_i{\cal A}_j^*$ by
Laplace's method, in the same way as in section
\ref{sectionprobatrans}. This involves expanding the argument of
the exponential (\ref{interferencephasestat}) in powers of small
mass differences. However, we should take care not to expand $m_j$
around $m_0$, as there is no guarantee, in the relativistic case,
that $\delta m_j=m_j-m_0$ is much smaller than $m_0$. This is
linked to the extreme sensitivity on $T$ of the factor
$\sqrt{T^2-L^2}$ in the relativistic case. Thus the masses $m_i$
and $m_j$ should be expanded around $\tilde m_0 =(m_i+m_j)/2$. A
new expansion parameter is defined by $\delta \tilde
m_j=m_j-\tilde m_0$. The parameters $\delta \tilde m_j$ are said
to be of order $\epsilon$.

Unfortunately, the value $T=\tilde T_0$ minimizing $\tilde
f_{ij}(T,L)$ to order $\epsilon^0$ cannot be computed exactly. An
approximate solution would be any $T$ satisfying approximately
$p_{cl,j}\cong p_0$ and the exact solution can be computed by
perturbation around it. If $T=\tilde E_0 L/p_0$ is chosen as the
approximate solution, with $\tilde E_0=\sqrt{p_0^2+\tilde m_0^2}$
(this solution satisfies $p_{cl,j}= p_0$ to order $\epsilon^0$),
the value of $\tilde T_0$, to order $\delta \tilde m_0^2=\tilde
m_0^2-m_0^2$, reads
$$
   \tilde T_0 = \frac{\tilde E_0 L}{p_0} \,+\,
   \frac{\tilde m_0^2 \delta \tilde m_0^2}{2p_0^3 E_0} \, (1-\rho) L \, ,
$$
where $\rho$ is defined by Eq.~(\ref{definitionrho}). The
expansions in $\delta \tilde m_0^2$ around $m_0$ and in $\delta
\tilde m_j$ around $\tilde m_0$ will be performed to the same
order. It can be checked that the value of $T$ minimizing $\tilde
f_{ij}(T,L)$ to order $\epsilon$ is still given by $T=\tilde T_0$
(the reason is that $\tilde m_0$ is the average of $m_i$ and
$m_j$).

The value of $\tilde f_{ij}(T,L)$ at its minimum reads
\begin{equation}
   \tilde f_{ij}(\tilde T_0,L) =
       \frac{\tilde m_0^2 (\delta m_{ij})^2 + (\delta \tilde m_0^2)^2}
            {8 \sigma_m^2 E_0^2}
   \;+\; \frac{E_0^4}{8 \tilde m_0^4 \sigma_{peff}^2} \,
       \frac{\ell^2}{1+\ell^2} \,
       (\rho(1-\tilde v_0^2)-1)^2 \,
       \left( \frac{\tilde m_0 \delta m_{ij}}{p_0} \right)^2
   \;+\; {\cal O}(\epsilon^3) \, ,
   \label{minfijstat}
\end{equation}
where $\ell=L/(v_0T^{disp})$ as before, $\sigma_m$ is defined by
Eq.~(\ref{omegam}) and $\tilde v_0=p_0/\tilde E_0$.

The value of the second derivative of $\tilde f_{ij}(T,L)$ with
respect to $T$ reads
\begin{equation}
   \frac{1}{2} \frac{d^2\tilde f_{ij}}{dT^2}(\tilde T_{ij},L) =
   \frac{2 v_0^2 \sigma_{peff}^2}{1+\ell^2}
   \;+\, {\cal O}(\epsilon) \, .
   \label{fijstatprime}
\end{equation}
This equation shows that the spatial width (linked to the time
width by a factor $v_0$) of the wave packet associated to the
oscillating particle increases linearly with $L$ for $\ell \gg 1$
(i.e.\ at a distance much larger than the dispersion length):
\begin{equation}
   spatial \; width \; = \sigma_{xeff} \, \sqrt{1+\ell^2}
   \sim \frac{\sigma_{peff}}{p_0} \, \frac{m_j^2}{E_0^2} \, L \, .
   \label{spacewidth}
\end{equation}
This expression agrees with the quantum-mechanical result
\cite{Kayser81,Kimbis}, derived by observing that the relation
\mbox{$\delta t=\frac{m^2}{p^2E}L\delta p$}, obtained from
$t=\frac{L}{v}$ (with $L$ kept fixed), leads to \mbox{$\delta
L=\frac{m^2}{E^2} \frac{\sigma_p}{p} L$}.

The value of the phase at $T=\tilde T_0$ is
\begin{equation}
   \tilde \phi_{ij}(\tilde T_0,L) = \frac{\tilde m_0\delta m_{ij}}{p_0} \, L
   =\frac{\delta m_{ij}^2}{2p_0} \, L \;+\; {\cal O}(\epsilon^2) \, ,
   \label{statphase}
\end{equation}
where as before $\delta m_{ij}^2=m_i^2-m_j^2$. The derivative of
the phase is given by
\begin{equation}
   \frac{d\tilde\phi_{ij}}{dT}(\tilde T_0,L) =
   \frac{\rho (1-\tilde v_0^2) +  \ell^2}{1+\ell^2} \,
   \frac{E_0 \delta m_{ij}}{\tilde m_0}
   \;+\; {\cal O}(\epsilon^2) \, .
   \label{statphaseprime}
\end{equation}
The second derivative of the phase is of order $(\delta \tilde
m_j)^2$ and can thus be neglected. Eqs.~(\ref{minfijstat}),
(\ref{fijstatprime}), (\ref{statphase}) and (\ref{statphaseprime})
can be compared with Eqs.~(\ref{minfijlaplace}),
(\ref{fijlaplaceprime}), (\ref{laplacephase}) and
(\ref{laplacephaseprime}). The relation $\delta m_{ij}=\delta
m_{ij}^2/2\tilde m_0$, valid for nearly degenerate masses, may be
used. From that comparison, it is clear that the final result will
be the same as Eq.~(\ref{averagelaplace}).

In conclusion, the flavor-mixing transition probability
(\ref{transitionproba}) for a propagation distance ${\bf
L}=L\,{\bf l}$ satisfying $L \gtrsim L^{disp}$ reads
\begin{eqnarray}
   && {\cal P}_{\alpha \to \beta}({\bf L}) \sim
   v_0N^2\sigma_{peff} \,
   \frac{g^2({\bf l})}{L^2} \,
   \sum_{i,j} V_{i\alpha} \, V_{\beta i}^{-1} \,
          V_{j\alpha}^* \, V_{\beta j}^{-1 \, *}
   \nonumber \\
   && \hspace{1.5cm} \times \exp \left(  -2 \pi i \frac{L}{L^{osc}_{ij}}
   - \frac{(\delta m_i^2)^2+(\delta m_j^2)^2}{16 \sigma_m^2 E_0^2}
   - 2 \pi^2 \left( \frac{\rho \, \sigma_{xeff}}{L^{osc}_{ij}} \right)^2
   - \left( \frac{L}{L^{coh}_{ij}} \right)^2
               \right) \, ,
   \label{averagephasestat}
\end{eqnarray}
where the relation $2\tilde m_0^2 (\delta m_{ij})^2 + 2 (\delta
\tilde m_0^2)^2= (\delta m_i^2)^2+(\delta m_j^2)^2$, valid for
nearly degenerate masses, has been used, and with $L^{osc}_{ij}$,
$L^{coh}_{ij}$ defined respectively by Eqs. (\ref{osclength}) and
(\ref{cohlength}). Recall that we can set $L_i^{disp}\cong
L_j^{disp}=L^{disp}$ in this regime, since decoherence occurs in
the transversal-dispersion regime if the mass eigenstates are not
nearly degenerate.

As in section \ref{sectionprobatrans}, the time interval $\Delta
T$ used to average is assumed to be large enough. At first sight,
we should have $v_0\Delta T \gtrsim\ell\sigma_{xeff}$ in the limit
$\ell\gg1$ but this condition is always violated at some large
distance. Actually, it is sufficient to suppose that $\Delta T
\gtrsim \frac{T^{osc}}{\gamma^2}$ ($\gamma=\frac{E_0}{\tilde m_0}$
is the usual Lorentz factor) so that the phase can freely
oscillate around its average value (see Eq.~(\ref{statphaseprime})
in the limit $\ell\gg1$). This last condition is easily satisfied.
Putting together this condition with the conditions on $\Delta T$
valid for the no-dispersion and transversal-dispersion regime
(Eqs.~(\ref{deltaTnodisp}) and (\ref{deltaTtransverse})), we
obtain (with $\tilde\sigma_{xeff}\sim\sigma_{xeff}$):
\begin{equation}
  \Delta T \gtrsim
  max\left(
  \frac{\sigma_{xeff}}{v_0},\frac{T^{osc}_{ij}}{\gamma^2}
  \right) \, .
  \label{deltaTboth}
\end{equation}
The relative strength of the constraints present in
Eq.~(\ref{deltaTboth}) is linked to the relative values of the
dispersion and coherence lengths:
$$
  \frac{\sigma_{xeff}}{v_0} \gtrless \frac{T^{osc}_{ij}}{\gamma^2}
  \hspace{5mm} \leftrightarrow \hspace{5mm}
  L^{disp}_{ij} \gtrless L^{coh}_{ij} \, .
$$

It is striking that two different methods of approximation, a
priori valid in different regimes, give the same oscillation
formula, Eqs.~(\ref{averagelaplace}) and (\ref{averagephasestat}).
The dispersion length $L^{disp}$ does not play any special role in
the final result. Each method is thus accurate enough to be
extended to the whole range of distances. However, it will be seen
in section~\ref{section6} that the physical interpretation depends
on the relative values of $L$ and $L_{i,j}^{disp}$.

\subsection{A shortcut}
\label{shortcut}

We now explain another method to derive the oscillation formula
(in the transversal- and longitudinal dispersion regime) which
does not require the notion of a dispersion length. Although this
method is shorter, it is not obvious how to apply it to unstable
oscillating particles. Moreover it does not make the wave packet
picture clearly apparent.

Returning to Eq.~(\ref{defAj}), we can choose to integrate on the
3-momentum before doing the energy integral. In particular, the
integration on the 3-momentum can be done with the help of the
Grimus-Stockinger theorem \cite{Grimus96}. Let $\psi({\bf p})$ be
a 3 times continuously differentiable function on ${\bf R}^3$ such
that $\psi$ itself and all its first and second derivatives
decrease at least like $1/{\bf p}^2$ for $|{\bf p}| \to\infty$.
Then, for a real number $A>0$,
$$
   \int d^3p \,
   \frac{\psi({\bf p}) \, e^{i \, \bf p \cdot L}}{A - {\bf p}^2 + i \epsilon}
   \; \stackrel{ L \to \infty }{\longrightarrow} \;
   - \frac{2\pi^2}{L} \, \psi(\sqrt{\!A}\,{\bf l}) \, e^{i\sqrt{\!A}L}
   + {\cal O}(L^{-3/2}) \, ,
$$
where $L=|{\bf L}|$ and ${\bf l}={\bf L}/L$. For $A<0$, the
integral decreases like $L^{-2}$.

The remaining energy integral in the amplitude (\ref{defAj}) can
be done by a saddle-point approximation \cite{Giunti98a}. However,
it is quicker to perform first the time average in the probability
(\ref{transitionproba}), which yields a delta function, and makes
one of the energy integrations trivial:
\begin{equation}
   \int dT \, {\cal A}_i {\cal A}_j^* = \frac{N^2}{L^2} \,
   \int dE \, \psi(E,q_i\,{\bf l}) \,
   \psi^*(E,q_j\,{\bf l}) \,
   e^{ i (q_i-q_j)L } \, ,
   \label{incohsumQFT}
\end{equation}
where $\psi(E,{\bf p})$ is the overlap function defined by
Eqs.~(\ref{overlapPD})-(\ref{overlapfP}) and
$q_j=\sqrt{E^2-m_j^2}$. $N^2$ absorbs numerical constants.
Actually the time interval of integration $\Delta T$ is finite, so
that the the delta function is only an approximation certainly
valid\footnote{As seen from Eq.~(\ref{deltaTboth}), this condition
can be weakened to $\Delta T\gtrsim
max(v_0^{-1}\sigma_{xeff},\gamma^{-2}T^{osc}_{ij})$.} for $\Delta
T\gtrsim T^{osc}_{ij}$.

Eq.~(\ref{incohsumQFT}) shows that the transition probability can
be interpreted as an incoherent sum (i.e.\ occurring in the
probability) over energy eigenstates: interference occurs only
between the components of $\psi(E,{\bf p})$ having the same energy
\cite{Sudarsky91}. In this way, the correspondence between models
with and without stationary boundary conditions is obvious: {\it
the time-integrated nonstationary probability is equivalent to the
energy-integrated stationary probability}. For example, the
oscillation formula obtained by Grimus and Stockinger with
stationary boundary conditions \cite{Grimus96} has the form of the
integrand in the right-hand side of Eq.~(\ref{incohsumQFT}). This
equivalence confirms that the stationary case can be obtained from
the more general nonstationary case in the limit of a vanishing
energy width. Note however that the stationary limit cannot be
realized in experiments and that the oscillation formula is always
averaged over the energy spectrum.

In section \ref{poleintegrations}, we have shown how a wave packet
can be associated with the amplitude ${\cal A}_i$, so that
oscillations can be seen, like in the quantum-mechanical
treatment, as the result of an interference between propagating
wave packets. Thus the equivalence (\ref{incohsumQFT}) shows that
this physical picture still holds in the case of stationary
boundary conditions, provided that an incoherent sum over the
energy is performed, contrary to what was claimed in
Ref.~\cite{Grimus99}. Therefore a wave packet picture can always
be associated with the oscillation formula (once the incoherent
energy average has been done), though this physical picture is
well hidden in the formalism using stationary boundary conditions.

This equivalence between the time-integrated nonstationary
probability and the energy-integrated stationary probability is
similar to the equivalence we have met, in quantum-mechanical
models, between intermediate wave packet models and stationary
approaches (see Eq.~(\ref{incohsum})). However, the question of
the equality of propagation times does not crop up in the quantum
field theory formalism. Recall that integrations over microscopic
space-time variables are included in the overlap function, with
the result that the phase depends only on the average time $T$ and
$L$.

Note that the time average on the probability in sections
\ref{osctransdisp} and \ref{osclongdisp} could also have been done
before the longitudinal-momentum integration in the amplitude,
apparently yielding a delta function which makes one of the
momentum integrals trivial. However this method is spoilt in 3
dimensions by the prefactor $|T-i\mu|^{-2}$. Instead of a delta
function, one obtains a delta function look-alike of width
$\mu^{-1}\sim \sigma_p^2/E_0$, introducing an additional momentum
uncertainty which is larger than the mass difference $\delta
m^2/E$ since $L^{osc}_{ij}\gtrsim p_0/\sigma_p^2$. For this
reason, it was preferable to avoid this shortcut in sections
\ref{osctransdisp} and \ref{osclongdisp} (though it yields the
same final answer as given by the following method). Moreover, it
was interesting for the physical interpretation to postpone the
time average, so as to obtain the explicit dependence of the
amplitude on time and distance as shown in
Eqs.~(\ref{amplitudelaplace}) and (\ref{amplitudephasestat}).

If the coordinate system is chosen so that ${\bf L}$ is oriented
along a coordinate axis, it is easy to rewrite the integral
(\ref{incohsumQFT}) as
\begin{equation}
   \int dT \, {\cal A}_i {\cal A}_j^*
   = \frac{N^2g^2({\bf l})}{L^2} \,
   \int dE \; e^{ i (q_i-q_j)L - f_i(E) - f_j(E) } \, ,
   \label{timeaverageshort}
\end{equation}
with the definitions $f_j(E)=f_{j\scriptscriptstyle P}(E) + f_{j\scriptscriptstyle D}(E)$ and
\begin{equation}
   f_{j\scriptscriptstyle P}(E) =
   \frac{ \left( \sqrt{E^2-m_j^2}-p_0 \right)^2}{4\sigma_{p \scriptscriptstyle P}^2}
   + \frac{ \left( E - E_0 - \left( \sqrt{E^2-m_j^2} - p_0 \right)
      v_{\scriptscriptstyle P} \right)^2}{ 4\sigma_{e\scriptscriptstyle P}^2 } \, ,
   \label{definitionfjPshort}
\end{equation}
where $v_{\scriptscriptstyle P}$ and $p_0$ are the components of ${\bf v}_{\scriptscriptstyle
P}$ and ${\bf p}_0$ along ${\bf L}$, while $E_0$ has been
redefined so as to absorb the transversal part of ${\bf p}_0\cdot
{\bf v}_{\scriptscriptstyle P}$. The definition of $f_{j\scriptscriptstyle D}(E)$ is similar.
The geometrical constraint $g({\bf l})$ is defined by
Eq.~(\ref{geom}).

Since the phase in Eq.~(\ref{timeaverageshort}) has no stationary
point, there is no problem in using Laplace's method to integrate
over the energy $E$. The integrand is maximal for
$$
   E_{ij} = E_0 \,+\, \rho \, \frac{\delta m_i^2 + \delta m_j^2}{4 E_0}
   \;+\;  {\cal O}(\epsilon^2) \, ,
$$
where the dimensionless number $\rho$ is defined by
Eq.~(\ref{definitionrho}), and $\epsilon\sim\delta m_i^2\sim\delta
m_j^2$.

The value of $f_i(E) + f_j(E)$ at the extremum reads
$$
   f_i(E_{ij}) + f_j(E_{ij}) =
   \frac{(\delta m_i^2)^2+(\delta m_j^2)^2}{16 \sigma_m^2 E_0^2}
   + 2 \pi^2 \left( \frac{\rho \, \sigma_{xeff}}{L^{osc}_{ij}} \right)^2 \,
   + \, {\cal O}(\epsilon^3) \, ,
$$
where $\sigma_{xeff}$,$\sigma_m$ and $L^{osc}_{ij}$ are defined by
Eqs.~(\ref{sigmaeff}), (\ref{omegam}) and (\ref{osclength}),
respectively. The value of the second derivative of $f_i(E) +
f_j(E)$ at the extremum reads
\begin{equation}
   \frac{1}{2} \frac{d^2(f_i+f_j)}{dE^2}(E_{ij}) =
   \frac{1}{2 v_0^2 \sigma_{peff}^2}
   \;+\, {\cal O}(\epsilon) \, .
   \label{fijshortprime}
\end{equation}
The expansion of the phase around the extremum reads
\begin{equation}
   \phi_{ij}(E) \cong - \frac{\delta m_{ij}^2}{2p_0} \, L \,
   +  \, \frac{\delta m_{ij}^2}{2p_0^2v_0} \, L \, (E-E_{ij}) \, .
   \label{shortcutphase}
\end{equation}
The second derivative of the phase is of order $\epsilon$ and can
be neglected with respect to the second derivative of $f_i+f_j$.
The approximation of the integral (\ref{timeaverageshort}) by
Laplace's method is now straightforward and yields
\begin{eqnarray}
   && \int dT \, {\cal A}_i {\cal A}_j^*
   = v_0N^2\sigma_{peff} \,
   \frac{g^2({\bf l})}{L^2} \,
   \nonumber \\ && \hspace{1.5cm}\times
   \exp \left(
   - 2 \pi i \frac{L}{L^{osc}_{ij}}
   - \frac{(\delta m_i^2)^2+(\delta m_j^2)^2}{16 \sigma_m^2 E_0^2}
   - 2 \pi^2 \left( \frac{\rho \, \sigma_{xeff}}{L^{osc}_{ij}} \right)^2
   - \left( \frac{L}{L^{coh}_{ij}} \right)^2
        \right) \, .
   \label{averageshortcut}
\end{eqnarray}
The linear superposition of the different partial transition
probabilities $\int dT \, {\cal A}_i {\cal A}_j^*$ gives the same
result as those obtained in the transverse- and
longitudinal-dispersion regimes (Eqs.~(\ref{averagelaplace}) and
(\ref{averagephasestat})). Though the computation is shorter, the
origin of the decoherence and localization terms is not as clear
(see section~\ref{section6}), because `wave packets' in
configuration space do not appear explicitly at any stage of the
calculation. Moreover the oscillation formula (\ref{probanodisp})
valid in the no-dispersion regime cannot be exactly reproduced.

\section{Analysis of the probability in the stable case}
\label{section6}

In section~\ref{section5}, the oscillation probability was
computed in the three distance ranges studied in
section~\ref{section4}, yielding the formulas (\ref{probanodisp}),
(\ref{averagelaplace}) and (\ref{averagephasestat}). In this
section, we analyze the different terms appearing in these
formulas and we explain their origin. This analysis will allow us
to answer most of the questions raised in section~\ref{section2}.

Before analyzing the oscillation formulas, let us sum up the
assumptions used in its derivation. The oscillating particle is
stable, and propagates in vacuum over a macroscopic distance $L$,
i.e.\ $L\gg 1/p_0$, where $p_0$ is the average momentum of the
particle. The velocity of the oscillating particle is arbitrary,
but the condition \mbox{$\delta m_{ij}^2 \ll E_0^2$} is assumed to
be satisfied. This means that nonrelativistic particles are
supposed to have nearly degenerate masses (if it were not the
case, oscillations would vanish anyway). The oscillation formula
has been derived for a scalar particle, but this assumption is not
very restrictive, since the spin structure factorizes from the sum
on the mass eigenstates as long as $\delta m_{ij}^2 \ll E_0^2$.

In order to obtain a time-independent formula, the transition
probability has been averaged over a time interval $\Delta T$
satisfying $\Delta T\gtrsim
max(v_0^{-1}\sigma_{xeff},\gamma^{-2}T^{osc}_{ij})$. We have also
supposed that the time separation $T^{sep}$ between the wave
packets, given by Eq.~(\ref{defTsep}), is smaller than the time
interval $\Delta T$. The dispersion has been taken into account.

The oscillation formula valid at large distance (i.e. in the
second and third regimes) will be studied first since it is the
generic experimental case.

\subsection{Oscillations at large distance}
\label{osclargedist}

\subsubsection{Oscillation formula at large distance}
\label{oscformulalargedist}

At large distance ($L\gtrsim p_0/\sigma_p^2$), the flavor-mixing
transition probability for a stable particle of arbitrary velocity
propagating over a distance ${\bf L}=L{\bf l}$, with dispersion
taken into account, is given in a very good approximation by
Eqs.~(\ref{averagelaplace}), (\ref{averagephasestat}) or
(\ref{averageshortcut}):
\begin{eqnarray}
  && {\cal P}_{\alpha \to \beta}({\bf L}) = \frac{N_g g^2({\bf l})}{L^2} \,
   \sum_{i,j} V_{i\alpha} \, V_{\beta i}^{-1} \,
          V_{j\alpha}^* \, V_{\beta j}^{-1 \, *}
   \nonumber \\
   && \hspace{1.5cm} \times \exp \left(  -2 \pi i \frac{L}{L^{osc}_{ij}}
   - \frac{(\delta m_i^2)^2+(\delta m_j^2)^2}{16 \sigma_m^2 E_0^2}
   - 2 \pi^2 \left( \frac{\rho \, \sigma_{xeff}}{L^{osc}_{ij}} \right)^2
   - \left( \frac{L}{L^{coh}_{ij}} \right)^2
               \right) \, ,
   \label{proba}
\end{eqnarray}
where $g({\bf l})$ is the geometrical factor defined by
Eq.~(\ref{geom}):
$$
   g({\bf l})
   = \exp \left(
                 - \frac{({\bf p}_0\times{\bf l})^2}{4\sigma_p^2}
          \right) \, ,
$$
with $\sigma_p$ defined by Eq.~(\ref{definitionsigmap}). Recall
that the function $g({\bf l})$ restricts the propagation to a cone
of axis ${\bf p}_0$ and angle $arcsin(\sigma_p/p_0)$ (see Fig.~6).
$N_g$ is a normalization constant determined by the conservation
of the probability, in the case of a stable oscillating particle:
$$
   \sum_\beta \int L^2 \, d\Omega \,
   {\cal P}_{\alpha \to \beta}({\bf L}) =1 \, .
$$
From Eq.~(\ref{proba}), we can see that $\sum_\beta {\cal
P}_{\alpha \to \beta}({\bf L})$ is independent of $L$, since
$V^{-1}=V^\dagger$ for stable particles (unless the indices
$\alpha,\beta$ are restricted to active flavors in the presence of
sterile flavors). Note that the unitarity relation is only
verified to order $\epsilon$ because of the approximations made in
the computations. Thus the normalization constant $N_g$ is fixed
by
\begin{equation}
   N_g \int d\Omega \, g^2({\bf l}) = 1 \, .
   \label{normNg}
\end{equation}
The question of the normalization of the probability will be
addressed in more detail in section \ref{normalization}. The
oscillation and coherence lengths are defined by
Eqs.~(\ref{osclength}) and (\ref{cohlength}), respectively:
\begin{equation}
   L^{osc}_{ij} = \frac{4\pi p_0}{\delta m_{ij}^2}
   \hspace{1cm} \mbox{and} \hspace{1cm}
   L^{coh}_{ij} = \frac{1}{\sqrt{2}\pi} \, \frac{p_0}{\sigma_{peff}}
                  L^{osc}_{ij} \, .
   \label{cohlengthbis}
\end{equation}
The effective width $\sigma_{xeff}$ is defined by
$\sigma_{peff}\sigma_{xeff}=1/2$, with $\sigma_{peff}$ given by
Eq.~(\ref{sigmaeff}):
\begin{equation}
   \frac{1}{\sigma_{peff}^2} =
   \frac{1}{\sigma_{p\scriptscriptstyle P}^2} + \frac{1}{\sigma_{p\scriptscriptstyle D}^2}
   + \frac{(v_0-v_{\scriptscriptstyle P})^2}{\sigma_{e\scriptscriptstyle P}^2}
   + \frac{(v_0-v_{\scriptscriptstyle D})^2}{\sigma_{e\scriptscriptstyle D}^2} \, ,
   \label{sigmaeffbis}
\end{equation}
The dimensionless parameter $\rho$ is defined by
Eq.~(\ref{definitionrho}):
\begin{equation}
   \rho = \sigma_{peff}^2
   \left(
   \frac{1}{\sigma_{p\scriptscriptstyle P}^2} + \frac{1}{\sigma_{p\scriptscriptstyle D}^2}
   - \frac{v_{\scriptscriptstyle P}(v_0-v_{\scriptscriptstyle P})}{\sigma_{e\scriptscriptstyle P}^2}
   - \frac{v_{\scriptscriptstyle D}(v_0-v_{\scriptscriptstyle D})}{\sigma_{e\scriptscriptstyle D}^2}
   \right) \, .
   \label{definitionrhobis}
\end{equation}
The mass width $\sigma_m$ is defined by Eq.~(\ref{omegam}):
\begin{equation}
   \frac{1}{\sigma_m^2} = \sigma_{peff}^2
   \left(
   \frac{1}{\sigma_p^2} \,
   \left( \frac{1}{\sigma_{e\scriptscriptstyle P}^2} + \frac{1}{\sigma_{e\scriptscriptstyle D}^2} \right)
   + \frac{(v_{\scriptscriptstyle P}-v_{\scriptscriptstyle D})^2}{\sigma_{e\scriptscriptstyle P}^2\sigma_{e\scriptscriptstyle D}^2}
   \right) \, .
   \label{omegambis}
\end{equation}

As will be made clear in the following sections, the oscillation
formula (\ref{proba}) reduces to the standard formulas
(\ref{standardproba}) and (\ref{standardphase}) (with the
additional property of $1/L^2$ geometrical decrease) if the
observability conditions $|{\bf p}_0\times{\bf
l}|\lesssim\sigma_p$, $L\ll L^{coh}_{ij}$ and $L^{osc}_{ij}\gg
S_{\scriptscriptstyle P,D}$ are satisfied.
\begin{figure}
\begin{center}
\includegraphics[width=10cm]{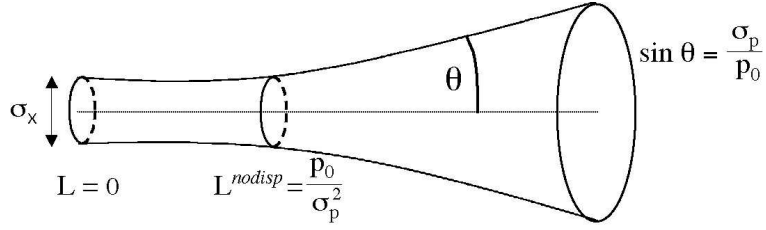}
\caption{The probability of detection is maximal within a cone at
large distance and within a cylinder at short distance.}
\end{center}
\end{figure}

\subsubsection{Oscillation phase}
\label{oscillationsubsection}

Each term of the sum appearing in Eq.~(\ref{proba}) oscillates
with a phase given by $2\pi L/L^{osc}_{ij}$. This phase is
independent of the details of production and detection up to order
$\epsilon$ (except the dependence on the momentum $p_0$, of
course). Its origin can be better understood by going back to the
expression of the probability before the time average is done.
\begin{enumerate}

\item In the transversal-dispersion regime, the phase is given by
Eq.~(\ref{defphaseij}). In the limit $T\ll T^{disp}$, the third
term of (\ref{defphaseij}) can be neglected and the phase reads
$$
   \phi_{ij}(T,L) \cong (E_i-E_j)T - (p_i-p_j)L \, .
$$
Using the expansions (\ref{expansionpj}) and (\ref{expansionEj}),
we can write
\begin{equation}
   E_jT-p_jL \cong E_0T-p_0L
   + \frac{\delta m_j^2}{2p_0} \, \left( L+\rho(v_0T-L) \right) \, .
   \label{phaseosc}
\end{equation}
The explicit expression for $\rho$ is not useful, as we shall see
below. The oscillating phase $\phi_{ij}(T,L)$ can now be written
\begin{equation}
   \phi_{ij}(T,L)
   \cong 2\pi \frac{L}{L^{osc}_{ij}} + 2\pi\rho \,
   \frac{v_0T-L}{L^{osc}_{ij}} \, .
   \label{phaseosc1}
\end{equation}
Since $|v_0T-L| \lesssim \sigma_{xeff}$ (see
Eq.~(\ref{amplitudelaplace})), the second term of the right-hand
side of Eq.~(\ref{phaseosc1}) is negligible if
\begin{equation}
   |\rho| \, \frac{\sigma_{xeff}}{L^{osc}_{ij}} \ll 1 \, .
   \label{phaseosc2}
\end{equation}
Such a constraint is indeed present in the third exponential term
of the probability (\ref{proba}), so that either the second term
of the right-hand side of Eq.~(\ref{phaseosc1}) does not
contribute to the phase, or the corresponding interference term in
the probability is negligible.

Since the leading term of the oscillating phase does not depend on
$\rho$, the phase is independent to order $\epsilon$ of the exact
values of $E_{i,j}$ and $p_{i,j}$. In other words the phase is
independent of the conditions of production and detection. The two
fundamental reasons for this independence are clearly seen in
Eq.~(\ref{phaseosc}):
\begin{enumerate}
   \item
   the particle is on-shell: $E_j=\sqrt{p_j^2+m_j^2}$;
   \item
   the particle is well localized in space-time:
   $|v_0T-L| \lesssim \sigma_{xeff}$.
\end{enumerate}

\item In the longitudinal-dispersion regime, the phase is given by
Eq.~(\ref{defphaseijtilde}). In the limit $T\gg T^{disp}$, the
second term of the right-hand side of Eq.~(\ref{defphaseijtilde})
can be neglected. The phase now reads
$$
   \tilde \phi_{ij}(T,L) \cong \delta m_{ij} \sqrt{T^2-L^2} \, .
$$
Using the expansions (\ref{statphase}) and (\ref{statphaseprime})
around the average propagation time $\tilde T_0$, the phase can be
written in the limit $\ell\gg 1$ as
\begin{equation}
   \tilde \phi_{ij}(T,L) \cong 2\pi \frac{L}{L^{osc}_{ij}}
   + \frac{E_0 \delta m_{ij}}{\tilde m_0} \, (T-\tilde T_0) \, .
   \label{oscilanalysis}
\end{equation}
The second term of this equation is small if the spatial spread of
the wave packet is smaller than $L^{osc}_{ij}/\gamma^2$, where
$\gamma$ is the usual Lorentz factor. Since the possible time
range is constrained by the width of the overlap function, i.e.\
$|T-\tilde T_0| \lesssim \ell \sigma_{xeff}/v_0$ (see
Eq.~(\ref{fijstatprime})), the second term of the right-hand side
of Eq.~(\ref{oscilanalysis}) is negligible if
\begin{equation}
   \frac{\sigma_{peff}}{p_0} \, \frac{L}{L^{osc}_{ij}} \ll 1 \, ,
\end{equation}
i.e.\ if $L \ll L^{coh}_{ij}$. Such a constraint is indeed present
in the last exponential term of the probability (\ref{proba}). If
it is satisfied, the oscillating phase is equal to $2\pi
L/L^{osc}_{ij}$, as usual. If not, the $ij$ interference term
vanishes and decoherence occurs.

\end{enumerate}

\subsubsection{Coherence length}
\label{subdecoh}

The last exponential term in the probability (\ref{proba}) shows
that the $ij$-interference term vanishes at a distance larger than
the coherence length $L^{coh}_{ij}$. This decoherence, predicted
by Nussinov \cite{Nussinov76}, has two possible origins, since
both the overlap function and the derivative of the phase
contribute to $exp(-(L/L^{coh}_{ij})^2)$. The physical explanation
depends on whether the coherence length is larger than the
dispersion length or not\footnote{As noted in section
\ref{sectionprobatrans}, the condition $L^{coh}\gtrsim L^{disp}$
can be written as $\frac{\delta m^2}{m^2} \lesssim
\frac{v_0^2\sigma_p}{p_0}$, which is equivalent to say that the
masses are nearly degenerate.}.
\begin{enumerate}

\item If the coherence length is smaller than the dispersion
length (here $L^{disp}=min(L^{disp}_i,L^{disp}_j)$), decoherence
takes place in the transversal-dispersion regime. In that case,
the decoherence term comes mainly from the function
$f_{ij}(T_{ij},L)$, i.e.\ from the overlap function. This is clear
from Eq.~(\ref{minfijlaplace}), with the dispersion neglected for
simplicity: $\ell \ll 1$. This decoherence arises from the
progressive separation of the wave packets, due to the different
group velocities $v_i$ and $v_j$: if T is large enough, both terms
$|v_iT-L|$ and $|v_jT-L|$ cannot remain small.

\item If the coherence length is larger than the dispersion length
(here $L^{disp}\cong L^{disp}_i \cong L^{disp}_j$), decoherence
takes place in the longitudinal-dispersion regime. In that case,
the decoherence term does not come anymore from the separation of
the wave packets since the packets spread out beyond the
dispersion length as quickly as they separate (see
Eq.~(\ref{spacewidth})). Still, decoherence arises from the
variation of the phase over the width of the overlap function (see
Eq.~(\ref{oscilanalysis})). The interference term is averaged to
zero by the time integral when the spatial spread of the `wave
packet' becomes larger than $L^{osc}_{ij}/\gamma^2$. A similar
mechanism has been observed in connection with neutron
interferometry \cite{Klein83}.

\end{enumerate}
All in all, the coherence length arises not only from the
separation of wave packets, as it is usually explained in the
literature: {\it in the case of nearly degenerate masses, it can
also originate in a too large dispersion of the wave packet in
comparison with the oscillation length} (Fig.~7).
\begin{figure}
\begin{center}
\includegraphics[width=13cm]{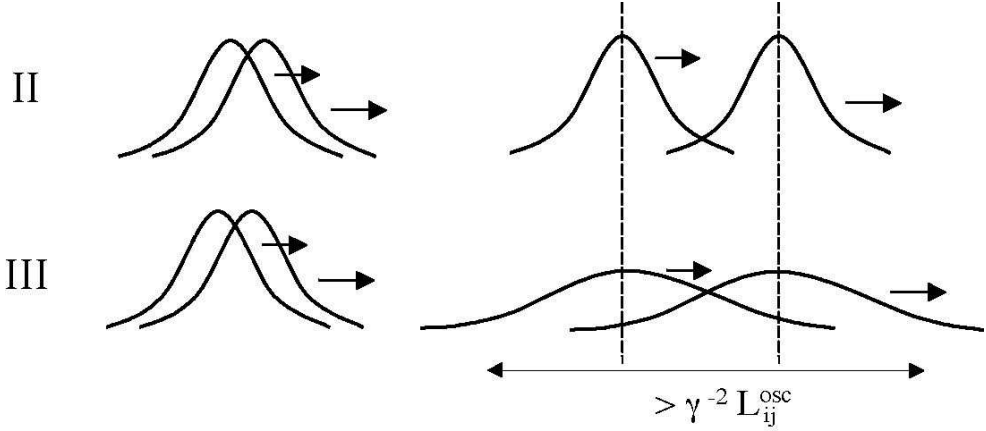}
\caption{In the transversal-dispersion regime (II), the coherence
length is due to the non-overlap of wave packets, whereas it is
due in the longitudinal-dispersion regime (III) to their
dispersion (the coherence length was not computed in the
no-dispersion regime).}
\end{center}
\end{figure}

Remark that the phenomena of separation of wave packets and
dispersion contribute to the coherence length in both regimes: if
approximations such as $\ell\ll 1$ (or $\ell \gg 1$) are not made,
terms proportional to $L^2$ come from $f_{ij}(T_{ij},L)$ (or
$\tilde f_{ij}(\tilde T_0,L)$) and from the squared derivative of
the phase, divided by the second derivative of $f_{ij}$ (or of
$\tilde f_{ij}$). Furthermore, the transition between the two
regimes is not visible in the oscillation formula, because the two
decoherence mechanisms smoothly blend.

Besides the above explanations in configuration space, the
coherence length can be also understood in energy-momentum space,
where it is due to the variation of the phase over the `wave
packet' width $\sigma_{peff}$ (see Eq.~(\ref{shortcutphase})). The
explanation in momentum space is thus simpler, but the two
mechanisms at the origin of the coherence length in configuration
space are more intuitive.

As dispersion is usually neglected in the literature, it is
interesting to derive the oscillation formula with this
approximation ($T_j^{disp}=\infty$). Such a computation in the
transversal-dispersion regime yields the same result as when
dispersion is taken into account (Eq.~(\ref{proba})), except that
the following substitution has to be made:
\begin{equation}
   L^{coh}_{ij} \to
   \frac{L^{coh}_{ij}}{|\rho(1-v_0^2)-1 |}
    \;\;\; (FALSE) \, .
   \label{falsecohlength}
\end{equation}
The incorrect multiplying factor has its origin in
Eq.~(\ref{expansionvj}): in the limit $T_j^{disp}=\infty$, the
coherence length term arises only from the difference between the
group velocities $v_i$ and $v_j$. However the factor
$|\rho(1-v_0^2)-1 |$ tends to 1 in the relativistic limit, so that
the substitution (\ref{falsecohlength}) becomes trivial in that
limit. This observation explains why our result (\ref{proba})
coincides with Eq.~(26) of Ref.~\cite{Giunti98a}, as the authors
of this article, while neglecting dispersion, consider only
relativistic neutrinos. Note however that even relativistic
neutrinos spread at large distances so that a calculation
neglecting dispersion such as in Ref.~\cite{Giunti98a} is only
valid for $L\lesssim L_j^{disp}$.

It is interesting to observe that the coherence length increases
when a long coherent measurement in time is performed at the
detector, even if the oscillating `wave packets' have separated
spatially \cite{Kiers96,Kiers98}. In that case, the energy
uncertainty at detection goes to zero, $\sigma_{e\scriptscriptstyle D}\to0$, so
that the effective width also goes to zero, $\sigma_{peff}\to0$,
and the coherence length becomes infinite,
$L^{coh}_{ij}\to\infty$.

After all these theoretical considerations, it must be said that
the decoherence at the level of the wave packet is irrelevant in
most experiments, since it is usually dominated by decoherence
effects originating in the energy spread of the beam. Different
situations are discussed in section \ref{coherenteffects}.
Finally, let us remark that the exponential decrease in $L^2$ of
this decoherence term is model-dependent. It results here from the
Gaussian approximation. However, the definition of the coherence
length is model-independent, apart from a multiplying constant.

\subsubsection{Localization}
\label{localization}

A third kind of term appearing in the flavor-mixing transition
probability (\ref{proba}) are {\it localization} terms, that is,
observability constraints imposing that the oscillation length
should be larger than the space-time uncertainty: $L^{osc}_{ij}
\gtrsim \sigma_{xeff}$. This condition can be rewritten as $\delta
m_{ij}^2/p_0 \lesssim \sigma_{peff}$, stating that oscillations
vanish if the energy-momentum measurements allow to distinguish
between the different mass eigenstates.

Two localization terms appear in the probability (\ref{proba}),
while one more is implicitly assumed when applying the Jacob-Sachs
theorem:
\begin{enumerate}

\item The term containing $\sigma_m$ can be rewritten
\begin{equation}
   \frac{(\delta m_i^2)^2+(\delta m_j^2)^2}{16 \sigma_m^2 E_0^2}
   =  \frac{(\delta m_{ij}^2)^2}{32 \sigma_m^2 E_0^2}
    + \frac{(\delta m_i^2+\delta m_j^2)^2}{32 \sigma_m^2 E_0^2}\, .
   \label{decomposition}
\end{equation}
The first term of the right-hand side of Eq.~(\ref{decomposition})
can be written as a localization term:
\begin{equation}
  L^{osc}_{ij} \gtrsim \sigma_x \, ,
  \label{local2}
\end{equation}
as $\sigma_m\sim v_0\sigma_p$ whether the stationary limit is
taken or not. The second term of the right-hand side of
Eq.~(\ref{decomposition}) is not a localization term and will be
discussed in section \ref{massconstraint}.

\item The term containing $\rho$ is also a localization term and
imposes that
\begin{equation}
    L^{osc}_{ij} \gtrsim |\rho| \, \sigma_{xeff} \, .
   \label{localconstraint}
\end{equation}
It is not obvious whether this constraint is stronger than the
previous one, i.e.\ whether it is possible that $|\rho| \,
\sigma_{xeff} \gg \sigma_x$. This situation might arise from an
energy uncertainty much smaller than the momentum uncertainty. In
that case, the definition (\ref{sigmaeffbis}) of the effective
width shows that $\sigma_{xeff}\gg\sigma_x$. Say, for example,
that the energy uncertainty at the detection goes to zero, i.e.\
$\sigma_{e\scriptscriptstyle D} \to 0$. With the help of
Eqs.~(\ref{sigmaeffbis}), (\ref{definitionrhobis}) and
(\ref{statlim}), we see that
\begin{equation}
   |\rho| \, \sigma_{xeff}
   \stackrel{\sigma_{e\scriptscriptstyle D} \to 0}{\longrightarrow}
   \frac{|v_{\scriptscriptstyle D}|}{\sigma_{e\scriptscriptstyle D}}
   \lesssim S_{\scriptscriptstyle D}  \, ,
   \label{statlimbis}
\end{equation}
where $S_{\scriptscriptstyle D}$ is the size of the macroscopic detection region.
Thus the localization term does not give a stronger constraint
than $L^{osc}_{ij}\gtrsim S_{\scriptscriptstyle D}$. This constraint is always
satisfied, as it is equivalent to the constraint obtained by
averaging the transition probability over the production region
(see section \ref{incoherenteffects}). Therefore, the coherence
length can be increased without bound by more accurate energy
measurements, contrary to what was claimed in
Refs.~\cite{Giunti98a,Giunti98b}. Note that this is not true if
the accuracy of the 3-momentum measurements is increased, as the
localization makes the oscillations vanish when the corresponding
spatial uncertainty becomes larger than the oscillation length. Of
course the opposite conclusions would be reached if experiments
measured time, not distance.

\item The contour integral used in the Jacob-Sachs theorem yields
a third localization constraint. The poles corresponding to the
mass eigenstates $m_i$ and $m_j$ cannot be both included in the
same contour integration unless $|\delta m_{ij}^2|/p_0 \lesssim
\sigma_p$. This third constraint is not stronger than the previous
ones and can be ignored.
\end{enumerate}

As was the case for the coherence length, there are two origins
for the localization term containing \nolinebreak $\rho$ (Fig.~8).
\begin{enumerate}

\item In the limit $\ell\ll1$ (transversal-dispersion regime),
this term comes from the variation of the phase over the width of
the wave packet (see Eq.~(\ref{phaseosc1})). If the oscillation
length is smaller than the width of the wave packet, the
interference term is averaged to zero by the time integration.

\item In the limit $\ell\gg1$ (longitudinal-dispersion regime),
this term arises from the overlap function (see
Eq.~(\ref{minfijstat})). Thus it comes from the separation of the
wave packets, which remains constant in the
longitudinal-dispersion regime.
\end{enumerate}

\begin{figure}
\begin{center}
\includegraphics[width=13cm]{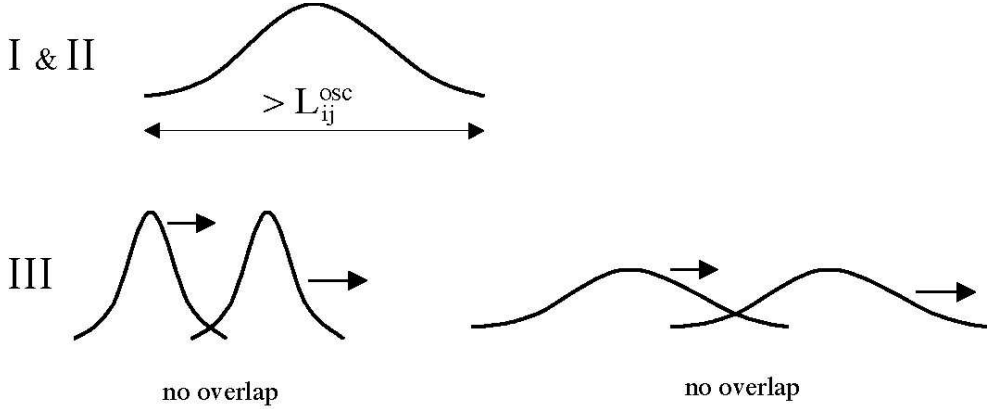}
\caption{In the no-dispersion and transversal-dispersion regimes
(I and II), the localization condition is due to the initial
spread of the wave packet, whereas it is due in the
longitudinal-dispersion regime (III) to their non-overlap.}
\end{center}
\end{figure}

Recall that the coherence length was shown to arise, in the
transversal-dispersion regime, from the wave packet separation
and, in the longitudinal-dispersion regime, from the variation of
the phase. Therefore, {\it the coherence-length and localization
conditions in the transversal-dispersion regime transform
respectively into localization and coherence-length conditions in
the longitudinal-dispersion regime}. The situation is summarized
in Table~1.
\begin{table}
\caption{Origins of the coherence length and localization terms.}
\vspace{5mm} \hspace{-5mm}
\begin{tabular}{|c|c|c|}
\hline
& Transversal-dispersion  regime & Longitudinal-dispersion regime \\
\hline
Coherence length  & decreasing overlap of wave packets & increasing dispersion of each wave packet \\
Localization & initial spread of each wave packet & constant overlap of wave packets  \\
\hline
\end{tabular}
\end{table}

\subsubsection{Energy-momentum conservation}
\label{massconstraint}

The last term of the oscillation probability (\ref{proba}) to be
discussed is the second term in the decomposition of
Eq.~(\ref{decomposition}). This term could be a matter of concern
since it does not vanish in the limit $m_i=m_j$. Note that this
kind of term is not specific to our computation. For example, it
would appear in the oscillation formula (26) of
Ref.~\cite{Giunti98a} if the terms $S_a(E_a)$ present in Eq.~(22)
of that article had been expanded beyond zeroth order in
$m_a^2/E_a^2$.

The second term in Eq.~(\ref{decomposition}) imposes that
\begin{equation}
  \frac{|m_i^2+m_j^2-2m_0^2|}{E_0} \lesssim \sigma_m \, .
  \label{onshell}
\end{equation}
As $m_0$ is related to the average in- and outgoing momentum by
$m_0^2=E_0^2-p_0^2$, condition (\ref{onshell}) means that the mass
eigenstates have to be on-shell with respect to $(E_0,p_0)$ within
the uncertainty $\sigma_m$. For example, this constraint is
impossible to satisfy in the mixing of relativistic and
nonrelativistic neutrinos if the available energy-momentum is such
that only the lightest neutrino can be produced. However it has no
effect on the oscillations in the two cases considered in the
present report, namely relativistic particles or nearly mass
degenerate particles. Condition (\ref{onshell}) should simply be
considered as expressing the conservation of energy-momentum. Such
kinematical constraints are usually not included in the
oscillation formula, though they rightly belong to it. A complete
computation of the transition probability should not only include
this energy-momentum constraint, but also similar terms arising
from the interaction amplitudes $M_{P,D}$, from the prefactor
resulting from the Gaussian integration and from the spin
structure of the propagator. Actually the masses of the
oscillating particle should be expected to appear not only through
mass differences but also through their absolute values.

\subsection{Oscillations at short distance}
\label{oscshortdistance}

At short distance ($L\lesssim p_0/\sigma_p^2$), the oscillation
formula is given by Eq.~(\ref{probanodisp}):
\begin{eqnarray}
  && {\cal P}_{\alpha \to \beta}({\bf L}) = N_{\tilde g} \,
  \exp \left( - \frac{(\tilde{\bf v}_0\times\tilde{\bf L})^2}
                     {2\sigma_x^2\,\tilde{\bf v}_0^2}
       \right)
  \sum_{i,j} V_{i\alpha} \, V_{i\beta}^* \, V_{j\alpha}^* \, V_{j\beta}
  \nonumber \\
  && \hspace{2cm} \times
  \exp \left(
      - 2\pi i\frac{L}{L_{ij}^{osc}}
      - \frac{(\delta m_i^2)^2+(\delta m_j^2)^2}{16\tilde\sigma_m^2E_0^2}
      - 2\pi^2\left( \frac{\tilde\rho\tilde\sigma_{xeff}}{L_{ij}^{osc}} \right)^2
      \right) \, ,
  \label{probanodispbis}
\end{eqnarray}
where $\tilde\sigma_{xeff}$ and $L_{ij}^{osc}$ are defined by
Eq.~(\ref{defsigmatilde}) and Eq.~(\ref{osclengthnodisp}),
respectively. $N_{\tilde g}$ includes all prefactors independent
of $L$ as well as normalization factors.

The normalization of the probability can be done in the same way
as in Eq.~(\ref{normNg}):
\begin{equation}
   N_{\tilde g} \int L^2 \, d\Omega \,
   \exp \left( - \frac{(\tilde{\bf v}_0\times\tilde{\bf L})^2}
                     {2\sigma_x^2\,\tilde{\bf v}_0^2}
       \right) = 1 \, .
   \label{normNgtilde}
\end{equation}
As in the case of oscillations at large distance, it can be argued
that the two last exponential terms in Eq.~(\ref{probanodispbis})
yield an observability condition not stronger than
$L^{osc}_{ij}\gtrsim S_{\scriptscriptstyle P,D}$ (see section
\ref{localization}). It could also be interesting to understand
the origin of the oscillation phase by going back to the
amplitude, before doing the time average. This discussion can be
found in Ref.~(\cite{Beuthe02}).

Actually, the oscillation formula (\ref{probanodispbis}) is very
similar to the formula valid at large distance
(Eq.~(\ref{proba})). There are three main differences. First there
is no coherence-length term in Eq.~(\ref{probanodispbis}), which
is due to the neglect of the terms beyond ${\cal O}(\epsilon^0)$
in the evaluation of $F(T_0)$ (see Eq.~(\ref{Fdiag})). Second, the
geometrical decrease in $1/L^2$ is lacking in
Eq.~(\ref{probanodispbis}), which is explained by the fact that
the dispersion of the oscillating particle `wave packet' is not
yet significant. Finally, the directional constraint present in
Eq.~(\ref{probanodispbis}) confines the propagation to a cylinder
of radius $r$ (with $\sigma_x\lesssim r\lesssim S_{\scriptscriptstyle P,D }$),
whereas the directional constraint present in Eq.~(\ref{proba})
restricts the oscillating particle propagation to a cone of angle
$arcsin(\sigma_p/p_0)$ (Fig.~6 in section
\ref{oscformulalargedist}). This different behavior is also a
result of the absence of dispersion for $L\lesssim
p_0/\sigma_p^2$.

In conclusion, the oscillation formula (\ref{probanodispbis})
reduces to the standard formulas (\ref{standardproba}) and
(\ref{standardphase}) if the observability conditions $|{\bf
v}_0\times{\bf L}|\lesssim |{\bf v}_0|\sigma_x$ and
$L^{osc}_{ij}\gg S_{\scriptscriptstyle P,D}$ are satisfied.

\subsection{Answers at last}
\label{answers}

In section~\ref{section2}, our review of field-theoretical
treatments brought to the fore several questions:
\begin{itemize}

\item Is there a limit in which the oscillation formula, derived
with stationary boundary conditions, can be obtained from the
oscillation formula derived with external wave packets?

\item Does an infinitely precise measurement of the energy lead to
an infinite coherence length, or is there decoherence anyway? This
question is inseparable from the first one.

\item Does a strong localization of the source lead to a
dependence of oscillation formula on mass prefactors?

\item Do `plane waves oscillations' exist?
\end{itemize}
The external wave packet model used in this report is of
sufficient generality to allows us to clarify and answer these
questions.

\subsubsection{Existence of a stationary limit}

Two important results concerning stationary boundary conditions
have been demonstrated. First, there is no contradiction between
models with stationary boundary conditions and those with external
wave packets. The former type of model can be obtained from the
latter in a smooth limit. Let us state again the gist of the
problem. Stationary boundary conditions, given by $v_{\scriptscriptstyle P,D}=0$
and $\sigma_{e\scriptscriptstyle P,D}=0$, lead to an infinite effective width
$\sigma_{xeff}$ and thus to an infinite coherence length
$L^{coh}_{ij}$ (see
Eqs.~(\ref{cohlengthbis})-(\ref{sigmaeffbis})). Hence, the
localization term depending on $\rho\sigma_{xeff}/L^{osc}_{ij}$
seems to diverge in Eq.~(\ref{proba}), with the result that the
$ij$-interference term seems to vanish in the stationary limit
$\sigma_{xeff}\!\to\!\infty$.  If it were true, it would be
impossible to increase without limit the coherence length by
performing long coherent measurements. This would be in
contradiction with stationary boundary condition models, which
have a zero energy uncertainty but an infinite coherence length.
However the product $\rho \sigma_{xeff}$ remains finite, as shown
by Eq.~(\ref{statlimbis}), so that the interference term does not
vanish. For example, the Grimus-Stockinger model \cite{Grimus96}
is obtained in the limit $\rho=0$, $\sigma_{xeff}\!\to\!\infty$
with $\rho\sigma_{xeff}=0$. The latter condition means that this
model can be recovered from the external wave packet model if the
stationary limit has the property $v_{\scriptscriptstyle P,D}/\sigma_{e\scriptscriptstyle
P,D}=0$ (see discussion after Eq.~(\ref{statlim})). The only
localization condition comes from the momentum spreads of the
source and detector (see Eq.~(\ref{local2})). Note also that the
amplitude ${\cal A}_j$ (see Eq.~(\ref{jacobsachsinteg})) takes in
the stationary limit the form of a plane wave, with a well-defined
energy and momentum (see
Eqs.~(\ref{expansionpj})-(\ref{expansionEj}) with $\rho\to0$).

The second important result is the equivalence (\ref{incohsumQFT})
between the time-integrated nonstationary probability and the
energy-integrated stationary probability. Since oscillations can
be described in the former model by interfering `wave packets',
this interpretation is also valid in the latter, contrary to what
was claimed in \cite{Grimus99}. Thus the incoherent superposition
of stationary amplitudes, which have a plane wave form, reproduces
the characteristics of wave packet propagation. Of course, this is
not true if the probability is not integrated over the energy
spectrum, but we have already argued that a source and a detector
with zero energy uncertainties do not constitute realistic
boundary conditions.

\subsubsection{Unbounded coherence length}

As shown by Kiers, Nussinov and Weiss \cite{Kiers96}, a more
precise measurement of the energy at detection increases the
coherence length. This feature is reproduced within the external
wave packet model: $L^{coh}_{ij} \sim \sigma_{xeff} \sim
1/\sigma_{e\scriptscriptstyle D}$ (see
Eqs.~(\ref{cohlengthbis})-(\ref{sigmaeffbis})). In the limit
$\sigma_{e\scriptscriptstyle D}\to0$, the external wave packet model becomes
equivalent to a stationary model (even if the source is
nonstationary), as the detector picks up wave packet's components
having exactly the same energy. Contrary to what was asserted in
Refs.~\cite{Giunti98a,Giunti98b}, oscillations do not vanish in
that limit since we argued above that $\rho \sigma_{xeff}$ remains
finite.

\subsubsection{No mass prefactors at large distance}

Contrary to what was claimed by Shtanov \cite{Shtanov98}, mass
prefactors never appear in front of the oscillating exponentials
present in the transition probability (\ref{proba}). Let us first
explain what these `mass prefactors' mean. Shtanov derives
oscillation formulas for neutrinos within a source-propagator
model in configuration space. Shtanov computes the convolution of
the neutrino propagator with a source, keeping only the phase in
the oscillating exponentials but neglecting the contributions
depending on the width $\sigma_x$ of the source. With this last
approximation, the space-time localization (i.e.\ the `wave packet
envelope') of the amplitude is lost.

Let us consider his model will be considered in a scalar version
for an easier comparison with our results. The propagator for a
scalar particle propagating from $x$ to $x'$ is asymptotically
($m_j\sqrt{(x'-x)^2}\gg1$) given in configuration space by
\begin{equation}
  {\cal A}_j\sim \frac{\sqrt{m_j}}{\left( (x'-x)^2 \right)^{3/4}} \,
  e^{-im_j\sqrt{(x'-x)^2} } \, ,
  \label{shtanovresult}
\end{equation}
where $(x'-x)^2$ is the Lorentz interval. The convolution of
Eq.~(\ref{shtanovresult}) with a monochromatic source yields
without problem the standard oscillation formula
(\ref{standardproba}). However, convolution with a strongly
localized source ($\sigma_{x,t}\lesssim1/E$) leads to an amplitude
that keeps its dependence on the mass prefactor $\sqrt{m_j}$. In
that case, the transition probability is not equivalent to the
oscillation formula (\ref{standardproba}), unless the masses
$m_{i,j}$ are nearly degenerate.

We now proceed to show that mass prefactors only appear at an
intermediate stage of the calculation, but not in the final
result. Note first that the amplitude (\ref{shtanovresult}) is in
correspondence with our amplitude (\ref{longdisp}), computed with
the stationary phase method, since the prefactor in
Eq.~(\ref{amplitudephasestat}) can be rewritten, for $T \gg
T^{disp}_j$, as
$$
  \frac{\sigma_{peff}}{T\sqrt{1+iT/T^{disp}_j}}
  \sim \frac{\sqrt{m_j}}{(T^2-L^2)^{3/4}} \, ,
$$
where the definition (\ref{dispersiontime}) of the dispersion time
has been used. This prefactor coincides with the prefactor in
Eq.~(\ref{shtanovresult}). However, the subsequent time average of
the transition probability completely cancels this dependence on
the mass, yielding Eq.~(\ref{averagephasestat}). This can be seen
by expanding the argument of the exponential in
Eq.~(\ref{amplitudephasestat}) around the average propagation
time. The width with respect to $T$ of the amplitude ${\cal A}_j$
is found to be equal to
$\frac{\sigma_{peff}T}{m_jv_0\gamma_{cl}^3}$ (where $\gamma_{cl}$
is the Lorentz factor associated with the velocity $v_{cl}=L/T$),
thereby providing a $m_j$-dependent factor that cancels the $m_j$
prefactor in the Gaussian integration.

Shtanov does not obtain such a result, since he does not perform
any time average on the probability. Besides, mass prefactors
would remain even if a time average were performed on the
probability: no `wave packet envelope' appears indeed in Shtanov's
amplitude, so that the cancellation mechanism explained in the
previous paragraph is not at work. For that reason, Shtanov
wrongly concludes that the mass prefactor remains if the source is
strongly localized. Another non-standard oscillation formula,
derived by Shtanov for mass eigenstates satisfying
$m_2\sqrt{(x'-x)^2}\ll1\ll m_1\sqrt{(x'-x)^2}$, reduces to the
standard formula when decoherence mechanisms are taken into
account. In conclusion, Shtanov's computations in configuration
space do not lead to new oscillation formulas.

\subsubsection{No plane waves}

Ioannisian and Pilaftsis \cite{Ioannisian99} derive a formula for
neutrino oscillations which exhibits a plane wave behavior if the
condition $L\lesssim p_0/\sigma_p^2$ is satisfied (with
$\sigma_x\ll L^{osc}_{ij}$, as usual). Their term `plane wave
oscillations' means that the oscillation amplitude ${\cal A}_j$
does not decrease as the inverse of the distance, that its phase
depends on the direction ${\bf L}$ as ${\bf p_j}\cdot {\bf L}$,
and that there is no geometrical constraint linking the average
3-momentum and the direction of observation. If it were true, the
oscillation length measured by a specific detector would depend on
the direction of the total momentum of the initial particles. The
plane wave condition is satisfied, for example, by taking
$\sigma_x\!\sim\!1\,$cm, $L\!\sim\!1000\,$km and
$p\!\sim\!1\,$GeV. These conditions might be verified for
atmospheric neutrinos.

In section \ref{sectionprobanodisp}, we have proved that the
detection probability of an oscillating particle, derived under
the condition $L\lesssim p_0/\sigma_p^2$, is negligible in
directions other than the average momentum whether the stationary
limit is taken or not. Therefore the oscillation phase has the
standard form given in Eq.~(\ref{standardphase}) and no `plane
wave oscillations' can be observed, contrary to the Ioannisian and
Pilaftsis' claim. Also, the absence of the $1/T$ (or $1/L$) factor
in Eq.~(\ref{nodispersion}), as noted by the same authors, is
easily understood by noting that the absence of dispersion entails
that the amplitude does not decrease with the distance (the
probability is maximal within a cylinder as pictured on Fig.~6 of
section \ref{oscformulalargedist}).

It is interesting to understand why a directional constraint is
missing for $L\lesssim p_0/\sigma_p^2$ in Ioannisian and
Pilaftsis' result, as this fact explains their  `plane wave
oscillations' prediction. At the end of their computations,
Ioannisian and Pilaftsis obtain an amplitude ${\cal A}_j$ whose
dominant term depends on $exp(ip_j|{\bf L'}|)$, where $|{\bf
L'}|=\sqrt{{\bf L'}^2}$ is the complex `norm' of a complex vector
${\bf L'}={\bf L}-2i\sigma_x^2{\bf p}_0$ (their results are
translated into our notation through the correspondence $q_j\to
p_j$, $\vec k\to{\bf p}_0$, $\delta l^2\to4\sigma_x^2$, $\vec
L\to{\bf L'}$, $\vec l\to{\bf L}$). For $L\ll p_0/\sigma_p^2$, the
quantity $|{\bf L'}|$ can be expanded and the argument of the
exponential reads to second order
$$
  ip_j|{\bf L}-2i\sigma_x^2{\bf p}_0| \cong
  2\sigma_x^2p_0p_j
  + i p_j\frac{{\bf p}_0\cdot{\bf L}}{p_0}
  -\frac{p_j}{4\sigma_x^2p_0}\,
  \left( L^2 - \frac{({\bf p}_0\cdot{\bf L})^2}{p_0^2} \right)
  \, .
$$
The two last terms were neglected in Ref.~\cite{Ioannisian99},
though $L\gg\sigma_x$. They lead to the following directional
constraint:
\begin{equation}
  \exp\left( - \frac{({\bf v}_0\times{\bf L})^2}{4\sigma_x^2v_0^2} \right)
  \, , \label{ipdirconstraint}
\end{equation}
where the factor 4 becomes a factor 2 when the amplitude is
squared. Therefore Ioannisian ad Pilaftsis' result includes a
directional constraint forbidding plane wave oscillations.

Note that the stationary limit assumption $|{\bf v}_{\scriptscriptstyle
P,D}|/\sigma_{e\scriptscriptstyle P,D}=0$ leads to $s_1=s_2$ (see
Eq.~(\ref{s12stat})), so that the constraint
(\ref{ipdirconstraint}) becomes equal to the one present in
Eq.~(\ref{probanodisp}). Thus the condition $|{\bf v}_{\scriptscriptstyle
P,D}|/\sigma_{e\scriptscriptstyle P,D}=0$ seems generic for models with
stationary boundary conditions, since it was also applied in the
case of the Grimus-Stockinger model (see discussion after
Eq.~(\ref{statlim}) and at the end of section
\ref{gaussianoverlap}).

In conclusion, the analysis of the transition probability
(\ref{probanodisp}) derived in the external wave packet model
under condition (\ref{ipcondition}) has shown that `plane wave
oscillations' do not exist. This negative result was confirmed by
a reexamination of the formula derived in
Ref.~\cite{Ioannisian99}.

\subsection{Conclusion}

Through the discussions of this section, the external wave packet
model has shown its power and generality. This model can either
reproduce in some limit the field-theoretical oscillation formulas
found in the literature, or allows to understand why they are
wrong. Thus the three categories of models described in section
\ref{review} (external wave packet models, stationary boundary
condition models,source-propagator models) really make one.
Furthermore, the external wave packet model has the advantage of
associating a clear physical picture to the different stages of
the calculations. For example, the role played by the dispersion
in the observability conditions (coherence length and
localization) could be cleared up in section \ref{osclargedist},
while the threshold $L=p_0/\sigma_p^2$ identified by Ioannisian
and Pilaftsis could be interpreted as indicating the onset of
dispersion.

\section{Oscillations of unstable particles}
\label{section7}

First, we study nonexponential corrections specific to mixed
unstable particles, and then examine in which respect the
oscillation formula obtained in the stable case is modified by the
decay term. Next, we establish the relation between our formalism
and the Wigner-Weisskopf-Lee-Oehme-Yang model. Finally, we apply
the formulas derived in this section to the $B^0\bar B^0$ system.

\subsection{Nonexponential corrections due to mixing}
\label{nonexponential}

If the oscillating particle is unstable, the flavor-mixing
amplitude (\ref{amplitot}) is a superposition of mass-eigenstate
amplitudes ${\cal A}_j$ given by Eq.~(\ref{defAj}). However these
formulas were derived on the assumption that the diagonalization
matrices $V$ appearing in Eq.~(\ref{amplitot}) are constant. This
hypothesis is only true in the limit of a negligible renormalized
self-energy. This approximation cannot be made for unstable
particles, since their decay arises from the imaginary part of
this self-energy. Actually, nonexponential corrections are always
important at large time for unstable particles (though until now
not observable), so it is worth checking whether the
energy-momentum dependence of the diagonalization matrix has a
measurable effect on the transition probability.

Nonexponential corrections to the propagation of an unmixed
particle of well-defined mass are studied in the Appendix, in the
context of the derivation of the Jacob-Sachs theorem. These
corrections have two origins. First, they are due to the bounded
character of the energy-momentum spectrum of the particle. Second,
they are due to multi-particle thresholds, if these are included
in the energy-momentum spectrum.

The first type of correction is easily calculated for mixed
propagators, since it has nothing to do with the diagonalization
matrices. The corrections can be evaluated separately for each
partial amplitude ${\cal A}_i$, and are negligible (see Appendix).
In order to compute the second type of corrections, the
energy-dependent diagonalization matrices are kept inside the
energy-momentum integral (\ref{defAj}). Let us consider the case
of $P^0\bar P^0$ mixing in the limit of no CP violation ($P$ is a
$K$ or a $B$ meson). Starting from Eq.~(\ref{fullpropag}), the
inverse propagator for the neutral meson system can be
parametrized by \cite{Sachs63}
\begin{equation}
   i G^{-1}(p^2) =
   \left(
   \begin{array}{cc}
   p^2 - m^2 - f_{00}(p^2)   &  - f_{0\bar0}(p^2)  \\
   - f_{\bar0 0}(p^2)        &  p^2 - m^2 - f_{00}(p^2)
   \end{array}
   \right) \, ,
   \label{propKK}
\end{equation}
where $m$ is the renormalized mass in the degenerate limit and
$-if_{\alpha\beta}(p^2)$ are the renormalized complex
self-energies. The diagonal elements are equal because of CPT
symmetry. In the limit of CP symmetry, the off-diagonal can be
taken to be equal: $f_{0\bar0}(p^2)\stackrel{CP}{=}f_{\bar0
0}(p^2)$.

This propagator can be diagonalized as follows:
$$
   -i \, G(p^2) =
   V^{-1}(p^2)
   \left(
   \begin{array}{cc}
   ( p^2 - m_1^2 - f_1(p^2) )^{-1}  &  0                          \\
   0                            &  ( p^2 - m_2^2 - f_2(p^2) )^{-1}
   \end{array}
   \right)
   V(p^2) \, .
$$
Thresholds of multi-particle states appear in the renormalized
self-energies $f_j(p^2)$ and in the diagonalization matrix
$V(p^2)$.

Let us define the function $I_{\beta\alpha}(T)$ by
$$
   I_{\beta\alpha}(T) = \int dE \, \psi(E,{\bf p}) \,
   G_{\beta\alpha}(p^2) \, e^{-iET} \, ,
$$
where $G_{\beta\alpha}(p^2)$ is the nondiagonal propagator in
flavor space and $\psi(E,{\bf p})$ is the overlap function defined
by Eq.~(\ref{recouvrement}). The Fourier transform of this
function with respect to the 3-momentum yields the amplitude
${\cal A}(\alpha \!\to\! \beta,T,{\bf L})$. Since we are
interested by nonexponential corrections to mixing, let us study
the effect of a threshold $p^2=b^2$ on the matrix element
$I_{0\bar0}(T)$. The corresponding propagator matrix element is
\begin{equation}
   G_{0\bar0}(p^2) =
     \frac{i \, V_{01}^{-1}(p^2) V_{1\bar0}(p^2)}{p^2 - m_1^2 -f_1(p^2)}
   + \frac{i \, V_{02}^{-1}(p^2) V_{2\bar0}(p^2)}{p^2 - m_2^2 -f_2(p^2)} \, .
   \label{G00}
\end{equation}
The function $I_{\beta\alpha}(T)$ is computed with the method
explained in the Appendix. The integration on the energy is done
by a contour integration including the two poles $z_{1,2}$
appearing in the right-hand side of Eq.~(\ref{G00}). The function
$I_{\beta\alpha}(T)$ is equal to the sum of the poles'
contribution $J$, the contributions $J_{1,2}$ of the
energy-momentum thresholds of the overlap function $\psi(E,{\bf
p})$, as well as the contribution $J_b$ of the multi-particle
threshold:
$$
   I_{0\bar0}(T) = J+J_1+J_2+J_b \, .
$$
$J$ is given by the residues of the two poles $z_j = m_j^2 - i m_j
\Gamma_j$:
$$
   J = V_{01}^{-1}(z_1) V_{1\bar0}(z_1) \, \frac{\pi\psi(z_1,{\bf
p})}{\sqrt{z_1 + {\bf p}^2}}
          \, e^{ -i \sqrt{ z_1 + {\bf p}^2} \, T }
        + V_{02}^{-1}(z_1) V_{2\bar0}(z_2) \, \frac{\pi\psi(z_2,{\bf
p})}{\sqrt{z_2 + {\bf p}^2}}
          \, e^{ -i \sqrt{ z_2 + {\bf p}^2} \, T } \, .
$$
Since CP violation is neglected, the propagation eigenstates are
the CP eigenstates $K_{1,2}\!\sim\!K^0\pm\bar K^0$, so that the
matrix $V$ is given at the pole by
\begin{equation}
   V(z_1) \cong V(z_2) \cong \frac{1}{ \sqrt{2} }
   \left(
   \begin{array}{rr}
   1 & 1 \\
   1 & -1
   \end{array}
   \right) \, .
   \label{matrixVpole}
\end{equation}
We shall take into account, in the evaluation of $J_b$, that these
expressions are not valid far from the poles. An estimate of $J$
is given by
\begin{equation}
   J \sim \frac{1}{m} \, \psi(m^2,{\bf p}) \, e^{-\Gamma_2 T/2}
   \left| \,
          e^{ - i \Delta m T - \Delta \Gamma T/2 }  - 1 \,
   \right| \, ,
   \label{grandeurJ}
\end{equation}
where $m$ is the mass in the degenerate limit, $\Delta m=m_1-m_2$,
$\Delta\Gamma=\Gamma_1-\Gamma_2$ and we have approximated $\sqrt{
m^2+{\bf p}^2 }\cong m$.

The contributions $J_1$ and $J_2$, due to the bounded spectrum of
the overlap function, can be computed separately on each term of
the right-hand side of Eq.~(\ref{G00}), exactly as in the
one-particle case. The result is likewise negligible.

Let us examine in detail the contribution of $J_b$.
Eq.~(\ref{exprJb}) of the Appendix shows that $J_b$ depends on the
difference between the complete propagator $G_{\beta\alpha}(z)$
and its analytic continuation $G_{\beta\alpha,II}(z)$:
\begin{equation}
    J_b = -i \, e^{ -i \sqrt{ b^2 + {\bf p}^2 } T }
   \int_0^\infty d\omega \, \psi(z(\omega),{\bf p}) \,
   \left( G_{0\bar0 \,,\, II}(z) - G_{0\bar0}(z) \right) \,
   e^{-\omega T} \, .
   \label{Jbnondiag}
\end{equation}
The difference $G_{II}(z) - G(z)$ can be computed with the
parametrization (\ref{propKK}). To first order in
$f_{\alpha\beta}$, we have
\begin{equation}
   G_{0\bar0 \,,\, II}(z) - G_{0\bar0}(z) =
   -i \left( f_{0\bar0 \,,\, II}(z) - f_{0\bar0}(z) \right)
    G_{0\bar0}(z) \, G_{0\bar0 \,,\, II}(z) \, .
   \label{differenceGGtwo}
\end{equation}
For large $T$, the dominant contribution to $J_b$ comes from the
$\omega$ values near zero, that is $z=b^2$, because of the
decreasing exponential. It is thus enough to know how the
integrand tends to zero near the threshold $z=b^2$. Just under the
real axis ($z=x-i\epsilon$), the spectral representation of the
self-energy given by Eq.~(\ref{spectralenergy}) entails that
$$
   f_{0\bar0 \,,\, II}(z) - f_{0\bar0}(z)
   = f_{0\bar0}(x+i\epsilon) - f_{0\bar0}(x-i\epsilon)
   = 2i \, {\cal I}\!m \, f_{0\bar0}(x+i\epsilon) \, .
$$
The value at the pole of the function $f_{0\bar0}(x+i\epsilon)$
can be related to experimentally known quantities with the help of
Eqs.~(\ref{propKK}) and (\ref{matrixVpole}):
$$
   {\cal I}\!m \, f_{0\bar0}(m^2+i\epsilon) \cong -
\frac{m\,\Delta\Gamma}{2} \, .
$$
As the main decay channels of $K^0$ and $\bar{K}^0$ are two-pion
decays, the nondiagonal self-energy ${\cal I}\!m \,
f_{0\bar0}(x+i\epsilon)$ has a two-particle threshold behavior:
$$
   {\cal I}\!m \, f_{0\bar0}(x+i\epsilon)
   \cong - \frac{m\,\Delta\Gamma}{2} \,
   \sqrt{ \frac{x-b^2}{m^2-b^2} } \, .
$$
Eq.~(\ref{differenceGGtwo}) can then be rewritten as
$$
   G_{0\bar0 \,,\, II}(z) - G_{0\bar0}(z)
   \cong - m \, \Delta \Gamma \, \sqrt{ \frac{z-b^2}{m^2-b^2} } \,
   G_{0\bar0}(z) \, G_{0\bar0 \,,\, II}(z) \, .
$$
Substituting $y=\omega T$ in Eq.~(\ref{Jbnondiag}) and expanding
the integrand in $1/T$ with the help of the $z(\omega)$
parametrization given by Eq.~(\ref{contourb}), we obtain
$$
   z = b^2 - 2 i \frac{y}{T} \, \sqrt{ b^2 + {\bf p}^2 }
       \;+\; {\cal O}\left( T^{-2} \right) \, .
$$
It follows that
$$
   J_b \sim T^{-3/2} \;
   \frac{ m^{3/2} \, \Delta \Gamma }{ (m^2 - b^2)^{5/2} } \;
   \psi(b^2,{\bf p}) \;
   e^{-i \sqrt{ b^2 + {\bf p}^2 } \, T} \, ,
$$
where we have approximated $b^2 + {\bf p}^2 \cong m^2$ and used
the fact that the self-energy is negligible at the threshold. Thus
$J_b$ is of the order of
\begin{equation}
   J_b \sim
   (QT)^{-3/2} \, \frac{ \Delta \Gamma }{Q}\, \frac{1}{m} \,
   \psi(b^2,{\bf p}) \, .
   \label{grandeurJb}
\end{equation}
It is now possible to compare the estimates of $J$ and $J_b$,
given by Eqs.~(\ref{grandeurJ}) and (\ref{grandeurJb}),
respectively.

\paragraph{At small times:}

$J_b \sim J$ if
$$
   (QT)^{-3/2} \, \frac{ |\Delta \Gamma| }{Q}
   \sim \bigg| \sin \frac{\Delta m \, T}{2} \bigg|
   \sim \frac{|\Delta m| \, T}{2} \, .
$$
It is known that $|\Delta \Gamma| \lesssim |\Delta m|$ (from the
experimental data in the case of the $K$ mesons \cite{Hagiwara02}
and from theoretical predictions in the case of $B$ mesons
\cite{Fleischer97}). Thus, $J_b \sim J$ if $\, T \sim  Q^{-1}$.
Since $Q$ is of the order of $0.2-1\,$GeV, for the $K$ and $B$
mesons, nonexponential corrections are dominant at small times for
$T \lesssim 10^{-24}\;$s. This time range is not observable.

\paragraph{At large times:}

$J \sim J_b$ if
$$
   e^{-\Gamma_L T/2}
   \sim (QT)^{-3/2} \, \frac{ |\Delta\Gamma| }{Q}
   \sim (\Gamma_L \, T)^{-3/2} \,
   \frac{ |\Delta\Gamma| \, \Gamma_L^{3/2} }{ Q^{5/2} } \, ,
$$
where $\Gamma_L=min(\Gamma_1,\Gamma_2)$, that is if
$$
   \Gamma_L \, T - 3 \, \ln (\Gamma_L \, T)
   \sim   5 \, \ln \left( \frac{Q}{\Gamma_L} \right)
        - 2 \, \ln \left( \frac{|\Delta\Gamma|}{\Gamma_L} \right) \, .
$$
For kaons, $Q\!\sim\!220\,$MeV so that $\Gamma_L \, T \sim 190$.
The amplitude is negligible at such large times so that
nonexponential corrections are not observable. For $B$ mesons,
$Q\!\sim\!1\,$GeV and $\Delta \Gamma/\Gamma_L$ can be estimated at
$4 \times 10^{-3}$ for the $B^0_d$ and at $10^{-1}$ for the
$B^0_s$ \cite{Fleischer97}, so that $\Gamma_L \, T \sim 168$ in
the former case and $\Gamma_L \, T \sim 162$ in the latter. Thus,
nonexponential corrections are not observable in either case.

The estimate (\ref{grandeurJb}) of nonexponential corrections to
the propagation of mixed neutral mesons is in agreement with the
theoretical formulas obtained by Chiu and Sudarshan (Eq.~(3.30) of
\cite{Chiu90}), and by Wang and Sanda (Eq.~(59) of \cite{Wang97}).
The authors of these two papers compute nonexponential corrections
in a quantum-mechanical framework and extend the
Wigner-Weisskopf-Lee-Oehme-Yang formalism
\cite{Lee57,Kabirww,Nachtmann} (see also \cite{Nowakowski99}).

In conclusion, nonexponential corrections can be neglected in the
propagation of quasi-stable mixed particles. Therefore, the
flavor-mixing amplitude (\ref{amplitot}) with constant matrices
$V$ can be used not only in the stable case, but also for an
unstable oscillating particle.

\subsection{Oscillation formula for unstable particles}
\label{sectionunstable}

The approximation of the amplitude ${\cal A}_j$ and of the time
average with Laplace's or the stationary phase methods can be
transposed for an unstable oscillating particle. The decay term
$\gamma_j(p)$ introduces a new small parameter, namely
$m_j\Gamma_j/p^2$. This parameter is of order $\epsilon=\delta
m_{ij}^2/2p^2$, or smaller, since the mean decay length
$L^{decay}_j=p/m_j\Gamma_j$ should be of the order, or larger,
than the oscillation length $L^{osc}_{ij}=4\pi p/\delta m_{ij}^2$,
or else the oscillations would not be observable. In parallel with
the discussion of section \ref{alternative}, the influence of this
new parameter on the transition probability shows up in the
argument of the exponential, through combinations with large
dimensionless coefficients.

Let us inspect the possible combinations of $\epsilon$ with the
available large parameters, where $\epsilon$ refers either to
$(pL^{osc})^{-1}$ or to $(pL^{decay})^{-1}$. First, if the
conditions of production and detection are not considered, only
one such coefficient is available, i.e.\ $pL$. This combination
gives the oscillation term \mbox{$exp(-2\pi iL/L^{osc})$} and the
usual exponential decrease of the probability in
\mbox{$exp(-L/L^{decay})$}. Next, the analysis of the propagation
with external wave packets adds a new parameter, i.e.\
$\sigma_{peff}$. The combinations $(\epsilon p/\sigma_{peff})^2$
and $(\epsilon \sigma_{peff} L)^2$ give the following
contributions to the argument of the exponential (with minus signs
omitted):
\begin{enumerate}
   \item
   Decay-independent terms, like the decoherence term, $(\frac{L}{L^{coh}})^2$,
   and the localization term, $(\frac{\sigma_{xeff}}{L^{osc}})^2$, which were
   already obtained in the stable case.
   \item
   Mixing-independent terms, like
   $\left( \frac{\sigma_{peff}}{p} \frac{L}{L^{decay}} \right)^2$ and
   $\left(\frac{\sigma_{xeff}}{L^{decay}}\right)^2$. The first is completely
   negligible in comparison with the exponential decrease at order $\epsilon$.
   The second imposes that $L^{decay} \gg \sigma_{xeff}$, which is true for
   quasi-stable particles.
   \item
   Mixing-decay cross terms, like
   $\frac{\sigma_{xeff}^2}{L^{osc}L^{decay}}$ and
   $\frac{\sigma_{peff}^2}{p^2}\frac{L^2}{L^{osc}L^{decay}}$.
\end{enumerate}
With the help of the condition ${\cal O}(L^{decay}) \gtrsim {\cal
O}(L^{osc})$, it can be seen that the new contributions, with
respect to the stable case, do not impose new constraints for the
observability of the oscillations. They are thus neglected in the
transition probability.

We have yet to check whether the oscillation phase is modified or
not by the widths $\Gamma_j$. Let us consider first the
transversal-dispersion regime (the no-dispersion regime can be
treated similarly). The position $p_j$ of the maximum is shifted
by the decay term $\gamma_j(p)$. The new maximum is the solution
of
$$
   f_j'(p) + \gamma_j'(p)=0 \, ,
$$
where $f_j(p)$ and $\gamma_j(p)$ are defined by
Eqs.~(\ref{definitionfj}) and (\ref{defgammaj}), respectively. It
can be computed as before to first order in $\epsilon$ ($\epsilon$
referring to $\delta m_j^2/2E_0^2$ or to $m_j\Gamma_j/2E_0^2$) and
is equal to
\begin{equation}
   p_j = p_j^{(0)} + \delta p_j^{(\Gamma)} \, ,
   \label{pjgamma}
\end{equation}
where $p_j^{(0)}$ is the solution for $\Gamma_j=0$, given by
Eq.~(\ref{expansionpj}), whereas $\delta p_j^{(\Gamma)}$ is the
contribution from the decay term, given by
\begin{equation}
   \delta p_j^{(\Gamma)} \lesssim p_0 \, \frac{\sigma_{peff}^2}{E_0^2} \,
   \frac{T}{T^{decay}_j} \, .
   \label{deltapjgamma}
\end{equation}
The decay time is defined by $T^{decay}_j=E_0/m_j\Gamma_j$.
Although the expansion parameter $m_j\Gamma_j/2E_0^2$ is
multiplied in the exponential by the large parameter $T$, the
shift $\delta p_j^{(\Gamma)}$ is small and the expansion
(\ref{pjgamma}) is valid as long as the propagation time $T$ is
not much larger than the decay time $T^{decay}_j$. The
corresponding energy $E_j=\sqrt{p_j^2+m_j^2}$ and velocity
$v_j=p_j/E_j$ are given by
\begin{eqnarray}
   E_j &=& E_j^{(0)} + \delta E_j^{(\Gamma)} \, ,
   \label{Ejgamma} \\
   v_j &=& v_j^{(0)} + \delta v_j^{(\Gamma)} \, ,
   \label{vjgamma}
\end{eqnarray}
where $E_j^{(0)}$ and $v_j^{(0)}$ are the solutions for
$\Gamma_j=0$, given by Eqs.~(\ref{expansionEj}) and
(\ref{expansionvj}), respectively, whereas the contributions from
the decay term are equal to \mbox{$\delta E_j^{(\Gamma)}=v_0
\delta p_j^{(\Gamma)}$} and \mbox{$\delta v_j^{(\Gamma)}=m_0^2
\delta p_j^{(\Gamma)} /E_0^3$}.

Apart from the modified values of $p_j$, $E_j$ and $v_j$, the
longitudinal momentum integration with Laplace's method yields the
same results as in section \ref{laplaceregimelongitud}. The phase
of the interference term ${\cal A}_i{\cal A}_j^*$ is thus given,
before the time average, by Eq.~(\ref{defphaseij}), but with the
modified values (\ref{pjgamma}), (\ref{Ejgamma}) and
(\ref{vjgamma}). It can be split in a $\Gamma-$independent part
\mbox{$\phi_{ij}^{(0)}(T,L)$}, and a $\Gamma-$dependent part
\mbox{$\phi_{ij}^{(\Gamma)}(T,L)$}. The latter is equal to
\begin{equation}
   \phi_{ij}^{(\Gamma)}(T,L) = (v_0T-L) \,
   (\delta p_i^{(\Gamma)} - \delta p_j^{(\Gamma)} ) \,
   \left( 1 - \frac{\ell^2}{1+\ell^2} \right) \, ,
   \label{defphaseijgamma}
\end{equation}
where $\ell=T/T^{disp}$ and the definition (\ref{dispersiontime})
of $T^{disp}$ has been used. Using the condition $|v_0T-L|
\lesssim \sigma_{xeff}$ (see Eq.~(\ref{amplitudelaplace})) and the
values of $p_{i,j}^{(\Gamma)}$ given by Eqs.~(\ref{pjgamma}) and
(\ref{deltapjgamma}), we have
$$
  \left| \phi_{ij}^{(\Gamma)}(T,L) \right| \lesssim
  v_0 \, \frac{\sigma_{peff}}{E_0} \,
  \left| \frac{T}{T^{decay}_i} - \frac{T}{T^{decay}_j} \right| \, ,
$$
which is much smaller than 1 in the observable domain ${\cal
O}(T)\lesssim {\cal O}(T^{decay})$. Therefore, the oscillation
phase is not modified by the width:
$$
   \left| \phi_{ij}^{(\Gamma)}(T,L) \right|
   \ll \left| \phi_{ij}^{(0)}(T,L) \right| \, .
$$

Let us now turn our attention to the longitudinal-dispersion
regime. Of course, the stationary point $p_{cl,j}$ is not shifted
by the decay term $\gamma_j(p)$, but the phase receives a
contribution from the derivative of the decay term. More
precisely, the phase (\ref{defphaseijtilde}) becomes
\begin{equation}
   \tilde \phi_{ij}(T,L) = \delta m_{ij} \, \sqrt{T^2-L^2}
   + \frac{\sigma_{peff}^2\,\ell}{1+\ell^2} \,
     \left(    \left( f_i'(p_{cl,i}) + \gamma_i'(p_{cl,i} ) \right)^2
             - \left( f_j'(p_{cl,j}) + \gamma_j'(p_{cl,j} ) \right)^2
     \right)
   \label{phaseijtildegamma}
\end{equation}
where $\ell=T/T^{disp}$ as above. The derivative of the decay term
is equal to
$$
   \gamma_j'(p_{cl,j}) = - \frac{p_0}{2E_{cl,j}^2} \, \frac{L}{L^{decay}_j} \, .
$$
The terms depending on $\Gamma_{i,j}$ are proportional to
$\sigma_{peff}\gamma_j'$. They will be small in the observable
domain ${\cal O}(T)\lesssim {\cal O}(T^{decay})$, so that they can
be neglected in the phase (\ref{phaseijtildegamma}).

Thus the oscillation phase is not modified by the decay term at
any distance. The result (\ref{proba}), obtained for a stable
oscillating particle, is only modified by the usual exponential
decrease $exp(-L/L^{decay}_{ij})$, with
\begin{equation}
   L^{decay}_{ij}=\frac{2p_0}{m_i\Gamma_i+m_j\Gamma_j} \, .
   \label{Ldecay}
\end{equation}

In conclusion, the flavor-mixing transition probability for a
long-lived unstable particle (i.e.\ satisfying $\sigma_{xeff}\ll
L^{decay}_{ij}$), having a relativistic or a nonrelativistic
velocity (but always with $\delta m_{ij}^2 \ll E_0^2$), and with
dispersion taken into account, is given for $L\gtrsim
p_0/\sigma_p^2$ by
\begin{eqnarray}
   && {\cal P}_{\alpha \to \beta}({\bf L}) = \frac{N_g g^2({\bf l})}{L^2} \,
   \sum_{i,j} V_{i\alpha} \, V_{\beta i}^{-1} \,
          V_{j\alpha}^* \, V_{\beta j}^{-1 \, *} \,
   \exp \left( -\frac{L}{L^{decay}_{ij}} \right) \nonumber \\
   && \hspace{1cm} \times \exp \left(  -2 \pi i \frac{L}{L^{osc}_{ij}}
   - \frac{(\delta m_i^2)^2+(\delta m_j^2)^2}{16 \sigma_m^2 E_0^2}
   - 2 \pi^2 \left( \frac{\rho \, \sigma_{xeff}}{L^{osc}_{ij}} \right)^2
   - \left( \frac{L}{L^{coh}_{ij}} \right)^2
               \right) \, .
   \label{probaunstable}
\end{eqnarray}
The definitions of the characteristic lengths and parameters
appearing in this formula are gathered below Eq.~(\ref{proba}).
For $L\lesssim p_0/\sigma_p^2$, the oscillation formula for stable
oscillating particles is similarly modified by the insertion of a
decay term into Eq.~(\ref{probanodisp}).

\subsection{Wigner-Weisskopf effective mass matrix}
\label{wweffmass}

Oscillating neutral mesons are usually described in quantum
mechanics with the Wigner-Weiss\-kopf method as extended by Lee,
Oehme and Yang \cite{Lee57,Kabirww,Nachtmann}. In this framework,
the evolution of a two-meson state satisfies a Schr\"odinger-like
equation with a complex (and non-Hermitian) effective mass matrix
$H$, which can be split into a Hermitian mass matrix $M$ and a
Hermitian decay matrix $\Gamma$, $H=M-i\Gamma$. It would be
interesting to establish a correspondence between this method and
the field-theoretical formula (\ref{probaunstable}).

First, let us simplify the oscillation formula
(\ref{probaunstable}). Note that the coherence length can be
neglected for particles such as the neutral $K$ or $B$ mesons,
since the oscillation length is comparable with the decay length:
$$
   \frac{L^{coh}_{ij}}{L^{decay}_{ij}} \sim \frac{p_0}{\sigma_{peff}} \,
   \frac{L^{osc}_{ij}}{L^{decay}_{ij}} \gg 1 \, .
$$
Next, the quasi-degeneracy of the mass eigenstates makes it
impossible to set up experimental conditions which destroy
oscillations. Thus the localization terms (containing the
parameters $\sigma_m$ and $\rho$) drop from
Eq.~(\ref{probaunstable}). These approximations lead to the
following simplified formula:
\begin{equation}
   {\cal P}_{\alpha \to \beta}({\bf L}) = \frac{N_g g^2({\bf l})}{L^2} \,
   \sum_{i,j} V_{i\alpha} \, V_{\beta i}^{-1} \,
          V_{j\alpha}^* \, V_{\beta j}^{-1 \, *} \,
   \exp \left( -\frac{L}{L^{decay}_{ij}} - 2 \pi i
\frac{L}{L^{osc}_{ij}} \right) \, .
   \label{probasimpli}
\end{equation}
The comparison of Eq.~(\ref{probasimpli}) with the
Wigner-Weisskopf-Lee-Oehme-Yang theory is more easily done at the
level of the amplitude. Whereas the true amplitude depends on both
$T$ and ${\bf L}$, the probability (\ref{probasimpli}) can,
equivalently, be generated from the following {\it effective
amplitude}
\begin{equation}
   {\cal A}_{eff}(\alpha\to\beta,{\bf L}) =
   \frac{\sqrt{N_g} g({\bf l})}{L} \, \sum_j V_{\beta j}^{-1} \,
   \exp \left(
              -i \left( m_j -i\frac{\Gamma_j}{2} \right) \frac{m_0 L}{p_0}
        \right)
   V_{j\alpha} \, ,
   \label{effectiveampli}
\end{equation}
where the masses are assumed to be nearly degenerate ($m_i\cong
m_j\cong m_0$). If $M$ is a diagonal matrix the diagonal terms of
which are given by $m_j-i\Gamma_j/2$, the effective amplitude can
be written as
\begin{eqnarray*}
   {\cal A}_{eff}(\alpha\to\beta,{\bf L})
   &=& \frac{\sqrt{N_g} g({\bf l})}{L} \;
   \left(
         V^{-1} \exp \left( -iM \, \frac{m_0L}{p_0}  \right) V
   \right)_{\beta\alpha}
   \\ &=& \frac{\sqrt{N_g} g({\bf l})}{L} \;
   \left(
         \exp \left( -iM_{flavor} \, \frac{m_0L}{p_0} \right)
   \right)_{\beta\alpha} \;\; ,
\end{eqnarray*}
where $M_{flavor}=V^{-1}MV$. The mass matrix $M_{flavor}$
corresponds to the effective Hamiltonian in the Lee-Oehme-Yang
theory. We recognize the factor $ m_0 L/p_0$ as being the
classical proper time of propagation, common to the two mass
eigenstates. However, let us again emphasize that this observation
does not justify the equal time prescription, which was explained
in section \ref{timespaceconversion}. It must not be forgotten
that $T$ and $L$ are the {\it average} propagation time and
length, and that interference takes place in a time and space
range defined by the width of the effective `wave packet'.

The effective oscillation amplitude confirms the soundness of the
reciprocal basis treatment in quantum mechanics explained in
section \ref{oscilwithplane}. Recalling that the diagonalization
matrix $V$ is related to the matrix $U$ diagonalizing the states
through $V=U^t$, the Hamiltonian (\ref{Hreciprocal}) can be
written as
\begin{equation}
   \hat H_{propag} =
   | \, \nu_\beta \!> \,
   \sum_j V^{-1}_{\beta j}\, e^{-i \lambda_j t} \, V_{j\alpha} \,
   <\! \nu_\alpha \, | \, ,
   \label{hathamilt}
\end{equation}
where $\lambda_j=m_j-i\Gamma_j/2$. If the time $t$ is taken to be
$m_0 L/p_0$, the matrix element $<\! \nu_\beta \, | \hat
H_{propag} | \, \nu_\alpha \!>$ is in correspondence with the
amplitude ${\cal A}_{eff}(\alpha\!\to\!\beta,{\bf L})$ given by
Eq.~(\ref{effectiveampli}). The field-theoretical method will thus
give the same results as the reciprocal basis formalism in the
case of nearly degenerate unstable states. Both formalisms are
easier to use that the cumbersome nonorthogonal basis
(\ref{orthogmasse}), since any transition amplitude can be simply
computed as a matrix product. Since the reciprocal basis formalism
has already been applied to the study of CP violation (see
\cite{Sachs63,Enz65,Alvarez-Gaume99,Branco,Silva00}), we shall
only give one example in section \ref{oscillBB} of the application
of formula (\ref{effectiveampli}) to the computation of a CP
asymmetry.

Let us end this section by an remark on the observability of
intermediate unstable particles. It is tempting to associate a
well-defined mass eigenstate with the exponential behavior in
$e^{-i \lambda_j t}$ in Eqs.~(\ref{effectiveampli}) or
(\ref{hathamilt}), but normalized mass eigenstates cannot be
defined if the mass matrix $M_{flavor}$ is not normal (see section
\ref{mixing}). The reason is that a final state can be produced by
all mass channels. In practice, it is sometimes possible to
maximize the probability of a channel (since the lifetimes
$\Gamma_j$ can be very different). In that case, one talks about
decays into a mass eigenstate represented by its reciprocal basis
vector ${}_{out}\!<\! \nu_j(0) \, |$ \cite{Branco}.

\subsection{Oscillations in the $B^0\bar B^0$ system}
\label{oscillBB}

$B^0\bar B^0$ oscillations were first measured in 1987
\cite{Albrecht87}, but the discovery of CP violation in this
system is very recent \cite{Abe01,Aubert01}. We compute here as an
example the following CP asymmetry \cite{Nir}:
\begin{equation}
   A_{CP}(B^0_d \to f,L) =
   \frac{ \Gamma\left( B^0_d(L) \to f\right) - \Gamma\left( \bar
B^0_d(L) \to f\right) }
        { \Gamma\left( B^0_d(L) \to f\right) + \Gamma\left( \bar B^0_d
(L) \to f\right) } \, ,
   \label{asymCP}
\end{equation}
where the index $f$ refers to a CP eigenstate. The propagation
eigenstates have nearly equal decay widths \cite{Fleischer97}, and
must be distinguished by their different masses. Let us define
$\Delta m_d \equiv m_{\scriptscriptstyle H} - m_{\scriptscriptstyle L}$, where the indices refer
to $B_L$, for {\sl Light}, and $B_H$, for {\sl Heavy}.

In the limit of a constant self-energy matrix $f_{\alpha\beta}$,
the diagonalizing matrix of the mixed propagator (\ref{propKK})
can be parametrized by
\begin{equation}
   V = \frac{v}{ \sqrt{2\sigma} }
   \left(
   \begin{array}{rr}
   \sigma & 1 \\
   \sigma & -1
   \end{array}
   \right)
   \hspace{1cm} \mbox{and} \hspace{1cm}
   V^{-1} = \frac{v^{-1}}{ \sqrt{2\sigma} }
   \left(
   \begin{array}{rr}
   1 & 1 \\
   \sigma & -\sigma
   \end{array}
   \right) \, ,
   \label{matriceV}
\end{equation}
where $\sigma^2=f_{\bar0 0}/f_{0\bar 0}$. Note that the phase of
this parameter depends on the phase convention chosen for flavor
states. The physics should be invariant \cite{Branco} under
\begin{equation}
   | B^0 \rangle \to e^{i\gamma} \, | B^0 \rangle
   \hspace{1cm} \mbox{and} \hspace{1cm}
   | \bar B^0 \rangle \to e^{i\bar\gamma} \, |\bar B^0 \rangle \, .
   \label{rephasing}
\end{equation}
Under the ket rephasing (\ref{rephasing}), $\sigma$ transforms as
$\sigma\to e^{i(\gamma-\bar\gamma)}\sigma$. This parameter is
related to the notations of \cite{Branco} through
$\sigma=-q_B/p_B$. The deviation of $|\sigma|$ from 1 parametrizes
the amount of CP violation in mixing, called `indirect CP
violation'. The constant $v$ in Eq.~(\ref{matriceV}) is an
arbitrary normalization, underlining the fact that the meson
cannot be considered as an asymptotic state. Its decay amplitude
should always be included in the full amplitude.

Theoretical calculations (\cite{Branco} and references therein)
show that $|\sigma|-1\sim{\cal O}(10^{-4})$, which is much smaller
than the present experimental uncertainty, so that only a small CP
violation occurs in the mixing. On the other hand, a much larger
CP violation is expected in the decays (`direct CP violation'),
since three quark generations are involved in processes such as
$B_d \to J/\psi K_S$. Thus, CP violation in mixing is usually
neglected for $B$ mesons (i.e.\ $|\sigma|=1$), which is the
opposite of what occurs with neutral kaons.

The computation of the asymmetry (\ref{asymCP}) requires the
knowledge of the amplitude ${\cal T}_f(L)$, corresponding to the
propagation of an initial $B^0$. Starting from the effective
oscillation amplitude (\ref{effectiveampli}), we can write ${\cal
T}_f(L)$ as
$$
   {\cal T}_f(L) \sim
   \left(
          {\cal M}(B^0 \!\to\! f ) \quad
          {\cal M}(\bar B^0 \!\to\! f )
   \right) \,
    V^{-1} \, \exp \left\{ -i M \frac{m_0L}{p_0} \right\} \, V \,
   \left(
   \begin{array}{c}
   {\cal C} \\ 0
   \end{array}
   \right) \, ,
$$
where the effective mass matrix $M$ is equal to
$diag(m_L-i\Gamma_L/2,m_H-i\Gamma_H/2)$. It will be useful to
define the parameter $\mu_f$ by
\begin{equation}
   \mu_f = \frac{ {\cal M}( \bar B^0 \to f ) }{ {\cal M}( B^0 \to f ) } \, .
   \label{defmu}
\end{equation}
Under the rephasing (\ref{rephasing}) of the kets, $\mu_f\to
e^{-i(\gamma-\bar\gamma)} \mu_f$. If $|\mu_f|\neq1$, `direct CP
violation' occurs in decay amplitudes to the state $f$. Whereas
the quantities $\sigma$ and $\mu_f$ are not invariant under the
rephasing (\ref{rephasing}), the product $\sigma\mu_f$ is
invariant. This quantity is related to the notations of
\cite{Branco} through $\sigma\mu_f=-\lambda_f$. If the final state
is a CP eigenstate and the transition dominated by a single CKM
amplitude (for example $B^0_d \!\to\! J/\psi K_S$), $\sigma\mu_f$
can be expressed in terms of the elements of the quark mixing
matrix, so that its measurement is very important for the
determination the CKM matrix elements.

The amplitude ${\cal T}_f(L)$ can be written as
\begin{equation}
   {\cal T}_f(L) \sim
   \frac{ {\cal C} {\cal M}(B_0 \!\to\! f)}{2} \,
   \left(
   ( 1 + \sigma\mu_f ) \,
   e^{ - i m_L \frac{m_0}{p_0}L - \frac{ m_0}{2p_0} \Gamma_L L}
   + ( 1 - \sigma\mu_f ) \,
   e^{- i m_H \frac{m_0}{p_0}L - \frac{ m_0}{2p_0} \Gamma_H L}
   \right) \, ,
   \label{ampli+-}
\end{equation}
where $p_0$ is the modulus of the total momentum of the final
pions.

We also need to compute the amplitude $\overline{ {\cal T}_f
}(L)$, corresponding to the propagation of an initial $\bar B^0$.
The same method yields
\begin{equation}
   \overline{ {\cal T}_f }(L)  \sim
   \frac{ {\cal C}^* {\cal M}(B_0 \!\to\! f)}{2\sigma} \,
   \left(
   ( 1 + \sigma\mu_f ) \,
   e^{ - i m_L \frac{m_0}{p_0}L - \frac{ m_0}{2p_0} \Gamma_L L}
   - ( 1 - \sigma\mu_f ) \,
   e^{- i m_H \frac{m_0}{p_0}L - \frac{ m_0}{2p_0} \Gamma_H L}
   \right) \, .
   \label{amplibar+-}
\end{equation}
With the insertion of Eqs.~(\ref{ampli+-})-(\ref{amplibar+-}) and
the approximations $|\sigma|=1$ and $\Gamma_H=\Gamma_L$, the
asymmetry (\ref{asymCP}) becomes
$$
   A_{CP}(B^0_d \!\to\! f,L) =
   A_{CP}^{dir}(B^0_d \!\to\! f) \, \cos \left(\Delta m_d
\frac{m_0L}{p_0}\right)
   + A_{CP}^{interf}(B^0_d \!\to\! f) \, \sin \left(\Delta m_d
\frac{m_0L}{p_0}\right) \, ,
$$
where direct CP violation, defined by
$$
   A_{CP}^{dir}(B^0_d \to f) =
   \frac{1-|\lambda_f|^2}{1+|\lambda_f|^2} =
    \frac{1-|\mu_f|^2}{1+|\mu_f|^2} \, ,
$$
has been separated from CP violation coming from the interference
between the mixing and the decay:
$$
   A_{CP}^{interf}(B^0_d \to f) =
     \frac{2 \, Im \,\lambda_f }{1+|\lambda_f|^2}
   =  - \frac{2 \, Im \,(\sigma \mu_f) }{1+|\mu_f|^2} \, .
$$
Three different kinds of CP violation have been met in this
section: indirect, direct and interference CP violations, signaled
by $|\sigma|\neq1$, $|\mu_f|\neq1$ and $Im(\sigma\mu_f)\neq0$,
respectively (see \cite{Branco} p.~78). If the final state is
$J/\psi \, K_S$, a theoretical study (see for example
\cite{Fleischer97}) shows that the CP violation parameter is in a
very good approximation equal to $\lambda_{\scriptscriptstyle J/\psi \, K_S} =
e^{-2i\beta}$, where $\beta$ is one of the angles of the unitary
triangle. The CP violation in the neutral kaon system that is
included in the final state can be neglected. The direct and
interference asymmetries become
$$
    A_{CP}^{direct}(B^0_d \to J/\psi \, K_S) = 0
   \quad \mbox{and} \quad
   A_{CP}^{interf}(B^0_d \to J/\psi \, K_S) = - \sin 2\beta \, .
$$
The quantities $\sin 2\beta$ and $|\lambda|$ were recently
measured by the BABAR \cite{Aubert01,Aubert02} and Belle
\cite{Abe01,Abe02} collaborations:
\begin{eqnarray*}
   && \sin 2\beta_{BABAR} = 0.741 \pm 0.067 \,(stat) \pm 0.033 \,(syst) \, , \\
   && \sin 2\beta_{Belle} = 0.719 \pm 0.074 \,(stat) \pm 0.035 \,(syst) \, \\
   && |\lambda_{BABAR}| = 0.948 \pm 0.051 \,(stat) \pm 0.017 \,(syst) \, .
\end{eqnarray*}
The value of the interference asymmetry is an indication of CP
violation in the $B$ system, whereas the value of $|\lambda|$ is
consistent with no direct CP violation.

\section{Selected topics}
\label{section8}

In section \ref{normalization}, we examine the question of the
normalization of the transition probability. We then discuss in
section \ref{cohincohdecoh} the relation between coherent and
incoherent effects, as well as the different estimates of the wave
packet size. Finally, the case of an unstable source is reviewed
in section \ref{unstablesource}.

\subsection{Normalization of the transition probability}
\label{normalization}

Until now, we have not explained why the time-averaged squared
modulus of the amplitude (see Eq.~(\ref{transitionproba})) can be
interpreted as the flavor-mixing transition probability. Actually,
the interpretation of the expression (\ref{proba}) as a transition
probability was a bit of a guess. As a matter of fact, the steps
between the amplitude ${\cal A}(\alpha\to\beta)=\sum_j V_{\beta
j}^{-1} {\cal A}_j V_{j\alpha}$ and a normalized event rate,
containing an expression which can be interpreted as a
flavor-mixing transition probability, are not straightforward. In
the case of the large-distance oscillation formula, the constant
prefactors $v_0N^2\sigma_{peff}$ appearing in
Eq.~(\ref{averagelaplace}) (or in Eqs.~(\ref{averagephasestat})
and (\ref{averageshortcut})) have been shoved into a constant
$N_g$, which was normalized in  Eq.~(\ref{normNg}) in order to
obtain a unitary evolution in the stable case. The normalization
of the short-distance oscillation formula (\ref{probanodisp})
proceeded in the same way. Whereas these prefactors can be
factorized from the sum on the mass eigenstates if \mbox{$\delta
m_{ij}^2 \ll E_0^2$}, this is not possible for a mixing of a
relativistic $m_i$ and a nonrelativistic $m_j$ mass eigenstate.

Cardall has made an attempt \cite{Cardall00} to go all the way to
a normalized event rate. Although he claims to resort to arbitrary
external wave packets, his model is equivalent to the Gaussian
external wave packet model, since he uses Gaussian approximations
to perform the integrals. We shall explain how Cardall's argument
can be applied to the external wave packet model developed in
sections \ref{section3} to \ref{section5}. Only the large-distance
case will be considered ($L\gtrsim p_0/\sigma_p^2$); the
short-distance case can be treated similarly.

The expression ${\cal A}(\alpha \!\to\! \beta,T,{\bf L}) {\cal
A}^*(\alpha \!\to\! \beta,T,{\bf L})$ (see Eq.~(\ref{amplitot}))
is proportional to a transition probability between one-particle
states. In order to obtain a formula applicable to experiments,
the external wave packets should be interpreted as densities of
particles. Cardall proposes three rules of correspondence:
\begin{enumerate}

\item Let us first examine the external wave packets normalization
constants $N_{P_{in}}$, $N_{P_{out}}$, $N_{D_{in}}$ and
$N_{D_{out}}$. They come from the overlap function and are
included in the factor $N^2$ appearing in the transition
probability (\ref{averagelaplace}). Eq.~(\ref{packetin}) shows
that $N_{P_{in}}$ is proportional to the modulus of the initial
wave function at production. The constant $N_{D_{in}}$ is
interpreted in the same way. The constants $N_{P_{in}}^2$ and
$N_{D_{in}}^2$ are then interpreted as the initial state particle
densities at production and detection, respectively:
$$
  N_{P_{in}}^2 \to
  \frac{d{\bf Q}}{(2\pi)^3 2E_{P_{in}}} \, f({\bf Q},x_{\scriptscriptstyle P})
  \hspace{1cm} \mbox{and} \hspace{1cm}
  N_{D_{in}}^2 \to
  \frac{d{\bf Q'}}{(2\pi)^3 2E_{D_{in}}} \, f({\bf Q'},x_{\scriptscriptstyle D}) \, ,
$$
where $f$ is the phase space density. The normalization constants
$N_{P_{out}}^2$ and $N_{D_{out}}^2$ are interpreted as particle
densities for the final state particles at production and
detection:
$$
  N_{P_{out}}^2 \to \frac{d{\bf K}}{(2\pi)^3 2E_{P_{out}}}
  \hspace{1cm} \mbox{and} \hspace{1cm}
  N_{D_{out}}^2 \to \frac{d{\bf K'}}{(2\pi)^3 2E_{D_{out}}} \, .
$$
Note that there may be more than one particle in the initial and
final states.

\item Let us now consider the production and detection widths.
Recall that the factor $N^2$ in Eq.~(\ref{averagelaplace})
contains a volume factor $V^2=(2^{-4}\pi^4\sigma_{p \scriptscriptstyle P}^{-3} \,
\sigma_{e\scriptscriptstyle P}^{-1}\sigma_{p \scriptscriptstyle D}^{-3} \, \sigma_{e\scriptscriptstyle
D}^{-1})^2$ coming from the overlap function (see
Eq.~(\ref{overlapP})). One factor $V$ is interpreted as an
integration on macroscopic space-time variables:
$$
  V \to d{\bf x}_{\scriptscriptstyle P} \, dx_{\scriptscriptstyle P}^0 \, d{\bf x}_{\scriptscriptstyle D} \,
  dx_{\scriptscriptstyle D}^0 \, .
$$
Since the production time is not measured, it is integrated over
in the event rate. This integration is equivalent to the time
average over the propagation time $T$ done in the present report.

\item Finally, the effective width $\sigma_{peff}$, appearing in
front of the exponential in Eq.~(\ref{averagelaplace}), is
interpreted as the result of an integration over the energy
spectrum of the oscillating particle. Furthermore, the second
volume factor $V$, the geometrical factor $g^2({\bf l})$ and the
term including $\sigma_m$ are interpreted, in the limit
$m_i=m_j\equiv \tilde m_0$, as a product of delta functions
constraining the direction of ${\bf L}$ and the energy of the
oscillating particle. More precisely,
$$
   v_0\sigma_{peff} V
   \exp \left(
   - \frac{{\bf p}_0^2 - ({\bf p}_0 \cdot {\bf l})^2}{2\sigma_p^2}
   - \frac{(\tilde m_0^2 - m_0^2)^2}{8\sigma_m^2E_0^2}
       \right)
   \to \pi^8 \int \frac{dE}{\sqrt{2\pi}} \,
   \delta^{(4)}(p-p_{\scriptscriptstyle P}) \, \delta^{(4)}(p-p_{\scriptscriptstyle D}) \, ,
$$
where $p=(E,\sqrt{E^2-\tilde m_0^2}\,{\bf l})$, and with the
approximations $p_{\scriptscriptstyle P}=p_{\scriptscriptstyle D}=p_0$, $v_{\scriptscriptstyle P}=v_{\scriptscriptstyle D}=0$,
so that the energy and momentum decouple. The relationship is
exact in the limit $\sigma_{p\scriptscriptstyle P}=\sigma_{p\scriptscriptstyle D}=0$ (with the
approximations for the momenta and velocities just mentioned).

\end{enumerate}
These three rules lead to the macroscopic event rate at the
detector at time $x_{\scriptscriptstyle D}^0$:
$$
   d\Gamma(x_{\scriptscriptstyle D}^0) = \int d{\bf x}_{\scriptscriptstyle P} \int d{\bf x}_{\scriptscriptstyle D}
   \int \frac{d{\bf Q}}{(2\pi)^3} \, f({\bf Q},x_{\scriptscriptstyle P})
   \int \frac{d{\bf Q'}}{(2\pi)^3} \, f({\bf Q'},x_{\scriptscriptstyle D}) \;
   d\Gamma({\bf Q},{\bf Q'},{\bf x}_{\scriptscriptstyle P},{\bf x}_{\scriptscriptstyle D}) \, ,
$$
with the constraint $x_{\scriptscriptstyle P}^0=x_{\scriptscriptstyle D}^0 - L/v_0$. The single
particle event rate is given by
$$
   d\Gamma({\bf Q},{\bf Q'},{\bf x}_{\scriptscriptstyle P},{\bf x}_{\scriptscriptstyle D}) =
   \int dE \, flux \times P_{mix} \times d\sigma \, .
$$
The $flux$ is the flux of oscillating particles of energy $E$,
produced at $x_{\scriptscriptstyle P}$ and detected at $x_{\scriptscriptstyle D}$. It includes a
phase space factor for final state particles at production, $\int
d{\bf K}$, as well as the delta function $\delta^{(4)}(p-p_{\scriptscriptstyle
P})$, the interaction vertex $|M_P(Q,K)|^2$, the geometrical
factor $1/L^2$, and a velocity factor $|v_0-v_{\scriptscriptstyle D}|$.

The factor $d\sigma$ is the cross section for the interaction of
particle of mass $\tilde m_0$ in the detector. It includes a phase
space factor for final state particles at production, $d{\bf K'}$
(not integrated over if these momenta are measured), as well as
the delta function $\delta^{(4)}(p-p_{\scriptscriptstyle D})$, the interaction
vertex $|M_D(Q',K')|^2$ and the M{\o}ller factor $|v_0-v_{\scriptscriptstyle
D}|^{-1}$.

Finally, $P_{mix}$ is identifiable as the flavor mixing transition
probability:
$$
   P_{mix} = \sum_{i,j} V_{i\alpha} \, V_{\beta i}^{-1} \,
          V_{j\alpha}^* \, V_{\beta j}^{-1 \, *} \,
   \exp \left(
    -2 \pi i \frac{L}{L^{osc}_{ij}}
    - \frac{(\delta m_{ij}^2)^2}{32 p_0^2}
      \left( \frac{v_0^2}{\sigma_m^2} + \frac{\rho^2}{\sigma_{peff}^2} \right)
    - \left( \frac{L}{L^{coh}_{ij}} \right)^2
         \right) \, .
$$
In comparison with ${\cal P}_{\alpha \to \beta}({\bf L})$, given
by Eq.~(\ref{proba}), the geometrical decrease and the
normalization factor, i.e.\ $N_g g^2({\bf l})/L^2$, are included
in the flux and in $d\sigma$. The approximation $m_0=(m_i+m_j)/2$
has also been made in each interference term.

In the case of a mixing of relativistic and nonrelativistic
particles, the flavor-mixing probability does not factorize from
the amplitudes of production and detection. The interference terms
are however negligible because decoherence occurs. The dependence
on the mass $m_j$ should be kept in the noninterference terms
\mbox{$\int dT |{\cal A}_j|^2$}, with the result that the
prefactor $v_0$ in (\ref{averagelaplace}) should be replaced by
$v_j$. Velocity-dependent prefactors were already derived in the
intermediate wave packet model of Giunti, Kim and Lee
\cite{Giunti91}, except that the one-dimensional treatment in that
article leads to a peculiar dependence in $v_j^{-1}$ (in our case,
the conversion of the $1/T^2$ prefactor into a $v^2_j/L^2$
prefactor yields an additional $v^2$, leading to the expected
linear dependence in $v_j$). Cardall's correspondence rules are
still valid, although the flux and the cross-section now depend on
the mass eigenstate. For example, $\tilde m_0$ and $v_0$ are
replaced respectively by $m_j$ and $v_j$ in the third rule. The
single particle event rate is replaced by
$$
   d\Gamma({\bf Q},{\bf Q'},{\bf x}_{\scriptscriptstyle P},{\bf x}_{\scriptscriptstyle D}) =
   \sum_j |V_{j\alpha}|^2 \, |V_{\beta j}^{-1}|^2 \,
   \int dE \, flux_j \times d\sigma_j \, .
$$
Mixings of relativistic and nonrelativistic neutrinos have been
studied by Ahluwalia and Goldman \cite{Ahluwalia97}, who identify
the third mass eigenstate with the $33.9\,$MeV particle suggested
by the KARMEN experiment \cite{Armbruster95}. Note that the
amplitude of production (and also of detection) should be computed
separately for the relativistic and nonrelativistic neutrinos.

\subsection{Coherence, incoherence and decoherence}
\label{cohincohdecoh}

Generally speaking, decoherence is said to occur in particle
oscillations if the interference terms in the transition
probability are averaged to zero by some mechanism. In that case,
the transition probability becomes independent of the distance. We
have seen in section~\ref{section6} that this phenomenon appears
in the large-distance flavor-mixing transition probability
(\ref{proba}) through the localization term and through the
coherence length. Localization conditions, such as $L^{osc}_{ij}
\gtrsim \sigma_x$, determine whether there is decoherence from the
start. If decoherence only occurs beyond a certain distance, the
threshold is called the coherence length. These two phenomena are
closely connected, since the coherence-length and localization
conditions in the transversal-dispersion regime transform into the
localization and coherence-length conditions in the
longitudinal-dispersion regime, respectively. In the end, both
mechanisms of decoherence originate in the wave packets widths of
the external particles.

Other effects lead to similar constraints on the oscillations.
They can be classified in {\it coherent} or {\it incoherent}
effects. In quantum field theory, a coherent effect has to be
taken into account in the amplitude, whereas an incoherent effect
is incorporated into the computation only at the level of the
probability \cite{Grimus99}. Actually, this distinction has to be
made only because approximations made in the computations often
make the intrinsic decoherence effects disappear. For example,
there are endless discussions about the energy and momentum
coherence in the plane wave treatment of oscillations, since this
approximation destroys all natural decoherence mechanisms. This
should be contrasted with the fact that coherence-length and
localization conditions appear explicitly in the oscillation
formula (\ref{proba}) obtained in the external wave packet model.
Moreover, decoherence between different energy components is
automatic in the external wave packet model (see section
\ref{shortcut}). Note that one should be careful not to confuse
the term `(in)coherent', referring here to a constraint applied to
the amplitude or to the probability, with the term
`(de)coherence', referring to the existence or disappearance of
oscillations.

\subsubsection{Incoherent effects}
\label{incoherenteffects}

First, the energy-momentum spread of the beam has to be taken into
account \cite{Bahcall69,Bilenky78,Bahcall}. For example, the
average of the oscillation term over a Gaussian momentum
distribution of width $\Delta p$ and mean value $\bar p$ gives
$$
   \int dp \;
   \exp \left(
     -2\pi i \frac{L}{L^{osc}_{ij}}
     - \frac{(p-\bar p)^2}{2(\Delta p)^2}
        \right)
   \sim \exp \left(
    -2\pi i \frac{L}{\bar L^{osc}_{ij}}
    -2\pi^2 \left(
      \frac{\Delta p}{\bar p} \,\frac{L}{\bar L^{osc}_{ij}}\right)^2
            \right) \, ,
$$
where $\bar L^{osc}_{ij}=4\pi \bar p/\delta m_{ij}^2$. A new
coherence length can be defined by
$$
   \bar L^{coh}_{ij} = \frac{1}{\sqrt{2}\pi} \, \frac{\bar p}{\Delta p} \,
   \bar L^{osc}_{ij} \, .
$$
This new coherence length can be obtained from the coherence
length (\ref{cohlengthbis}), derived in the external wave packet
model, by substituting the beam spread $\Delta p$ for the
effective width $\sigma_{peff}$. This result could be expected
from the discussion of section \ref{shortcut}, by extending the
incoherent sum over the energy from a $\sigma_{peff}$ range to a
$\Delta E=\Delta p/v_0$ range.

Second, the  macroscopic propagation distance $L$ is not perfectly
known \cite{Gribov69,Bilenky78}. A Gaussian average over the
macroscopic region of production gives
$$
   \int dL \; \exp \left( -2\pi i \frac{L}{L^{osc}_{ij}}
   - \frac{(L-\bar L)^2}{2(\Delta L)^2} \right)
   \sim \exp \left(
    -2\pi i \frac{L}{L^{osc}_{ij}}
    -2\pi^2 \left( \frac{\Delta L}{L^{osc}_{ij}} \right)^2
             \right) \, ,
$$
where $\bar L$ is the average propagation distance and $\Delta L$
is the size of the source. Thus a new localization condition has
to be satisfied:
$$
   L^{osc}_{ij} \gtrsim \Delta L \, .
$$
This condition is similar to the localization condition
(\ref{localconstraint}) which comes from a coherent effect. The
only thing to do is to substitute the size of the source $\Delta
L$ for the effective width $\sigma_{xeff}$. As noted several times
(\cite{Dicke,Comsa83,Rich93,Kiers96,Stodolsky98}, and references
therein), it is impossible, in stationary cases, to distinguish
wave packets from an incoherent plane wave superposition with the
same energy-momentum spectrum.

This observation leads directly to the generalization of the
quantum field computation of the transition probability already
done with external Gaussian wave packets. Arbitrary wave packets,
with space width and momentum widths not minimizing the
uncertainty, lead to an oscillation probability similar to
Eq.~(\ref{proba}) (or Eq.~(\ref{probaunstable}) if the particle is
unstable), except that the relation
$\sigma_{xeff}\sigma_{peff}=1/2$ does not hold anymore. This
result can be understood as follows. If the arbitrary external
wave packets are decomposed in Gaussian wave packets, the
amplitude ${\cal A}_i$ becomes a superposition of Gaussian
amplitudes. These do not interfere with each other if their phases
are very different. Thus each Gaussian amplitude will mostly
interfere with itself, from which an incoherent superposition
follows.

In practice, the size of the  region of production is usually much
larger than the size of the wave packets. Similarly, the
energy-momentum spread of the beam is usually much larger than the
energy-momentum spread of the wave packet. Incoherent effects are
thus very often dominant. At worst, the sizes of the regions of
production and detection and of the energy-momentum spectrum are
determined by the characteristics of the wave packet. Similar
decoherence mechanisms were found by Gabor \cite{Gabor56} in
connection with electron-interference experiments. This author
obtains three decoherence factors, coming from an average over the
size of the source, from an average over the energy spectrum, and
from the path difference between the interfering beams. The two
first effects have been explained in this section, and the last
effect can be related to the wave packet separation in particle
oscillations.

\subsubsection{Coherent effects}
\label{coherenteffects}

Recall that coherent effects are constraints on the oscillation
process, which should be applied at the level of the amplitude.
Most coherent effects can be expressed through their influence on
the sizes of the external wave packets associated to the
production and detection of the oscillating particle.
Unfortunately, the estimate of a wave packet size is not an easy
matter and no consensus exists on whichever evaluation method is
the best.

Let us consider first a solar neutrinos. Only the cases of line
spectra ($pep$ or ${}^7Be$) deserve careful thought, since the
energy average always dominates coherent effects for continuum
spectra (except if a detector with a extremely high energy
resolution is invented). The most commonly discussed constraint on
the wave packet size comes from the {\it pressure broadening},
that is, the interruption of coherent emission due to collisions
of the emitting atoms. Nussinov \cite{Nussinov76}, Loeb
\cite{Loeb89}and Kim and Pevsner \cite{Kimbis} estimate the wave
packet size of the parent nuclei at $\sigma_x\!\sim\!10^{-6}\,$cm,
whereas Krauss and Wilczek \cite{Krauss85} propose
$\sigma_x\!\sim\!10^{-4}\,$cm. Kiers, Nussinov and Weiss
\cite{Kiers96} claim that the small wave packets of captured
electrons give a stronger constraint
$\sigma_x\!\sim\!6\times10^{-8}\,$cm. However, none of these
estimates gives a momentum width $\sigma_p$ as large as the
$1\,$keV energy spread of the solar neutrino line spectra
\cite{Krauss85,Pakvasa90,Bahcall94}. The main contribution to this
energy spread comes from the thermal energy spread of the captured
electron \cite{Krauss85}, as well as from the Doppler shift due to
the thermal motion of the emitting nucleus \cite{Pakvasa90}. These
effects cannot be modelized at the level of the amplitude and
belong thus to incoherent effects. Finally, Malyshkin and Kulsrud
\cite{Malyshkin00} compute the effect of Coulomb collisions on the
solar neutrino flux. They obtain a quantity $\sigma_a$, equivalent
to the wave packet width, and consistent with Nussinov and Loeb's
estimates. While Malyshkin and Kulsrud agree that pressure
broadening can be neglected with respect to the line width, they
claim that it could be relevant to the decoherence in the case of
the continuous solar spectra, for a detector of very high
resolution and a very long oscillation length. This conclusion is
in contradiction with the increase of the coherence length with
the resolution of the detector: the coherence length is mainly
determined by the energy resolution of the detector, if it is
smaller than the wave packet size at the source. All these results
show that coherent effects seem to be irrelevant to solar
neutrinos.

As regards neutrinos from supernovae, the source wave packet width
has been estimated by Anada and Nishimura at
$\sigma_x\!\sim\!10^{-14}\,$cm for neutrinos from the supernova
core \cite{Anada88}, and at $\sigma_x\sim10^{-9}\,$cm for
neutrinos from the neutrino sphere \cite{Anada90}. Since
$p\!\sim\!10\,$MeV, core neutrinos decohere before oscillating
($p/\sigma_p\!\sim\!0.1$), whereas sphere neutrinos might
oscillate if the oscillation length is not too short
($p/\sigma_p\!\sim\!10^4$). However, the incoherent momentum
spread ($p/\Delta p\!\sim\!1-10$) destroys any interference effect
left, unless the oscillation length is, by chance, comparable to
the supernova-Earth distance \cite{Reinartz85}.

Another type of neutrino source is a radioactive nucleus in an
atomic lattice, which is the case for reactor neutrinos. The wave
packet size can be estimated by Rich \cite{Rich93} and Grimus and
Stockinger \cite{Grimus96} at $\sigma_x\!\sim\!10^{-10}\,$m. The
wave packet size of the emitted electrons is larger and can be
neglected \cite{Kimbis}. With the neutrino energy around $1\,$MeV,
decoherence occurs beyond $\frac{p}{\sigma_p}\!\sim\!10^3$
oscillation lengths, so that there are no coherent effects
relevant for laboratory experiments. Of course, oscillations may
vanish because of the incoherent energy average.

Coherent effects may also appear because of the finite lifetime of
the source $\tau_{decay}$, which interrupts the classical emission
of the wave train and limits the size of the wave packet to
$\sigma_x\sim c\tau_{decay}$ \cite{Kiers96}. This effect could be
relevant to atmospheric neutrinos and in accelerators. For
quasi-stable sources, the constraint $L^{osc}_{ij}\gtrsim\sigma_x$
can be very stringent. For example, in the case of $\pi\to\mu\nu$,
the length of the wave packet is bounded by $c\tau_{decay}\!\sim\!
7.8\,$m and could be macroscopically large. However this argument
is only valid as long as the decay point of the source is not
observed at all. As emphasized by Kayser \cite{Kayser81}, the
detection of a final state, such as the muon in $\pi\to\mu\nu$,
can localize the decay point to a precision much better than
either $c\tau_{decay}$ or the macroscopic size of the production
region. An extreme example is given by the neutrinos from the
$\beta$ decay of a nucleus with $\tau_{decay}\!\sim\!1\,$sec. The
observation of the $\beta$ particle allows to pin down the decay
point to a precision much better than either
$c\tau_{decay}\!\sim\!10^5\,$km or the production region, for
example a nuclear reactor. Thus the relation between the decay
time of the source and the wave packet size of the oscillating
particle is not direct. The decay time only puts an upper bound to
the wave packet length. Some papers dealing with the finite
lifetime of the source are reviewed in section
\ref{unstablesource}.

Since the coherence length depends not only on the characteristics
of the source, but also on those of the detector, the minimal wave
packet size in the detection process must also be checked. It can
be roughly estimated at $\sigma_x\!\sim\!10^{-10}\,$m, that is,
$\sigma_p\!\sim\!10^{-3}\,$MeV \cite{Grimus99}. In all cases, this
momentum width is not larger than the energy spread and can be
neglected.

As regards K and B mesons, particle decay takes place before
decoherence occurs:
$$
   \frac{L^{coh}_{ij}}{L^{decay}_{ij}} \sim \frac{p_0}{\sigma_{peff}} \,
   \frac{\Gamma_i + \Gamma_j}{2|m_i-m_j|}
   \sim \frac{p_0}{\sigma_{peff}} \gg1 \, ,
$$
where $L^{coh}_{ij}$ and $L^{decay}_{ij}$ are given by
Eqs.~(\ref{cohlengthbis}) and (\ref{Ldecay}), respectively.

To sum up, coherent effects are most likely irrelevant with
respect to incoherent effects. This discussion also shows that, in
practice, we are far from being able to increase the coherence
length by more accurate energy measurements.

\subsection{Unstable source}
\label{unstablesource}

In principle, the instability of the source can be taken into
account in a field-theoretical model by considering the source as
another internal line of the global Feynman diagram describing the
process. The difficulty, of technical nature, consists in
integrating on both propagators, with a constraint on the decay
point which can be either nonexistent, or very stringent.

The case of an unstable source decaying in flight has been
considered by Campagne in a field-theoretical model
\cite{Campagne97}. Besides the usual condition
$L^{osc}_{ij}\gtrsim\sigma_{x\scriptscriptstyle P,D}$, this author obtains a new
localization condition $L^{osc}_{ij} \gtrsim L^{decay}_{\scriptscriptstyle P}$,
where $L^{decay}_{\scriptscriptstyle P}$ is either the decay length $p/m\Gamma$
of the unstable source, or the length of the decay tunnel, if the
latter is shorter. However, this treatment is not a real
improvement on the one proposed by Rich, who had derived earlier
the same results in the framework of time perturbation theory in
quantum mechanics \cite{Rich93}. It is not satisfactory that
macroscopic quantities, such as the lengths of the source, of the
target and of the pion decay tunnel, are treated in
\cite{Campagne97} on the same footing as microscopic quantities.

The case of an unstable source at rest has been considered in
detail by Grimus, Mohanty and Stockinger \cite{Grimus99,Grimus00}.
They use their previous field-theoretical model \cite{Grimus96},
modified by a quantum-mechanical Wigner-Weisskopf approximation,
in order to take into account the finite lifetime of the source.
Besides the already known condition
$L^{osc}_{ij}\gtrsim\sigma_{x\scriptscriptstyle P,D}$, they obtain a new
localization condition:
\begin{equation}
   \frac{\sigma_{\scriptscriptstyle P}}{m_{\scriptscriptstyle P}\Gamma_{\scriptscriptstyle P}}
   \lesssim \frac{1}{4\pi} \, L^{osc}_{ij} \, ,
   \label{sfc}
\end{equation}
where $\sigma_{\scriptscriptstyle P}$, $m_{\scriptscriptstyle P}$ and $\Gamma_{\scriptscriptstyle P}$ are
respectively the wave packet momentum width, the mass, and the
decay width of the unstable parent particle. In brief, the
unstable source should not move on distances larger than the
oscillation length during its lifetime, or else the oscillations
vanish. They find that this condition is most likely satisfied in
the experiments LSND \cite{Athanassopoulos96,Athanassopoulos98}
and KARMEN \cite{Armbruster95,Armbruster02}. The same authors have
also found a new coherence length, $L^{coh}_{ij}=4E^2/\delta
m_{ij}^2\Gamma_{\scriptscriptstyle P}$, which is however completely irrelevant,
since it is much larger than the coherence lengths discussed
previously ($\Gamma_{\scriptscriptstyle P} \ll \sigma_{\scriptscriptstyle P}$ for all weakly
unstable particles). The widths $\sigma_{x\scriptscriptstyle P,D}$ can be
estimated at $10^{-2}\,$MeV for LSND and KARMEN \cite{Grimus00},
so that the condition $L^{osc}_{ij}\gtrsim\sigma_{x\scriptscriptstyle P,D}$ is
also satisfied.

A slightly different model for $\pi\to\mu\nu$ is proposed by
Dolgov \cite{Dolgov00}. This author computes the transition
amplitude by coupling the neutrino propagator to external wave
packets at the source, but not at the detector (source-propagator
model). The finite lifetime of the source is taken into account
with a Wigner-Weisskopf approximation, like in the articles
discussed above \cite{Grimus99,Grimus00}, but without the
restriction of a source at rest. Dolgov considers first the case
where the decay point of the muon is not registered (but its
energy-momentum is perfectly known) and computes an amplitude of
spatial width $\gamma=p_\pi/m_\pi\Gamma_\pi$. The oscillating
phase is given by
$$
   \phi_{ij}(T,L) = 2\pi \frac{L}{L^{osc}_{ij}}
               + \alpha \, \frac{L-v_0T}{L^{osc}_{ij}}
   \hspace{1cm} \mbox{with} \hspace{1cm}
   \alpha = 2\pi \frac{{\bf v}_\pi \cdot {\bf v}_0}
                      {{\bf v}_0^2 - {\bf v}_\pi \cdot {\bf v}_0} \, ,
$$
where ${\bf v}_\pi$ and ${\bf v}_0$ are the velocities of the pion
and neutrino, respectively. With the help of the constraint
$0\lesssim v_0T-L\lesssim\gamma$, the second term of the phase is
seen to be negligible if $L^{osc}_{ij}\gg \alpha\gamma$. For a
source at rest, $\alpha=0$ and this localization condition
vanishes. When the muon decay point is registered, Dolgov obtains
an oscillation phase equal to the standard result as long as the
localization condition $\sigma_\pi\ll L^{osc}_{ij}$ is satisfied
(where $\sigma_\pi$ is the size of the wave packet of the initial
pion). It would be interesting to study intermediate situations
where the muon is registered with a space-time uncertainty and to
compute the time-independent oscillation probability.

This brief account of the treatment of oscillations from an
unstable source shows that there is still work to do, especially
regarding decays in flight.

\section{Correlated oscillations}
\label{section9}

\subsection{Introduction}

Experiments where two correlated mesons oscillate together are
very interesting for the study of CP violation
\cite{Carter80,Carter81,Branco} and can provide tests of the
Einstein-Podolsky-Rosen (EPR) effect
\cite{Apostolakis98,Bertlmann98,Bertlmann99,Foadi99,Pompili00}. At
the present time, the process $\phi(1020)\to K^0\overline{K^0}$ is
studied by the experiment DA$\Phi$NE at Frascati
\cite{Adinolfi00}. The process $\Upsilon(4s)\to B^0\overline{B^0}$
is studied at $B$ factories such as KEKB at Tsukuba \cite{Abe02}
or BABAR at Stanford \cite{Aubert02,Babar}, where asymmetric
collisions allow the measurement of the $B$ pathlengths. In
principle, correlated oscillations could occur with particles not
conjugated to each other, for example the lepton could oscillate
between $e$ and $\mu$ and the neutrino between $\nu_e$ and
$\nu_\mu$ in the process $\pi\!\to\! l\nu$. However, the charged
lepton masses are too different for such oscillations to be
observable (but note that if the masses were close enough to allow
oscillations, it would be impossible to identify the flavor).

In this section, we develop the formalism appropriate for
correlated oscillations and we apply it to oscillations of
correlated mesons. We then disprove a claim of a non-standard
oscillation length. The question of the oscillation of recoil
particles, like $\Lambda$ in $\pi p\!\to\!\Lambda K$ or $\mu$ in
$\pi\!\to\!\mu\nu$, can be examined in the same framework,
allowing us to answer the fifth question posed in section
\ref{fivequestions}.

The treatment of one-particle oscillations with the external wave
packet model of section~\ref{section3} can be easily extended by
associating a propagator with each oscillating particle. Consider
the two successive processes:
\begin{eqnarray}
  (e^+e^-)(p_{\scriptscriptstyle P}) &\to& {\cal R} \, X \, ,
   \\
   {\cal R} &\to& P_1 \,  P_2
   \to f_1(p_{\scriptscriptstyle D_1})  f_2(p_{\scriptscriptstyle D_2}) \, ,
   \label{process}
\end{eqnarray}
where $p_{\scriptscriptstyle P}$ and $p_{\scriptscriptstyle D_{1,2}}$ are the average momenta of
the source ${\cal R}$ and of the final states, respectively. We
would like to apply the external wave packet model to the second
process. In principle this model should be modified to include the
decay width of the source ${\cal R}$, since one external particle
at an interaction point is not sufficient to localize it. However
the decay widths of the $\phi(1020)$ and $\Upsilon(4s)$ are large,
so that their decay point will be close to their production point.
The latter can be localized with the $e^\pm$ external wave
packets. In the case of a small decay width, it will be seen below
that the correlated oscillations become independent of the
production point.

Note that the initial flavors of the oscillating particles cannot
be observed at the source and must be summed over. For example,
$B^0\bar B^0$ and $\bar B^0 B^0$ are both produced in the
$\Upsilon(4s)$ decay. In contradistinction to the one-particle
oscillation case, this sum does not destroy the oscillations,
because of the flavor correlation between the oscillating
particles. Indeed, two mesons of the same flavor cannot be
observed in the above process at the same time in the center of
mass frame (EPR effect).

Let us first write the amplitude corresponding to the process
(\ref{process}). The masses of the first particle are noted
$m_{1a}$, $m_{1b}$ etc., while the masses of the second particle
are noted $m_{2i}$, $m_{2j}$ etc. The two sets coincide if $P_1$
and $P_2$ are conjugated mesons, but it is not necessarily the
case. The average production and detection points are noted
$x_{\scriptscriptstyle P}$ and $y_{\scriptscriptstyle D_1}$, $y_{\scriptscriptstyle D_2}$, respectively. The
partial amplitude, corresponding to the propagation of the
eigenstates of mass $m_{1a}$, from $x_{\scriptscriptstyle P}$ to $y_{\scriptscriptstyle D_1}$,
and of mass $m_{2i}$, from $x_{\scriptscriptstyle P}$ to $y_{\scriptscriptstyle D_2}$, can be
written as
$$
   {\cal A}_{a,i} =
   \int d^4p_1 \int d^4p_2 \; \psi(p_1,p_2) \,
   G_{1a}(p_1^2) \, G_{2i}(p_2^2) \;
   e^{ -ip_1 \cdot (y_{\scriptscriptstyle D_1}-x_{\scriptscriptstyle P})
       -ip_2 \cdot (y_{\scriptscriptstyle D_2}-x_{\scriptscriptstyle P}) } \, ,
$$
with the overlap function given by
$$
   \psi(p_1,p_2) = N \, \psi_{\scriptscriptstyle P}(p_1^0+p_2^0,{\bf p}_1+{\bf p}_2) \,
   \psi_{\scriptscriptstyle D_1}(p_1^0,{\bf p}_1) \, \psi_{\scriptscriptstyle D_2}(p_2^0,{\bf p}_2) \, .
$$
The production overlap function $\psi_{\scriptscriptstyle P}$ is defined by
Eq.~(\ref{overlapP}). The detection overlap function $\psi_{\scriptscriptstyle
D_1}$ is defined by the same equation, where ${\bf p}_{\scriptscriptstyle D_1}$,
$E_{\scriptscriptstyle D_1}$ and ${\bf v}_{\scriptscriptstyle D_1}$ have been substituted to
${\bf p}_{\scriptscriptstyle P}$, $E_{\scriptscriptstyle P}$ and ${\bf v}_{\scriptscriptstyle P}$,
respectively. The definition of $\psi_{\scriptscriptstyle D_2}$ is similar.

The integration on the energies $p_{1,2}^0$ can be done with the
help of the Jacob-Sachs theorem (\ref{integIT}), yielding
\begin{equation}
   {\cal A}_{a,i} \sim
   \int d^3p_1 \int d^3p_2 \, \psi(z_1,z_2,{\bf p}_1,{\bf p}_2) \,
   e^{\,-\,i \, \sqrt{ z_1 + {\bf p}_1^2 } \, T_1
      \,+\, i \, {\bf p}_1 \cdot {\bf L}_1} \;
   e^{\,-\,i \, \sqrt{ z_2 + {\bf p}_2^2 } \, T_2
      \,+\, i \,  {\bf p}_2 \cdot {\bf L}_2 } \, ,
   \label{doubleampli}
\end{equation}
where $T_{1,2}=t_{\scriptscriptstyle D_{1,2}}-t_{\scriptscriptstyle P}$, ${\bf L}_{1,2}={\bf
y}_{\scriptscriptstyle D_{1,2}}-{\bf x}_{\scriptscriptstyle P}$ are the average propagation time
and distance of $P_i$, and $z_i$ is the pole of the propagator
$G_i(p^2)$.

Since our aim is not to prove the conservation of energy-momentum
between the initial state $\phi(1020)$ and the final states
$f_{1,2}$, we set
$$
   {\bf p}_{\scriptscriptstyle D_1} + {\bf p}_{\scriptscriptstyle D_2} = {\bf p}_{\scriptscriptstyle P}
   \hspace{1cm} \mbox{and} \hspace{1cm}
   E_{\scriptscriptstyle D_1} + E_{\scriptscriptstyle D_2} = E_{\scriptscriptstyle P} \, .
$$
As before, it will be useful to define reference masses $m_1$ and
$m_2$ through $m_{1,2}^2=E_{\scriptscriptstyle D_{1,2}}^2 - {\bf p}_{\scriptscriptstyle
D_{1,2}}^2$. The velocities ${\bf v}_{1,2}$ are defined by ${\bf
v}_{1,2}={\bf p}_{\scriptscriptstyle D_{1,2}}/E_{\scriptscriptstyle D_{1,2}}$ (they should not
be confused with ${\bf v}_{\scriptscriptstyle D_{1,2}}$, which appear in the
overlap function and refer to the velocities of the detection
regions, as explained in section \ref{gaussianoverlap}).

\subsection{Factorization}
\label{factorization}

The momentum integrations in the amplitude (\ref{doubleampli}) can
be easily evaluated, either with Laplace's method or with the
stationary phase method, provided that the energy-momentum
correlation is not too stringent at the source. More precisely,
the integrations on ${\bf p}_1$ and ${\bf p}_2$ can be done
independently if the energy uncertainty at the source is larger
than the energy uncertainties at the detection points (recall that
the momentum width is always larger than the energy width). For
example, the production widths for the $\phi(1020)$ decay and the
$\Upsilon(4s)$ decay can be estimated by their average decay
widths, which are $4.26\,$MeV and $14\,$MeV, respectively
\cite{Hagiwara02}. Both widths are larger than typical detector
uncertainties. In that case, the energy-momentum width at the
source can be neglected and the overlap function factorizes in
one-particle overlap functions:
\begin{equation}
   \psi(z_1,z_2,{\bf p}_1,{\bf p}_2) \cong
   \psi_1(z_1,{\bf p}_1) \, \psi_2(z_2,{\bf p}_2) \, .
   \label{overlapfactor}
\end{equation}
Thus the resulting amplitude factorizes in one-particle
oscillation amplitudes. The rest of the computation proceeds as in
the one-particle case, except when the detection times are
measured: the time average is then done only once, on the
production time.

If the energy-momentum uncertainty is smaller at the production
point than at the detection points, energy-momentum correlations
are introduced at the source so that the amplitude cannot be
factorized. However, the oscillation formula is not expected to be
modified, as long as the energy-momentum uncertainty at the source
is larger than the mass difference between the interfering mass
eigenstates:  $\sigma_{p\scriptscriptstyle P} \gtrsim \delta m_{ij}^2/p{\scriptscriptstyle
D_{1,2}}$. This condition resembles the localization condition
derived in the one-particle oscillation case, which stated that
oscillations vanish if the uncertainty on the position of the
source is larger than the oscillation length. This similarity is
misleading: it will be seen correlated oscillations do not vanish
if $\sigma_{p\scriptscriptstyle P} \lesssim \delta m_{ij}^2/p{\scriptscriptstyle D_{1,2}}$,
since the knowledge of the source energy-momentum is not
sufficient to ascertain which mass eigenstates are produced. This
is because the energy-momentum of the source is shared between two
intermediate particles, instead of one as in the one-particle
oscillation model presented in section~\ref{section3}. Moreover,
the position and time of the production process can be
reconstructed, in principle, from the final states
characteristics, so that the oscillation pattern is not washed out
by an ill-defined production point. The discussion of the above
constraint is somewhat academic, as it is satisfied for the
experimentally studied processes, $\phi\to K^0\overline{K^0}$ and
$\Upsilon(4s)\to B^0\overline{B^0}$. Nevertheless, the question of
whether the violation of this constraint changes the oscillation
formula will be examined, partly as a matter of principle, and
partly because it will be useful for the discussion of the recoil
oscillation conundrum.

\subsection{Energy-momentum correlation at the source}

There are two reasons to study more carefully energy-momentum
correlations at the source. First, this analysis yields a
correlated localization condition, arising from the source. In
particular, it allows to check explicitly that a zero momentum
width at the source does not wash out the oscillations. Second, it
will be useful for the examination of the so-called recoil
oscillations. Unfortunately the treatment of the full
3-dimensional case is involved, because the correlation at the
source might link a variation in the longitudinal momentum of one
particle, with a variation in the transversal momentum of the
other. If the transversal dispersion is neglected, the integrals
can in principle be evaluated with Laplace's method in three
dimensions, but the results are lengthy. For that reason, only the
collinear case will be presented. Momenta, velocities and lengths
can take positive or negative values, with the sign indicating the
direction.

Since the stationary phase and Laplace's method give the same
results (see section~\ref{section5}), we choose the latter for the
longitudinal momenta integrations. Besides, the dispersion will be
neglected. Recall that the masses of the first particle are noted
$m_{1a}$, $m_{1b}$ etc., while the masses of the second particle
are noted $m_{2i}$, $m_{2j}$ etc. Suppose that the overlap
function $\psi$ is maximal for $p_1=p_{1a}$ and $p_2=p_{2i}$. As
before, these momenta can be computed by expanding the argument of
the overlap function in small mass differences. The expansion
parameters are given by $\delta m_{1a}^2=m_{1a}^2-m_1^2$ and
$\delta m_{2i}^2=m_{2i}^2-m_2^2$. At first order in mass
differences, the average momenta of the particles are
\begin{eqnarray}
   p_{1a} &=& p_{\scriptscriptstyle D_1}
   \,+\, \frac{\sigma_{xeff3}^2 c_2 - \sigma_{xeff2}^2 c_1}{4\Delta} \, ,
   \label{extremum1} \\
   p_{2i} &=& p_{\scriptscriptstyle D_2}
   \,+\, \frac{\sigma_{xeff3}^2 c_1 - \sigma_{xeff1}^2 c_2}{4\Delta} \, ,
   \label{extremum2}
\end{eqnarray}
where
$\Delta=\sigma_{xeff1}^2\sigma_{xeff2}^2-\sigma_{xeff3}^4\geq0$.
In momentum space, the effective widths read
\begin{eqnarray}
   \frac{1}{\sigma_{peff1,2}^2} &=&
   \frac{1}{\sigma_{p\scriptscriptstyle P}^2} + \frac{(v_{1,2}-v_{\scriptscriptstyle P})^2}{\sigma_{e\scriptscriptstyle P}^2}
   + \frac{1}{\sigma_{p\scriptscriptstyle D_{1,2}}^2}
   + \frac{(v_{1,2}-v_{\scriptscriptstyle D_{1,2}})^2}{\sigma_{e\scriptscriptstyle D_{1,2}}^2} \, ,
   \nonumber \\
   \frac{1}{\sigma_{peff3}^2} &=&
   \frac{1}{\sigma_{p\scriptscriptstyle P}^2}
   + \frac{(v_1-v_{\scriptscriptstyle P})(v_2-v_{\scriptscriptstyle P})}{\sigma_{e\scriptscriptstyle P}^2} \, .
\end{eqnarray}
They are related to the effective widths in configuration space by
$\sigma_{peff1,2}\sigma_{xeff1,2}=1/2$. Their name is justified
below by their appearance as second order coefficients in the
expansion of the overlap function. These expressions can be
compared to the effective width (\ref{sigmaeff}) in the
one-particle oscillation case. The mass differences are included
in the constants $c_k$:
$$
   c_k = \frac{v_k-v_{\scriptscriptstyle D_k}}{\sigma_{e\scriptscriptstyle D_k}^2} \, \delta_k
   + \frac{v_k-v_{\scriptscriptstyle P}}{\sigma_{e\scriptscriptstyle P}^2} \, ( \delta_1 + \delta_2) \, ,
$$
with $\delta_1=\delta m_{1a}^2/2E_{\scriptscriptstyle D_1}$ and $\delta_2=\delta
m_{2i}^2/2E_{\scriptscriptstyle D_2}$.

If the overlap function is noted $\psi=N\exp(-f(p_1,p_2))$, the
expansion of the argument of the exponential around its extremum
can be written as
$$
   f(p_1,p_2) =
   f(p_{1a},p_{2i})
   + \frac{(p_1-p_{1a})^2}{4\sigma_{peff1}^2}
   + \frac{(p_2-p_{2i})^2}{4\sigma_{peff2}^2}
   + \frac{(p_1-p_{1a})(p_2-p_{2i})}{2\sigma_{peff3}^2} \, .
$$

The evaluation in one dimension of the integral
(\ref{doubleampli}) by Laplace's method yields
\begin{eqnarray}
   \hspace{-1cm} &&
   {\cal A}_{a,i} \sim \exp \left( - f(p_{1a},p_{2i}) \right) \,
   \exp \left(
   -iE_{1a}T_1+ip_{1a}L_1-\frac{m_1\Gamma_1T_1}{2E_{1a}}
   -iE_{2i}T_1+ip_{2i}L_1-\frac{m_2\Gamma_2T_2}{2E_{2i}}
          \right)
    \nonumber \\ \hspace{-1cm} && \times
    \exp \frac{-1}{4\Delta} \bigg(
         \sigma_{xeff2}^2(v_{1a}T_1\!-\!L_1)^2 +
         \sigma_{xeff1}^2(v_{2i}T_2\!-\!L_2)^2
           - 2\sigma_{xeff3}^2(v_{1a}T_1\!-\!L_1)(v_{2i}T_2\!-\!L_2)
                            \bigg) \, ,
   \label{doublelaplace}
\end{eqnarray}
where $v_{1a}=p_{1a}/E_{1a}$, with $E_{1a}=\sqrt{p_{1a}^2+m_a^2}$.
The velocity $v_{2i}$ and the energy $E_{2i}$ are similarly
defined. Once more `wave packets' can be associated with the
oscillating particles $1$ and $2$, but they are correlated by the
finite width $\sigma_{xeff3}$. The phase can be expanded around
$p_{\scriptscriptstyle D_{1,2}}$ and $m_{1,2}$. For example,
\begin{equation}
   \phi_{1a} = E_{1a}T_1-p_{1a}L_1 \cong E_{\scriptscriptstyle D_1} T_1 - p_{\scriptscriptstyle D_1} L_1
   + (v_1 T_1-L_1) \left( \delta p_{1a} + \frac{\delta m_{1a}^2}{2p_1} \right)
   + \frac{\delta m_{1a}^2}{2p_1} \, L_1 \, ,
   \label{doublephase}
\end{equation}
where $\delta p_{1a}=p_{1a}-p_1$. The phase difference between two
amplitudes vanishes but for the oscillation term already
encountered in the previous sections, provided the wave packet is
sufficiently localized, so that the term proportional to
$v_1T_1-L_1$ in (\ref{doublephase}) is negligible.

At the extremum, the argument of the overlap function reads
\begin{equation}
   f(p_{1a},p_{2i}) =  \frac{\alpha \delta_1^2 + \beta \delta_2^2
                             + 2\gamma \delta_1 \delta_2}{64\Delta} \, ,
   \label{doublelocal}
\end{equation}
and yields a localization condition, that is, the mass differences
cannot be too large compared with the energy-momentum widths. The
coefficients $\alpha$ and $\gamma$ read
\begin{eqnarray*}
   \alpha &=&
     \frac{1}{\sigma_{p\scriptscriptstyle P}^2\sigma_{e\scriptscriptstyle P}^2}
     \left(  \frac{1}{\sigma_{\scriptscriptstyle D}^2}
           + \frac{(v_2-v_{\scriptscriptstyle D_1})^2}{\sigma_{e\scriptscriptstyle D_1}^2}
           + \frac{(v_2-v_{\scriptscriptstyle D_2})^2}{\sigma_{e\scriptscriptstyle D_2}^2}
     \right)
   \\
   &+& \left(  \frac{1}{\sigma_{p\scriptscriptstyle D_2}^2}
           + \frac{(v_2-v_{\scriptscriptstyle D_2})^2}{\sigma_{e\scriptscriptstyle D_2}^2} \right)
     \left(  \frac{1}{\sigma_{p\scriptscriptstyle P}^2\sigma_{e\scriptscriptstyle D_1}^2}
           + \frac{1}{\sigma_{e\scriptscriptstyle P}^2\sigma_{p\scriptscriptstyle D_1}^2}
           + \frac{(v_{\scriptscriptstyle P}-v_{\scriptscriptstyle D_1})^2}
                   {\sigma_{e\scriptscriptstyle P}^2\sigma_{e\scriptscriptstyle D_1}^2}
     \right)
   + \frac{1}{\sigma_{p\scriptscriptstyle D_1}^2\sigma_{e\scriptscriptstyle D_1}^2\sigma_{peff2}^2} \, ,
   \\
   \gamma &=& \frac{1}{\sigma_{p\scriptscriptstyle P}^2\sigma_{e\scriptscriptstyle P}^2}
     \left(  \frac{1}{\sigma_{\scriptscriptstyle D}^2}
           + \frac{(v_1-v_{\scriptscriptstyle D_1})(v_2-v_{\scriptscriptstyle D_1})}{\sigma_{e\scriptscriptstyle D_1}^2}
           + \frac{(v_2-v_{\scriptscriptstyle D_2})(v_1-v_{\scriptscriptstyle D_2})}{\sigma_{e\scriptscriptstyle D_2}^2}
     \right)
   + \frac{(v_1-v_{\scriptscriptstyle D_1})(v_2-v_{\scriptscriptstyle D_2})}
          {\sigma_{p\scriptscriptstyle P}^2\sigma_{e\scriptscriptstyle D_1}^2\sigma_{e\scriptscriptstyle D_2}^2}
   \\
   &+& \frac{1}{\sigma_{e\scriptscriptstyle P}^2}
     \left(  \frac{1}{\sigma_{p\scriptscriptstyle D_1}^2}
           +  \frac{(v_1-v_{\scriptscriptstyle D_1})(v_{\scriptscriptstyle P}-v_{\scriptscriptstyle
D_1})}{\sigma_{e\scriptscriptstyle D_1}^2}
     \right)
     \left(  \frac{1}{\sigma_{p\scriptscriptstyle D_2}^2}
           +  \frac{(v_2-v_{\scriptscriptstyle D_2})(v_{\scriptscriptstyle P}-v_{\scriptscriptstyle
D_2})}{\sigma_{e\scriptscriptstyle D_2}^2}
     \right) \, ,
\end{eqnarray*}
where $\sigma_{p\scriptscriptstyle D}^{-2}=\sigma_{p\scriptscriptstyle D_1}^{-2}+\sigma_{p\scriptscriptstyle
D_2}^{-2}$. The coefficient $\beta$ is obtained from $\alpha$ by
the exchange of the indices $1\leftrightarrow2$.

What happens when the energy-momentum uncertainty at the source
goes to zero? Three features of the amplitude
(\ref{doublelaplace}) must be examined: the localization condition
(\ref{doublelocal}) given by the function $f(p_{1a},p_{2i})$, the
value of the phase and the `wave packet' effect on the time
average. In the limit $\sigma_{p\scriptscriptstyle P}\!\to\!0$ (so that
$\sigma_{e\scriptscriptstyle P}\!\to\!0$ too), the argument of the overlap
function becomes at the extremum
$$
   \lim_{\sigma_{p\scriptscriptstyle P}\to 0} f(p_{1a},p_{2i}) =
     \frac{\left(
                  (v_2\!-\!v_{\scriptscriptstyle D_1})\delta_1 + (v_1\!-\!v_{\scriptscriptstyle D_1})\delta_2
           \right)^2}
         {4(v_1\!-\!v_2)^2\sigma_{e\scriptscriptstyle D_1}^2}
   + \frac{\left(
                  (v_2\!-\!v_{\scriptscriptstyle D_2})\delta_1 + (v_1\!-\!v_{\scriptscriptstyle D_2})\delta_2 \right)^2}
          {4(v_1\!-\!v_2)^2\sigma_{e\scriptscriptstyle D_2}^2}
   +\frac{(\delta_1 + \delta_2)^2}{4(v_1\!-\!v_2)^2\sigma_{p\scriptscriptstyle D}^2}
$$
Thus a small, or even zero energy-momentum uncertainty at the
source, does not destroy the oscillations. This result confirms
our expectations, since a zero width at the source does not give
information on which mass eigenstates propagate.  On the other
hand, the function $f(p_{1a},p_{2i})$ diverges when either
$\sigma_{p\scriptscriptstyle D_1}^2$ or $\sigma_{p\scriptscriptstyle D_2}^2$ goes to zero,
yielding the expected localization conditions at the detectors.

In the limit of a zero energy-momentum width at the source, wave
packets associated to oscillating particles are of infinite
extent. This phenomenon was expected, since the time of production
$T_{\scriptscriptstyle P}$, included in $T_1$ and $T_2$, becomes ill-defined.
However the average on the time of production does not destroy the
oscillations, at least if the detection times are measured. The
explanation follows. In the limit $\sigma_{p\scriptscriptstyle P}\!\to\!0$,
Eqs.~(\ref{extremum1}) and (\ref{extremum2}) show that the
quantities $\delta p_{1a}=p_{1a}-p_{\scriptscriptstyle D_1}$ and $\delta
p_{2i}=p_{2i}-p_{\scriptscriptstyle D_2}$ become
$$
   \lim_{\sigma_{p\scriptscriptstyle P}\to 0} \delta p_{1a}
   = -\lim_{\sigma_{p\scriptscriptstyle P}\to 0} \delta p_{2i}
   = - \frac{\delta_1+\delta_2}{v_1-v_2} \, .
$$
Note that these values can be directly obtained, in the
one-dimensional case, from energy-momentum conservation at the
source. The phase of the amplitude (see Eq.~(\ref{doublephase}))
can then be written as
\begin{equation}
   \lim_{\sigma_{p\scriptscriptstyle P}\to 0} ( \phi_{1a}+\phi_{2i}) =
    \delta_1 \, \frac{L_1-L_2-v_2(T_1-T_2)}{v_1-v_2}
   +\delta_2 \, \frac{L_1-L_2-v_1(T_1-T_2)}{v_1-v_2} \, ,
   \label{limdoublephase}
\end{equation}
where the term of order zero in the mass differences has been
dropped. Since $T_1-T_2=t_{\scriptscriptstyle D_1}-t_{\scriptscriptstyle D_2}$ and
$L_1-L_2=y_{\scriptscriptstyle D_1}-y_{\scriptscriptstyle D_2}$, the phase is independent of the
production point. However the production point is implicit in the
phase (\ref{limdoublephase}) when the classical limit is taken.
The substitution of the classical relations
$L_{1,2}=v_{1,2}T_{1,2}$ shows indeed that the expressions
multiplying $\delta_1$ and $\delta_2$ are the propagation times
$T_1$ and $T_2$, respectively. In other words the space-time
coordinates of the production point can be reconstructed from the
coordinates of detection and the velocities of the oscillating
particles, with the exception of the $v_1=v_2$ case (recall that
velocities of the same sign have the same direction). While
complete computations in three dimensions are complicated, it can
be seen that the phase difference becomes independent of the
production coordinates, in the limit of zero energy-momentum
uncertainty at production:
\begin{equation}
   \lim_{\sigma_{p\scriptscriptstyle P}\to 0} (\phi_{1a}+\phi_{2i}-\phi_{1b}-\phi_{2j}) =
     \delta E_{1ab} (T_1-T_2)
   - \delta {\bf p}_{1ab} \cdot ({\bf L}_1-{\bf L}_2) \, .
   \label{limphaseproba}
\end{equation}

\subsection{No recoil oscillations}

It has been claimed that particles produced together with mixed
states also oscillate because of the energy-momentum recoil. For
example, the $\Lambda$ baryon, in the process $\pi^- p\!\to\!
\Lambda K^0$, is supposed to be in a superposition of two
energy-momentum eigenstates, so that its detection probability
should oscillate in space \cite{Srivastava95a}. Furthermore the
oscillation frequency of the kaon is modified by the momentum
recoil against the $\Lambda$. In another example, muons produced
in $\pi\!\to\!\mu\nu$ are supposed to oscillate in space, because
of the momentum recoil against the mixed state of the neutrino
\cite{Srivastava95c,Srivastava98}. As explained in section
\ref{timespaceconversion}, recoil oscillations arise in
quantum-mechanical models if different propagation times are
associated to the different mass eigenstates. Recoil oscillations
are unacceptable from basic principles, since the oscillation of
the detection probability of a non-mixed state means that
probability is not conserved at all distances. On the contrary,
when several mass eigenstates are mixed, the sum of the detection
probabilities of the different mass eigenstates is always equal to
1 for a given propagation distance.

This assertion has been questioned in several papers in the
framework of quantum mechanics. These refutations involve either
the consideration of the different proper times associated with
the oscillating particle and the recoil particle
\cite{Lowe96,Zralek98,Burkhardt99}, or the use of classical
trajectories \cite{Dolgov97}, or the use of energy-momentum
conservation for the average energy-momenta of the propagating
wave packets \cite{Zralek98,Nauenberg99}. The source-propagator
approach of Shtanov \cite{Shtanov98} is not a real improvement on
the arguments of Ref.~\cite{Dolgov97}, since Shtanov treats
configuration space variables, which have a microscopic role, as
classical macroscopic variables.

Quantum field theory allows to understand which quantum-mechanical
explanations are correct {\it and} important. The correlated
oscillation model of the previous section is easily adapted to the
treatment of recoil oscillations. Suppose that the first particle
is in a superposition of several mass eigenstates $m_{1a}$,
whereas the second particle has only one mass eigenstate $m_2$ (it
is the so-called recoil particle). Since the phase of the
amplitude (\ref{doublelaplace}) depends only on the average
propagation times $T_{1,2}$ and distances $L_{1,2}$, which are
common to the different mass eigenstates, no recoil oscillations
will be observed as long as these quantities are well-defined.
This can be checked explicitly if $\sigma_{p\scriptscriptstyle
P}\gtrsim\sigma_{p\scriptscriptstyle D_{1,2}}$, as the overlap function
factorizes in that case. Thus the only case requiring a careful
examination is when the uncertainty on the position of the source
is larger than the oscillation length, i.e.\ when $\sigma_{p\scriptscriptstyle
P}\lesssim \delta m_{1ab}^2/2p_1$. This condition is not satisfied
in the process $\pi^- p\!\to\!\Lambda K^0$. The momentum
uncertainty of the proton, in this fixed-target experiment, is
around 3 keV, i.e.\ much larger than the kaon mass difference of
$3 \times 10^{-9}\,$keV \cite{Lipkin99}. Therefore the problem of
recoil oscillations  does not arise in the $\Lambda K^0$ complex.

In contradistinction to the case of the $\pi^-p$ collision, it is
possible to consider a $\pi$ decay with a sharp momentum. Its
theoretical minimal uncertainty is given by its $2.5
\times10^{-8}\,$eV decay width, which is smaller than the neutrino
mass differences. This case has been treated, with essentially
correct quantum-mechanical arguments, by Dolgov, Morozov, Okun and
Schepkin \cite{Dolgov97} and within a model coupling the $\mu$ and
$\nu$ propagators with a source wave packet by Dolgov
\cite{Dolgov00}. On the one hand, if only the neutrino is
observed, the detailed study of the one-particle oscillation case
in section~\ref{section5} has shown that neutrino oscillations
occur, provided that the momentum width associated with the muon
is larger than the inverse oscillation length. On the other hand,
if only the muon is observed, no oscillations in the probability
of muon detection occur. The reason is that the sum over the
neutrino flavors makes the transition probability (\ref{proba})
independent of the distance (apart from the geometrical decrease
in $L^{-2}$):
$$
   \sum_\alpha \int L^2 \, d\Omega \,
   {\cal P}_{\alpha \to \beta}({\bf L} ) =1 \, .
$$
Furthermore oscillations also vanish after integration over the
unknown source decay coordinates. This point was discussed in
section \ref{unstablesource}.

Finally, the only case left is the detection of both muon and
neutrino, coming from the decay of a pion endowed with a sharp
momentum. As explained above, this process can be treated like a
correlated oscillation, but with only one mass eigenstate $m_2$
for the muon. The formula (\ref{limphaseproba}) shows that the
oscillation of the probability is independent of the production
point, so that the average over the production point does not have
any effect on the phase. Moreover the oscillation term can be
expressed as depending only on the detection point of the
neutrino, and on the reconstructed decay point (see
Eq.~(\ref{limdoublephase}) with $\delta_2=0$):
$$
   \lim_{\sigma_{p\scriptscriptstyle P}\to 0} ( \phi_{1a}+\phi_{2}) =
    \delta_1 \, \frac{L_1-L_2-v_2(T_1-T_2)}{v_1-v_2} \, .
$$
The substitution of the classical relations
$L_{1,2}=v_{1,2}T_{1,2}$ in the above equation shows indeed that
the phase difference depends only on the classical propagation
time of the neutrino:
$$
  \lim_{\sigma_{p\scriptscriptstyle P}\to 0} ( \phi_{1a}+\phi_{2}) =
  \delta_1 T_1
  = \frac{\delta m_{1a}^2L_1}{2p_{\scriptscriptstyle D_1}} \, .
$$
In short the probability to detect both muon and neutrino can be
interpreted as the result of a neutrino oscillation alone. Recall
that the above relation can be obtained in a simple way in one
dimension by energy-momentum conservation at the source.

As regards the experimental data, oscillations of recoil particles
have not been probed (in the case of the $\pi\!\to\!\mu\nu$, it
would only be possible for a neutrino oscillation length shorter
than the muon decay length). However the neutral kaon mass
difference obtained from strangeness oscillation experiments is
consistent with the one obtained from regeneration experiments
\cite{Lowe96}.

\subsection{Oscillations of correlated mesons}

We compute here the amplitude associated to the correlated
oscillations of a pair of neutral mesons produced in $e^+e^-$
annihilation. The phenomenology of correlated $K$ and $B$ mesons
is discussed in \cite{Dunietz87,Buchanan92,Hayakawa93,Branco} and
in \cite{Carter80,Carter81,Bigi87,Fleischer97,Branco},
respectively.

Each meson oscillates between its components $P_L\!-\!P_H$, before
decaying into final states $f_1(k_1)$ and $f_2(k_2)$ at spacetime
points $y_1$ and $y_2$:
$$
   e^+e^- \to {\cal R}(q) \to P^0\bar P^0 \to f_1(k_1) f_2(k_2) \, ,
$$
where $q$, $k_1$ and $k_2$ are the corresponding energy-momenta.

Since each final state can be produced by either $P^0$ or $\bar
P^0$, the two amplitudes arising from the exchange of $P^0$ and
$\bar P^0$ as intermediate states must be coherently added. The
resonance quantum numbers of $\phi$ and $\Upsilon$(4s),
$J^{PC}=1^{--}$, are conserved by the strong interactions at the
source, so that the pair $P^0\bar P^0$ is in an antisymmetric
state under $P$ and under $C$. Thus the relative sign of the two
contributions to \mbox{${\cal R}\!\to\!f_1f_2$} is negative
\cite{Lipkin68}.

The total amplitude before antisymmetrization is given by
Eq.~(\ref{doubleampli}). In section \ref{factorization}, it was
shown that the energy-momentum correlation at the source can be
neglected for the resonances $\phi$ and $\Upsilon$(4s), so that
the overlap function factorizes as in Eq.~(\ref{overlapfactor}).
The amplitude for correlated oscillations can thus be approximated
by the product of two amplitudes, each corresponding to a single
oscillation. Furthermore, it was shown in section \ref{wweffmass}
that the oscillation amplitude, in the case of nearly degenerate
unstable particles, can be replaced by the effective amplitude
(\ref{effectiveampli}) depending only on the distance $L$. In the
end the antisymmetrized amplitude to detect a $f_1$ at distance
$L_1$ and a $f_2$ at distance $L_2$ can be written as
\begin{eqnarray*}
   {\cal T}_{f_1f_2} &\sim& {\cal M}_P \,
   \left(
   \left( {\cal M}_{01} \;\; {\cal M}_{\bar 01} \right) \,
    V^{-1} \, e^{ -i M \frac{m_0L_1}{p_1} } \, V \,
   \left(
   \begin{array}{c}
   \!1\! \\ \!0\!
   \end{array}
   \right)
   \right)
   \left(
   \left({\cal M}_{02} \;\; {\cal M}_{\bar 02} \right) \,
    V^{-1} \, e^{ -i M \frac{m_0L_2}{p_2} } \, V \,
   \left(
   \begin{array}{c}
   \!0\! \\ \!1\!
   \end{array}
   \right)
   \right)
   \\ & &
   - \;\; \mbox{same expression with
            $\left( \begin{array}{c} \!1\! \\ \!0\! \end{array} \right)
             \leftrightarrow
             \left( \begin{array}{c} \!0\! \\ \!1\! \end{array} \right)
            $} \, ,
\end{eqnarray*}
where ${\cal M}_P={\cal M}({\cal R} \to  P^0\bar P^0)$. The matrix
$M$ is defined by
\mbox{$M=diag(m_L-i\Gamma_L/2,m_H-i\Gamma_H/2)$}, $m_0$ is the
mass of the kaon in the degenerate limit and $p_j$ are the norms
of the 3-momenta of the final states $f_j$. The diagonalization
matrix $V$ is given by Eq.~(\ref{matriceV}). The amplitudes ${\cal
M}_{0j}$ and ${\cal M}_{\bar 0j}$ stand for \mbox{${\cal M}(P^0
\!\to\! f_j)$} and \mbox{${\cal M}(\bar P^0 \!\to\! f_j)$},
respectively.

The amplitude can be written as
\begin{eqnarray*}
   {\cal T}_{f_1f_2} &\sim& {\cal M}_P {\cal M}_{01} {\cal M}_{02} \,
   \frac{(1+\sigma\mu_{f_1})(1+\sigma\mu_{f_2})}{2\sigma}
   \\ &\times&
   \left( -\eta_{f_2} \,
   e^{ -i \left( m_L -\frac{i}{2}\Gamma_L \right) \frac{m_0L_1}{p_1}
        -i \left( m_H -\frac{i}{2}\Gamma_H \right) \frac{m_0L_2}{p_2}}
            +\eta_{f_1} \,
   e^{ -i \left( m_H -\frac{i}{2}\Gamma_H \right) \frac{m_0L_1}{p_1}
        -i \left( m_L -\frac{i}{2}\Gamma_L \right) \frac{m_0L_2}{p_2}}
   \right) \, ,
\end{eqnarray*}
where $\mu_f$ is defined by Eq.~(\ref{defmu}) and
$\eta_f=(1-\sigma\mu_f)/(1+\sigma\mu_f)$. Note that the amplitude
vanishes for identical final states ($f_1=f_2$ and $p_1=p_2$) and
identical propagation distances ($L_1=L_2$), as expected
\cite{Day61}. In the center-of-mass frame, the equality of momenta
$p_1=p_2\equiv p$ implies that the interference term oscillates
like
\begin{equation}
   \cos \left( \frac{m_0 ( m_{\scriptscriptstyle H} -m_{\scriptscriptstyle L} )}{p} \, (L_1-L_2) \right) \, .
   \label{oscilKK}
\end{equation}
This oscillation formula coincides with the quantum-mechanical
result obtained with the equal time prescription, whereas the
different time prescription leads to an oscillation length shorter
by a factor 2 \cite{Srivastava95b}. As already explained in
section \ref{timespaceconversion}, the quantum field treatment
shows that neither prescription is meaningful, since the
interference takes place over a space and time range, determined
by the effective `wave packet' width. We can only say that, in the
end, the different mass eigenstates have the same {\it average}
propagation time or distance, so that the oscillation formula will
agree with the one obtained with the identical time prescription.
As regards the experimental data, Kayser \cite{Kayser97a} has
shown, in the case of the process $\Upsilon\!\to\!
B^0\overline{B^0}$, that the mass difference obtained from the
oscillation formula (\ref{oscilKK}) is in agreement with the mass
difference extracted from single $B$ oscillations.

\section{Summary and outlook}

Although the plane wave derivation of the vacuum oscillation
formula has often been criticized in the literature, it is still
used in most articles and textbooks. There are two reasons for
this. First, the oscillation formula obtained in this way is
believed to be correct by most physicists, in spite of the
numerous inconsistencies present in its derivation. Second, the
other approaches are not felt to be completely satisfying, with
the consequence that more sophisticated treatments are not
considered worth the effort.

Let us first point out that we do not claim that the oscillation
formula obtained with the plane wave approach should be revised.
However we have argued that this approach becomes unacceptable
when all its inconsistencies are added up: the perfect knowledge
of the momentum precludes spatial oscillations, observability
conditions (such as $\sigma_x \lesssim L^{osc}$ or $L\lesssim
L^{coh}$) are not taken into account, flavor states are
ill-defined, unstable oscillating particles cannot be consistently
described in that model. Furthermore, the plane wave derivation
requires two prescriptions: the classical propagation condition
($|vt-x|\ll t$) and the equal time prescription ($t_i=t_j$).
Whereas the first prescription can be justified in a
quantum-mechanical wave packet approach, the second prescription
can only be proved right in a field-theoretical treatment. This is
an important argument in favor of the latter approach, since the
different time prescription ($t_i\neq t_j$) leads to oscillation
formulas differing significantly from the standard result. We have
also shown that an energy-momentum prescription (such as $E_i=E_j$
or $p_i=p_j$) is not necessary to derive the oscillation formula.

Once the necessity of a more sophisticated approach is understood,
we are faced with the choice between two main methods: on the one
hand, a quantum-mechanical treatment associating wave packets with
the propagating mass eigenstates and, on the other hand, a
field-theoretical treatment where the oscillating particle is
considered as an internal line of a Feynman diagram. The first
possibility is rich in physical insights (such as the existence of
a coherence length) but not satisfactory in many respects: flavor
states are still ill-defined, the equal time prescription is still
needed, the nonrelativistic limit is problematic if the mass
eigenstates are not nearly degenerate, the size of the wave packet
is hard to estimate, the coherence length is not well-defined in
the nonrelativistic limit,  the case of a vanishing energy
uncertainty is not included, and finally the treatment is
inadequate for unstable particles. Whereas most of these problems
are solved in the interacting wave packet model of Giunti
\cite{Giunti02b}, this last model requires quantum field theory to
compute the interactions and is as complex as the external wave
packet model.

Derivations of the oscillation formula resorting to
field-theoretical methods are not very popular, although they seem
the only way out. The first reason is that they are thought to be
very complicated. The second reason is that the existing quantum
field computations of the oscillation formula do not agree in all
respects. The aim of our report was to counter both objections.
The first of these is easily refuted by noting that the
oscillation formula for a stable particle can be derived in a very
simple field-theoretical model, the Kobzarev {\it et al.} model
\cite{Kobzarev82}, in which the source and the detector are
approximated by infinitely heavy nuclei. This is the simplest
model of all in which the oscillation formula can be consistently
derived. The second objection could only be countered by a
detailed and complete computation of the oscillation formula with
all approximations carefully considered.

In this report, we have shown that all existing field-theoretical
treatments (with the noteworthy exception of the Blasone-Vitiello
approach \cite{Blasone01b}) can be included in the so-called
external wave packet model, where the oscillating particle is
described as an internal line of a Feynman diagram and propagates
between a source and a detector localized with in- and outgoing
wave packets. In particular, we have paid attention to the case of
a vanishing energy uncertainty (stationary limit) and its wave
packet interpretation, so that we could prove that the
Grimus-Stockinger model \cite{Grimus96} is a subcase of the
external wave packet model. Two other limits have been considered.
First, we have proved that oscillations near the source can be
described by the standard oscillation formula, contrary to what
was claimed by Ioannisian and Pilaftsis \cite{Ioannisian99}.
Second, we have also shown that oscillations far from a
well-localized source are also described by the standard
oscillation formula, contrary to Shtanov's claim \cite{Shtanov98}.
In order to include all these cases in our formalism, it has been
necessary to evaluate the amplitude with two approximation
schemes: Laplace's and stationary phase methods. As a result, the
propagation range was divided into three regimes, distinguished by
the dispersion (or spreading in space-time) of the amplitude. In
the first regime, near the source, the dispersion is negligible
and Laplace's method is sufficient. This case corresponds to
Ioannisian and Pilaftsis' limit. In the second regime, the
transversal dispersion becomes important, so that both
approximation methods need to be used. In the third regime, far
from the source, the longitudinal dispersion becomes important and
the stationary phase method is sufficient.  This case corresponds
to Shtanov's limit. We have also discussed another method of
computation (in energy-momentum space) valid for stable
oscillating particles, as it clears up in which sense interference
occurs between same energy states.

The computation of the intrinsic decoherence at the wave packet
level has yielded two well-known observability conditions for
oscillations. However, our physical explanation differs from the
standard one by taking into account the dispersion of the
amplitude, so that the origin of decoherence depends on the
distance at which it occurs. The first condition of observability
of oscillations, $L^{osc}_{ij}\gg\sigma_x$, is either due to the
initial spread of the associated `wave packets', or to their
constant overlap at large distance. Similarly the existence of a
coherence length (yielding the second observability condition
$L\ll L^{coh}_{ij}$) is either due to the separation of `wave
packets', or to their dispersion. Our 3-dimensional treatment has
also yielded a third observability condition, $|{\bf
p}_0\times{\bf l}|\lesssim\sigma_p$ (modified into $|{\bf
v}_0\times{\bf L}|\lesssim v_0\sigma_x$ at short distance) which
has the obvious geometrical interpretation of constraining the
propagation within a cone (respectively a cylinder at short
distance). It is now clear that the coherence length tends to
infinity in the stationary limit, as noted by Kiers, Nussinov and
Weiss \cite{Kiers96}.

The oscillation of unstable particles, whether mesons or
neutrinos, has been analyzed in the same framework, with the
result that the oscillation formula is only modified, as expected,
by the well-known exponential decay term. Nonexponential
corrections to the oscillation formula have been estimated for the
first time in quantum field theory. Finally we have given a
field-theoretical treatment of correlated oscillations, with the
aim of disproving the existence of recoil oscillations.

Our work confirms that the standard vacuum oscillation formula
given by Eqs.~(\ref{standardproba}) and (\ref{standardphase}) is
correct if the observability conditions mentioned above are
satisfied. This fact is rather remarkable, in the light of the
numerous inconsistencies which were present in the plane wave
derivation of this formula. It can be explained by the following
observations:
\begin{itemize}
\item the energy and momentum present in the phase are linked by
the on-shell condition, and the average propagation time and
length are linked by a constraint equivalent to a wave packet
localization in space-time. For these reasons, the $E_jT$ and
$p_jL$ components of the phase cancel but for the usual
oscillating factor $\frac{\delta m^2L}{2p}$;

\item only same energy components interfere because of the time
average, so that the time-averaged transition probability computed
in the external wave packet model is equivalent to the
energy-integrated transition probability computed with plane
waves;

\item from the previous argument, decoherence effects arising from
external wave packets cannot be distinguished from decoherence
effects arising from averages over the production (and detection)
region and over the energy spectrum; since the former effects are
dominated by the latter, they have no influence on the oscillation
formula.
\end{itemize}
For ten years, new experimental results in neutrino physics have
stimulated the research on the theory of particle oscillations,
leading to the development of numerous (and sometimes conflicting)
field-theoretical approaches. In this report, we have tried to
unify these treatments and to extend them to oscillations of $K$
and $B$ mesons. We indeed believe that the neutrino oscillation
formula is much strengthened by considering it in the same
framework as the mesonic oscillation formula, whose parameters
have been confirmed by other methods such as regeneration in
matter. We hope to have convinced the reader that
field-theoretical models provide, on the one hand, the most secure
foundation to the oscillation formula and give, on the other hand,
a good physical understanding of oscillation phenomena, whatever
the particles involved.

One case needing further investigation is the decay in flight of
the source of the oscillating particle, which is important for
atmospheric neutrinos and some laboratory neutrino experiments. It
would also be interesting to apply a field-theoretical treatment
to cascade decays, such as $B\!\to\!J/\psi K\!\to\!J/\psi f_K$,
where double-flavor oscillations occur. These processes are useful
for the determination of the signs of the $B_HB_L$ mass and
lifetime differences \cite{Azimov90,Kayser97b,Branco,Quinn00}.

The field-theoretical approach faces its biggest challenge in the
description of neutrino oscillations in matter. A first task
consists in the rederivation of the MSW oscillation formula (see
\cite{Mannheim88,Notzold88,Pal89,Cardall99} for a few attempts).
Another difficult task is to describe the nonstationary evolution
of neutrinos in supernovae or in the early universe, where
interaction rates are in competition with the flavor oscillation
period \cite{Raffelt,Cardall01}.

The turn of the century marks a new golden age for CP violation
experiments, thanks to the $B$ factories, as well as the golden
era of neutrino oscillation experiments. These favorable auspices
should be a great stimulation for further research on the
theoretical foundations of oscillations.

\section*{Acknowledgments}

I am deeply indebted to Jeanne De Jaegher for a thorough reading
of the manuscript, which led to lengthy discussions on the quantum
mechanics of oscillations, as well as the rewriting of some
sections. Discussions with Jean Pestieau, Andreas Veithen, Jacques Weyers, Carlo
Giunti and Christian Cardall are gratefully acknowledged. I thank
Giuseppe Vitiello and Massimo Blasone for having organized in
Vietri a workshop dedicated to the theoretical foundations of particle oscillations.
It is a pleasure to thank Michel and Brita Beuthe for their
comments as well as for travel support. I must also express my
gratitude to Karel and C\'ecile De Jaegher and to Attilio Rivoldini
for their computer-related assistance.
Finally, I thank the Universit\'e Catholique de Louvain and
its Institut de Physique Th\'eorique for financial support.

\section*{Appendix: the Jacob-Sachs theorem}
\addcontentsline{toc}{section}{Appendix}

Let $\psi(E,{\bf p})$ be a function which is distinct from zero
only within certain bounds:
$$
   \psi(E,{\bf p}) \neq 0 \;\; \mbox{for} \;\;
   0<M_1^2<p^2<M_2^2 \;\; \mbox{and} \;\;  E\geq 0 \, ,
$$
with $p^2 = E^2-{\bf p}^2$. On this interval $\psi(E,{\bf p})$ is
taken to be infinitely differentiable. Let the function $I(T)$ be
defined by
$$
   I(T) = \int dE \, \psi(E,{\bf p}) \, G(p^2) \, e^{-iET} \, ,
$$
where the function $G(p^2)$ is the complete scalar propagator in
momentum space.

We are going to prove that the function $I(T)$ has the following
asymptotic behavior:
\begin{equation}
   I(T) \stackrel{ T \to \infty }{\longrightarrow}
   \frac{\pi Z}{\sqrt{z_0 \!+\! {\bf p}^2 }} \,
   \psi(\sqrt{z_0+{\bf p}^2},{\bf p}) \, e^{-i\sqrt{z_0+{\bf p}^2} \, T} \, ,
   \label{integIT}
\end{equation}
where $z_0$ is the pole of the integrand and $Z$ is the residue.
This section follows \cite{Jacob61}.

It will be easier to work with an overlap function depending
explicitly on $p^2$, so we write $\psi(p^2,{\bf p})$. We choose to
implement the assumption of compact domain of $\psi(p^2,{\bf p})$
by
\begin{eqnarray*}
   \psi(p^2,{\bf p}) &=& (p^2 - M_1^2)^n \, (p^2 - M_2^2)^n \,
   \Omega(p^2,{\bf p})
   \quad \mbox{for} \quad
   0 < M_1^2 < p^2 < M_2^2 \;\; (n>0) \, ,
   \\
   \psi(p^2,{\bf p}) &=& 0 \quad \mbox{otherwise} \, .
\end{eqnarray*}
The symmetrical behavior at $M_1^2$ and $M_2^2$ is chosen only to
simplify the algebra. The function $\Omega$ is taken to be
analytic over the domain of interest.

Under the change of variable $z=p^2$, the function $I(T)$ becomes
$$
   I(T) = \frac{1}{2} \int_{M_1^2}^{M_2^2} dz \,
   (z \!+\! {\bf p}^2 )^{ -\frac{1}{2} } \, \psi(z,{\bf p}) \, G(z) \,
   e^{ -i \sqrt{ z + {\bf p}^2 } T } \, .
$$
All functions in the integrand are analytically continued in the
complex plane. For example, the real propagator is considered as
the limit of an analytic function of the complex variable $z$:
$$
   G(p^2) = \lim_{z\to p^2 +i\epsilon} G(z) \, .
$$

The full propagator associated to a scalar field can be written in
the {\it Schwinger's spectral representation} \cite{Schwinger60}
as
\begin{equation}
   G(z) = \frac{i}{z-M_0^2-\Pi(z)} \, ,
   \label{spectralrepres}
\end{equation}
where $M_0$ is the bare mass of the scalar field and $\Pi(z)$ is
an analytic function defined by
\begin{equation}
   \Pi(z) = z \, \int^{\infty}_{b^2} \, ds \, \frac{\sigma(s)}{z-s} \, ,
   \label{spectralenergy}
\end{equation}
which can be interpreted as the self-energy of the scalar field.
The positive real function $\sigma(x)$ satisfies $\sigma(x)=0$ for
$x\leq b^2$. $b^2$ is the invariant mass of the lightest
multi-particle state in interaction with the scalar field
corresponding to the propagator. This spectral representation has
the advantage of being in direct correspondence with the full
propagator (Eq.~(\ref{fullperturbpropag})), obtained in
perturbation theory by an infinite sum over the self-energy
insertions. If $x$ is on the real axis, it can be checked that
$$
   \Pi(x+i\epsilon) - \Pi(x-i\epsilon) = -2\pi i x \sigma(x) \, .
$$
The self-energy can be written as
\begin{equation}
   \Pi(x\pm i\epsilon) = \delta M^2 + u(x) \mp i v(x) \, ,
   \label{selfenergy}
\end{equation}
where $v(x)=\pi x \sigma(x)$ and $\delta M^2$ is an infinite
constant which is absorbed in the renormalized mass $M^2$:
$$
   M^2 = M_0^2 + \delta M^2 \, .
$$
Note that $M$ is not the physical mass since it is not the pole of
the propagator.

The propagator $G(z)$ has branch points at the thresholds
corresponding to the multi-particle states. The first branch point
is at $z\!=\!b^2$. We assume that the other branch points are
above $M_2^2$. The analytic continuation of $G(z)$ into the second
Riemann sheet \cite{Brown} is such that the analytically continued
function $G_{II}(z)$ just below the real axis is equal to the
original function just above the real axis. The analytically
continued function $\Pi_{II}(z)$ of the function $\Pi(z)$ is
defined in the same way, $\Pi_{II}(x-i\epsilon)=\Pi(x+i\epsilon)$,
so that
\begin{equation}
   G_{II}(z) = \frac{i}{ z - M_0^2 - \Pi_{II}(z) } \, .
   \label{spectralcont}
\end{equation}
$G_{II}(z)$ has a pole at \mbox{$z_0 \!=\! m^2 - i m \Gamma$}.

Different cases have to be considered, according to the stability
or instability of the particle, and according to the positions of
the pole $z_0$ and threshold $b^2$. The pole is assumed to be
inside the contour (otherwise the particle cannot be emitted).

\subsection*{Unstable particle, with $b < M_1 < m < M_2$}

The path of integration is shown on Fig.~9.
\begin{figure}
\begin{center}
\includegraphics[width=7cm]{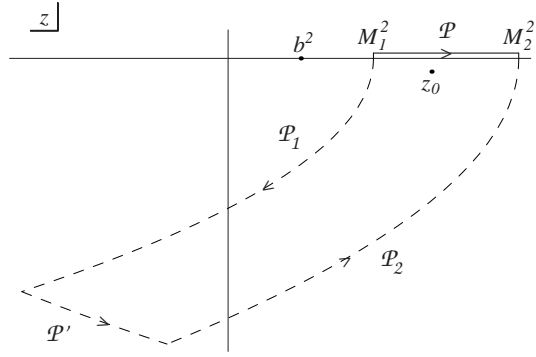}
\caption{Integration contour if $b < M_1 < m < M_2$.}
\end{center}
\end{figure}
The analytic expressions of the paths ${\cal P}_j$ read
$$
   {\cal P}_j : \, z =
   \left(
         -i \omega + \sqrt{ M_j^2 + {\bf p}^2 }
   \right)^2
   - {\bf p}^2
   \quad (j=1,2) \, ,
$$
where $\omega$ ranges from zero to $\omega_\infty$ on ${\cal P}_1$
and from $\omega_\infty$ to zero on ${\cal P}_2$. The analytic
expression of the path ${\cal P'}$ reads
$$
   {\cal P'} : \, z =
   \left(
         -i \omega_\infty + \sqrt{ M^2 + {\bf p}^2 }
   \right)^2
   - {\bf p}^2 \, ,
$$
where $M$ ranges from $M_1$ to $M_2$, with the limit
$\omega_\infty \to \infty$ understood.

The integral $I(T)$ is equal to
$$
   I(T) = J + J_1 + J_2 + J' \, .
$$
$J$ is the contribution of the pole $z_0$, whereas $J_1$, $J_2$
and $J'$ are the contributions of ${\cal P}_1$, ${\cal P}_2$ and
${\cal P'}$, respectively. Their analytic expressions read
\begin{enumerate}

   \item
   Contribution of the pole:
   $$
     J = Z \, \pi \, \left( z_0 + {\bf p}^2 \right)^{ - \frac{1}{2} } \,
         \psi(z_0,{\bf p}) \;
         e^{-i \sqrt{ z_0 + {\bf p}^2 } \, T} \, .
   $$
   Setting $\Delta M \cong |m-M_{1,2}|$, a rough estimate of $J$ is given
   by $J \sim m^{2n-1} \, (\Delta M)^{2n} \, e^{-\Gamma T/2}$.

   \item
   Contribution of the sickle-shaped paths:
   $$
     J_j = i (-1)^j \,
           e^{-i \sqrt{ M_j^2 + {\bf p}^2 } \, T}
           \int_0^\infty d\omega \, \psi \left( z(\omega),{\bf p} \right) \,
           G_{II}\left( z(\omega) \right) \, e^{-\omega T} \, .
   $$
   At large $T$, the dominant contribution to the integral comes from the
   $\omega$ values near zero, because of the decreasing exponential. Since the
   integrand tends to zero with $\omega$ ($\psi(M_j^2,{\bf p})=0)$), the
   asymptotic behavior of the integral depends on the way the integrand
   tends to zero. Setting $y=\omega T$ and expanding in $1/T$, we obtain:
   $$
     J_j \sim  \left( M_2^2 \!-\! M_1^2 \right)^n \,
               \left( M_j^2 \!+\! {\bf p}^2 \right)^{n/2} \,
               \Omega \left( M_j^2,{\bf p} \right) \,
               G_{II}\left( M_j^2 \right)
               \, T^{-(n+1)} \,
               e^{-i \sqrt{ M_j^2 + {\bf p}^2 } \, T} \, .
   $$
   The corrections to this formula are of order $(\Delta M T)^{-(n+2)}$,
   where \mbox{$\Delta M \approx M_j \!-\! m$}, that is, they are of the order of
   the uncertainty on the particle mass.
   It is assumed that the function \mbox{$\Omega (z,{\bf p})$} diverges more
   slowly than \mbox{$\exp (-\omega T)$} on the paths ${\cal P}_j$ as $\omega$
   tends to infinity. This assumption is true for a wide class of functions, in
   particular for Gaussians, whereas their path integral diverges on half-circles
   at infinity. A rough estimate of $J_j$ is given by
   $J_j \sim m^{2n-1} \, (\Delta M)^{n-1}\, T^{-(n+1)}$.

   \item
   Contribution from the path at infinity:
   $$
     J' = e^{-\omega_\infty T} \int_{M_1^2}^{M_2^2} dM
         \frac{M}{ \sqrt{M^2 \!+\! {\bf p}^2 } } \,
         \psi \left( z(M) \right) \,
         G_{II} \left( z(M) \right) \,
         e^{ -i \sqrt{ M^2 + {\bf p}^2 } \, T } \, .
   $$
   If $\Omega$ satisfies the same conditions at infinity as above,
   $J' \!\sim\! \exp(-\omega_\infty T)$ and tends to zero as
   $\omega_\infty \to \infty$.

\end{enumerate}
In conclusion, the contribution of the pole is a decreasing
exponential in $T$, whereas the contributions due to the bounded
character of the energy spectrum decrease in inverse powers of
$T$.

At small $T$, $J_j/J \sim (\Delta M \, T)^{-(n+1)}$, which is
nonnegligible for $\Delta M \, T \lesssim 1$. Below that value,
the asymptotic evaluation of $J_j$ is not valid anymore, because
terms in $(\Delta M T)^{-(n+2)}$ have been neglected.

First, let us consider weakly decaying particles. In the case of
the $K_S^0$, the mass is measured with a precision of $\Delta M
\approx 10^{-2}\;$MeV. Thus nonexponential corrections will be
important for $T \lesssim 10^{-19}\;$s, which is not observable
since the $K_S^0$ lifetime is $0.89 \, \times \, 10^{-10}\,$s. In
the case of the $B^0$, $\Delta M \approx 2\,$MeV, so that
nonexponential corrections will be important for $T \lesssim
10^{-22}\,$s, which is not observable since the $B^0$ lifetime is
$1.29 \, \times \, 10^{-12}\,$s. Next, let us next consider
resonances. In the example of $\Delta(1232)$, $\Delta M \approx
2\,$MeV, so that nonexponential corrections are important for $T
\lesssim 10^{-22}\,$s, which is large compared to the inverse
width equal to $5 \, \times \, 10^{-24}$ s. Thus the propagation
of resonances can never be modelized by the contribution of the
pole alone.

In the case of the weakly decaying particles, power law
corrections are also important at large times and dominate if
$$
   \Gamma T \, \gtrsim \, 2(n+1) \, \ln(\Delta M T) \,
   = \, 2(n+1) \, (\ln(\Gamma T)+\ln(\Delta M/\Gamma)) \, .
$$
For $K_S^0$, the two-particle threshold is characterized by
$n=1/2$ and $\Delta M / \Gamma \sim {\cal O}(10^{10})$, so that
the threshold of the nonexponential behavior is given by $\Gamma T
\gtrsim 69$, at which time the amplitude will be much too small to
be observable. For $B^0$, taking $n=1/2$ and $\Delta M / \Gamma
\sim {\cal O}(10^{9})$, we obtain $\Gamma T \gtrsim 62$, with the
same conclusion as above.

To sum up the analysis of the propagation of a weakly decaying
particle (with no multi-particle thresholds included in the
spectrum), the function $I(T)$ is very well approximated in the
observable time domain by Eq.~(\ref{integIT}).

\subsection*{Unstable particle, with $M_1 < b < m < M_2$}

In this case, the multi-particle threshold is included in the
energy spectrum. It generates new power law corrections to the
amplitude.

The path of integration is shown on Fig.~10.
\begin{figure}
\begin{center}
\includegraphics[width=7cm]{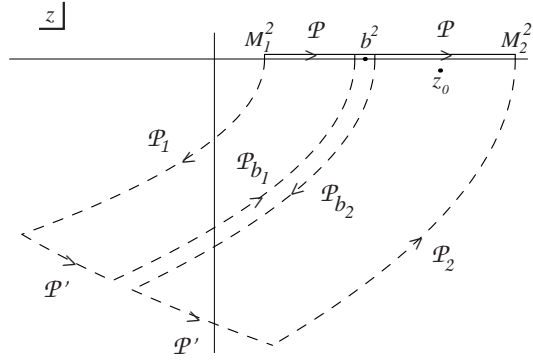}
\caption{Integration contour if $M_1 < b < m < M_2$.}
\end{center}
\end{figure}
The contributions of the paths ${\cal P}_j$ are the same as in the
previous case, except that  $G(z)$ replaces $G_{II}(z)$ on ${\cal
P}_1$, since this path lies on the first Riemann sheet. The
asymptotic value of $J_j$ as $T \!\to\! \infty$ does not change.
The contributions of the paths ${\cal P}_{b_j}$ do not cancel each
other out, since they do not lie on the same Riemann sheet. Their
sum reads
\begin{eqnarray}
   J_b &=& \frac{1}{2} \int_{ {\cal P}_b } dz \,
   (z \!+\! {\bf p}^2 )^{ -\frac{1}{2} } \, \psi(z,{\bf p}) \,
   \left( G_{II}(z) - G(z) \right) \,
   e^{ -i \sqrt{ z + {\bf p}^2 } T }
   \nonumber \\
   &=& -i \, e^{ -i \sqrt{ b^2 + {\bf p}^2 } T }
   \int_0^\infty d\omega \, \psi(z(\omega),{\bf p}) \,
   \left( G_{II}(z) - G(z) \right) \,
   e^{-\omega T} \, .
   \label{exprJb}
\end{eqnarray}
The analytic expression of the paths ${\cal P}_b$ reads
\begin{equation}
   {\cal P}_b : \, z =
   \left( -i \omega + \sqrt{ b^2 + {\bf p}^2 } \right)^2 - {\bf p}^2 \, ,
   \label{contourb}
\end{equation}
where $\omega$ ranges from zero to $\omega_\infty$, with the limit
$\omega_\infty\to\infty$ understood. The function $I(T)$ is equal
to
$$
   I(T) = J+J_1+J_2+J_b+J' \, .
$$
The asymptotic behavior of $J_b$ is studied in the same way as for
$J_j$. As before, the dominant contribution to $J_b$, for large
$T$, comes from $\omega$ values near zero. Again, we set $y=\omega
T$ and expand the integrand in $1/T$. Recall that $G_{II}(z)$ is
defined by the analytic continuation of $G(x+i\epsilon)$ below the
cut, $G_{II}(x-i\epsilon)=G(x+i\epsilon)$. Its discontinuity
through the cut can be computed with Eqs.~(\ref{spectralrepres}),
(\ref{selfenergy}) and (\ref{spectralcont}):
$$
   G_{II}(x-i\epsilon) - G(x-i\epsilon)  =
   \frac{2v(x)}{ \left( x - M^2 - u(x) \right)^2 + v^2(x) }\, .
$$
In the example of $K^0$, the self-energy is given at first order
in $g^2$ by the pion bubble diagram (where $g$ is the coupling
constant between a kaon and two pions), yielding
$$
   v(z) = \frac{g^2}{4\pi} \sqrt{ 1- b^2/z } \, ,
$$
where $b=2 m_\pi$ is the two-pion threshold. The function $u(z)$
is also of order $g^2$.

The evaluation of $J_b$ with the same asymptotic method as used
for $J_j$ gives
$$
   J_b = -i \, (-2i\pi)^\frac{1}{2} \, T^{-3/2} \, \frac{g^2}{4\pi b} \;
   \frac{ \left( b^2 + {\bf p}^2 \right)^{1/4} }
        { \left( b^2 - M^2 - u(b^2) \right)^2 } \;
   \psi(b^2,{\bf p}) \;
   e^{ -i \sqrt{ b^2 + {\bf p}^2 } T } \, .
$$
The corrections to this formula are of order \mbox{$(QT)^{-5/2}$},
where \mbox{$Q \equiv M-b$} is the energy release on decay of the
unstable particle. Noting that cutting rules give
\mbox{$v(m^2)=m\Gamma$}, the coupling constant $g^2$ can be
replaced by its expression in function of $\Gamma$, $m$ and $Q$. A
rough estimate of $J_b$ is then given by
$$
   J_b \sim
   (QT)^{-3/2} \, \frac{\Gamma}{Q} \, \frac{1}{m} \, \psi(b^2,{\bf p}) \, .
$$
At small $T$, the ratio $J_b/J$ is of the order of
$$
   \frac{J_b}{J} \sim (Q T)^{-3/2} \, \frac{\Gamma}{Q} \, ,
$$
which is much smaller than 1 if $QT \gg 1$ and $\Gamma/Q \ll 1$.
This is the case for weakly decaying particles. For example,
$\Gamma/Q \approx 10^{-14}$ for $K_S^0$. It is not true for
resonances. For example, $\Gamma/Q \approx 0.8$ for
$\Delta(1232)$. Below the value $QT \approx 1$, the asymptotic
computation of $J_b$ is not valid anymore since corrections in
$(QT)^{-5/2}$ have been neglected.

At large $T$, the power law contribution $J_b$ dominates the pole
contribution if
$$
   \Gamma T - 3 \ln(\Gamma T) \gtrsim 5 \ln(Q/\Gamma) \, ,
$$
so that the nonexponential time thresholds are given by $\Gamma_S
T \gtrsim 165$ for $K_S^0$, $\Gamma_L T \gtrsim 202$ for $K_L^0$
and \mbox{$\Gamma_{L,H} T \gtrsim 157$} for $B^0_{L,H}$ (in the
last case, we have taken $Q=1\,$MeV). Thus nonexponential effects
are not observable at large times for weakly decaying particles.
In the case of resonances, the contribution of $J_b$ is always of
the same order or larger than the pole contribution. Resonances do
not propagate macroscopically, since their width is of the same
order as the typical energy of the processes.

To sum up the analysis of the propagation of a weakly decaying
particle, including the multi-particle thresholds, the function
$I(T)$ is very well approximated in the observable time domain by
the Eq.~(\ref{integIT}).

\subsection*{Stable particle, with $M_1 < m < M_2 < b$}

This case can be examined like the first one, except that the pole
is real, so that the result is given by Eq.~(\ref{integIT}), with
$\Gamma=0$. There is no exponential decrease of the amplitude.
Power law corrections are negligible at large $T$.

\subsection*{Stable particle, with $M_1 < m < b < M_2$}

This case can be examined like the second one, except that the
pole is real. The result is again given by Eq.~(\ref{integIT}),
with $\Gamma=0$. The ratio $J_b/J$ is of the order of \mbox{$J_b/J
\sim (m T)^{-3/2} g^2/Q^2$}, where $g$ is the coupling constant
with the particles produced at the threshold. The contribution of
$J_b$ is too small to be observed at small $T$ in the current
experiments, and is negligible at large $T$.

\def\baselinestretch{1}

\end{document}